\documentclass[ALICE,manyauthors]{cernphprep}
\usepackage[utf8]{inputenc}
\usepackage[comma,square,numbers,sort&compress]{natbib}
\usepackage{hyperref}
\usepackage{lineno}
\usepackage{xspace}
\usepackage[table]{xcolor}
\usepackage{booktabs}
\usepackage{siunitx}
\usepackage[normalem]{ulem} 
\DeclareSIUnit\barn{b}
\DeclareSIUnit{\invbarn}{b^{-1}}

\usepackage[littleEndian,numberBitsAbove,numberFieldsTwice]{bitpattern} 
\DeclareSIUnit{\sps}{sps}
\DeclareSIUnit\permille{\text{\textperthousand}}

\DeclareSIUnit\clight{\text{\ensuremath{c}}}

\usepackage[T1]{fontenc}
\usepackage{orcidlink}

\newcommand{\answered}[1]   {\textcolor{purple}{\textit{#1}}}
\renewcommand{\answered}[1]{}

\newcommand{\pp}           {pp\xspace}

\newcommand{\XeXe}         {\mbox{Xe--Xe}\xspace}
\newcommand{\PbPb}         {\mbox{Pb--Pb}\xspace}

\newcommand{\pPb}          {\mbox{p--Pb}\xspace}

\newcommand{\snn}          {\ensuremath{\sqrt{s_{\mathrm{NN}}}}\xspace}
\newcommand{\pt}           {\ensuremath{p_{\rm T}}\xspace}

\newcommand{\dEdx}         {\ensuremath{\textrm{d}E/\textrm{d}x}\xspace}

\newcommand{\nineH}        {$\sqrt{s}~=~0.9$~Te\kern-.1emV\xspace}
\newcommand{\seven}        {$\sqrt{s}~=~7$~Te\kern-.1emV\xspace}
\newcommand{\twoH}         {$\sqrt{s}~=~0.2$~Te\kern-.1emV\xspace}
\newcommand{\twosevensix}  {$\sqrt{s}~=~2.76$~Te\kern-.1emV\xspace}
\newcommand{\five}         {$\sqrt{s}~=~5.02$~Te\kern-.1emV\xspace}
\newcommand{\twosevensixnn}{$\sqrt{s_{\mathrm{NN}}}~=~2.76$~Te\kern-.1emV\xspace}
\newcommand{\fivenn}       {$\sqrt{s_{\mathrm{NN}}}~=~5.02$~Te\kern-.1emV\xspace}

\newcommand{\GeVc}         {Ge\kern-.1emV/$c$\xspace}
\newcommand{\MeVc}         {Me\kern-.1emV/$c$\xspace}
\newcommand{\TeV}          {Te\kern-.1emV\xspace}
\newcommand{\GeV}          {Ge\kern-.1emV\xspace}
\newcommand{\MeV}          {Me\kern-.1emV\xspace}
\newcommand{\GeVmass}      {Ge\kern-.2emV/$c^2$\xspace}
\newcommand{\MeVmass}      {Me\kern-.2emV/$c^2$\xspace}

\newcommand{\Osq}           {\rm{O}$^2$\xspace}

\begin{document}

\begin{titlepage}
\PHyear{2023}       
\PHnumber{009}      
\PHdate{27 January}  

\title{ALICE upgrades during the LHC Long Shutdown 2}
\ShortTitle{ALICE LS2 upgrades}   

\Collaboration{ALICE Collaboration\thanks{See Appendix~\ref{app:collab} for the list of collaboration members}}
\ShortAuthor{ALICE Collaboration} 

\begin{abstract}
A Large Ion Collider Experiment (ALICE) has been conceived and constructed as a heavy-ion experiment at the LHC. During LHC Runs~1 and 2, it has produced a wide range of physics results using all collision systems available at the LHC.
In order to best exploit new physics opportunities opening up with the upgraded LHC and new detector technologies, the experiment has undergone a major upgrade during the LHC Long Shutdown~2 (2019--2022). This comprises the move to continuous readout, the complete overhaul of core detectors, as well as a new online event processing farm with a redesigned online-offline software framework.
These improvements will allow to record Pb--Pb collisions at rates up to \SI{50}{\kilo\hertz}, while ensuring sensitivity for signals without a triggerable signature.
\end{abstract}
\end{titlepage}

\setcounter{page}{2} 

\tableofcontents
\clearpage
\listoffigures
\clearpage
\listoftables

\section{Introduction}

A Large Ion Collider Experiment (ALICE) was proposed, conceived, and built to study the properties of the quark--gluon plasma (QGP) in heavy-ion collisions at the Large Hadron Collider (LHC) at CERN~\cite{Aamodt:2008zz}. 
The design was driven by the requirement to reconstruct tracks at high multiplicity in central \PbPb collisions and to provide particle identification over a wide range in transverse momentum (\pt).
In LHC Runs~1 and 2, the ALICE~1 apparatus was used to record and analyse hadronic collisions ranging from \pp to \PbPb~\cite{Abelev:2014ffa}. 
The measurements have provided new insights in the properties of the quark--gluon plasma as well as several other aspects of the strong interaction. A comprehensive review of this scientific output was reported in Ref.~\cite{ALICE:2022wpn}.
During the Long Shutdown~2 (2019 -- 2021), major upgrades have led to the new experimental setup, ALICE~2, extending the physics capabilities of the experiment for Runs~3 and 4.

\subsection{Motivation}

The main objectives of the upgrades in Long Shutdown 2 (LS2) are to significantly improve the capabilities of ALICE to probe the QGP with heavy-flavour quarks, and to enable completely new measurements of the thermal emission of dielectron pairs. In addition, the upgrades significantly improve the precision of measurements in several other areas, such as jet quenching phenomena
probing the interactions of high-energy partons, the production of light nuclei, momentum correlations of hadrons to determine the interaction potentials of unstable particles, and the study of collective effects in collisions of protons with high multiplicity.
To gain access to these areas of physics a two-fold approach was taken by improving the pointing resolution and increasing the readout rate capabilities of the entire system to collect larger data samples.
A thinner and lighter inner tracker with the first layer closer to the interaction point improves the pointing resolution by a factor of 3 in the transverse direction and a factor 6 in the longitudinal direction. 
This provides more effective suppression of backgrounds in the reconstruction of decays of heavy-flavour mesons and baryons as well as in the dielectron emission measurements. 
The increase of the readout rate from below \SI{1}{\kilo\hertz}
to \SI{50}{\kilo\hertz} for \PbPb{} collisions leads to improved statistical precision for all measurements, even in the presence of large backgrounds. 
The improvements in pointing resolution and readout rate
will also enable the measurement of thermal dilepton production in \PbPb{} collisions, as well as a number of new measurements of heavy-flavour production, which were out of reach of the ALICE detector in Runs~1 and 2.

\subsection{Experimental setup}

\begin{figure}
  \centering
\includegraphics[width=.96\textwidth]{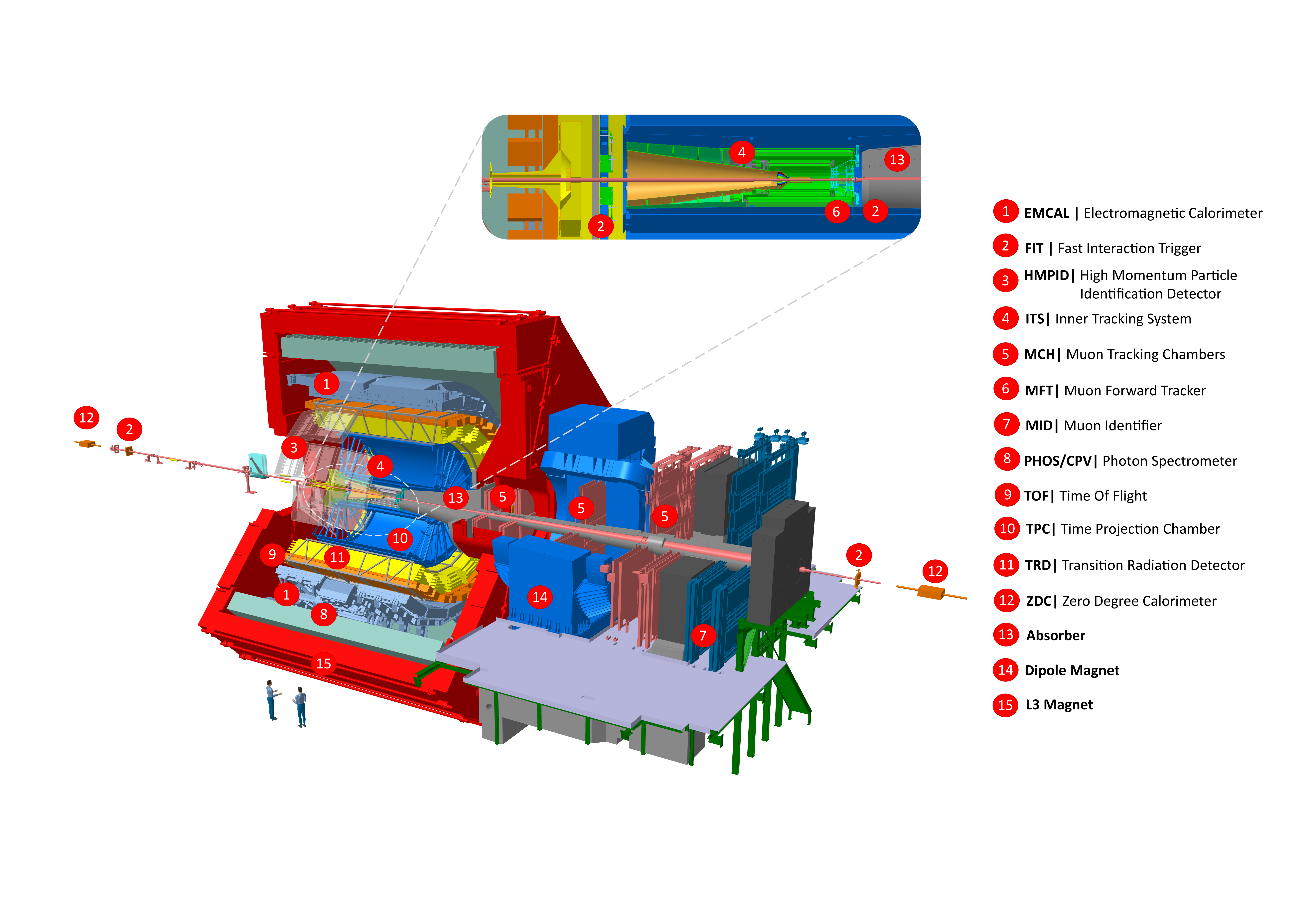}
  \caption[ALICE~2 detector systems]{ALICE 2 detector systems (see legend and text for details).}
  \label{fig:intro:alice_run3}
\end{figure}

The experimental setup consists of a central barrel contained in a solenoidal
magnet ($B = \SI{0.5}{\tesla}$) and a forward muon system with a dipole magnet
providing a total bending power of \SI{3}{\tesla\metre}, see Fig.~\ref{fig:intro:alice_run3}. 
The central barrel detector system is designed for efficient tracking in the high track-density environment of heavy-ion collisions, covering transverse momenta from \SI{\sim 100}{\mega\eV\per\clight} to \SI{\sim 100}{\giga\eV\per\clight} with excellent hadron and electron identification capabilities.

Until the end of Run~2, the Inner Tracking System (ITS) which is crucial for the extrapolation of tracks to the primary vertex, consisted of two layers of Silicon Pixel Detectors (SPD), two layers of Silicon Drift Detectors (SDD), and two layers of Silicon Strip Detectors (SSD)~\cite{Aamodt:2008zz}.
The readout rate of the full ITS was limited to \SI{1}{\kilo\hertz}.
The ITS was replaced with a new detector (ITS2), based on seven layers of ALPIDE monolithic active pixel sensors (MAPS),
which provides better pointing resolution thanks to its reduced distance to the interaction point and better position resolution.
It is also able to handle the hit densities resulting from \PbPb collisions at \SI{50}{\kilo\hertz} interaction rate.

In the radial direction, the ITS is followed by the Time Projection Chamber (TPC) extending from \SI{0.85}{\metre} to \SI{2.5}{\metre} in radius over a length of \SI{5}{\metre}. 
With the multiwire proportional chambers used in ALICE~1, the ion backflow into the drift region had to be suppressed by active gating, which in turn limited the readout rate to about \SI{700}{\hertz} for Pb--Pb collisions.
This limitation is removed in the upgraded TPC by employing readout chambers based on Gas Electron Multiplier (GEM) foils that reduce the ion backflow and resulting space charge in the TPC to a level that can be corrected for while operating the detector with \PbPb interaction rates up to \SI{50}{\kilo\hertz}.

The Transition Radiation Detector (TRD) (extending from \SIrange[range-phrase={ \text{to} }]{2.8}{3.5}{\metre} in radius) provides additional space points for tracking, which are also used to determine the size of the distortions due to space charge effects in the TPC, as well as  \dEdx measurements for particle identification, and the detection of transition radiation for electron identification.
The readout electronics were upgraded to minimise the data volume and to reduce the dead time to allow data taking at high interaction rates.

The subsequent Time-of-Flight detector (TOF) allows the identification of hadrons over a wide momentum range and electrons at low momentum.
Besides consolidation work on the front-end electronics, the readout was upgraded to handle the increased interaction rates.

A large part of the acceptance in the central barrel is covered by electromagnetic calorimeters.
The ElectroMagnetic Calorimeter (EMCal) is realised as Pb-scintillator sampling calorimeters with avalanche photon detector (APD) readout, whereas the PHOton Spectrometer (PHOS) uses PbWO$_4$ crystals with APD readout.
All calorimeters have undergone maintenance and improvements of the readout electronics.

The High Momentum Particle Identification Detector (HMPID) is a ring-imaging Cherenkov detector that adds hadron identification capabilities at large transverse momenta over a limited acceptance. 
A part of the system was equipped with additional absorbers to facilitate a measurement of the interaction cross section of light antinuclei.
Also here, the readout electronics were upgraded to improve the rate capability.

The muon detectors cover the forward pseudorapidity range $-4.0 < \eta < -2.5$  and use a system of absorbers to remove hadrons and identify muons.
The background of secondary muons from pion and kaon decays in the muon system is small at high \pt, thanks to the so-called `muon plug' absorber, which is placed at $z = 90$ cm from the interaction point. The main muon detector stations use multiwire proportional chambers (muon tracking chambers, MCH), and resistive plate chambers (muon identifier, MID), both of which were equipped with new front-end electronics. Following Run 2, and as a new addition to the muon detectors in ALICE 2, the Muon Forward Tracker (MFT) consists of tracking stations with the ALPIDE silicon pixel sensors that are installed in front of the muon plug to improve mass resolution and pointing resolution for the detection of secondary charmonia and muons from B-meson decays. 

A set of forward detectors form a Fast Interaction Trigger (FIT), which is used for triggering, event selection and determination of the collision time. The FIT system consists of two arrays of fast Cherenkov radiators placed on both sides of the interaction point (FT0), complemented with 3 sets of scintillator detectors.
The interaction trigger is provided by the FT0 together with a large azimuthally segmented scintillator detector placed on the opposite side of the muon detectors, which is also used to determine the 
reaction-plane orientation in \PbPb{} collisions. Two additional scintillator detectors, FDD, are placed on opposite sides of the interaction point at large distances to cover $4.7 < \eta < 6.3$ and $-6.9 < \eta < -4.9$ to select diffractive and ultra-peripheral collisions with rapidity gaps. The FIT detector replaces the T0, V0 and AD detectors, which had similar functionalities in ALICE~1~\cite{Aamodt:2008zz}.

The Zero-Degree Calorimeters (ZDC) are installed at $\approx \SI{100}{\meter}$ on either side of the interaction point to help determine the centrality and event plane orientation. The readout electronics of the ZDC were upgraded to increase the readout rate to match the rest of the system.

In addition to the interventions outlined here, significant consolidation work has been performed on several subsystems which are described in the sections on individual detector systems below. 

Furthermore, the readout infrastructure was completely renewed to support the continuous readout of the core detectors.
The raw data from the detectors are mostly transmitted through optical links and received by First Level Processors (FLPs), where the data are assembled to time frames for further processing. 
A dedicated farm of Event Processing Nodes (EPN) was installed at the experiment site for the online reconstruction of all collisions.
The output of this synchronous reconstruction is stored on mass storage systems and is used for an asynchronous reconstruction stage with improved calibration. 
The output of the latter is then used for physics analysis.
A new common software framework, O$^2$, was developed for online and offline reconstruction as well as the physics analysis.

\subsection{Data samples}

\begin{figure}
  \centering
  \includegraphics[width=.48\textwidth]{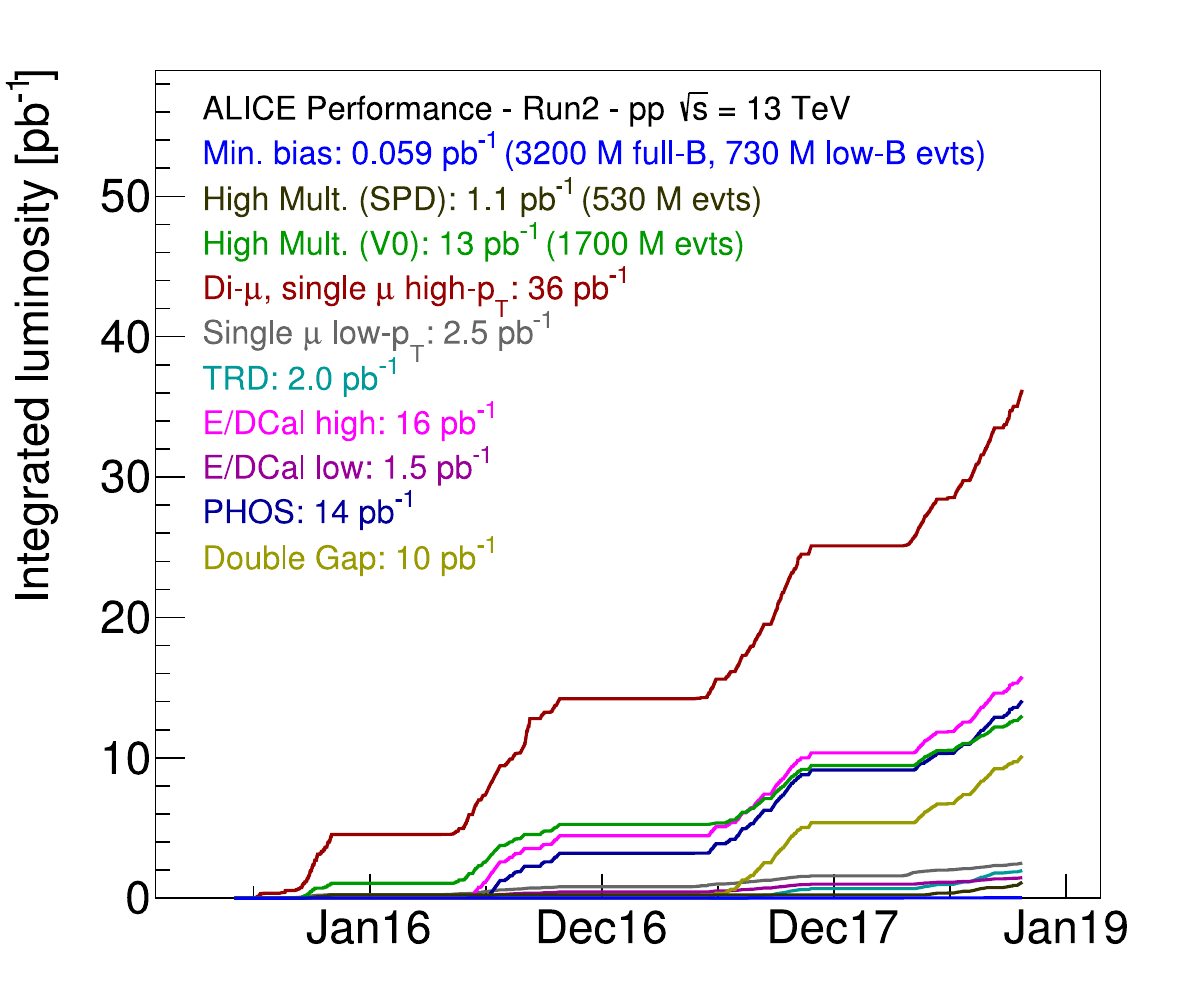}
  \includegraphics[width=.48\textwidth]{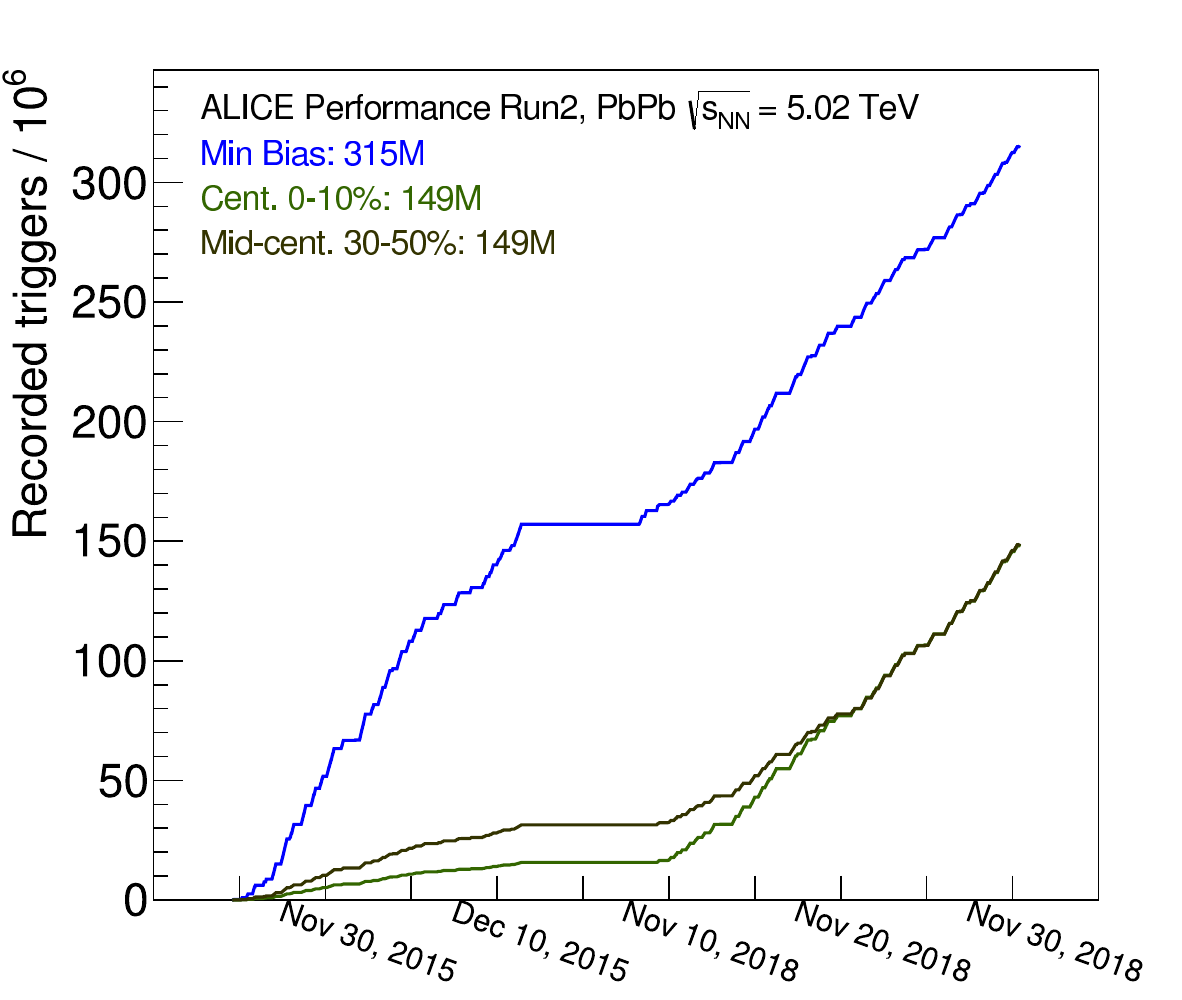}
  \caption[Accumulation of integrated luminosity]{Accumulation of integrated luminosity over time for different trigger types in pp (left) and \PbPb (right) collisions during LHC Run~2.}
  \label{fig:lumi_run2}
  \end{figure}

During LHC Runs~1 and 2, data were recorded with pp, \pPb{}, \XeXe{}, and \PbPb{} collisions at a variety of collision energies.
The collision  and readout rates were tuned to limit pile-up in pp collisions and to keep the total space charge generated in the gas amplification in the TPC readout chambers to manageable levels. 
Typical collision rates were up to \SI{8}{\kilo\hertz} for \PbPb{} collisions and around \SI{200}{\kilo\hertz} for \pp collisions. 
To make optimal use of the different readout rate capabilities across the detector systems, clusters of detectors were read out at different rates. The central barrel detectors were read out at a rate of \SIrange{500}{600}{\hertz},
while the cluster with 
the forward muon detectors together with V0, T0 and the silicon pixel layers for event characterisation were read out at a slightly higher rate. For specific triggers, such as coincidence triggers between the forward muon detectors and the calorimeters in the central barrel, the full detector was read out.
During \pp and \pPb{} data taking a `fast cluster' containing all barrel detectors except the SDD
was used in order to double the effective TPC readout rate. 
Figure~\ref{fig:lumi_run2} shows the luminosities accumulated during Run~2 with different trigger conditions.

For Runs~3 and 4, it is planned to record \pp and \PbPb data at interaction rates of \SIrange{0.5}{1}{\mega\hertz} and \SI{50}{\kilo\hertz}, respectively.
This will allow us to inspect integrated luminosities of \SI{200}{\pico\invbarn} and \SI{13}{\nano\invbarn}, respectively.
  
\subsection{Outline}

In this article, the upgrades made to ALICE during the LHC Long Shutdown~2 are discussed.
The next Chapter~\ref{chap:chips} presents the readout system design, the common readout unit and the integrated circuits (ASICs), that were conceived, designed and produced for the upgrades of multiple detector systems. Chapter~\ref{chap:dets} presents the upgrades of the inidividual detector systems in detail. 
Chapter~\ref{chap:integration} details the mechanical integration of the detector components within ALICE and the interfaces with the LHC. 
In Chapter~\ref{chap:readout} the trigger system, the readout chain, as well as the synchronous and asynchronous processing stages are discussed. 
The expected performance of the upgraded detector and reconstruction is reported in Chapter~\ref{sec:performance}. 
Chapter~\ref{chap:concl} comproses of a conclusion with prospects for the LHC Run~3 and a brief outlook on the future ALICE upgrade plans.

\section{System design and common developments}
\label{chap:chips}

A series of developments have been pursued commonly for multiple systems. Foremost, the readout chain was redesigned for all detectors (Sec.~\ref{sec:system}). A common readout unit was developed for the readout of the detectors (Sec.~\ref{sec:cru}). The ALICE Pixel Detector (ALPIDE) chip is designed and used for both the inner tracking system and the muon forward tracker (Sec.~\ref{alpide:subsection:alpide}). The SAMPA is used as front-end chip for the time projection chamber and the muon systems (Sec.~\ref{sec:sampa}).

\subsection{System design}
\label{sec:system}
In nominal operating conditions (\SI{50}{\kilo\hertz} interaction rate for Pb--Pb) each TPC drift time period of \SI{\sim 100}{\us} will contain on average 5~Pb--Pb events. It was therefore decided to use a continuous, untriggered readout strategy, combined with online data compression for the upgraded readout and data acquisition system.

\begin{figure*}[!t]
  \begin{center}
    \includegraphics[width=1.\textwidth]{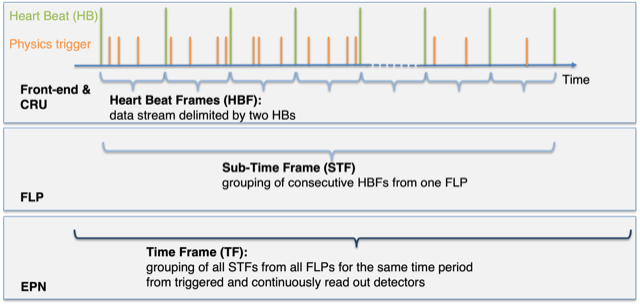}
  \end{center}
  \caption[Time frame and heartbeat frame structure]{Time frame and heartbeat frame structure in continuous and triggered mode. HeartBeat (HB) triggers are issued in continuous and triggered modes to all upgraded detectors. Physics triggers can be sent to upgraded detectors in triggered mode and are sent to non-upgraded detectors in all modes. HBF and TF rates are programmable with the following nominal values; HBF: 1 every orbit, $\sim$89.4 µs/$\sim$10 kHz, TF: 1 TF every 128 HBFs/$\sim$11 ms/$\sim$100 Hz.
  }
  \label{fig_ro:tf_hbf_structure}
\end{figure*}

In order to synchronise the continuous data stream across all readout and processing branches, the data stream is divided in so-called time frames (TF) of a nominal length of 128 LHC orbits (\SI{\sim 11}{ms}).
Each TF is subdivided in heartbeat frames (HBF) with a length corresponding to an orbit of \SI{\sim 89.4}{\us}. 
Figure~\ref{fig_ro:tf_hbf_structure} illustrates this structure. 
 For commissioning and calibration runs, for which the data throughput exceeds nominal conditions, all detectors also support triggered mode, in which only data from selected interactions are retained by the readout electronics. 
In addition, a subset of legacy detectors has not been upgraded to continuous readout and will operate in triggered mode only. 
For these detectors, as well as for dedicated runs, minimum bias triggers based on the fast interaction trigger detector (FIT) and the PHOS, EMCAL and TOF are distributed.
In both the continuous and triggered readout mode, the detector data are time stamped with a precision of an LHC bunch crossing of 25 ns; data belonging to a HBF are grouped together into HBF packets.

\begin{figure*}[hbtp]
  \begin{center}
    \includegraphics[width=1.\textwidth]{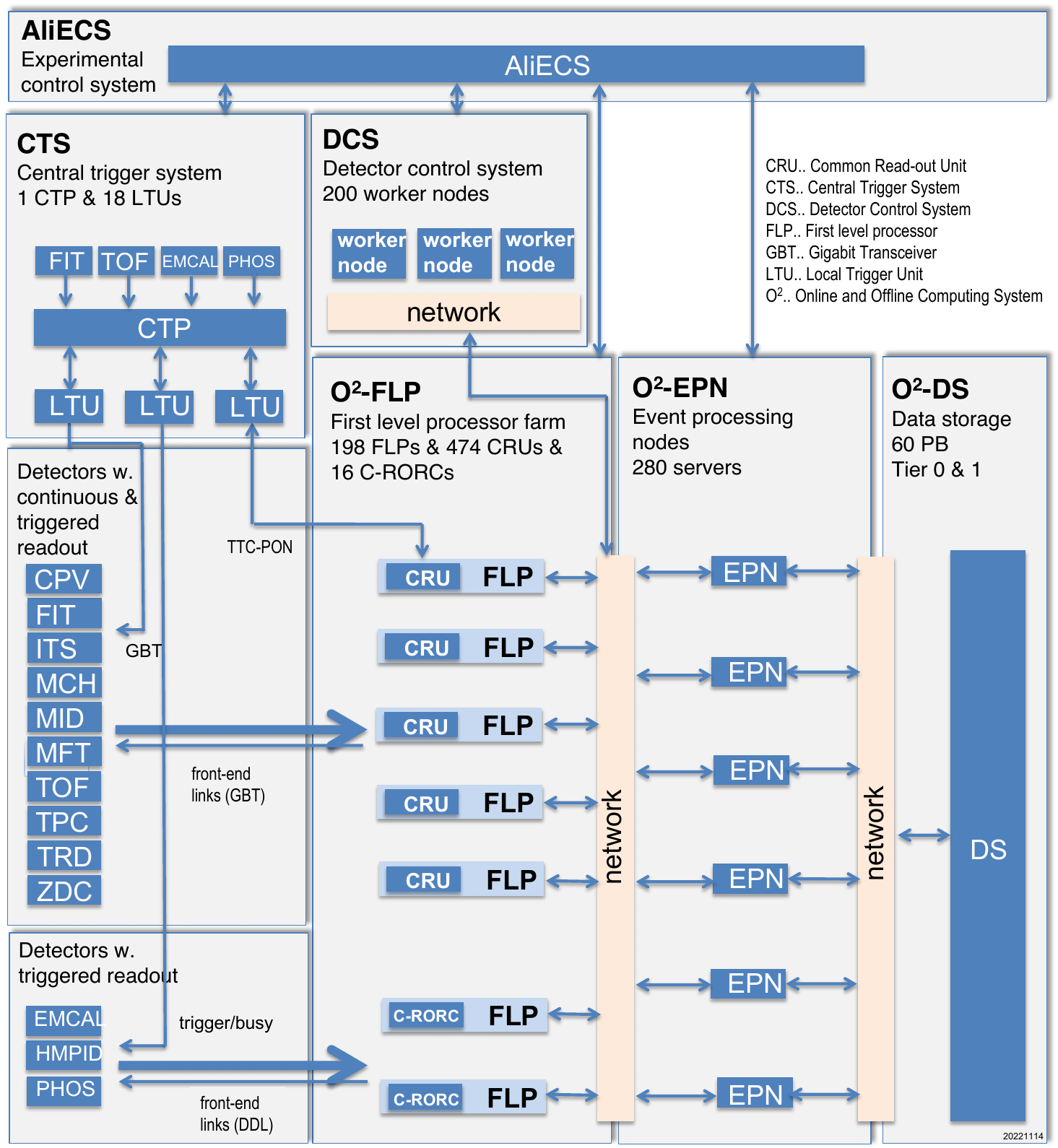}
  \end{center}
  \caption[ALICE readout architecture]{ALICE readout and control system architecture.}
  \label{fig_ro:alicesystemblock}
\end{figure*}

The upgraded ALICE system architecture is shown in Fig.~\ref{fig_ro:alicesystemblock}.
The Common Readout Units (CRU) are standardised PCIe FPGA-based optical I/O processor modules used by all upgraded detectors for data readout and configuration, see Sec.~\ref{sec:cru}. Data taking is governed by the Central Trigger System (CTS) which distributes timing and trigger signals. The CTS features a two-staged distribution system consisting of one central trigger processor (CTP) and up to 18~active distribution units, the local trigger units (LTU), one for each subdetector. The CTP-LTU and LTU-CRU connections are implemented using bidirectional TTC-PON links~\cite{Taylor:592719,Mendes:2017aok}.
The standard timing and trigger signal distribution path goes from the CTS via the detector-specific CRUs to the detector front-ends via bidirectional radiation tolerant GBT links~\cite{tpc:moreira2009}.
Trigger signals are distributed with three different latencies referred to as LM (level -1 at \SI{425}{\nano\second}), L0 (level 0 at \SI{1200}{\ns}), and L1 (level 1 at \SI{6100}{\ns}).
Detectors that require latency-critical trigger signals receive them additionally on a direct path from the CTS, which is located in the cavern, to the detector front-ends on GBT links.
A second group of detectors do not support continuous readout and require a trigger signal indicating the presence of an interaction with a latency of \SI{1.6}{\us}. 
Some detectors continue to be read out via legacy readout cards (C-RORC~\cite{ref_CRORC}) following a hardware trigger signal to initiate the readout. 
They receive the clock and trigger signals via the legacy TTC system~\cite{Taylor:592719,Mendes:2017aok}. 
For more details on the CTS, see Sec.~\ref{sec:cts}.

The Online \& Offline processing farm (\Osq{}) contains the first level processors (FLP) and event processing nodes (EPN). The detector front-ends send the data via GBT-based links to the CRUs and C-RORCs located in the FLPs.
Depending on the detector implementation, the readout data are reformatted or compressed either in the front-ends, the CRUs, or in the FLPs. The FLPs prepare Sub-Time Frames (STF) by merging all HBFs of one TF of the connected detector. Note that for most detectors the data is distributed over several FLPs.
The FLPs ship the STFs of all subdetectors via a network to the EPN farm where they are merged into TFs. In order to compress the data to be stored, the \Osq{} system performs a first synchronous online reconstruction pass, converts the data into compressed time frames (CTF) and sends them to the  storage system from where it is accessed asynchronously for further processing.
In total, a raw data throughput of 3.4 TB/s is processed in a continuous manner by the readout system. After zero suppression and data compression in the front-ends, the CRUs, and the FLPs, a data throughput of 635 GB/s is processed by the data network and the EPN farm. 

The detectors are configured via the detector control system (DCS) which is connected to the detector front-ends via the CRU. The experiment control system (ECS) governs the entire data taking process via direct network connections to the central systems (DCS, CTS, FLP, EPN). 

\subsection{Common readout unit}
\label{sec:cru}

For all upgraded detectors the Common Readout Unit (CRU) serves as interface between detector front-end links, the \Osq{} FLP processors, the CTS and DCS. The CRUs are custom developed FPGA-based Gen 3 PCI Express plug-in cards installed in the FLPs. 
The card (named PCI40) was originally developed for LHCb~\cite{ro:PCI40} and has requirements fully compatible with ALICE. ALICE adopted the PCI40 for its CRU and joined the qualification and test effort and has developed firmware for use in the experiment.

The CRU hardware features up to 48 high-speed, bidirectional, 10\,Gb/s optical links using 12-lane Minipod parallel optical transmitters (AFBR-812 and AFBR-822) and receivers from Avago/Broadcom. They are accessible from the CRU front-panel through MTP (Multi-Fiber Termination Push-on) optical ribbon cable connectors and establish the interface to the detector front-end electronics using the GBT protocol~\cite{tpc:moreira2009} implemented in the FPGA. GBT links are the result of a common development for all LHC experiments to provide a radiation tolerant transmission chip (GBTx, SCA) and optical transceiver set (VTTx, VTRx) to be used on the detector front-end cards communicating with the data aquisitioning and detector control systems via optical links.
The GBTx provides a data bandwidth of 4.48\,Gb/s in wide-bus mode, or 3.2\,Gb/s in GBT mode depending on whether forward error correction with superior correction capability for radiation induced transmission errors is activated. The slow control adapter ASIC (SCA) is an auxiliary chip compatible with the GBTx. Connected to a GBTx it allows the control of ADCs and digital IOs. VTTx and VTRx are radiation tolerant dual optical transmitter and transceiver components compatible with the GBTx ASIC.

The number of data links used for each CRU and the use of the forward error correction are adapted to the subdetector needs. In most detector implementations, 24 data links are connected to one CRU.
Table~\ref{tab:flp_farm_links} in Sec.~\ref{sec:flp} shows the number of CRUs, the number of readout links and the data throughput into and out of one CRU for each subdetector. 

For the GBT downlink to the detector-front ends carrying the timing and trigger signals as well as the configuration data up to 320\,Mb/s are available from the GBTx ASIC on a configurable number of pins. The single word transmission protocol (SWT) has been developed to provide the front-end designers with a common configuration data framework.

One of the two CRU SFP+ optical transceivers is used to connect the CRUs to the CTS system via bidirectional TTC-PON~\cite{Mendes:2017aok} links. The TTC-PON link allows the distribution of timing and trigger signals with constant latency from the CTS to the CRU over passive optical splitters with a bandwidth of up to 9.6\,Gb/s. The links carry the LHC clock with a jitter below 20\,ps (rms) and synchronise all 474 CRUs and the connected detectors to each other. The upstream link from all CRUs to the CTS carry detector buffer status information, see Sec.~\ref{sec:ctp_interfaces}.

The 16-lane (x16) PCI Express card edge connector provides the interface between the CRU and the ALICE \Osq{} FLPs, in which up to 3 CRUs are installed. The interface achieves \SI{\sim 90}{Gb/s} sustainable data throughput from the CRU to the memory of the FLP computers~\cite{cru_fw}.
Depending on subdetector implementation, the CRU FPGA forwards data that has already been formatted and compressed in the detector front-end, or performs detector-specific formatting, compression and base line reconstruction. In both cases, the data stream to the FLP consists of data packets compatible with the HBF structure (see Sec.~\ref{sec:system}).
A central FPGA firmware framework provides the interfaces to CTS, FLP, CRU and the subdetectors. Subdetector-dependent functionality, such as link decoding, adding HBF structure, compression or data processing is added via a dedicated user logic (UL) firmware plug-in to the central FPGA firmware. A detailed description of the CRU firmware design can be found in~\cite{cru_fw}.

\begin{figure*}[hbtp]
  \begin{center}
    \includegraphics[width=1.\textwidth]{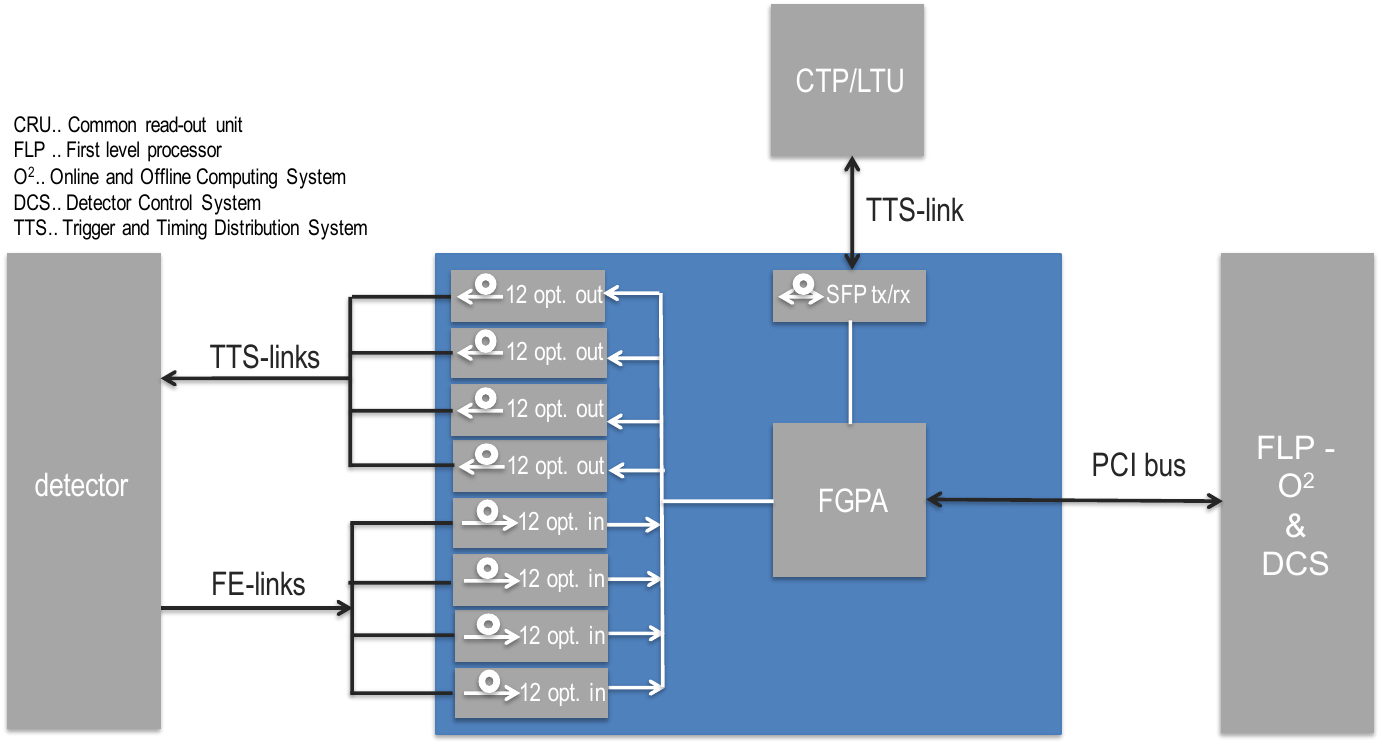}
  \end{center}
  \caption[Block diagram of the common readout unit]{Block diagram of the common readout unit (CRU): the CRU forms the interface between the first-level processors (via PCIe), the central trigger system (via TTS), and the detectors (via TTS and FE).}
  \label{fig_ro:cru_block}
\end{figure*}

\begin{figure*}[hbtp]
  \begin{center}
    \includegraphics[width=1.\textwidth]{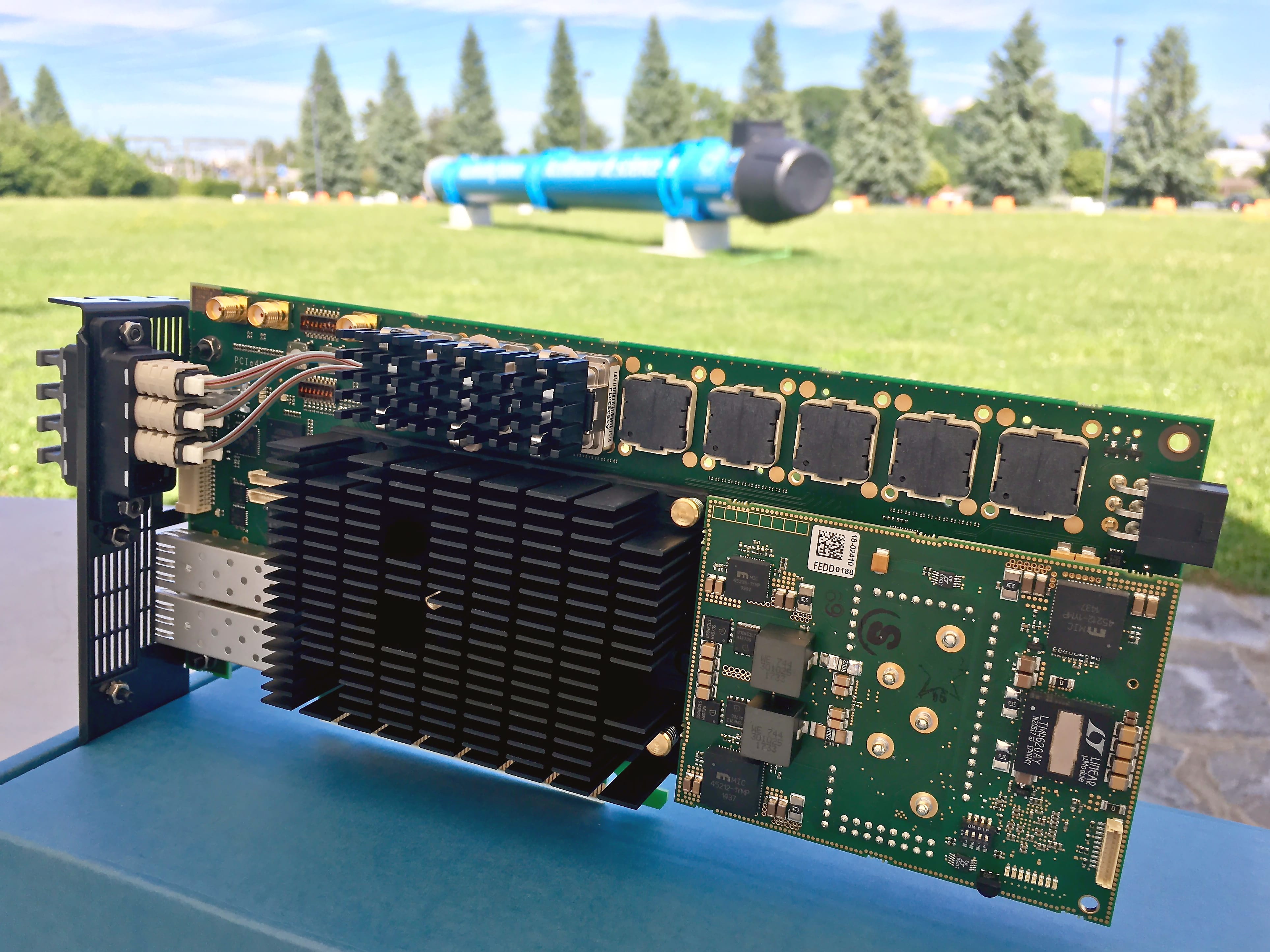}
  \end{center}
  \caption[Picture of a common readout unit]{Picture of a CRU; bottom left, FPGA cooling radiator, bottom right, power mezzanine; top row; 3 out of 8 Minipods installed; top left, fiber optics cable to MPO connector on front panel; bottom left, SFP transceivers.}
  \label{fig_ro:cru_photo}
\end{figure*}

 Figure~\ref{fig_ro:cru_block} shows the block diagram of the module functionality and Fig.~\ref{fig_ro:cru_photo} shows a photograph of the CRU card.

\subsection{The ALPIDE Chip}
\label{alpide:subsection:alpide}

\subsubsection{Technology, Sensing, Pixels}
\label{alpide:subsubsection:technology}

The ALPIDE chip~\cite{AGLIERIRINELLA2017583} is a Monolithic
Active Pixel Sensor (MAPS)~\cite{Snoeys2014167} implemented in a \SI{180}{\nm} CMOS technology for imaging
sensors provided by TowerJazz (Tower Semiconductor since March 2022)~\cite{Senyukov2013115}.
It was
designed for the upgrade of the Inner Tracking System (ITS2) 
to meet the requirements summarized in
Table~\ref{alpide:table:requirements}.

\begin{figure}
  \centering
  \includegraphics[width=0.5\linewidth]{{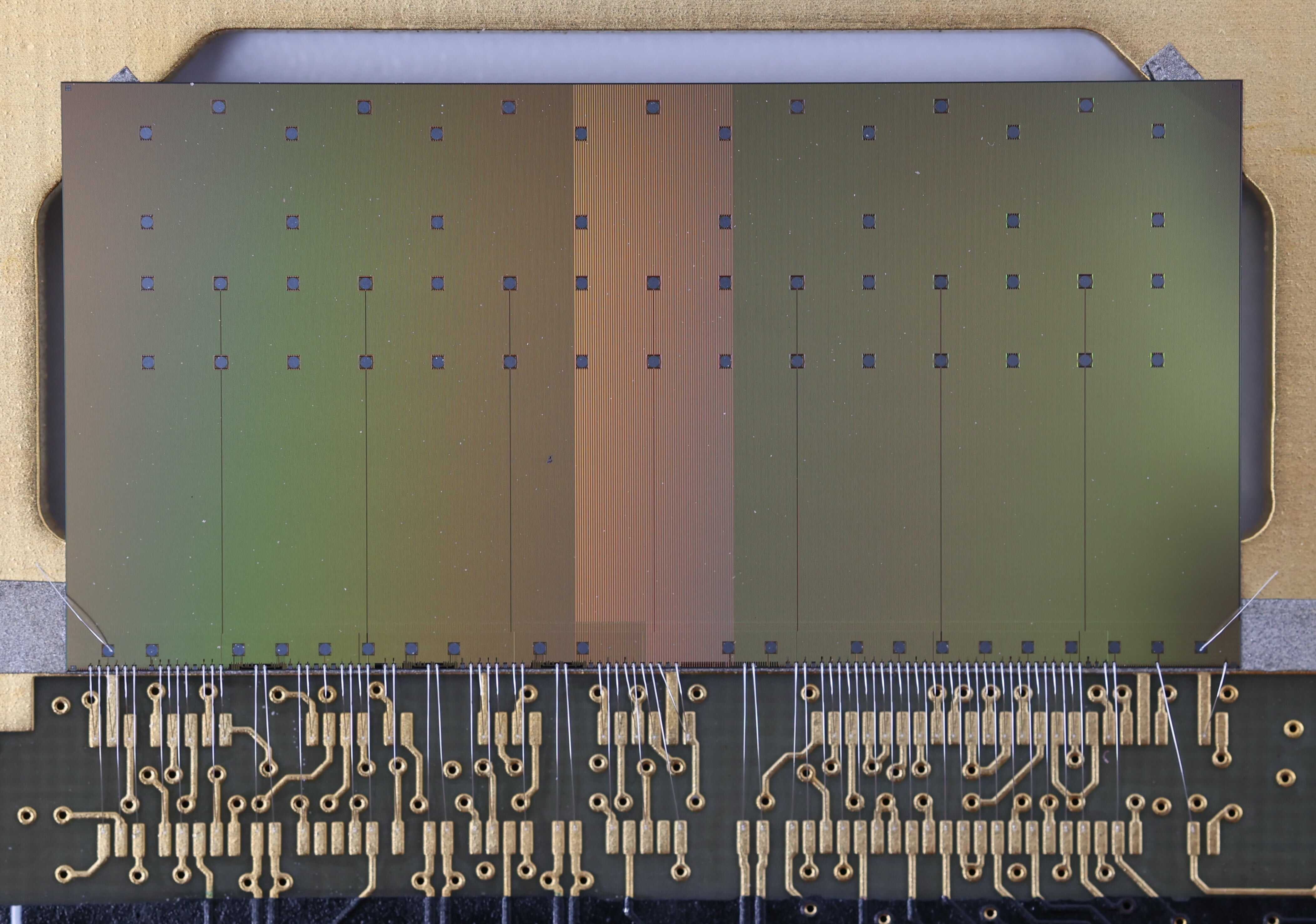}}
  \caption[Picture of ALPIDE]{Photograph of the ALPIDE chip on a test carrier.}
  \label{alpide:figure:ALPIDE-photo}
\end{figure}

The ALPIDE chip (Fig.~\ref{alpide:figure:ALPIDE-photo}) measures 15~mm by 30~mm and includes a matrix of
512$\times$1024 sensing pixels, each one measuring $\SI{29.24}{\um} \times \SI{26.88}{\um}\ (z\times r\varphi)$.
Analog biasing, control, readout and interfacing functionalities are
implemented in a peripheral region of ${\rm 1.2\times30~mm^2}$
(Fig.~\ref{alpide:figure:ALPIDE-architecture}).

\begin{figure}
  \centering
  \includegraphics[width=0.7\linewidth]{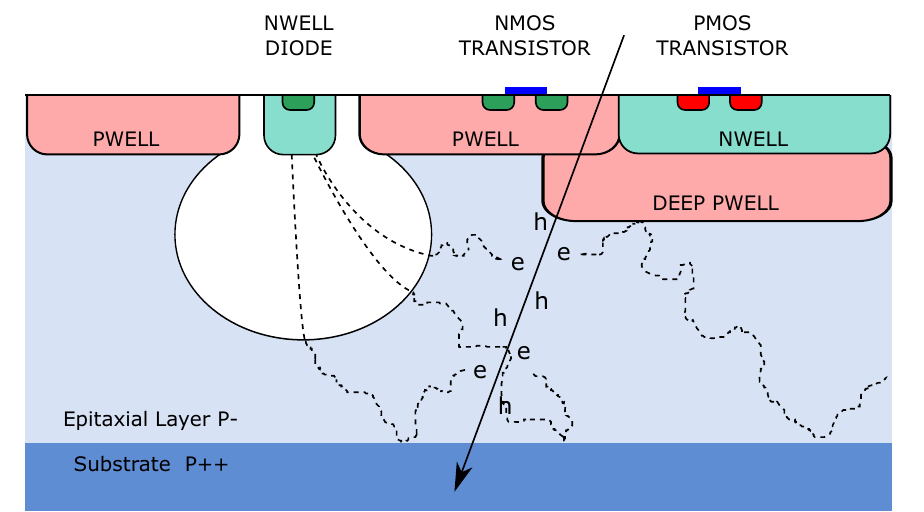}\caption[Cross section of ALPIDE pixel cell]{Schematic cross-section of a pixel cell.}
  \label{alpide:figure:cross-section}
\end{figure}

The ALPIDE chips are fabricated on substrates with a high-resistivity
($>1~{\rm k\Omega\cdot cm}$) epitaxial layer on p-type substrate.
Typical values for the thickness of the epitaxial layer are in the
range between 18 and \SI{30}{\um}.
Figure~\ref{alpide:figure:cross-section} illustrates that a charged particle crossing the sensor liberates charge carriers in the material.
The electrons released in the epitaxial layer can diffuse
laterally while they remain vertically confined  by potential
barriers at the interfaces with the overlying p-wells and the
underlying p-type substrate.
The signal sensing elements are n-well diodes (\SI{\sim 2}{\um}
diameter). Their area is typically 100 times smaller than the pixel
cell area.
The electrons that reach the depletion volume of a diode (or carriers
that are released directly inside it) induce a current signal at the
input of the pixel front-end.

The manufacturing process also provides a deep p-well layer that can
be used to shield the epitaxial layer from the n-wells of the pmos
transistors.
These would otherwise compete with the sensing diodes in collecting the
electrons, strongly impairing the charge collection.
This feature permits the use of full CMOS circuits, including pmos
transistors, in the active area.

A reverse bias voltage can be applied to the substrate.  This increases the
depletion volume around the n-well collection diodes and reduces the
capacitance of the input junction.
All these aspects contribute to increasing the S/N ratio.

\begin{table}
  \centering
  \begin{small}
    \begin{tabular}{lcc}
      \hline
      Parameter & Inner Barrel & Outer Barrel \\
      \hline
      Chip dimensions [${\rm mm \times mm}$] & \multicolumn{2}{c}{${\rm 15\times30}$} \\
      Silicon thickness  [\si{\um}]              &   50          & 100                   \\
      Spatial resolution [\si{\um}]              &     5          &   10~\emph{(5)} \\
      Detection efficiency            & \multicolumn{2}{c}{${\rm >99\%}$}  \\
      Fake-hit probability [${\rm evt^{-1}pixel^{-1}}$]   & \multicolumn{2}{c}{${\rm <10^{-6}}$~($<<\mathit{{10^{-6}}}$)}  \\
      Integration time [\si{\us}]       & \multicolumn{2}{c}{ ${\rm <30}~\mathit{(10)}$ }  \\
      Power density  [${\rm mW/cm^2}$]   & ${\rm <300}~\mathit{(\sim47)}$ &
      ${\rm <100}~\mathit{(\sim35)}$  \\
      TID radiation hardness\textsuperscript{*} [krad]             & 270                   &  10  \\
      NIEL radiation hardness\textsuperscript{*} {[}${\rm 1~MeV~n_{eq}/cm^2}${]}                  & ${\rm 1.7\times10^{12}}$ &  ${\rm 1\times10^{11}}$  \\
      Readout rate, Pb--Pb interactions {[kHz]}        &  \multicolumn{2}{c}{${\rm 100}$} \\ 
      \hline
    \end{tabular}
  \caption[Requirements for pixel sensor]{General requirements for the pixel sensor chip for the upgrade of the ALICE inner tracking system. In cases where the actual ALPIDE performance is significantly better than the requirements, the actual performance is indicated in parenthesis and italics. (*)~Radiation load integrated over 6 years of operation.}
  \label{alpide:table:requirements}
  \end{small}
\end{table}

\begin{figure}
  \centering
  \includegraphics[width=0.9\linewidth]{{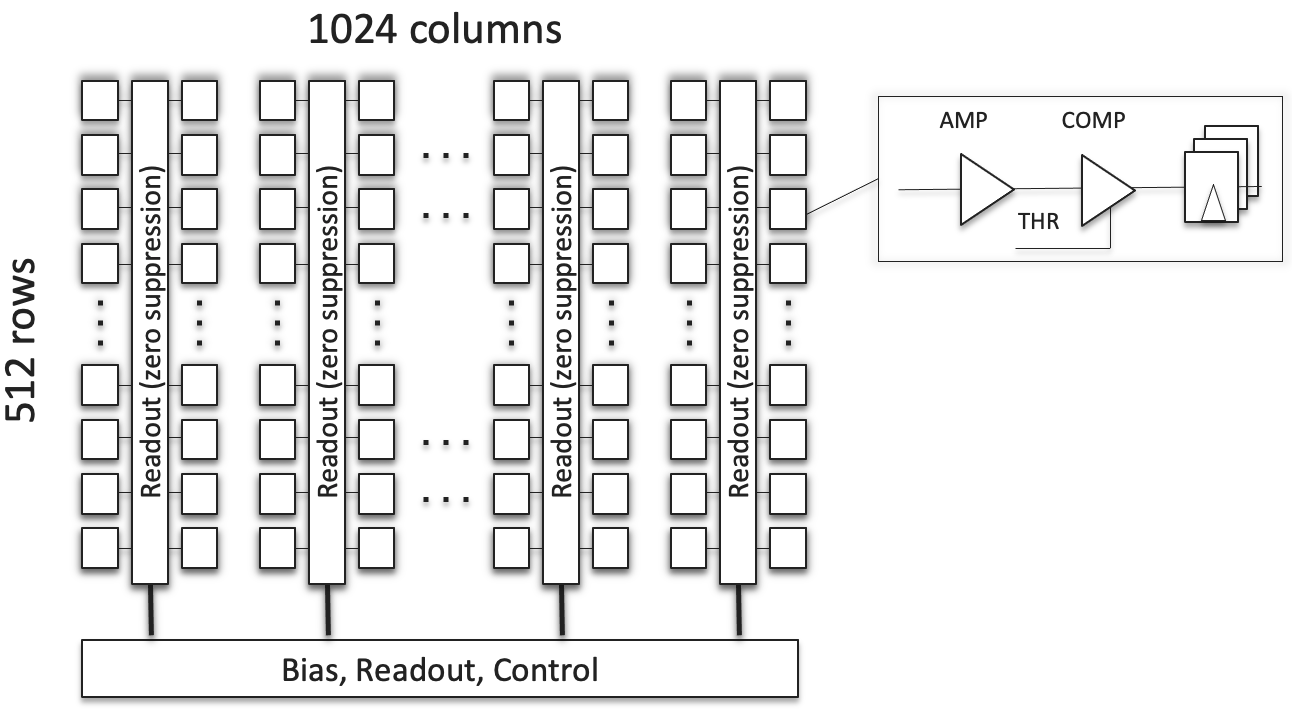}}
  \caption[ALPIDE architecture]{Architecture of the ALPIDE chip.}
  \label{alpide:figure:ALPIDE-architecture}
\end{figure}

\subsubsection{Analog Front-End and Discriminator}
\label{alpide:subsubsection:front-end}

Each pixel cell contains a sensing diode, a front-end amplifier and
a shaping stage, a discriminator and a digital section
(see the insert in Fig.~\ref{alpide:figure:ALPIDE-architecture}).
The digital section includes a multi-event buffer with three hit storage registers 
and a pixel mask register.

In every pixel, there is a pulse injection capacitor for injection of
test charge into the input of the front-end.
A digital-only pulsing mode is also available, directly forcing the
setting of the in-pixel memory cells, substituting the
latching of a discriminated pulse.
The analog and digital pulsing patterns are fully programmable.
These features are used routinely for testing and calibration.

The front-end and the discriminator are continuously active.
They feature a non-linear response and their transistors are biased in
weak inversion. The total power consumption of the pixel cell is \SI{40}{nW}.
The small signal gain of the front-end is \SI{4}{mV/{\it e}},
the equivalent noise charge is \SI{3.9}{{\it e}}, while
the minimum threshold is below \SI{100}{{\it e}}.
The typical value of the capacitance of the sensing diode is \SI{2.5}{fF}.
The input capacitance of the front-end is below \SI{2}{fF}.
The output of the front-end has a peaking time of the order of \SI{2}{\micro\second}, while the discriminated pulse has a typical duration of \SIrange{5}{6}{\micro\second}.
The front-end and the discriminator act as an analogue delay line.
This allows operating the chip in triggered mode when, as it happens in ALICE,
the latency of the incoming trigger is comparable with the peaking time of the
front-end.
A common threshold level is applied to all the pixels.

The latching of the discriminated hits in the storage
registers is controlled by global {STROBE} signals.
A pixel hit is stored into one of three in-pixel latch cells if
a STROBE pulse is applied to the pixel while the
output of the front-end is above threshold.
The generation of the internal STROBE signals can be either triggered
by an external command or optionally initiated
by an internal sequencer.
The duration of the STROBE pulses is programmable.
Two major operating modes are supported. In triggered mode the STROBE and
the frame readout are triggered externally from an event synchronous command.
In continuous mode the strobe is asserted periodically and for a duration almost equal to the period.
The event frames are continuously integrated and read out.

\subsubsection{Matrix and Readout}
\label{alpide:subsubsection:matrix-readout}

The readout of the frame data from the matrix is zero-suppressed and
is executed by an array of circuits named \emph{priority encoders} (Fig.~\ref{alpide:figure:ALPIDE-architecture}).
The priority encoder provides to the periphery the address of the
first pixel with a hit in its double column, selecting it
according to a hardwired topological priority.

During one hit transfer cycle a pixel with a hit is selected, its
address is encoded and transferred to the periphery and finally the
in-pixel memory element is reset.
The address of the next pixel with a hit in the double column is then
calculated.
This cycle is repeated until the addresses of all pixels initially
presenting a valid hit at the inputs of a priority encoder have been
transferred to the periphery and all the hit storage registers in the double column have been reset.

Each priority encoder is a fully combinatorial circuit and it is
steered by sequential logic in the periphery during the
readout of a matrix frame.
It is implemented in a very narrow region between the pixels, extending
vertically over the full height of the columns.
There is no free running clock distributed in the matrix and there is
no signaling activity if there are no hits to read out.
The average energy needed to encode the address of a hit pixel is of
the order of \SI{100}{\pico\joule}.
Power is consumed proportionally to the readout rate and to the
average hit occupancy of the frames.
The readout of the matrix consumes around \SI{3}{\milli\watt} under
normal conditions.
The priority encoders also implement the buffering and distribution of
readout and configuration signals to the pixels.

\begin{figure}
  \centering
  \includegraphics[width=0.7\linewidth]{{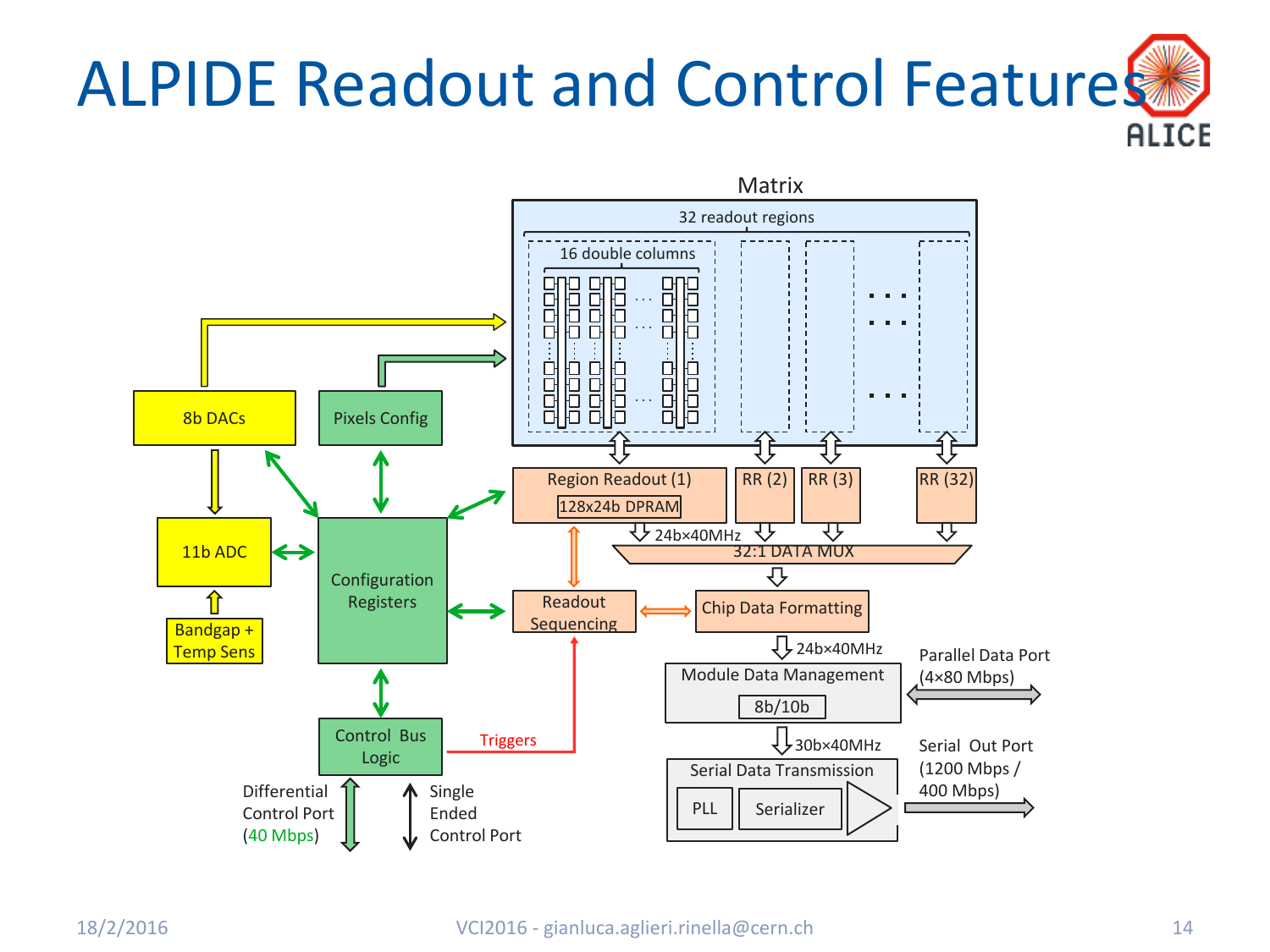}}
  \caption{Block diagram of the ALPIDE chip.}
  \label{alpide:figure:chip-block-diagram}
\end{figure}

The 512 double columns and the corresponding priority encoders are
functionally grouped in 32 regions (512$\times$32
pixels), each of them with 16 double columns being read out by 16
priority encoder circuits (Fig.~\ref{alpide:figure:chip-block-diagram}).
There are 32 corresponding region readout units in
the chip periphery, each one executing the readout of a region.
They steer the priority encoders, latch the encoded pixel hit address,
perform additional data reduction and formatting and
buffer the hit data into memories.
The 16 double columns inside each region are read out sequentially,
while the 32 regions are read out in parallel.
The data from the 32 region readout units are assembled
and formatted by a top readout unit module.

Data can be transmitted on two different readout ports.
The largest capacity data readout interface is a \SI{1.2}{\giga b\per\second} serial data port with differential signaling.
The serial transmission is 8b/10b encoded, therefore the maximum data
throughput is \SI{960}{\mega b\per\second}.
The serial port can optionally operate at reduced line rates (\SI{600}{\mega b\per\second} or 
\SI{400}{\mega b\per\second}).
A bidirectional parallel data port with single-ended
signaling is also available, with a capacity of \SI{320}{\mega b\per\second}.
This port enables the implementation of an inter-chip data transfer and relaying protocol designed 
to integrate multi-chip modules without additional external devices.
This is used in the modules of the ITS2 outer barrel.

The ALPIDE chip has custom control interfaces.
There are a differential control port supporting bidirectional (half duplex)
serial signaling at \SI{40}{\mega b\per\second} on differential links and 
a second single ended control port.
The two control interfaces and the dedicated internal logic allow interconnecting multiple chips 
on a module and control them via the differential interface of only one of the chips acting as 
hub of the control bus.
The control bus is also used to distribute broadcast commands and synchronization messages to the
chips, most notably the trigger commands.

The periphery of the chip contains fourteen 8--bit analog DACs for the
biasing of the pixel front-ends. The analog section of
the periphery also contains a band-gap reference and a temperature
sensing circuit.
An ADC with 11~bit resolution is available for monitoring and
testing purposes, and can probe the outputs of the DACs, the analog
as well as digital supply voltages, the band-gap voltage and the temperature sensor.

\subsubsection{Features for integration of ITS2 modules}
\label{alpide:subsubsection:module-integration-features}

\begin{figure}
  \centering
  \includegraphics[width=0.9\linewidth]{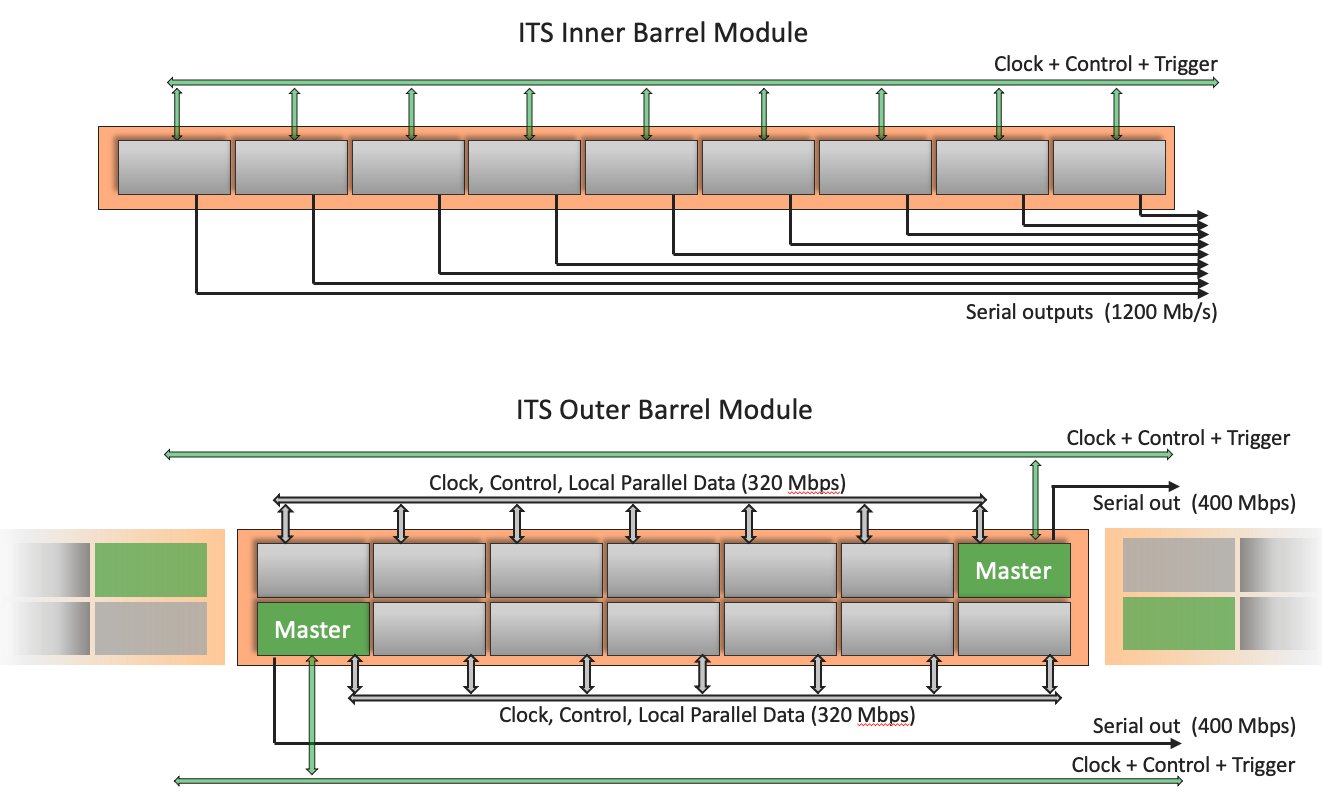}
  \caption{Diagrams of the ITS2 inner barrel and outer barrel modules.}
  \label{alpide:figure:its-modules}
\end{figure}

The ALPIDE chip has specific design features to enable the integration 
of multi-chip detector modules, to
minimise the electrical wiring between modules and off-detector
electronics and to provide common interfaces across the ITS2 staves.
Two different hybrid modules built with ALPIDE sensors are used in the
upgraded ALICE ITS2 (Fig.~\ref{alpide:figure:its-modules}): one in the three innermost layers constituting the inner barrel 
and the other in the staves of the remaining layers of the outer barrel (see also Sec.~\ref{sec:its:stavemodules}).

The ITS2 inner barrel module includes nine ALPIDE chips.  
They share a common differential control and clock distribution buses.  
Each chip transmits its own data
off-detector at maximum line rate (\SI{1.2}{\giga b\per\second}) on point-to-point high speed serial
links.

The ITS2 outer barrel module contains fourteen chips, arranged in two
subgroups of seven.  One chip in each group, called {\it master}, 
acts as control hub and data relaying chip.  
Only the master chips communicate with the
external electronics through differential clock and control busses shared between multiple modules and through point-to-point differential wire-line links for the transmission of data.  

Each of the master chips connects to six neighbouring chips, 
forwards them to the main clock and bridges
the control
transactions on electrical interconnects that are local to the
modules.

The chips neighbouring the master use a shared parallel local bus to transfer 
their data to the master.  
The master chip relays the data from the slave chips on the
serial output port driving the point-to-point links.  
In this configuration the master
chips transmit data on the serial data port using a lower bit rate
(\SI{400}{\mega b\per\second}).

Grouping of data from neighbouring chips and transmitting at lower rate
are possible in the ITS2 outer barrel layers given the lower
occupancy.  In addition to reducing the total number of copper links,
this scheme achieves a significant reduction of the power consumption given that
only one out of seven line drivers is maintained active.

Differential copper wire lines directly connect the ALPIDE chips to
the off-detector electronics. These links reach a length of \SI{8}{\meter}.  The
electrical receivers and transmitters on the ALPIDE chips were
designed and tailored to the electrical and protocol levels to operate
with these long interconnects.

\subsubsection{Power consumption}
\label{alpide:subsubsection:power-consumption}

The ALPIDE sensor chip has three power supplies: one analog domain, one digital domain 
and a power supply dedicated to the Phase-Locked Loop (PLL) of the high speed serial data transmitter.  
The power consumption of the analog section, dominated by the analog front-ends, 
is typically \SI{24}{mW}.
The digital power consumption includes the pixel digital sections, readout modules, 
peripheral circuits, and I/Os.
It depends strongly on the configuration and operating 
conditions. 
In the nominal conditions of the ALICE ITS2, it is about \SI{130}{mW}.

The output serial links are driven by a data transmission unit including
a PLL, a fast serializer and a line driver stage with a typical power consumption 
of about \SI{52}{mW}.
The data transmission unit is enabled in all the chips of the ITS2 inner barrel.
In the outer barrel modules it is active only in the master chips and disabled in the 
remaining chips, that is only 1 out of 7 sensors consumes this extra power.  

The power dissipation density
is about \SI{47}{mW/cm^2} in the ITS2 inner barrel
modules and around \SI{35}{mW/cm^2} in the outer barrel modules.

The readout of the matrix and the digital periphery consume power in
proportion to the clocking frequency, the readout rate and the pixel occupancy.
In less demanding applications not requiring the high speed links and 
the full rate capabilities, the power consumption can be reduced considerably
with various techniques including slowing down or suspending the primary clock 
and using the single ended I/Os to read out data at low rates.

\subsubsection{Results from the experimental characterization in laboratory and beam tests}

The ALPIDE chip and its prototype predecessors have been characterised with
an extensive test program including laboratory tests and a series of beam tests.
A summary of key results is given in this section. 
The full set of results and details on the methodologies will be presented 
in a separate paper.

The laboratory measurements were based on the sensible usage of
the built-in test pulse charge injection circuitry and on the
systematic analysis of threshold scans and noise measurements.
These allowed thorough characterisation the distributions of 
pixel thresholds, the fractions of pixels requiring masking, and 
the residual fake-hit rate after masking. Their dependencies on
operating conditions were accurately established.

A set of ALPIDE prototype sensors were characterised in beam tests to
quantify detection performance and hit-position resolution.
The samples under test were located at the center of 
beam telescopes acting as precision trackers. 
The telescopes were themselves constructed
with ALPIDE sensors and had six detection planes: three
upstream and three downstream of the Device Under Test (DUT).
The measurements were based on reconstructing particle tracks in 
the telescope and projecting them onto the DUT plane.
The presence of a matching cluster, its size in
pixels and its centroid were the basis for measuring the detection
efficiency and the hit-position resolution. 

\begin{figure}
  \centering
  \includegraphics[width=\linewidth]{{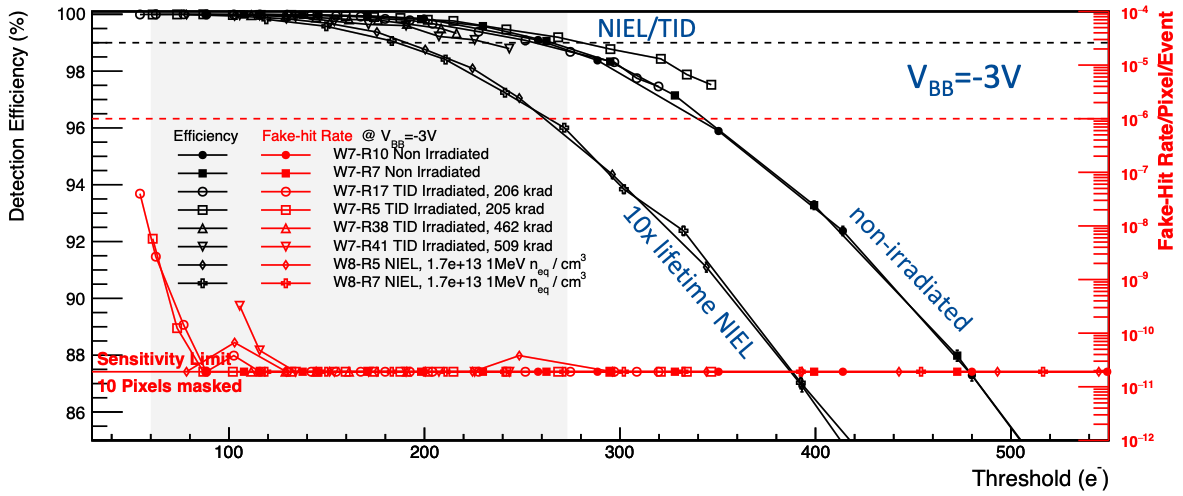}}
  \caption[ALPIDE detection efficiency and fake-hit rate]{ALPIDE sensor chip detection efficiency and fake-hit rate
    vs global threshold setting. Beam test results (\SI{6}{\giga
      eV/\it{c}} pions, orthogonal incidence). ALPIDE substrate reverse bias:~\SI{-3}{\volt}.}
  \label{alpide:figure:detection-eff-fhr}
\end{figure}

Figure~\ref{alpide:figure:detection-eff-fhr} provides a
summary of the beam test results on the detection efficiency and the
fake-hit rate as a function of the global threshold setting.  Data of
eight different samples are shown, including two non-irradiated DUTs
and pairs of devices exposed to increasing doses of ionising radiation
before the tests.  The samples exposed to the TID level of \SI{200}{\kilo
  rad} (75\% of the lifetime dose) received also a combined Non-Ionising Energy Loss (NIEL)
fluence at a level corresponding to 1.3 times the fluence expected
over the total lifetime. The Total Ionising Dose (TID) level of \SI{500}{\kilo rad} (190\%
of the lifetime dose) includes a combined fluence that is 3.2 times
the lifetime fluence.  Two devices were also irradiated with neutrons
for a cumulated non-ionizing energy loss of ${\rm
  1.7\times10^{13}~[1~MeV~n_{eq}/cm^2]}$, corresponding to ten times the
fluence expected over the full detector lifetime.

The results show a large operating margin for the threshold setting 
between 50 and 250 electrons, providing a detection
efficiency above 99\% and a fake-hit rate that is several orders of
magnitude smaller than the required value of $10^{-6}$
fake-hit probability per pixel per frame.  The ALPIDE sensor proved to be
extremely well performing in terms of noise.  Masking the ten most
noisy pixels out of the 524288 in the matrix (less than 0.002\%)
resulted in a residual fake-hit noise level below the sensitivity of
these experiments ($2\times10^{-11}$).

\begin{figure}
  \centering
  \includegraphics[width=0.9\linewidth]{{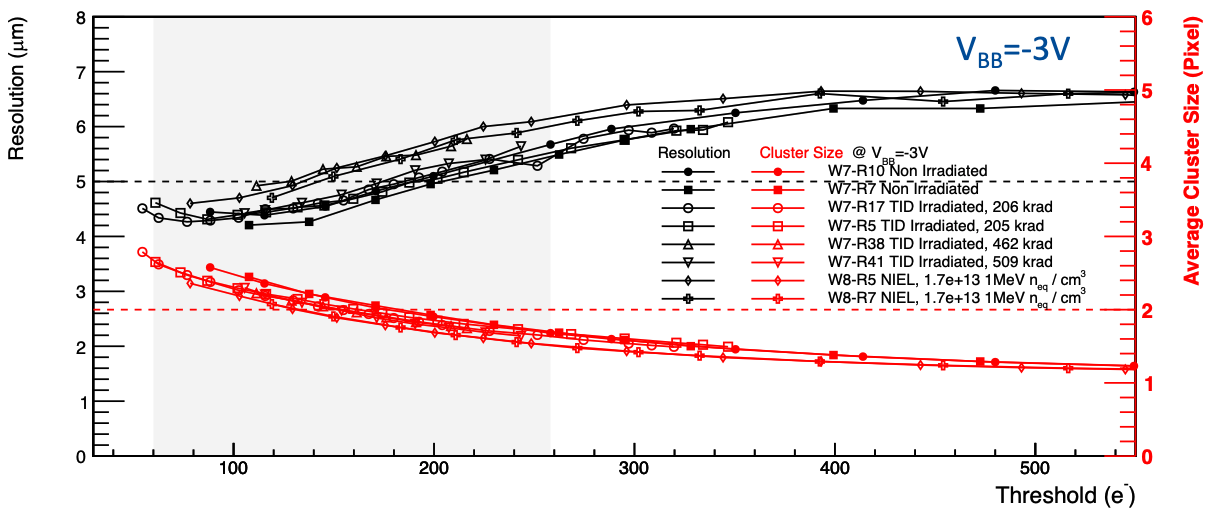}}
  \caption[ALPIDE position resolution and cluster size]{ALPIDE chip hit-position resolution and average
    cluster size as a function of global threshold setting. Beam test results with 
    \SI{6}{\giga eV/\it{c}} pions with perpendicular incidence. ALPIDE substrate reverse
    bias:~${\rm -3~V}$.}
  \label{alpide:figure:resolution-cluster-size}
\end{figure}

Figure~\ref{alpide:figure:resolution-cluster-size} shows
the beam test results on the hit-position resolution (black markers and lines, upper band) and the average
cluster size (red markers and lines, lower band) as a function of the global threshold.  The data sets
refer to the same samples of Fig.~\ref{alpide:figure:detection-eff-fhr}.  
The hit-position resolution is better than \SI{6}{\micro m} for thresholds 
below 300 electrons and better than \SI{5}{\micro m} for a threshold below 140
electrons.  As expected the average cluster size depends on the
threshold setting, due to the cutting on shared charge diffusing into pixels
adjacent to the seed pixel. It ranged between 1.5 and 2.5 pixel hits
in the range of interest.

The tests also showed that the chip-to-chip performance
variations were negligible, that the chips with combined TID and NIEL
irradiation performed similarly to the non-irradiated chips and that
sufficient operational margin was present also in the samples with a
NIEL dose ten times larger than the one expected over the lifetime.

\subsection{SAMPA}
\label{sec:sampa}

The SAMPA~\cite{Sam:SAMPA_IEEE} is a 32-channel custom front-end ASIC for the readout of gaseous detectors and specifically for the ALICE Muon Chambers (MCH, Sec.~\ref{sec:mch}) and Time Projection Chamber (TPC, Sec.~\ref{sec:tpc}). Each of the 32 channels contains a Charge Sensitive Amplifier (CSA) and a 10-bit 20\,MSample/s ADC. The digitised data of all 32 channels is made available on serial links as either a raw data stream or preprocessed by an internal Digital Signal Processor (DSP), supporting both the continuous and triggered readout of the upgraded ALICE system. The SLVS serial output links support 320\,Mb/s and are compatible with the input links (e-links) on the serial transceiver ASIC (GBTx) of the GBT-links used in ALICE for data transmission between the detectors and the CRUs. Depending on the data transfer rate needed in the application, the SAMPA data can be routed via a programmable number of up to 11 serial links.

\begin{figure}
\begin{centering}
\includegraphics[width=\textwidth]{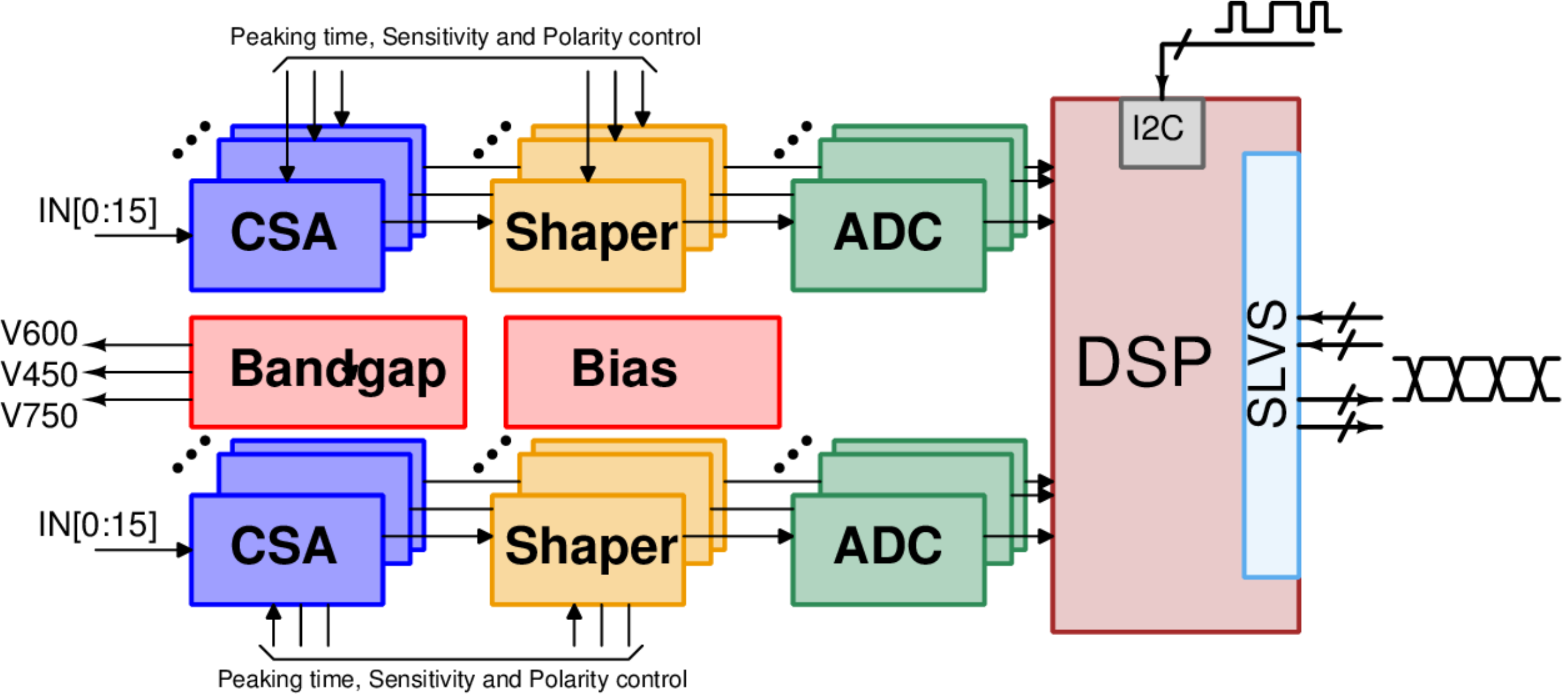}
\par
\end{centering}
\caption{Block diagram of the SAMPA ASIC.}
\label{Sam:fig:GlobalDiag}
\end{figure}

The block diagram of the SAMPA ASIC is shown in Fig.~\ref{Sam:fig:GlobalDiag}. The front-end is composed of a cascade connection of a CSA (Charge Sensitive Amplifier), a differential semi-Gaussian pulse shaper and an Analog-to-Digital Converter (ADC). The CSA and the pulse shaper convert signals into a semi-Gaussian pulse with an amplitude proportional to the total charge injected on the input. SAMPA was designed and fabricated in 130\,nm CMOS technology and it operates at a nominal supply voltage of 1.25\,V. 
In order to adapt the SAMPA to its two applications in the MCH and TPC, the sensitivity, polarity and peaking time of the front-end can be adjusted via external pins. SAMPA supports positive and negative polarity of the input charge and has three different gain modes with different sensitivity and peaking time: 20\,mV/fC@160\,ns, 30\,mV/fC@160\,ns for the TPC and 4\,mV/fC@300\,ns for the MCH. Table~\ref{tab:sampa_perf} summarizes the main characteristics and performance of the SAMPA.

\begin{table}
\caption{SAMPA key parameters.}
\begin{tabular}{llll}
\hline
  Parameter          				& MCH      				& TPC                  			&    \\
  \hline
Input polarity     				& pos      					& neg                  			&    \\
Input charge linear range	  		& 500\,fC 					& 100\,fC and 67\,fC   			&    \\
Sensor capacitance 				& 40-80\,pF 				& 12-25\,pF             			&    \\
Gain               					& 4\,mV/fC  				& 20\,mV/fC and 30\,mV/fC 		 &	\\
Gain channel-to-channel variation	& 1.5\,\%					& 1.5\,\%					&  	\\
Gain linearity					& 0.5\,\% up to 85\,\% of range	& 0.5\,\% up to 85\,\% of range	&  	\\
Channel-to-channel cross talk		& $<$0.3\,\%				&  $<$0.2\,\%				&  	\\
Noise 						& 2000\,$e^-$ @ 60 pF		& 600\,$e^-$ @ 12\,pF		&  	\\
Peaking time					& 330\,ns					& 170\,ns					&  	\\
Baseline return					& $<$550\,ms				& $<$500\,ns					 &	\\
ADC sampling rate max. 20\,MSa/s	&10\,MSa/s				& 5\,MSa/s					 &	\\
ADC ENOB					& $>$9.2					&  $>$9.2				&  	\\
ADC INL						&  $<$ 1\,LSB (abs.)			&  $<$ 1\,LSB (abs.)	&  	\\

\hline
\end{tabular}

\label{tab:sampa_perf}
\end{table}

Analog front-end reference voltages (nominal values 450\,mV, 600\,mV, and 750\,mV) are generated internally with temperature compensation and can be adjusted via configuration registers. The ADC requires an external voltage reference of \SI{1.1}{\volt}.
The DSP eliminates signal perturbations, distortion of the pulse shape, offsets, and signal variations due to changes in the environment. 
An I2C interface allows setting control registers.

\subsubsection{CSA and shaper}

\begin{figure}
\begin{centering}
\includegraphics[width=\textwidth]{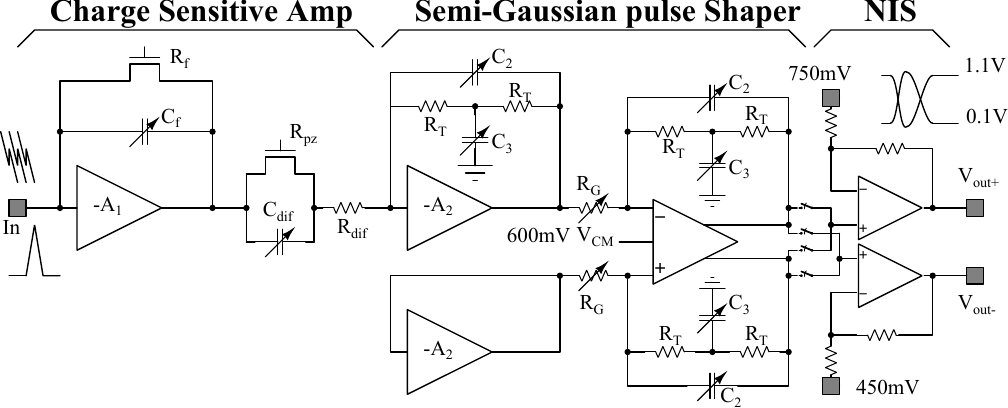}
\par\end{centering}
\caption{Block diagram of the front-end implemented in the SAMPA ASIC.}
\label{Sam:fig:DiagramFE}
\end{figure}

The SAMPA front-end is composed of a positive/negative polarity CSA with a capacitive feedback 
$\rm C_f$ and a resistive feedback $\rm R_f$ connected in parallel, converting the input charge signal ($Q$) into a voltage step signal proportional to $Q/\rm C_f$. The discharge resistor ($\rm R_f$) provides baseline restoration and reduces pile-up effects in the CSA (Fig.~\ref{Sam:fig:DiagramFE}).
A pole-zero cancellation resistor ($\rm R_{pz}$) eliminates the undershoot generated by the long time constant of the output step signal of the CSA.
The step signal is fed to a band-pass filter constituted by a first order high-pass filter $\rm C_{dif}R_{dif}$ (differentiator) and a two bridged-T second order low-pass filters (integrator). After that, a non-inverting stage (NIS) generates a semi-Gaussian output pulse with an amplitude proportional to the input charge.
The amplifier of the first shaper is a scaled-down version of the CSA amplifier.
In order to provide the second shaper with a differential mode input, a copy of the first shaper is included.
This copy is connected in unity gain configuration to minimize its noise contribution. The second shaper consists of a fully differential amplifier with a Miller configuration and a common-mode feedback network. It has the same functionality as the first shaper and implements two other poles and a zero creating a CR-(RC)$^4$ semi-Gaussian shaper together with the differentiator and the first shaper stage.

The gain of the front-end is controlled by $\rm R_G$, which is an array of parallel resistances that are switched by configuration registers. 
The peaking time of the semi-Gaussian shaper is adjusted for each operation mode (160\,ns and 300\,ns) by external configuration control of an array of parallel capacitors.
These front-end configurations are performed with transmission gates used as low resistance switches. 

\subsubsection{ADC}

The 32-channel 10-bit SAMPA ADC features a sampling frequency of up to 20\,MSa/s defined by an external clock. The MCH and the TPC use the SAMPA with \SI{10}{\mega\hertz} and \SI{5}{\mega\hertz} sampling clock, respectively.

\begin{figure}[htbp]
\begin{centering}
\includegraphics[width=\textwidth]{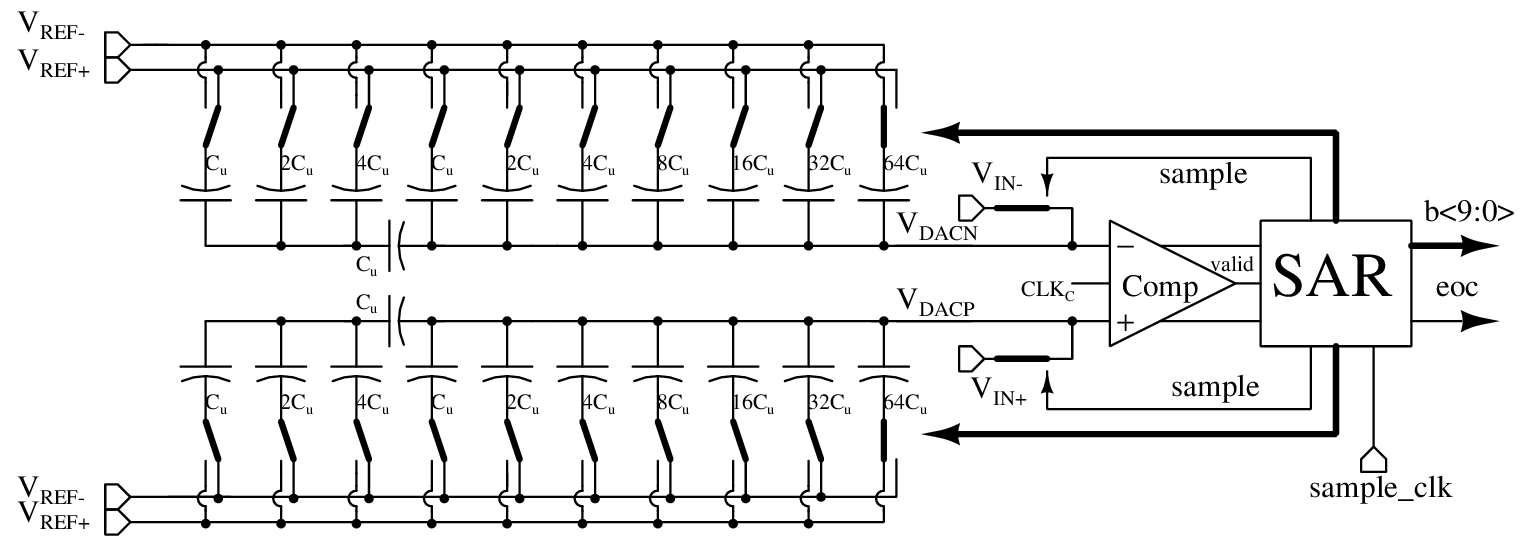}
\par\end{centering}
\caption{Block diagram of the SAMPA SAR ADC.}
\label{Sam:fig:SampaSAR}
\end{figure}

The SAMPA ADC is based on the successive-approximation register (SAR) architecture~\cite{sam:SAR}. It is shown in Fig.~\ref{Sam:fig:SampaSAR}. A differential capacitive DAC is implemented with the split capacitor topology. Top-plate sampling with MSB (Most Significant Bit) preset to achieve full-range sampling is used. A switching strategy with low energy dissipation per cycle is used. 

\subsubsection{DSP and readout}
\label{sampa:subsection:DSP}

\paragraph{Direct ADC Serialization}
In the Direct ADC Serialization (DAS) mode, the SAMPA sends out the unmodified raw data stream from all 32 ADC channels via 11 serial links, bypassing the readout processor and DSP.
In this mode, most of the digital circuitry is powered down via clock gating, keeping active only the communication links. 
10 SLVS links are used to send the 10-bit data samples of each ADC channel. The $11^{\rm th}$ link is used to provide a synchronisation clock.
Optionally, a split mode can be activated, such that data from ADC channels zero to 15 (16 to 31) are transmitted on serial links zero to four (five to nine). 
This allows connection of an odd number of SAMPAs to an even number of serial transmitters, as in the case of the TPC readout, where five SAMPAs are connected to two GBTx transmitter chips. 

\begin{figure} [h]
\centering
\includegraphics[width=\textwidth]{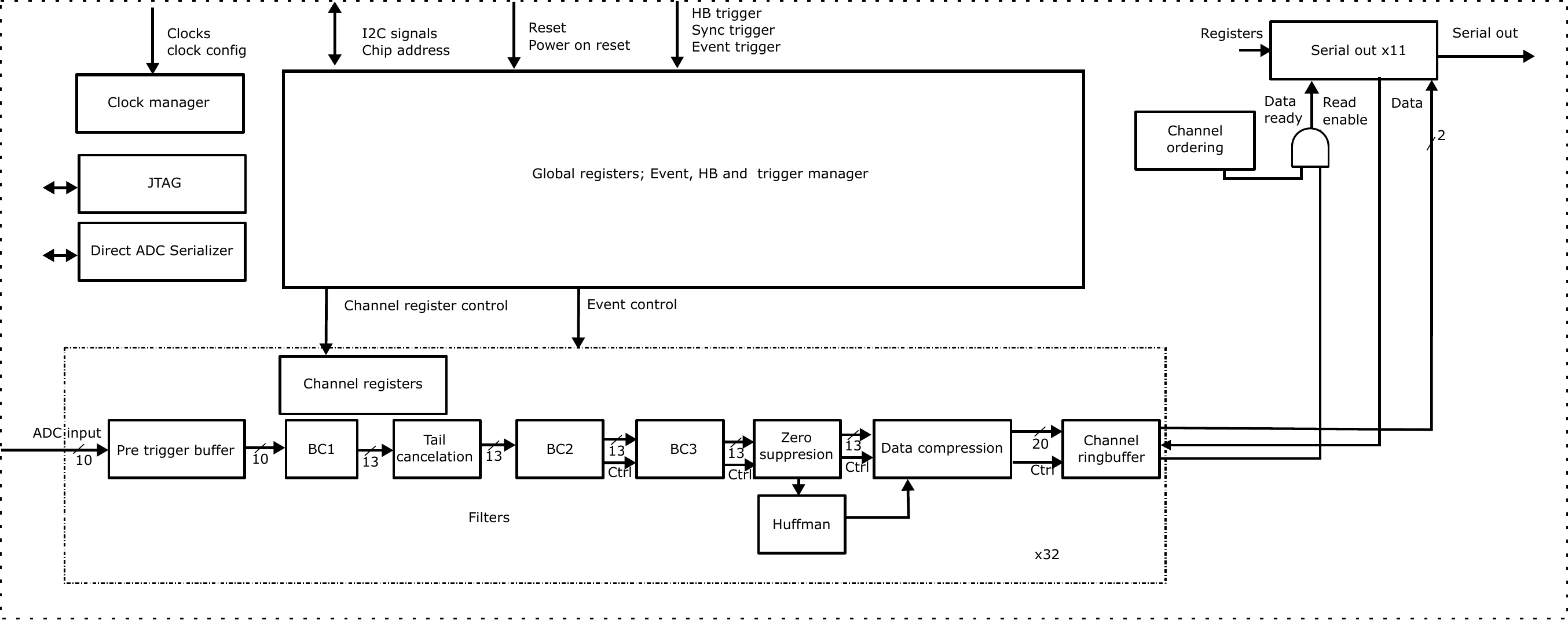}
\caption[Diagram of SAMPA DSP]{Diagram of digital signal processing chain in the SAMPA.}
\label{Sam:fig:DSPdiag}
\end{figure}

\paragraph{DSP}
The SAMPA DSP (Fig.\ref{Sam:fig:DSPdiag}) implements fully parallel data processing on the 32 channels and supports both continuous and triggered readout operation.
When in DSP mode, the data coming from the ADC are received via the pre-trigger buffer
with programmable depth of up to 192 10-bit words per channel. In triggered operation, the pre-trigger buffer delays the data and allows the collection of the detector signal samples before the arrival of the trigger signal.

The pre-trigger buffer is followed by a section of several configurable pipelined digital filters for signal conditioning. 
The filter blocks are:
\begin {itemize}
\item The \textbf{baseline correction 1} subtracts a given pedestal value for a fixed time after a trigger and applies a configurable Infinite Impulse Response (IIR) filter to correct slow fluctuations of the baseline.
\item The \textbf{tail cancellation} corrects long tails via a Digital Shaper, using a cascade of four fully configurable first order IIR filters which also can be used as general low-pass or band-pass filter.
\item The \textbf{baseline correction 2 and 3} offer a moving average Finite Impulse Response (FIR) filter and a non-linear slope-based filter.
\end{itemize}

The SAMPA is equipped with 3.2\,Gb/s output data bandwidth to extract the full raw data stream for up to 10\,MSa/s. This feature is used in the TPC application.
For the MCH application, data preprocessing in the SAMPA is used by applying a zero-suppression algorithm, removing all data below a threshold configurable for each channel. In addition, a cluster sum algorithm is available, where instead of delivering time and amplitude of each sample above threshold, consecutive active samples are added up to clusters in time and only the sum of the values and the time of arrival are delivered. 
The data are formatted for transmission in either continuous or triggered mode. A hamming code protected output buffer handles data size fluctuations and distributes the data to the activated serial links.

\subsubsection{Physical implementation and packaging}
\label{sampa:subsection:pack}

\begin{figure} [h]
\centering
\includegraphics[width=0.423\textwidth]{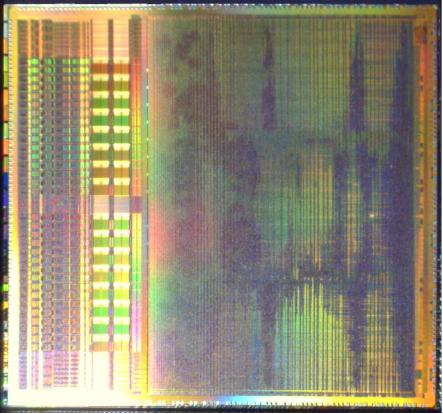}
\includegraphics[width=0.40\textwidth]{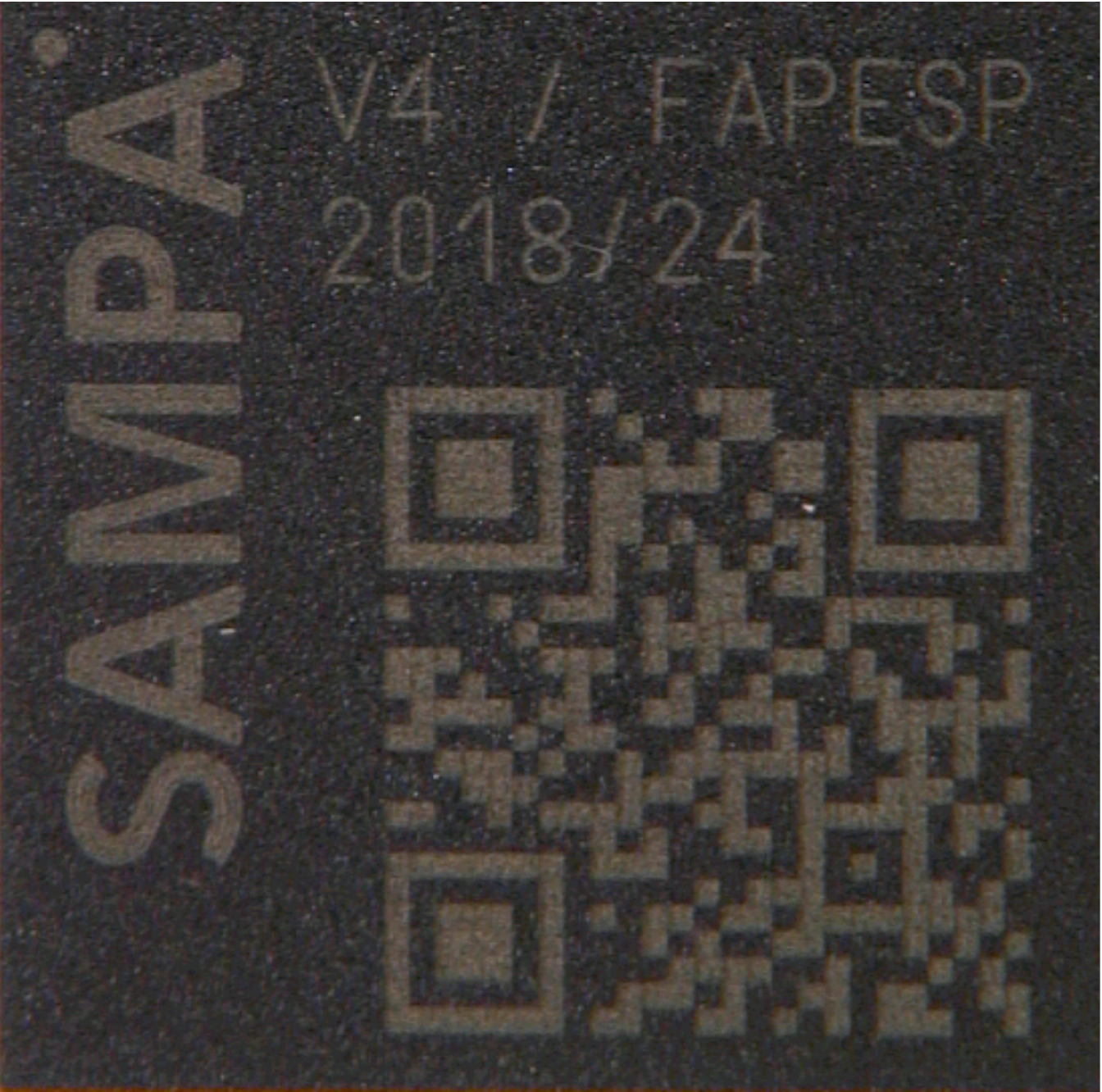}

\caption[SAMPA chip]{SAMPA bare die (left) and TFBGA packaged chip (right).}
\label{Sam:fig:die_chip}
\end{figure}

The SAMPA ASIC die is 8.9\,mm wide and 9.5\,mm long with 350 flip chip bond pads. As visible in the left panel of Fig.~\ref{Sam:fig:die_chip}, only a minor part of the die is devoted to the analog circuits (left on the picture), while the largest fraction contains the digital blocks, with part of the area being occupied by the buffer memories. During the implementation, special care was taken to isolate the power domains of the different circuits. There are five different power domains: CSA, Shaper and Output Buffer, ADC, core digital logic and SLVS IO drivers.
The SAMPA features a 372 ball ${\rm 15\,mm \times 15\,mm}$, 1.2\,mm thick, Thin Fine-pitch Ball Grid Array (TFBGA) package with 0.65\,mm ball pitch in order to be compatible with the MCH integration requirements. 
The high number of available balls allowed multiple connections to VDD and GND pads, reducing the inductive and resistive loss. The package includes filtering capacitors for ADC power connections and for on-chip ADC reference voltages. 
A QR-code on the SAMPA package (right panel in Fig.~\ref{Sam:fig:die_chip}) encodes the wafer lot-ID and a unique chip serial number, allowing the identification and tracking of each ASIC.

\subsubsection{SAMPA performance and tests}

\begin{figure}
\centering
\includegraphics[width=0.85\textwidth]{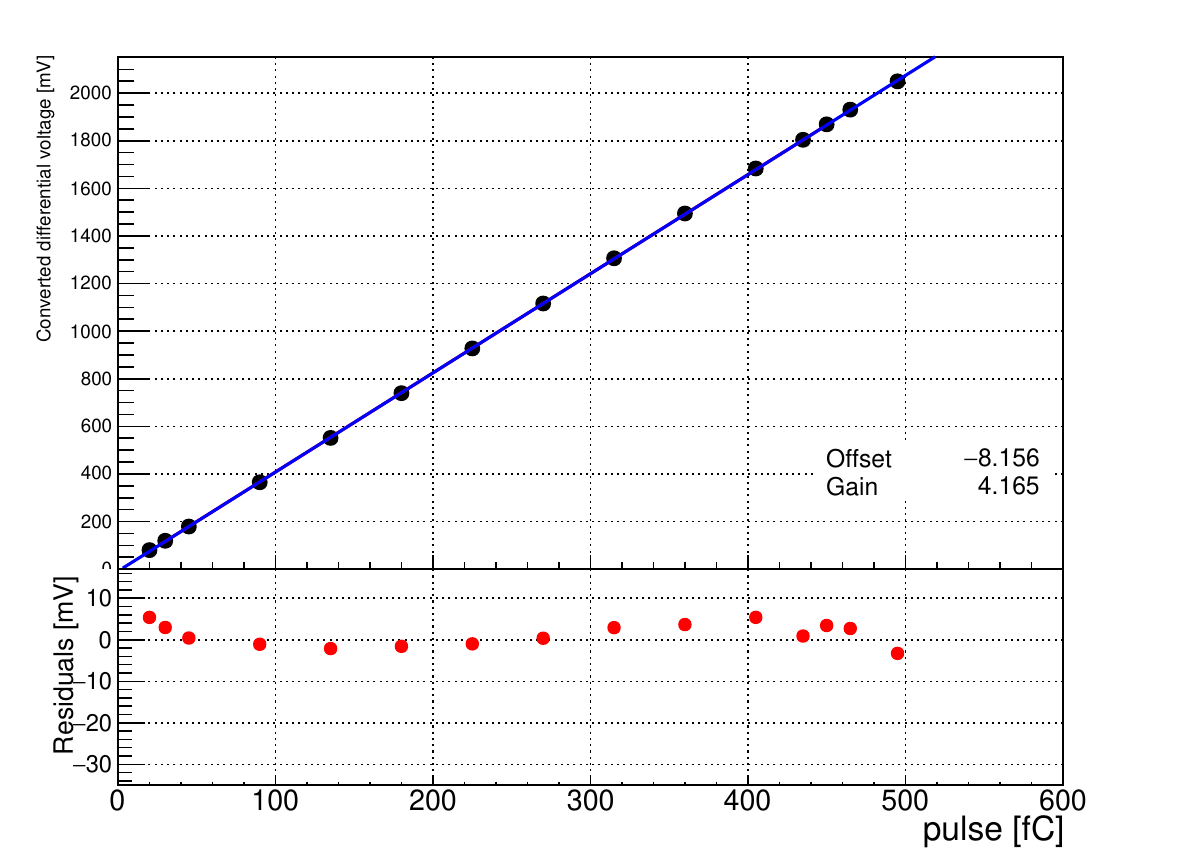}
\caption{Example response curve for 4\,mV/fC configuration. }
\label{Sam:fig:figLinear}
\end{figure}

The main specifications and performance of the SAMPA are listed in Tab.~\ref{tab:sampa_perf} above.
Figure~\ref{Sam:fig:figLinear} shows the response curve for the 4\,mV/fC gain setting.

Robustness measurements of the CSA against saturation in case of multiple consecutive signals were performed, showing that an average current of at least 30\,nA can be sustained for 60\,$\mu$s without significant baseline shift, indicating that the SAMPA can stand this charge rate indefinitely. 

The SAMPA functionality was verified successfully against the highest expected radiation load of 2.1\,kRad. Robustness of the SAMPA against single event upsets and single event latchups for an expected maximum flux of high-energy hadrons of 3.4\,kHz/cm$^2$ has been verified. An upper limit for the SEL cross section of 10$^{-7}$cm$^2$ for ions with a linear energy transfer of 16\,MeV~cm$^2$~mg$^{-1}$ has been measured.

80000 SAMPAs have been tested using a robotic test system to verify the functionality of the digital blocks and that the output baseline, noise, gain and peaking time are in a narrow intervals around the nominal values. The SAMPA mass production yield was 79.6\,\%. A breakdown of the rate of different kinds of failures can be found in Table~14 of~\cite{ALICETPC:2020ann}.

\section{Detector systems}
\label{chap:dets}

In the following subsections, each of the ALICE detector systems is
presented, with emphasis on the upgrades that were
installed during LHC Long Shutdown 2. Each system is presented in a separate subsection, starting from the inner tracking system, the muon forward tracker and the time projection chambers, which have undergone the most significant changes.

\subsection{Coordinate system}

The gloabel reference coordinate system used in ALICE is a right handed system with the $z$ axis point along the beam line, in the direction away from the muon arm, the $y$-axis pointing vertically up, and the $x$ axis pointing horizontally towards the center of the LHC. The nominal interaction point is the origin of the coordinate system. The two sides of the detector along the beam axis are refferred to as the C side, where the muon arm is positioned, and the A side, where FV0 is positioned.

\subsection{Inner Tracking System}

\label{sec:its:intro}
The new Inner Tracking System (ITS2)~\cite{ITS-TDR} uses the ALPIDE
sensor (described in Sec.~\ref{alpide:subsection:alpide}) and
represents the largest-scale application of Monolithic Active Pixel
Sensors (MAPS) in a high-energy physics experiment.
The main goal of the ITS upgrade is to improve the precision of the reconstruction of the primary vertex as well as of decay vertices
originating from heavy-flavour hadrons, and the performance in the detection of low-\pt particles. Additionally, readout rates of \SI{50}{\kHz} in Pb–Pb and \SI{400}{\kHz} in pp collisions are required.
In order to achieve this performance, the following key improvements were made in comparison with the previous ITS:
\begin{itemize}
    \item Granularity increased for all layers with pixel sensors with a cell size of $\SI{29.24}{\um} \times \SI{26.88}{\um}$. The number of layers for the inner barrel was increased from two to three, raising the total number of layers from six to seven.
    \item New beam pipe with a central beryllium section with an outer radius
      reduced from $\SI{28}{\mm}$ to $\SI{18}{\mm}$  (see Chapter~\ref{chap:integration}).
    \item Innermost detector layer moved closer to the interaction point, from $\SI{39}{\mm}$ to $\SI{22.4}{\mm}$ .
    \item Material budget reduced to $0.36\% \;X_0$ per layer for the innermost layers and limited to $1.10\%$ $X_0$ per layer for the outer layers. 
\end{itemize}
Table~\ref{tab:Table1} reports a list of main parameters of the old ITS1, used in Runs 1 and 2, and of the new ITS2.
The new design improves the tracking efficiency and momentum resolution at low \pt as well as the impact-parameter resolution by a factor of three and five in the r$\varphi$- and z-coordinate, respectively, at a \pt of 500~\MeVc~\cite{Aamodt:2008zz}.
\begin{table}[h]
    \centering
    \caption{Comparison of main detector parameters of the previous ITS1 and the new ITS2.}
   \begin{small}
    \begin{tabular}{ccc}
    \hline
    & \textbf{ITS1} & \textbf{ITS2} \\
        \hline
Technology       & Hybrid pixel, strip, drift &  MAPS \\
No. of layers    & 6                  & 7 \\
Radius           & 39--430 mm          & 22--395 mm \\
Rapidity coverage & $\mid\eta\mid\leq$ 0.9 & $\mid\eta\mid\leq$ 1.3 \\
Material budget / layer       & 1.14\% X$_0$               & inner barrel: 0.36\% X$_0$\\
& & outer barrel: 1.10\% X$_0$ \\
Pixel size       & 425 ~\si{\um} $\times$  50 ~\si{\um}  & 27 ~\si{\um} $\times$ 29 ~\si{\um} \\
Spatial resolution (r$\varphi \times$ z) & 12 ~\si{\um} $\times$ 100 ~\si{\um} & 5 ~\si{\um} $\times$ 5 ~\si{\um} \\
Readout          & Analogue (drift, strip), Digital (Pixel) & Digital \\
Max rate (Pb--Pb) & 1~\si{\kHz}                      & 50~\si{\kHz} \\
\hline

    \end{tabular}
    \label{tab:Table1}
    \end{small}
\end{table}

\begin{figure} [h]
\centering
\includegraphics[width=0.85\textwidth]{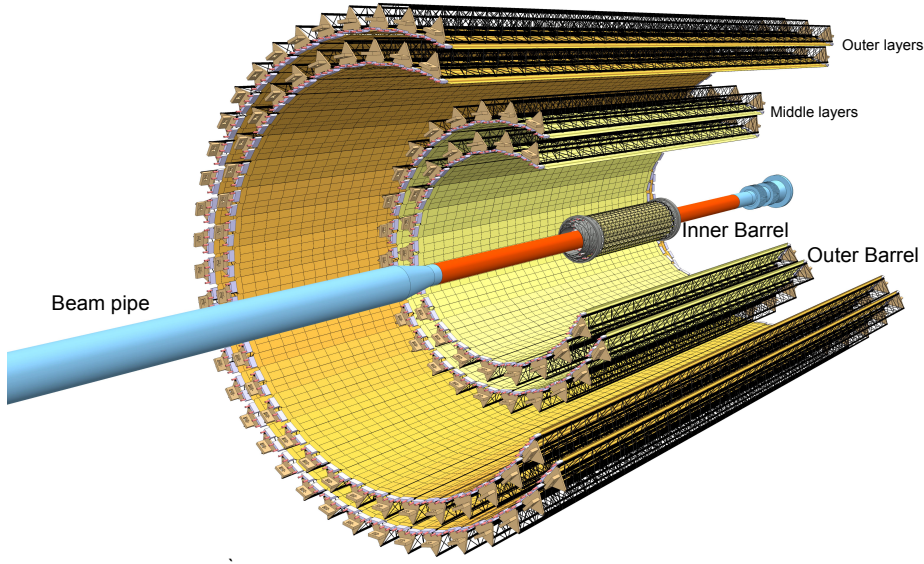}
\caption[ITS2 layout]{Schematic layout of the ITS2. The three innermost layers form the inner barrel, the middle and outer layers form the outer barrel.}
\label{fig:fig1}
\end{figure} 
An overview of the ITS2 structure is shown in Fig.~\ref{fig:fig1}. The detector is grouped into the inner barrel (IB) consisting of the three innermost layers, and the outer barrel (OB) arranged in two double layers. 
The radial position of each layer (listed in Table~\ref{tab:Table2}) was optimized to achieve the best performance in terms of pointing resolution, \pt resolution, and tracking efficiency in the high track-density environment of Pb–Pb collisions. The pseudorapidity coverage of the detector is $|\eta| < 1.22$ for the most luminous $90\%$ of the interaction region, i.e.\ for interaction vertices located in the range of approximately $\pm \SI{10}{\cm}$ around the nominal interaction point along the beam axis, see also Fig.~\ref{fig:ITSvertex}. The total surface area of the sensors is \SI{\sim 10}{\square\metre} instrumented with about 12.5~billion pixels with binary readout. The detector is operated at room temperature (20$^\circ$C to 25$^\circ$C), which is stabilized by water cooling. The radiation load at the innermost layer is expected to be 270~krad of Total Ionising Dose (TID) and $1.7$ $10^{12}$ 1 MeV n$_\mathrm{eq}$/cm$^2$ of Non-Ionising Energy Loss (NIEL), for 10 years of ALICE running, met by the ALPIDE sensors
(see Sec.~\ref{alpide:subsection:alpide}). In order to meet the material budget requirements, the silicon sensors are thinned down to \SI{50}{\um} and \SI{100}{\um} in the inner and outer barrel, respectively.

\begin{table}[t]
\centering
\caption{Main layout parameters of the new ITS2.}
\begin{small} 
\begin{tabular}{c|ccccc}
\hline
Layer no. &	Average&	Stave&	No. of &No. of &Total no.\\
&	radius&length&staves&HICs/	&of chips\\
&	(mm)&(mm)& &stave	&\\

\hline
0	&23 & 271 & 12 & 1&108\\
1	&31 & 271 & 16	& 1&144\\
2	&39 & 271 & 20	& 1&180\\
3	&196 & 844 & 24 & 8	&2688\\
4	&245 & 844 & 30 & 8	& 3360\\
5	&344 & 1478	&42 & 14 &	8232\\
6	&393 & 1478	& 48 & 14 &	9408\\

\hline
\end{tabular}
\label{tab:Table2}
\end{small}
\end{table}

\subsubsection{Stave modules}
\label{sec:its:stavemodules}
\begin{figure}
\centering
\includegraphics[width=1.\textwidth]{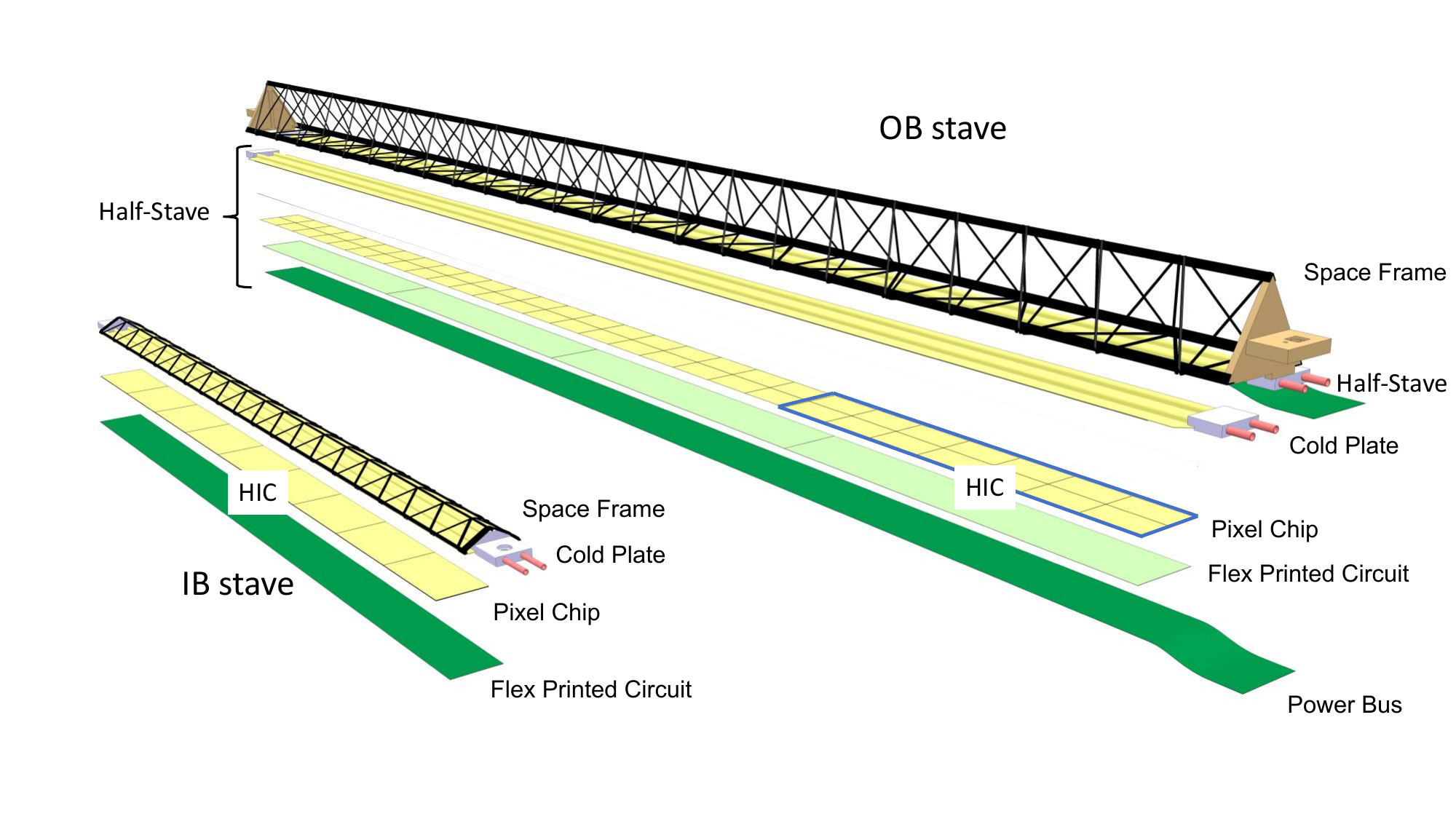}
\caption[ITS2 stave layout]{Layout of the staves of the inner and outer barrels.}
\label{fig:fig3}
\end{figure}

\begin{figure}
\centering
\includegraphics[width=0.90\textwidth]{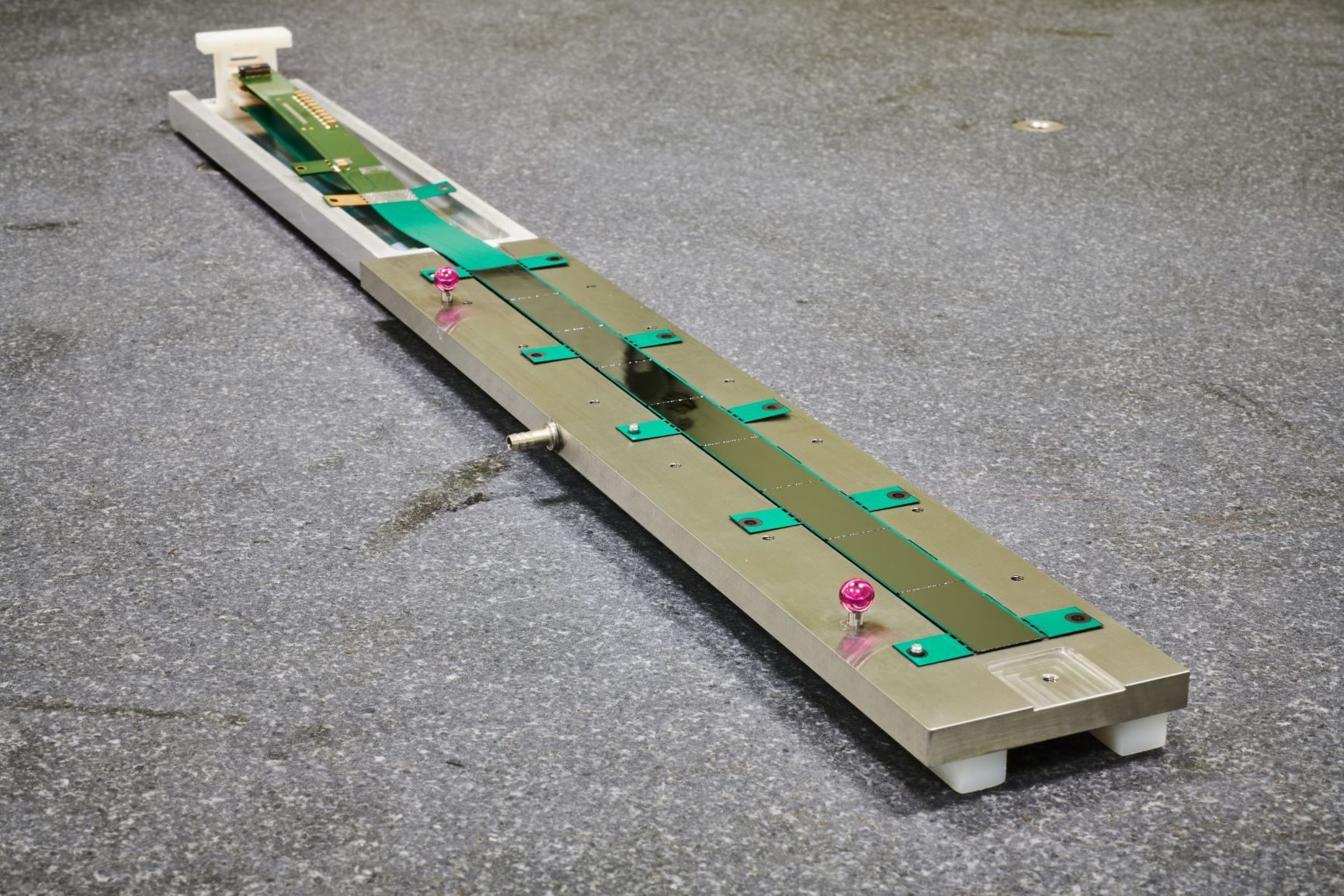}
\caption[ITS2 hybrid integrated circuit]{Inner HIC seen from the sensor side. The green tabs are used for fixing and handling, and are removed before mounting the HIC.}
\label{fig:fig4}
\end{figure}

\begin{figure}
\centering
\subfigure[Outer barrel HIC, top view. ]{\includegraphics[width=0.9\textwidth]{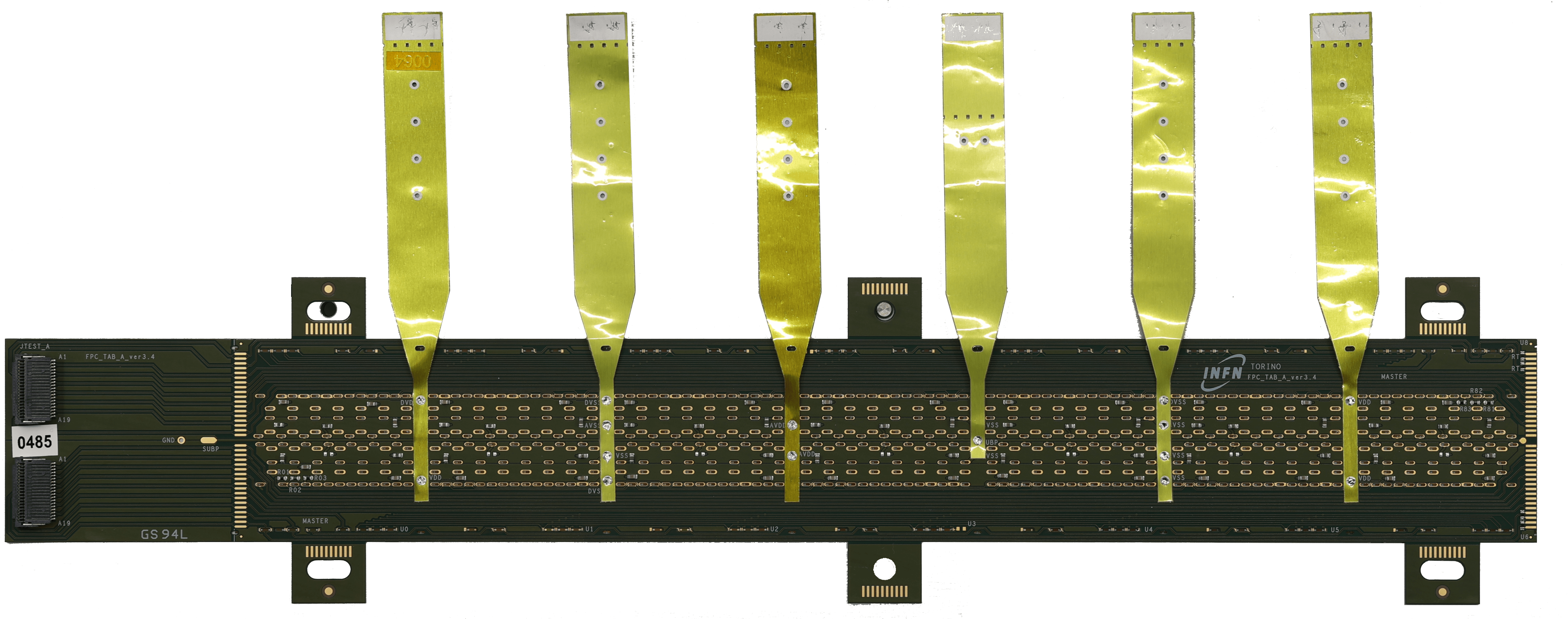}}
\subfigure[Outer barrel HIC, sensors view ]{\includegraphics[width=0.92
\textwidth]{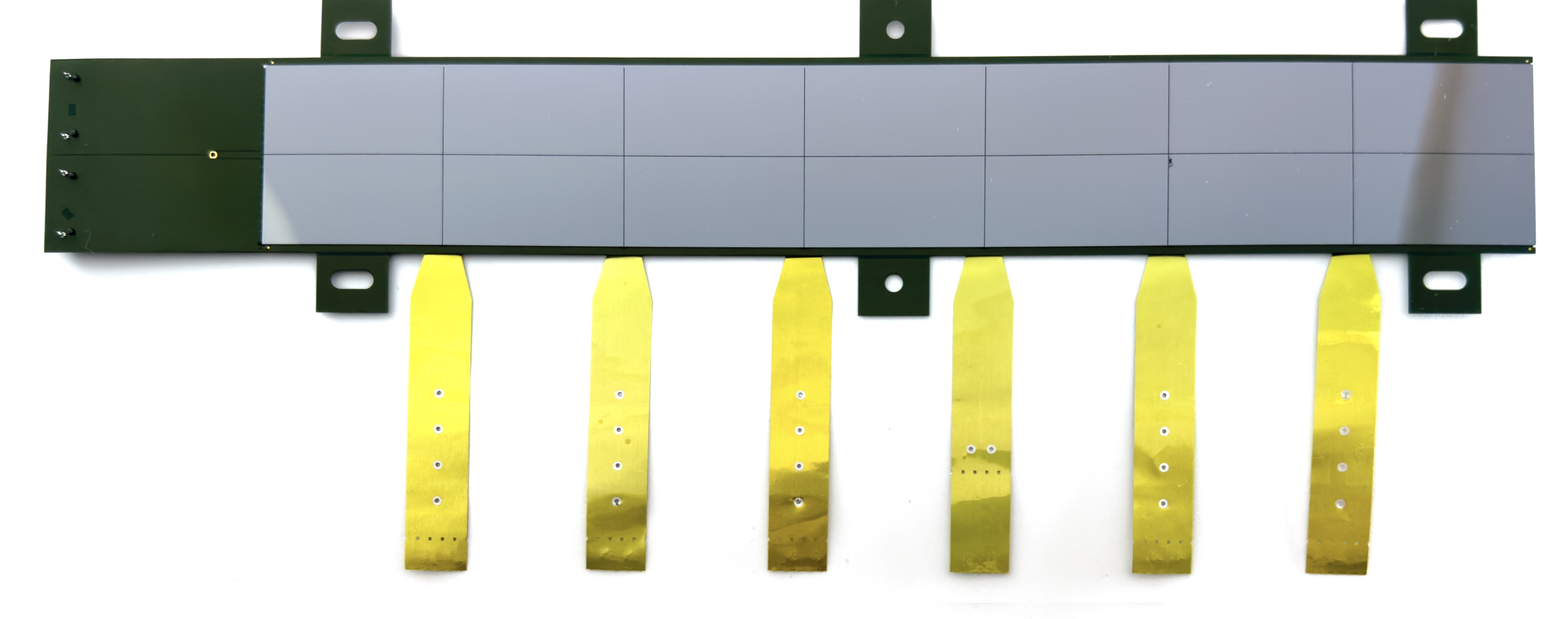}}
\caption[ITS2 outer barrel HIC]{Top view and bottom view of an outer barrel HIC. The yellow cables connect the HIC to the power bus.}
\label{fig:fig5}
\end{figure}

The basic detector unit, called stave, consists of the following elements (Fig.~\ref{fig:fig3}):
\begin{itemize}
  \item \textbf{Hybrid Integrated Circuit (HIC)}:  an assembly of a polyimide Flexible Printed Circuit (FPC) on which a number of pixel chips, namely 9 and 14 for the inner and outer barrel staves, respectively, and some passive components, are bonded. Figures~\ref{fig:fig4} and~\ref{fig:fig5} show photos of an inner and outer barrel HIC.
      \item \textbf{Coldplate}: a carbon fibre sheet with high thermal conductivity with embedded polyimide cooling pipes, which is either integrated within the space frame (for the inner barrel staves) or attached to the space frame (for the outer barrel staves).
    \item \textbf{Space Frame}: a carbon fiber truss-like support structure providing the mechanical support and the necessary stiffness to the assembly of HICs on cold plates.
\end{itemize}
The HICs are glued to the cold plate: 1 HIC for the inner barrel and 8 and 14 HICs, for the middle and outer layers, respectively. The cold plate is in thermal contact with the pixel chips to remove the generated heat.
For the inner barrel, each staves consists of a single HIC+cold plate assembly. In the outer barrel, staves are further segmented in azimuth in two half-staves. Each half-stave extends over the full length of the stave and consists of a cold plate on which four or seven modules (HICs) are glued depending on the length of the stave. 

\paragraph{Hybrid Integrated Circuit} 
As shown in Fig.~\ref{fig:fig3}, the inner barrel HIC includes one row of 9 sensors, whereas the outer barrel HIC comprises two rows of 7 sensors each, as visible in Fig.~\ref{fig:fig5}, bottom.
The HICs consist of an assembly of ALPIDE chips glued to an FPC, which provides the connection to analogue and digital power rails as well as p-well and substrate bias voltages. Differential pairs of traces of \SI{100}{\um} width and spacing are used to distribute control and clock signals in a local bus and to read out individual pixel sensors. 
In the inner-barrel staves, all 9~sensors share the same aluminium power bus on the FPC, while in the outer-barrel staves, a dedicated aluminium power bus extends over all FPCs of the half-stave and provides analogue and digital power as well as ground connections.
The baseline powering scheme is based on a conservative parallel connection: all chips in a HIC are directly connected to the analogue and digital power planes of the FPC, which are in turn fed by the power bus serving the half-stave. The electrical connection to the HICs is made by means of thin aluminium cables soldered onto the HIC, as visible in Fig.~\ref{fig:fig5}. To minimise the material budget, aluminium was chosen as conductor (having a radiation length of \SI{8.9}{\cm} compared to \SI{1.44}{\cm} for copper)
for the FPCs of the inner barrel. 
Since the resistivity of aluminium is 1.5 times larger than that of copper, the thickness of the power lines must be correspondingly increased.
A thickness of \SI{25}{\um}  ensures a voltage drop below \SI{50}{\milli\volt} over the full length, as well as an attenuation suitable for signal transmission up to 1.2 Gbps.
Polyimide Upilex-S75 was selected as substrate because it has a small thermal expansion coefficient (0.01\% at 200$^\circ$C) and therefore provides good dimensional stability during the aluminium coating by sputtering in vacuum. The material budget requirements for the outer barrel FPC are less severe and allow for a more standard production procedure, using copper-clad Pyralux, with a substrate of \SI{75}{\um} and \SI{18}{\um} metal layer. With the external power bus, the thinner copper traces are compatible with voltage drop requirements and the readout rate of 400~Mbps, sufficient for the lower occupancy of the outer layers. The power bus in the outer barrel is connected to the HICs via short 

\begin{figure}
    \centering
    \hfill
    \includegraphics[height=0.6\textheight]{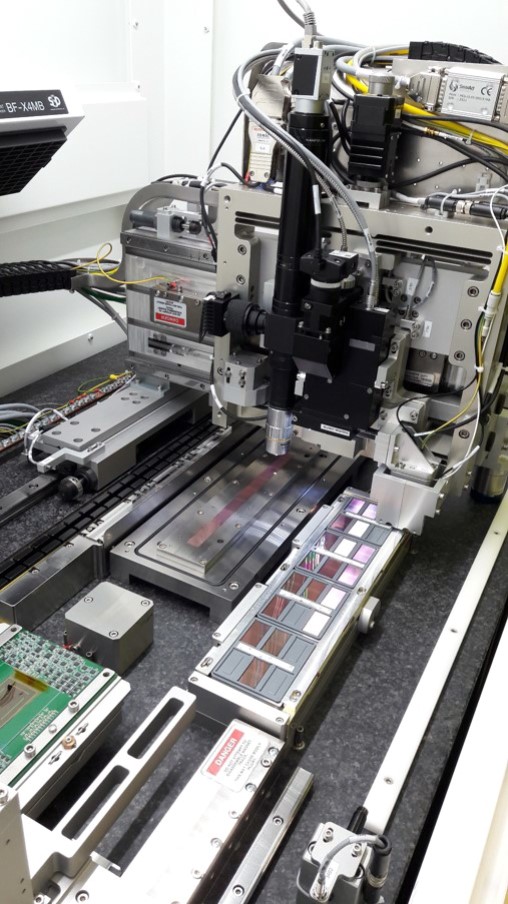}
    \hfill
    \includegraphics[height=0.6\textheight]{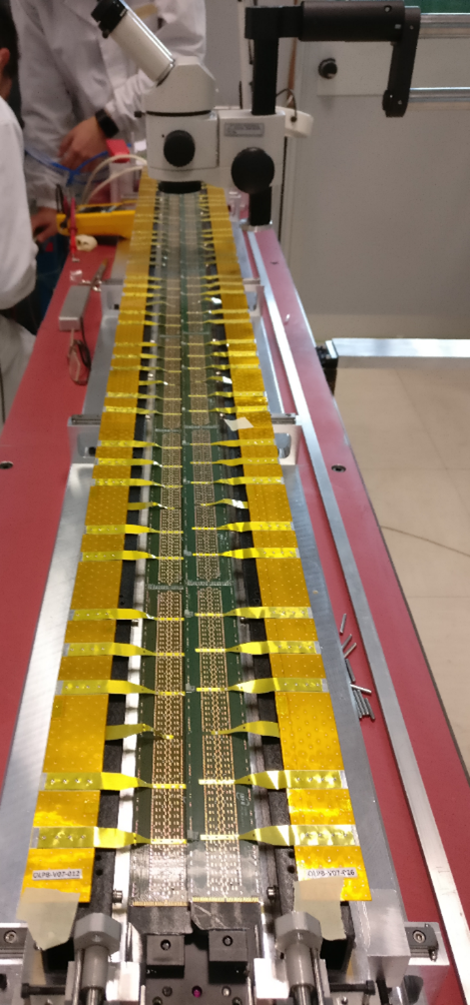}
    \hfill
    \\[1cm]
    \includegraphics[width=1.0\textwidth]{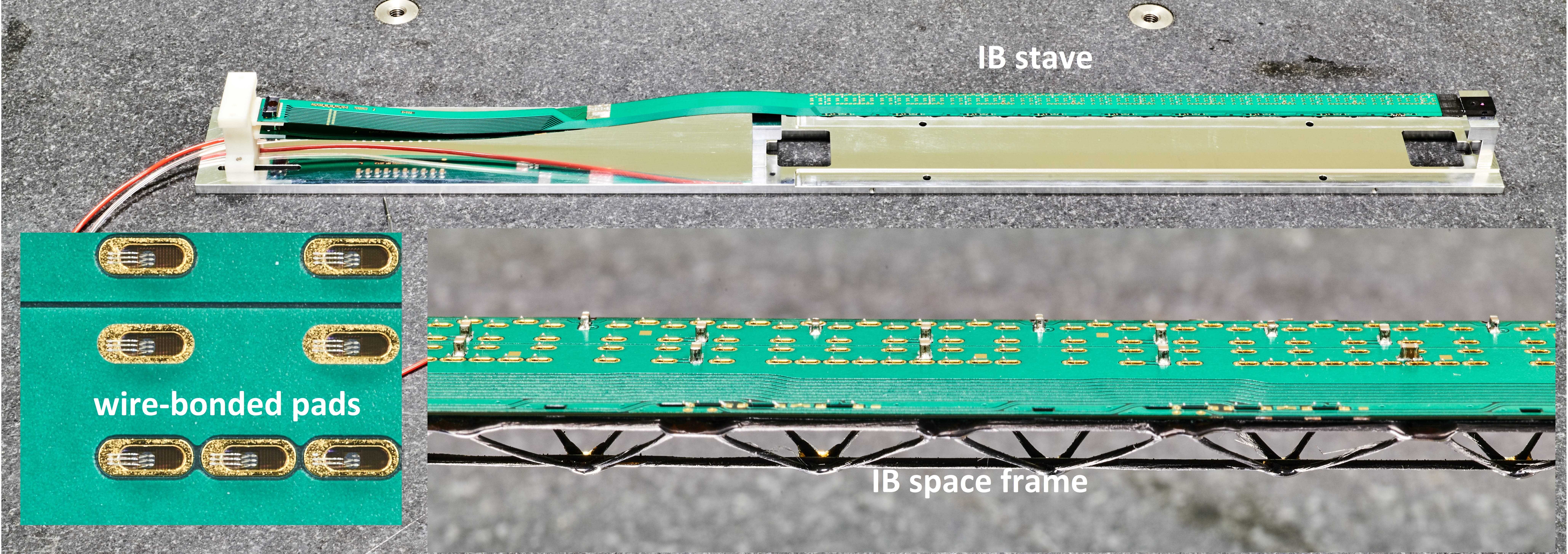}
    \caption[Pictures of the ITS2 assembly]{Pictures of the ITS2 assembly. Top left: A view of the MAM used for chip inspection and HIC assembly. Top right: Photo of an outer barrel stave, with power bus cables opened on the two sides. Bottom: Photo of an inner barrel stave, with detailed views shown in the insert.}
    \label{fig:fig7}
\end{figure}

The main requirements for the chip to FPC interconnection are: (i) compact module layout with minimal dead area; (ii) highly reliable and stable mechanical connection; (iii) high quality, low inductance electrical connection. A custom made automatic Module Assembly Machine (MAM),
named ALICIA, supplied by IBS-Precision Engineering, see top left panel of Fig.~\ref{fig:fig7}, implements electrical testing, dimension measurement, integrity inspection and alignment for assembly, was used to achieve a reproducible accuracy and the required production speed at the various HIC assembly sites.

Using a stencil manufactured in an adhesive film (\SI{90}{\um} thick), very precise spots of Araldite 2011 (\SI{0.6}{mm} diameter, 160 per chip) were applied on the FPC clamped on a gripper jig. After the chips were aligned by the MAM onto the vacuum chuck with a position accuracy of better than \SI{5}{\um} and a spacing of \SI{150}{\um}, the FPC was positioned precisely on top. Shims of \SI{50}{\um} were used to ensure a sufficient gap for the glue and variations related to tolerances of tooling (planarity: $\pm \SI{10}{\um}$) and components (FPC thickness: $\pm \SI{10}{\um}$; chip thickness: $\pm \SI{5}{\um}$). The assembly procedure was validated by mechanical tests where on average a pull strength of \SI{44}{N/chip} and a peel strength of \SI{3}{N} were measured. The electrical interconnection used a novel approach of wire bonding through the FPC vias (Fig.~\ref{fig:fig7} bottom left panel). In order to account for the clearance necessary for the wedge bonding tool, the FPC vias have an oblong shape (1.2 mm $\times$ 0.4 mm); in addition \SI{300}{\um} interconnection pads were implemented on the top surface. Wire bonding was performed using \SI{25}{\um} aluminium wire (three wires per connection); a typical pull force of \SI{11}{cN} with a standard deviation of \SI{0.8}{cN} was measured per wire.

\begin{figure}[h]
\centering
\includegraphics[width=0.80\textwidth]{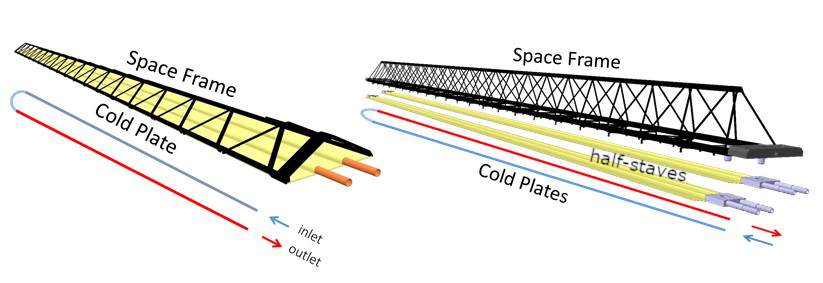}
\caption[Space frame and cold plate cooling scheme]{Space frame and cold plate cooling scheme.
(Left) Inner Barrel; (Right) Outer Barrel}
\label{fig:fig9}
\end{figure}

\clearpage
\paragraph{Space frame and cold plate} 
The layout of the ITS2 stave mechanics and cooling consists of a space frame and one or two cold plates. A large effort was devoted to the design of the lightest possible mechanical supports to maintain the silicon sensors in an accurate position while providing the cooling to remove the heat dissipated by the sensors. A novel technology was developed to directly embed polyimide pipes inside the cold plate, an assembly of highly thermally conductive carbon fibre laminate (see Fig.~\ref{fig:fig9}). 

\begin{figure}[h!]
\centering
\includegraphics[width=0.49\textwidth]{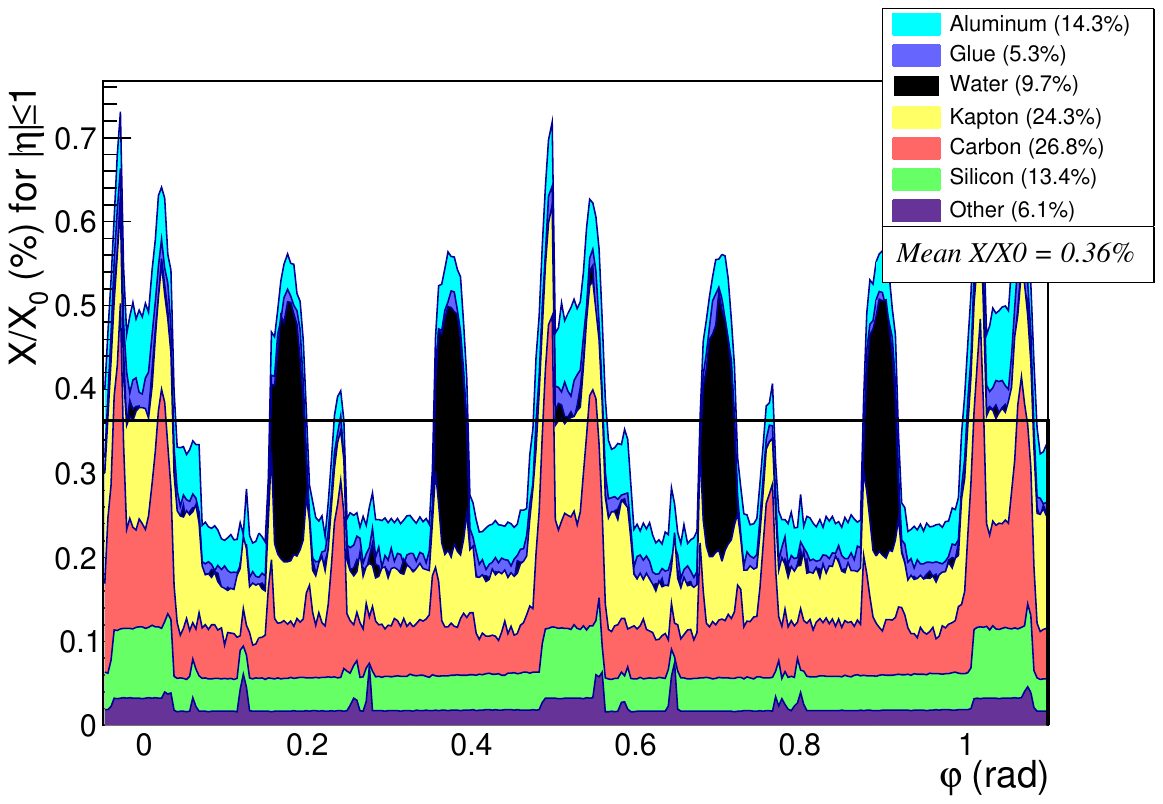}
\hfill
\includegraphics[width=0.49\textwidth]{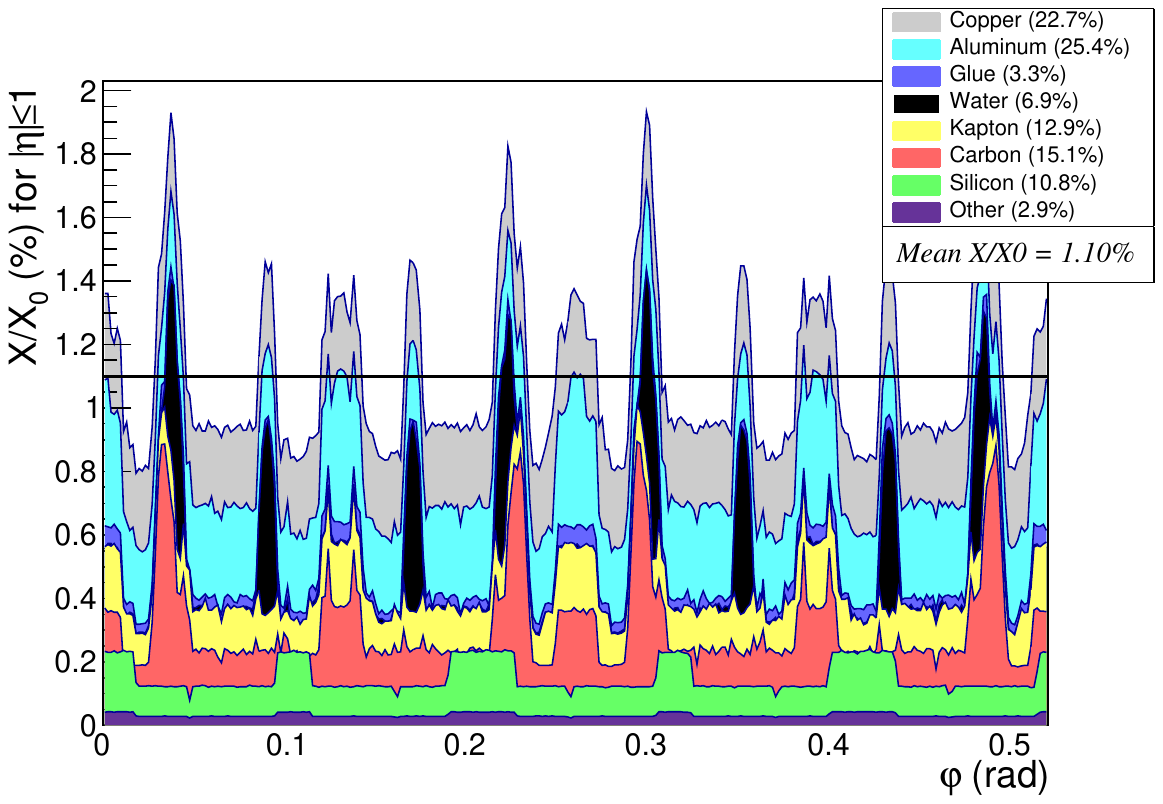}
\caption[Material composition of ITS2 staves]{Azimuthal distribution of single contributions to the material budget of an inner (left) and outer (right) barrel stave layer. The relative contribution of each component to the total material budget is quoted.
}
\label{fig:fig11}
\end{figure}

For mechanical stability, the cold plate is stiffened by the space frame, a light filament-wound carbon structure with a triangular cross section. The implementations of the cold plate and space frame differ in the inner and outer barrel to satisfy different geometrical and thermal constraints (Fig.~\ref{fig:fig9}). In order to guarantee electrical insulation, a Parylene coating was applied on the cold plates (Parylene volume resistivity is \SI{1.4e17}{\ohm\cm}, measured on a \SI{25}{\um} thick layer). 
Such a structure provides a highly efficient heat dissipation with single-phase liquid flow up to a power density of \SI{0.5}{W/cm^2} produced by the silicon chips glued on top of it. As can be seen in Fig.~\ref{fig:fig11}, this solution meets the quite stringent requirement of a very low material budget, in particular for the inner barrel stave ($0.36$\%~$X_0$/layer). The same solution for the integration of the cooling pipes was adopted for the outer barrel staves.
The material budget requirements for the much larger outer barrel layers are less stringent and the different layout corresponds to an average budget of $1.10$\%~$X_0$/layer (Fig.~\ref{fig:fig11}).

\subsubsection{Global support mechanics and services}
\begin{figure}[h!]
\centering
\includegraphics[width=0.9\textwidth]{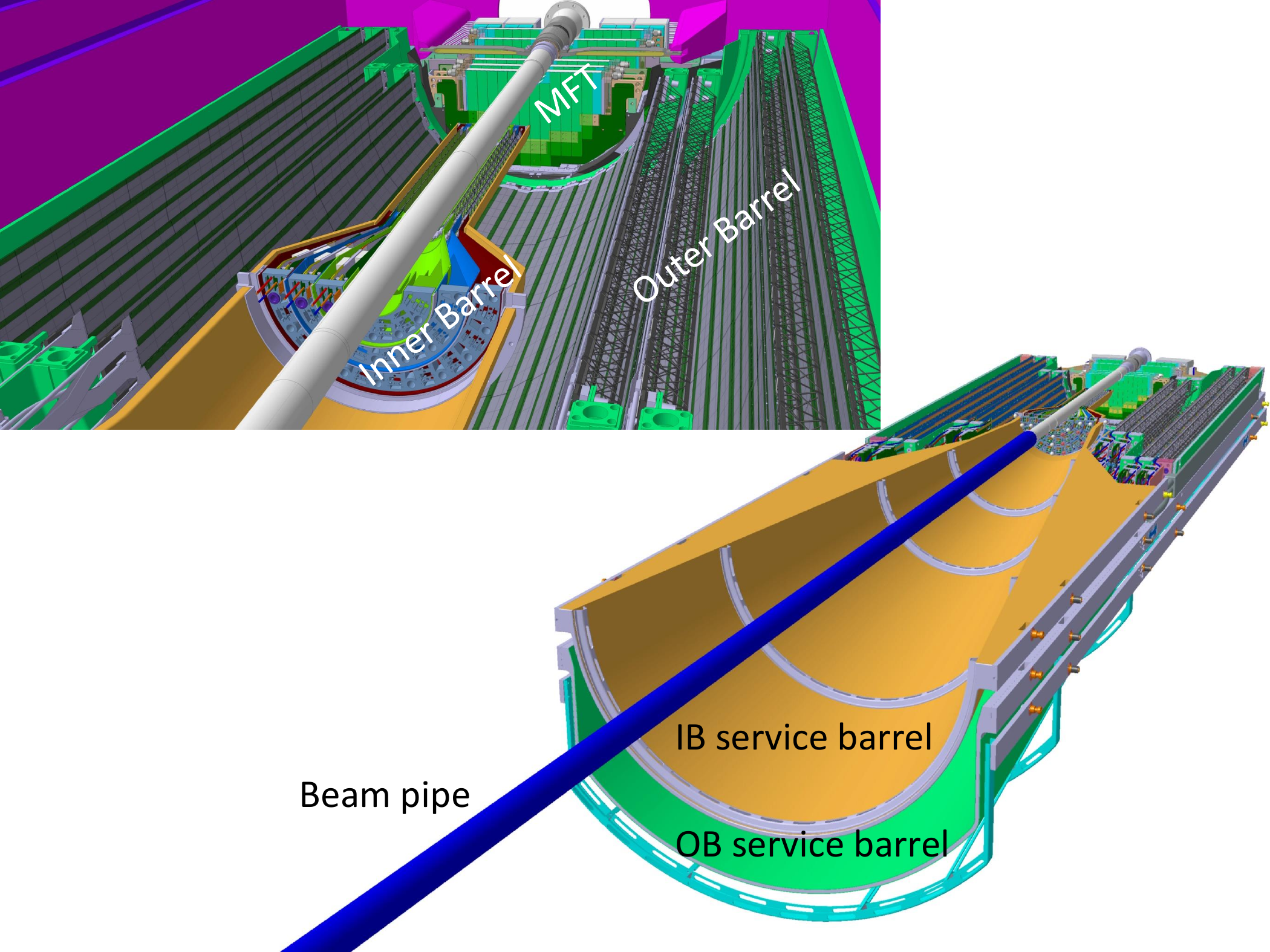}
\hfill
\caption[ITS2 support structures]{Overview of the mechanical structure of the ITS2. The upper panel shows the Inner Barrel, the Outer Barrel staves and the MFT in the back. The lower panel shows the IB and OB conical structural shells supporting the respective services.}
\label{fig:IBsupp}
\end{figure} 

In addition to the requirements of minimising the material in the sensitive region and ensuring high accuracy of the relative position of the detector sensors, discussed in the previous section, the ITS2 mechanical structure fulfils the following design criteria:
\begin{itemize}
\item{provide an accurate position of the detector with respect to the TPC and the beam pipe;}
\item{locate the first detector layer at a minimum distance to the beam pipe wall;}
\item{ensure thermo-mechanical stability over time;}
\item{ensure accessibility for maintenance and inspection.}
\end{itemize}
Also, the design of the support of the detector and services has to take into account the requirements set by the integration of the new Muon Forward Tracker (MFT) and the Fast Interaction Trigger system (FIT), which are installed very close to the ITS2. The main mechanical support structures of the ITS2 are shaped in two barrels made from carbon fiber composite. The inner shell supports the three innermost layers, while the outer shell supports the outer four layers.
Each barrel is divided into top and bottom halves, which are installed sequentially around the beam pipe. Each barrel is composed of a detector section and a service section, as shown in Fig.~\ref{fig:IBsupp}. The staves are housed in the detector barrel and are connected to the readout  and power systems via signal and power cables which are routed through the service barrel to the ALICE miniframe. The pipes that connect the on-detector cooling system to the cooling plant in the cavern are also routed through the service barrels.

\textbf{Detector support structure} The main structural components of the detector barrels are the end-wheels and the cylindrical and conical structural shells. Two light composite end-rings provide the reference plane for the fixation of the two extremities of each stave. The position of the staves in the reference plane is given by a ruby sphere, matching an insert in the mechanical connectors at both extremities. This system ensures accurate positioning, within a few \si{\um}, during the assembly and provides the possibility to dismount and re-position the stave with the same accuracy in case of maintenance. Finally, the staves are clamped by a bolt. The end-wheels on the A-side also provide the feed-through for the services.

An outer cylindrical structural shell connects the opposite end-wheels of the barrel and avoids that external loads are transferred directly to the staves. As shown in Fig.~\ref{fig:IBsupp}, in order to minimise the material budget in the detection area, the following design choices were adopted:
\begin{itemize}
\item{The inner barrel is conceived as a cantilever structure supported at one end outside the outer barrel acceptance;}
\item{The outer barrel has no intermediate mechanical structures between the four detection layers within the detector acceptance.}
\end{itemize}
The design of the outer barrel allows the separate assembly inside the TPC of all half-layers, which then are combined sequentially, starting from the outermost layer. The mechanical connection between the two double layers is provided by two conical structural shells located at the extremities of the detection area (Fig.~\ref{fig:IBsupp}).
All the barrel support structures are attached to the cage (Fig.~\ref{cage}), acting as the main supporting element inside the TPC bore.

\subsubsection{Readout and powering systems}
The readout and powering systems are composed of 192 identical readout units and 142~power boards, and have complete control over all sensor operations, including power management, triggering, data readout and slow control.
One of the major goals for the ITS upgrade was to minimize the detector material budget, which for the inner layers is on average 0.36\% $X_0$ per layer~\cite{ITS-TDR}. Reducing the sensor power consumption implies softer cooling requirements, and hence decreasing the passive mass in the system. Transferring data from the sensors to the front-end electronics represents a significant part of the total power budget. 
To reduce the power consumption, intermediate conversions between electrical and optical layers were avoided.
Therefore, the high-speed transceiver on the sensor~\cite{Mager:2016} directly drives the differential line connecting it to the front-end electronics, using the shortest possible path in order to achieve the target bit-rate of 1.2 Gb/s within the power budget~\cite{Szczepankiewicz:2016}. Consequently, both readout units and power boards are located within the ALICE L3 magnet, in a magnetic field of \SI{\sim 0.5}{T} and exposed to the radiation environment. All system components were validated for these conditions~\cite{Schambach:2018}.

The matrix of $1024 \times 512$ pixels of the ALPIDE sensor is digitally read out through serial links at 1.2 Gb/s or 400 Mb/s in the inner and outer barrel, respectively. 
Because of the higher occupancy, the sensors in the inner barrel are read out individually whereas those in the outer barrel are read out in groups of seven using the master-slave mode described in Sec.~\ref{alpide:subsection:alpide}.
As mentioned in Sec.~\ref{sec:its:stavemodules}, the middle and outer layers share the HIC as a common building block and are identical from the readout point of view. Such a HIC is composed of two rows of seven sensors, where each row implements the aforementioned master-slave topology.

\subsubsection{The readout system}
\label{sec:its:readout}
\begin{figure}
\centering
\includegraphics[width=0.80\textwidth]{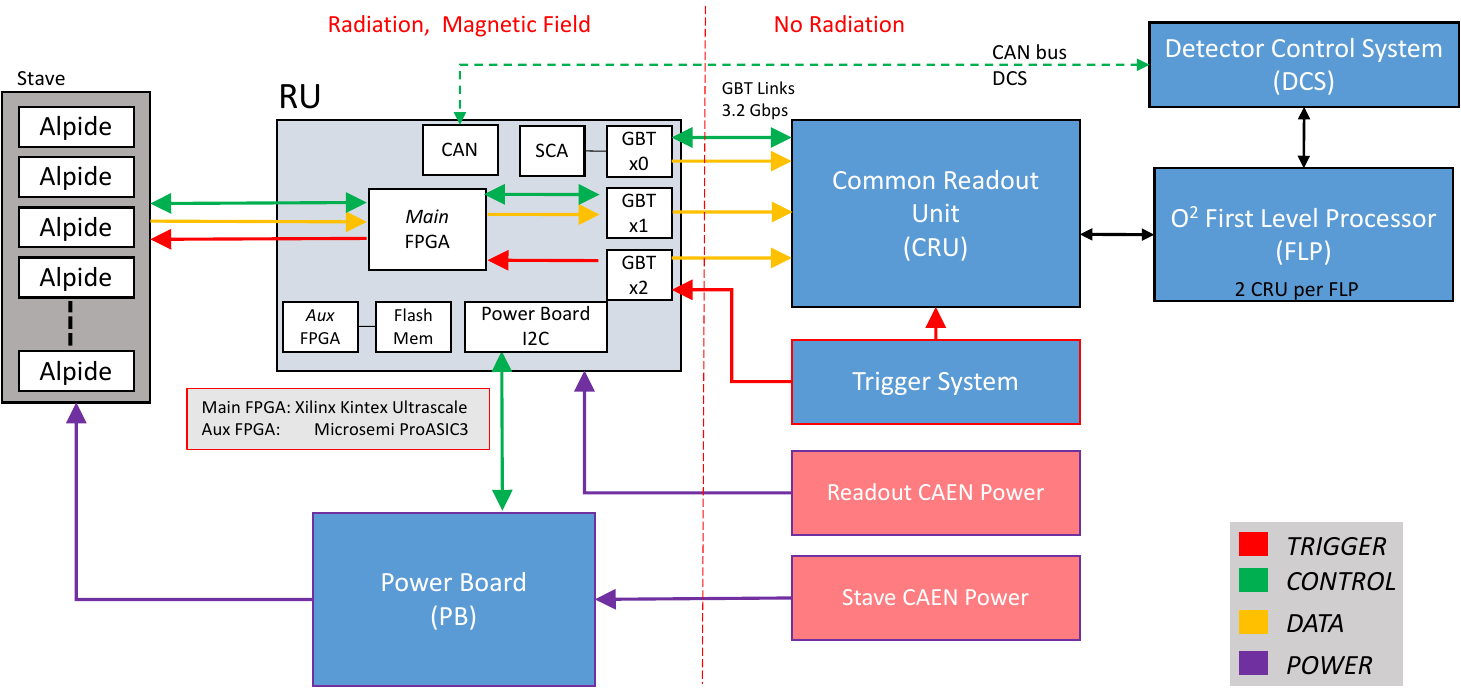}
\caption[Readout unit design]{Schematic design and interconnections of the ITS2 readout unit.}
\label{fig:RU_diagram}
\end{figure}

To maximize the modularity of the system, the readout electronics are organized in 192 autonomous readout units, one for each stave. As schematically shown in Fig.~\ref{fig:RU_diagram}, the readout units provide control and trigger and read the high-speed data lines from the ALPIDE chips. 
The readout units are identical, and only the I/O connections layer in the firmware adapts to the connected stave. Each readout unit connects to a common readout unit (CRU), see Sec.~\ref{sec:cru}, through the optical Versatile Link~\cite{VTRX:2009}, which is custom made by CERN and provides a \SI{3.2}{\giga\byte\per\second} radiation-tolerant physical transport layer including forward data correction. One downlink is used to transmit the slow-control commands from the counting room to the readout units, while up to three upstream links from each readout unit can transmit in parallel to achieve the needed readout bandwidth. A separate link is used to receive the trigger from the Central Trigger Processor (CTP).

The maximum necessary data bandwidth is determined by the collision system, the interaction rate, and the characteristics (pixel size, noise, etc.), and positioning of the sensors. 
The maximum bandwidth of \SI{1.2}{\giga\bit\per\second} suffices for the readout of Pb--Pb collisions at interaction rates up to \SI{100}{\kilo\hertz}.
For the middle and outer layers, the required bandwidth is lower, even though the sensors are read out in groups of seven sensors in master-slave mode. For those layers, the data link is configured for a maximum bandwidth of \SI{400}{\mega\bit\per\second} between the master chips and the readout unit. Table~\ref{tab:Table3} gives an overview of the data flow and bandwidth in the different parts for each layer. Each link represents a direct connection between a master sensor and a readout unit, while the payload is the actual available bandwidth for the data once the 8b/10b transmission encoding overhead is accounted for.

\begin{table}[b]
\centering
\caption{Summary of the ITS2 readout connections and payload capacity.~}
\begin{small} 
\begin{tabular}{ccccccccc}
\hline
Layer	& Staves	& Links	            & Links         	& Link      	& Bandwidth     	& Payload       	& Bandwidth         	& Payload  \\
    	&       	& per stave	        & bandwidth	        & payload	    & per stave	        & per stave	        & per layer	            & per layer \\
        &           &                   & [Gb/s]	        & [Gb/s]	    & [Gb/s]	            & [Gb/s]	        & [Gb/s]	            & [Gb/s]\\
\hline
0	    &12	        &9	                &1.2	            &0.96	        &10.8	                &8.6	            &129.6	                &104\\
1	    &16	        &9	                &1.2	            &0.96	        &10.8	                &8.6	            &172.8	                &138\\
2	    &20	        &9	                &1.2	            &0.96	        &10.8	                &8.6	            &216.0	                &173\\
3	    &24	        &16	                &0.4	            &0.32	        &6.4	                &5.1	            &153.6	                &123\\
4	    &30	        &16	                &0.4	            &0.32	        &6.4	                &5.1	            &192.0	                &154\\
5	    &42	        &28	                &0.4	            &0.32	        &11.2	                &9.0	            &470.4	                &376\\
6	    &48	        &28	                &0.4	            &0.32	        &11.2	                &9.0	            &537.6	                &430\\
\hline
Total	&192		&                   &                   &               &                       &                   &1872	                &1498\\

\hline
\end{tabular}
\label{tab:Table3}
\end{small}
\end{table}

The ITS2 can operate in two modes, triggered and continuous. In triggered mode, pixel hits on the sensors are latched into the sensor memory and then transmitted to the readout boards only if a trigger command arrives within a few microseconds after the event that generated them. In continuous mode, data are always recorded and transmitted, segmented in time frames of programmable duration, where all the events within the same time frame share the same timestamp.

The total bandwidth per stave was adapted to match the capacity of 3 CERN Versatile Link upstream connections per readout unit, with aggregate payload bandwidth of 9.6~Gb/s (3$\times$3.2 Gb/s). The system would saturate only at \SI{200}{\kilo\hertz} of Pb--Pb collisions.
A detailed description of the components and operating modes of the readout units can be found in Ref.~\cite{Schambach:2018}.

\subsubsection{The powering system}
The ITS ALPIDE sensors require 1.8~V analogue and digital power rails; a reverse bias can be applied to the sensor substrate. Power regulation happens through linear regulators mounted on custom power boards, which are housed in the same rack as the corresponding readout unit. The power boards have built-in I2C~\cite{Philips:1981} interconnection to monitor their functional parameters in real time, including the sourced currents and voltages. The readout unit has full control over the power board through the I2C bus: power sequence, monitoring and tuning is therefore managed from the counting room through the same Versatile Link used to manage the data acquisition and read the sensors, as shown in Fig.~\ref{fig:RU_diagram}.

As described in Sec.~\ref{sec:its:stavemodules}, staves are connected to the powering system via aluminium power buses. The inner barrel staves have the main power conductors integrated in the FPC that also carries the signal and control lines, while a separate power bus and bias bus are required for each half stave of the middle and outer layers.
In the inner staves all nine sensors share the same power bus, while in the middle and outer layers the power delivery path is (half-)separate for each HIC (see Fig.~\ref{fig:fig7}). The power boards can drive up to 16 analogue and digital power rails, providing supply for a full outer barrel stave, which is composed of 14 HICs. 
The main power supplies are CAEN mainframes (EasyCrates) populated with 61 A3009B radiation tolerant CAEN power modules complemented by 4 CAEN A2518 LV modules for the reverse substrate bias located in the racks in the cavern and in the counting rooms, respectively. 

The power board consists of two power units, which contain 16 low voltage and 8 reverse-substrate bias channels. The two power units feature independent I2C control interfaces and output connectors. They are based on radiation tolerant LDO regulators, shunt resistors, overcurrent protection circuitry, current and voltage measuring circuitry, and remote voltage-setting circuitry.

A power unit can power an inner-barrel stave, a middle-layer stave or an outer-layer half-stave. In order to allow every readout unit to control the power units of the attached detector segment, channels have to stay unused. A total of 142 power boards are used to supply the entire ITS2.

\subsubsection{Component production, detector assembly, and commissioning on surface}
\begin{figure}[h!]
\centering
\includegraphics[width=1.0\textwidth]{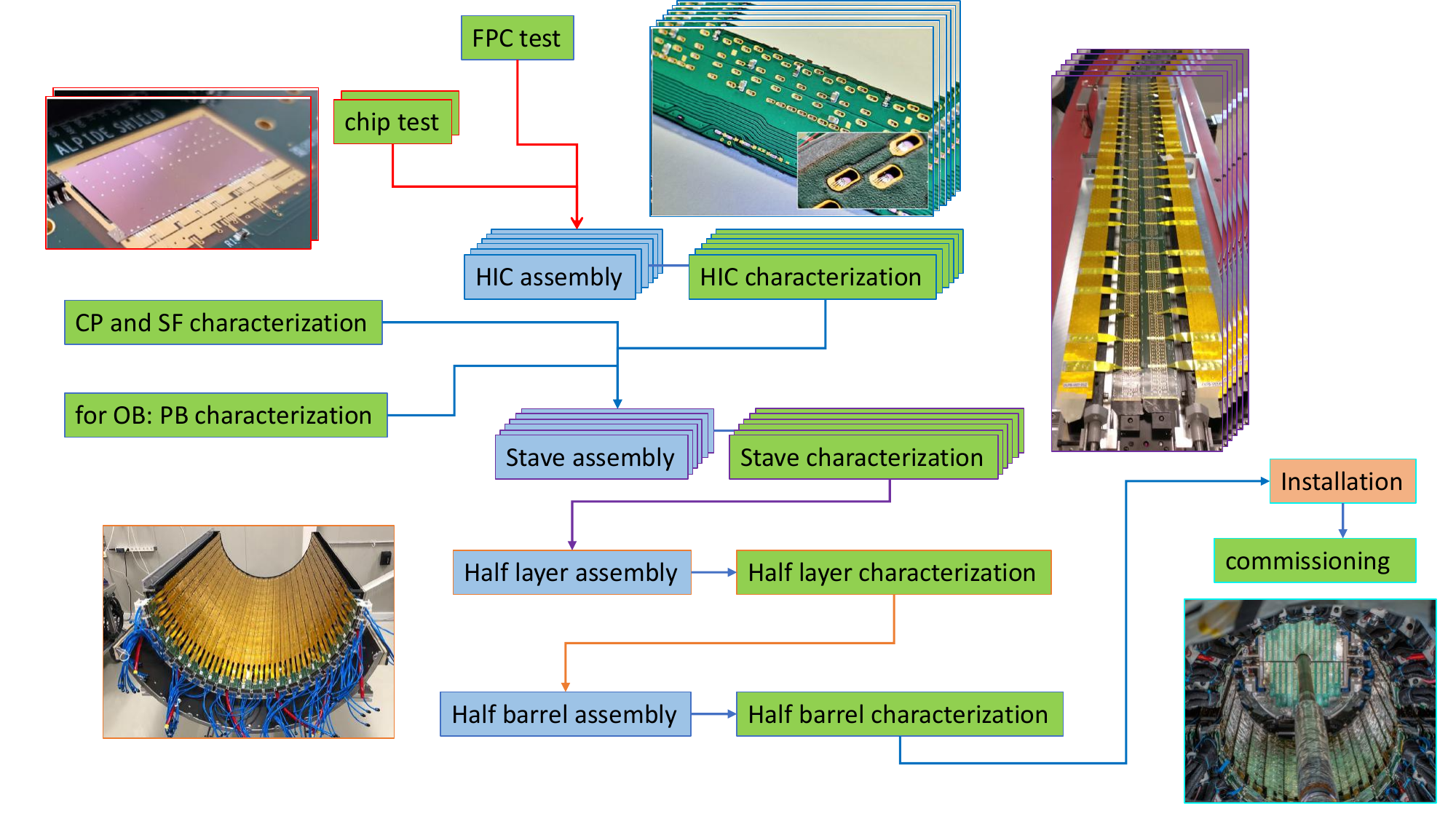}
\caption{Schematic description of the stave production workflow.}
\label{fig:Production_diagram}
\end{figure} 

The assembly of HICs and staves, and their characterization was distributed over 13 sites spread across North America, Europe and Asia. 
The workflow is schematically represented in Fig.~\ref{fig:Production_diagram}.

\begin{figure}[h!]
\centering
\includegraphics[width=1.0\textwidth]{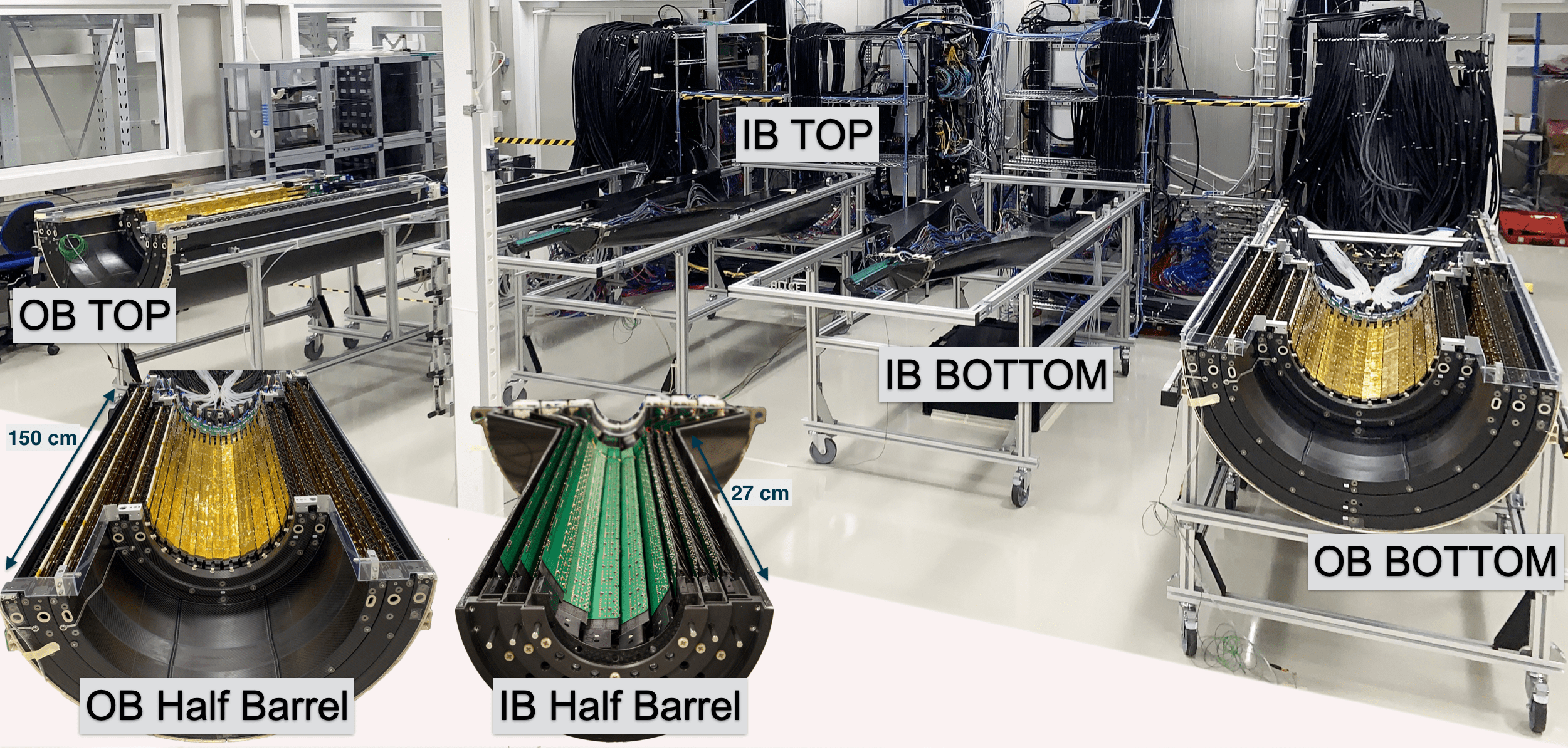}
\caption[ITS2 in the clean room during on-surface commissioning]{ITS2 in the clean room during on-surface commissioning. The lower left shows a zoomed-in view of the half barrels of the outer barrel (OB) and inner barrel (IB) type.
}
\label{fig:fig18}
\end{figure}

The produced staves were transported to CERN where a large clean room was built to allow assembly of the full detector as well as on-surface commissioning activities before the installation in the ALICE cavern. 
Staves were assembled into layers by mounting them on the support structures. The layers were assembled into four separate half-barrels, two for the inner and outer barrels each. Each half-barrel was then connected to the readout units, power boards, and the cooling system. 
Throughout the year 2020, the full detector was under commissioning on the surface, with the half-barrels located next to each other to facilitate stepwise integration (Fig.~\ref{fig:fig18}).

\begin{figure}
  \centering
  \includegraphics[width=0.7\textwidth]{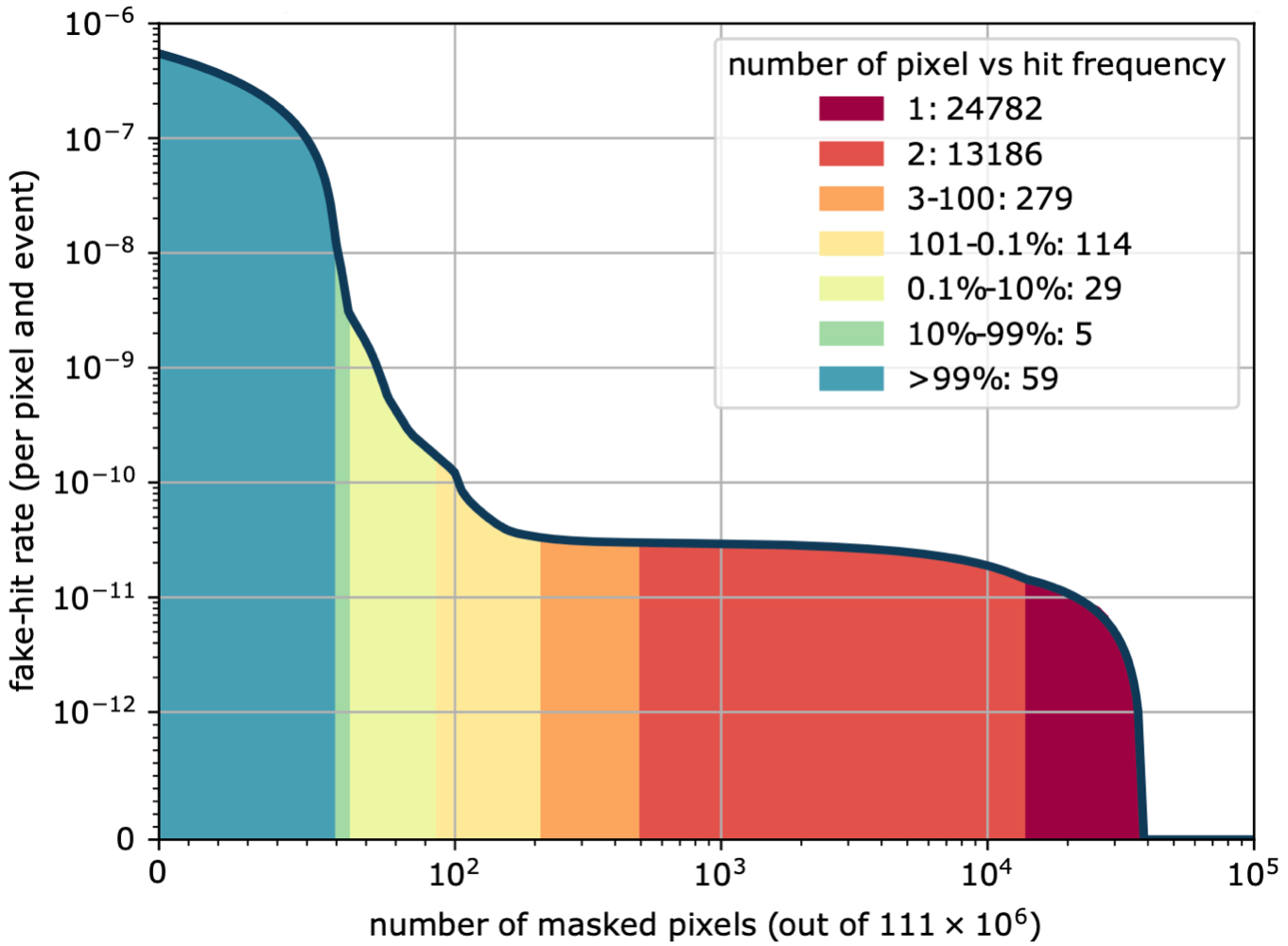}
  \caption[The fake-hit rate of an inner half-barrel]{Fake-hit rate of an inner half-barrel as a function of the number of masked pixels. \label{fig:IBfhr}}
\end{figure}

During commissioning, the detection efficiency, readout performance, and noise levels were characterised.
As an example of the results obtained during the commissioning, the fake-hit rate measured for one inner-barrel half barrel is reported here. Randomly distributed triggers were sent to all the chips in the layer and the generated hits were registered. These hits are due to noise as well as to cosmic rays crossing the detector in coincidence with the trigger. In the inner barrel, the fake-hit rate was found to be dominated by roughly hundred pixels per half-barrel as shown in Fig.~\ref{fig:IBfhr}~\cite{Reidt:2021Vertex}, where colors indicate how often a pixel fired in \SI{15e6}{events} acquired at a trigger rate of \SI{50}{kHz} using a charge threshold of \SI{100}{\elementarycharge^{-}}, e.g. there were 24782~pixels which fired once in the sample. Masking these pixels leads to a fake-hit rate of \SI{1e-10}{/ pixel \per event}. This is significantly better than the target value of \SI{1e-6}{\per pixel \per event }~\cite{ITS-TDR}. The majority of pixels which remain after this masking show one or two hits, which is consistent with the expected rate from cosmic muons.

\subsubsection{Detector calibration}

The calibration procedure for the ITS2 consists of two main steps, namely the identification and masking of noisy pixels and the optimization of the in-pixel discriminator thresholds. 
The calibration has to be performed on a regular basis both to determine the best operating point of the detector and to measure its performance for the chosen operating point at regular intervals. The majority of the calibration scans is based on the injection of analogue or digital pulses into the single pixels, most notably the measurements of the pixel thresholds and their tuning to optimal values.
The scan is executed completely by the DCS software, which controls sequencers on the readout units that trigger pulses to the chip.
The main challenge in the threshold calibration is related to the high data rate generated by 12.5 billion channels.  A threshold scan of the full detector with 50 charge injection points and 50 hits per point results in about 3 $\times 10^{13}$ hits or approximately 100 TB of raw hit data. If the scan is performed as fast as the on-detector bandwidth allows, this data will be collected in slightly less than 1 hour, resulting in a data rate of 20--30 GB/s. However, a full scan is generally used as a reference and it is not needed on a daily basis. In fact, to ensure a good threshold calibration of the detector it is sufficient to pulse about 1\% of the pixels; such a scan can be completed in under 5 minutes.

In order to process the larger amount of data in a timely way, the analysis is performed in a distributed manner on the EPNs (see Sec.~\ref{sec:epn}), making use of the ALICE DPL (see Chapter~\ref{chap:readout}). 
Once the pixel thresholds are properly tuned, the next calibration step is performed, namely the detection and masking of the noisiest pixels. In general, the fraction of noisy pixels masked is below 0.1 \textperthousand{} which leads to an overall fake-hit rate of about 10$^{-8}$ hits/event/pixel on average, well below the required 10$^{-6}$ hits/event/pixel by design. 

After a successful calibration, the addresses of the pixels to be masked and the ALPIDE register settings for threshold tuning are sent off to be stored in the configuration database using a dedicated WinCC panel in the DCS.

\subsubsection{Installation and global commissioning}
\begin{figure}
  \centering
  \includegraphics[width=0.9\textwidth]{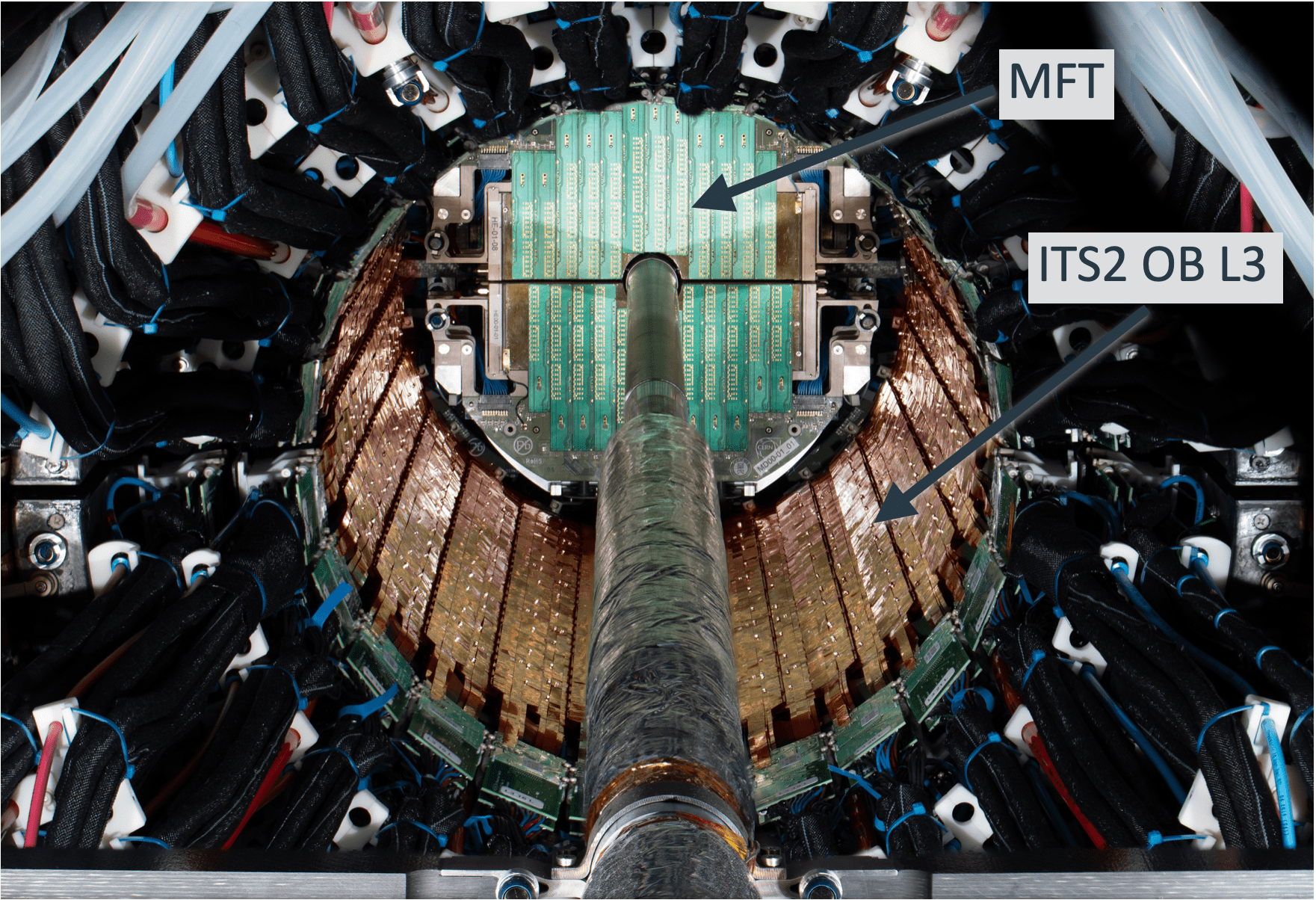}
  \includegraphics[width=0.9\textwidth]{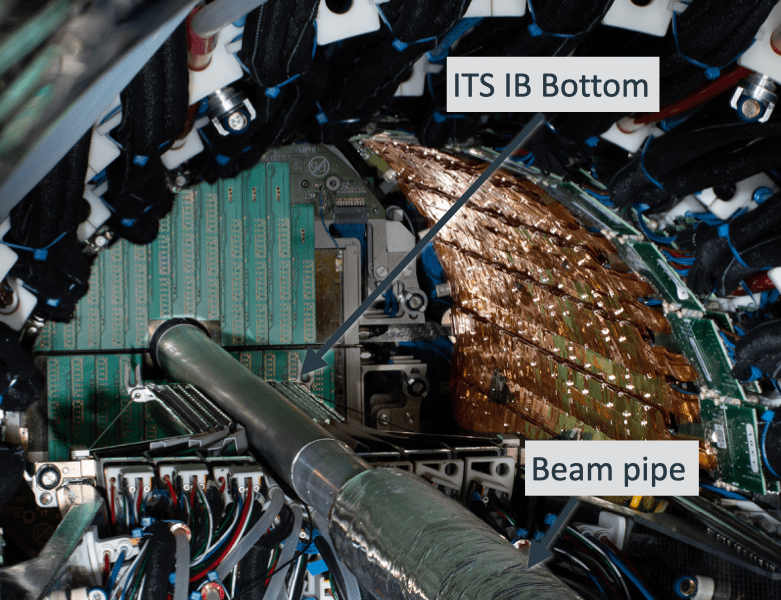}
  \caption[ITS2 installation]{Top: Outer barrel surrounding the beam pipe with the Muon Forward Tracker (MFT) in the background. \label{fig:OBinstallation} Bottom: ITS2 inner barrel bottom half-barrel next to the beam pipe, outer barrel and MFT in the background. \label{fig:IBinstallation}}
\end{figure}

The installation of the ITS2 in ALICE started in January 2021, when the services were transferred from the on-surface commissioning hall to the experimental cavern. 
The detector was installed on rails in the so-called "cage" (see Sec.~\ref{chap:integration}), hosting the beam pipe, ITS2 and MFT inside the ALICE TPC. 
Several insertion tests were performed on the surface to optimise the procedures and to identify potential interferences. For the final installation inside the experimental apparatus, up to six cameras where used to continuously monitor key contact points during insertion, since the clearance between staves of top and bottom barrels is of the order of a millimeter. Furthermore, surveys were carried out to determine the exact position of the detector elements and the beam pipe and visualize them with the help of three-dimensional scans carried out earlier on the surface. At each step of the insertion process, CAD models of the insertion process were compared with the camera images to verify the positioning of the detector elements.
The process started in March 2021, with the outer barrel installation. Figure~\ref{fig:OBinstallation} (top) shows layer 3 in position in ALICE, around the beam pipe. The visible surface is covered by the power distribution bus. After the installation, the outer barrel was thoroughly tested before the inner barrel was inserted into its final position in close vicinity to the beam pipe, in May 2021. Figure~\ref{fig:IBinstallation} (bottom) shows the bottom half of the inner barrel in its final position at around \SI{1}{mm} distance from the beam pipe.
 The full detector was installed without damaging any component. 
 \subsubsection{First results from global commissioning}
 After connection and verification of the detector and its services, the focus was set on the central system integration and on gaining experience with the final framework for the operation of the detector.
 \begin{figure}
  \centering
  \includegraphics[width=0.7\textwidth]{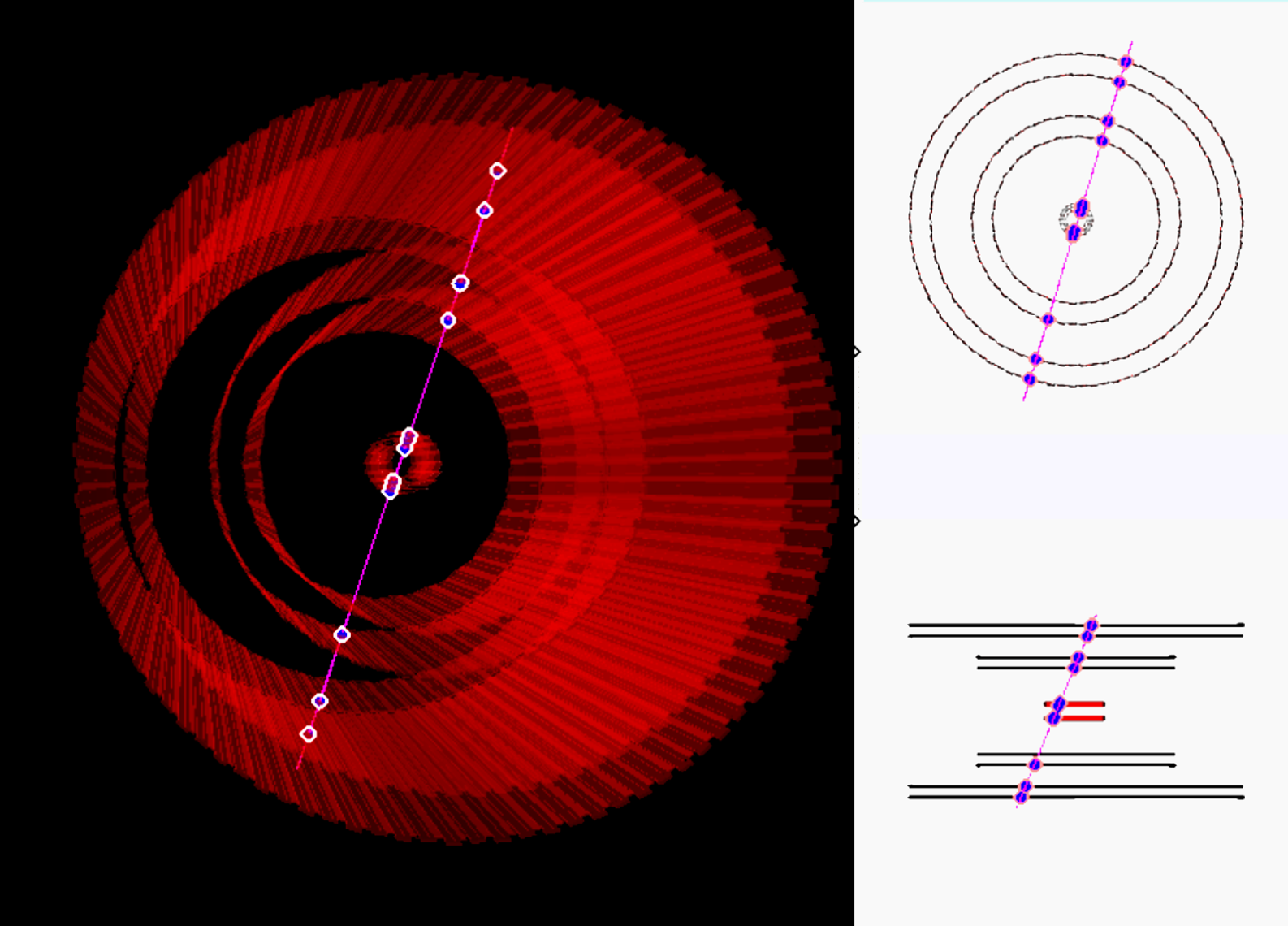}
  \caption[Cosmic muons in ITS2]{Event display of a cosmic muon traversing all layers of ITS2 twice, no magnetic field.\label{fig:ITScosmic}}
\end{figure}
First cosmic muon tracks traversing the full detector, like the one shown in Fig.~\ref{fig:ITScosmic}, could be acquired using continuous integration without a dedicated trigger signal. These tracks were found by matching three hit points in the layers 4 to 6 of the outer barrel and requiring another hit point in the inner barrel, at a rate of about \SI{0.02}{Hz}, while those traversing only the outer barrel were more frequent (\SI{0.5}{Hz}).
The commissioning campaign continued throughout the year and allowed to optimise the detector control system (DCS), the calibration procedures, and readout parameters. 
During the pilot beams with pp collisions at injection energy (\SI{450}{\giga\eV}) in October 2021, reconstructed tracks from the ITS2 were used to determine the position of the primary vertex, as shown in Fig.~\ref{fig:ITSvertex}, and continuously monitor it online in the QC plots. 
\begin{figure}
  \centering
  \includegraphics[width=0.7
  \textwidth]{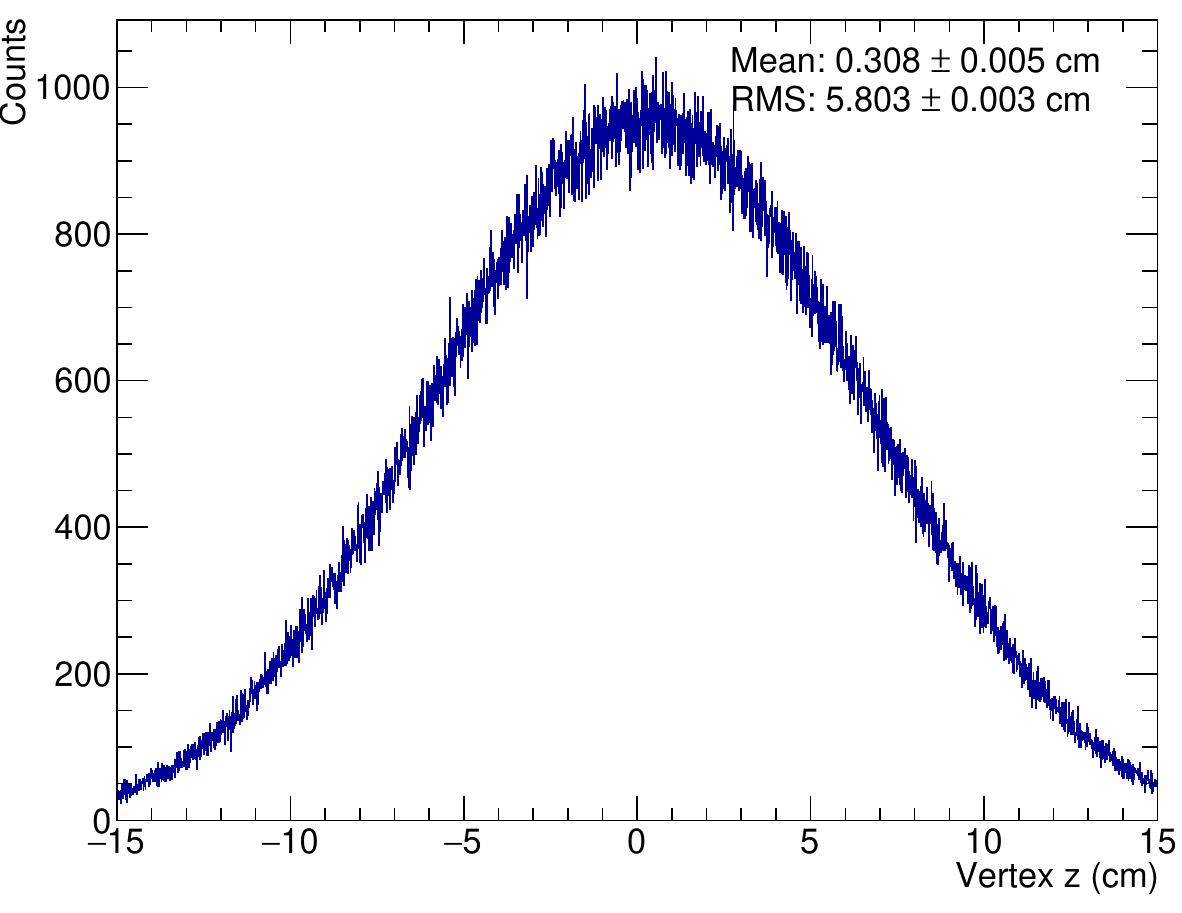}
  \caption[Primary vertices reconstructed by ITS2]{Longitudinal distribution of the primary vertex positions from ITS2 tracks reconstructed online during the LHC pilot beam in October 2021.  \label{fig:ITSvertex}~}
\end{figure}
The primary vertex position is one example of the quantities that are continuously monitored by the online quality control system of the O$^2$ analysis framework to assess the quality of the data acquired with the ITS2.

\clearpage
\subsection{Muon Forward Tracker}
\begin{figure}
\centering
\includegraphics[height=4.5cm]{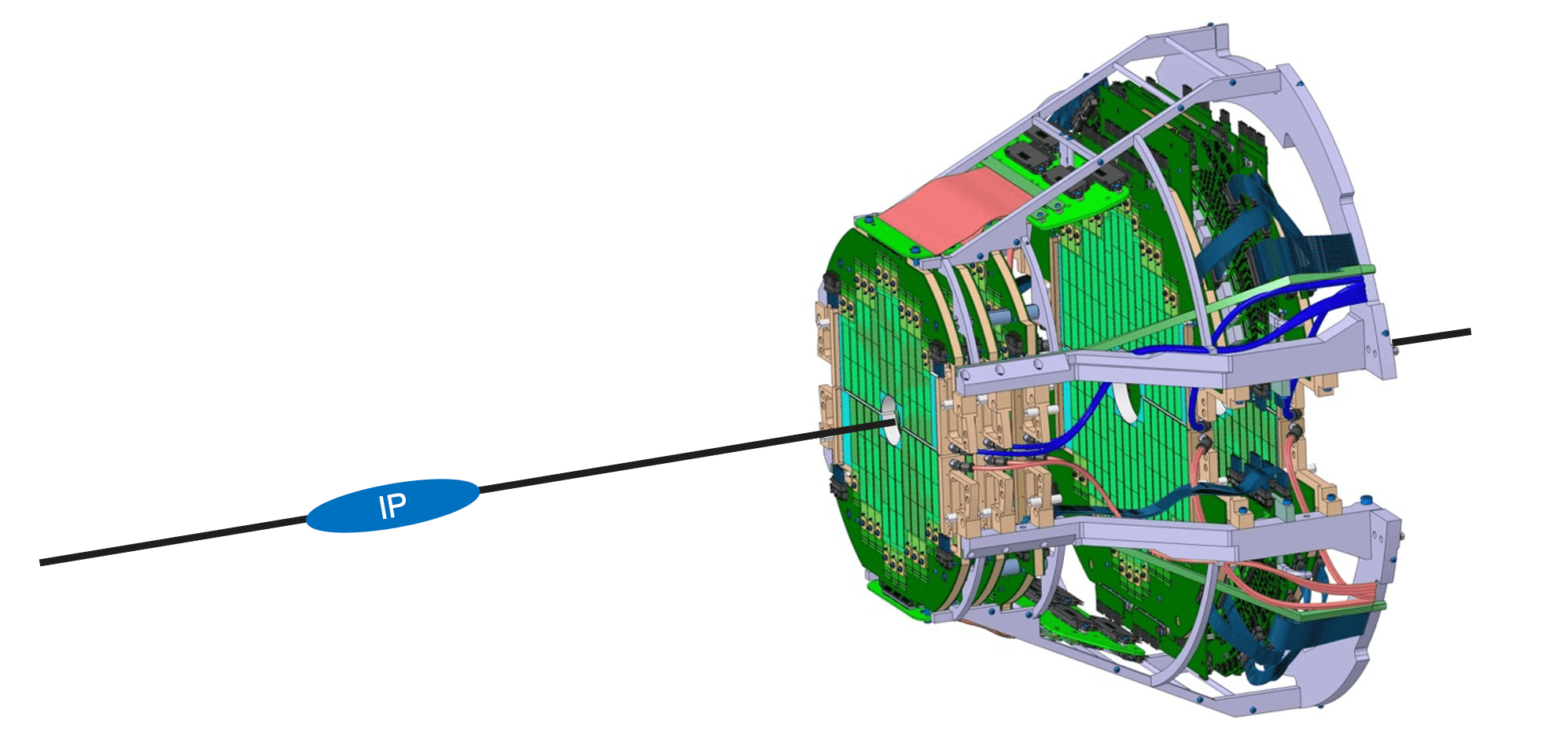}
\hfill
\includegraphics[height=4.5cm]{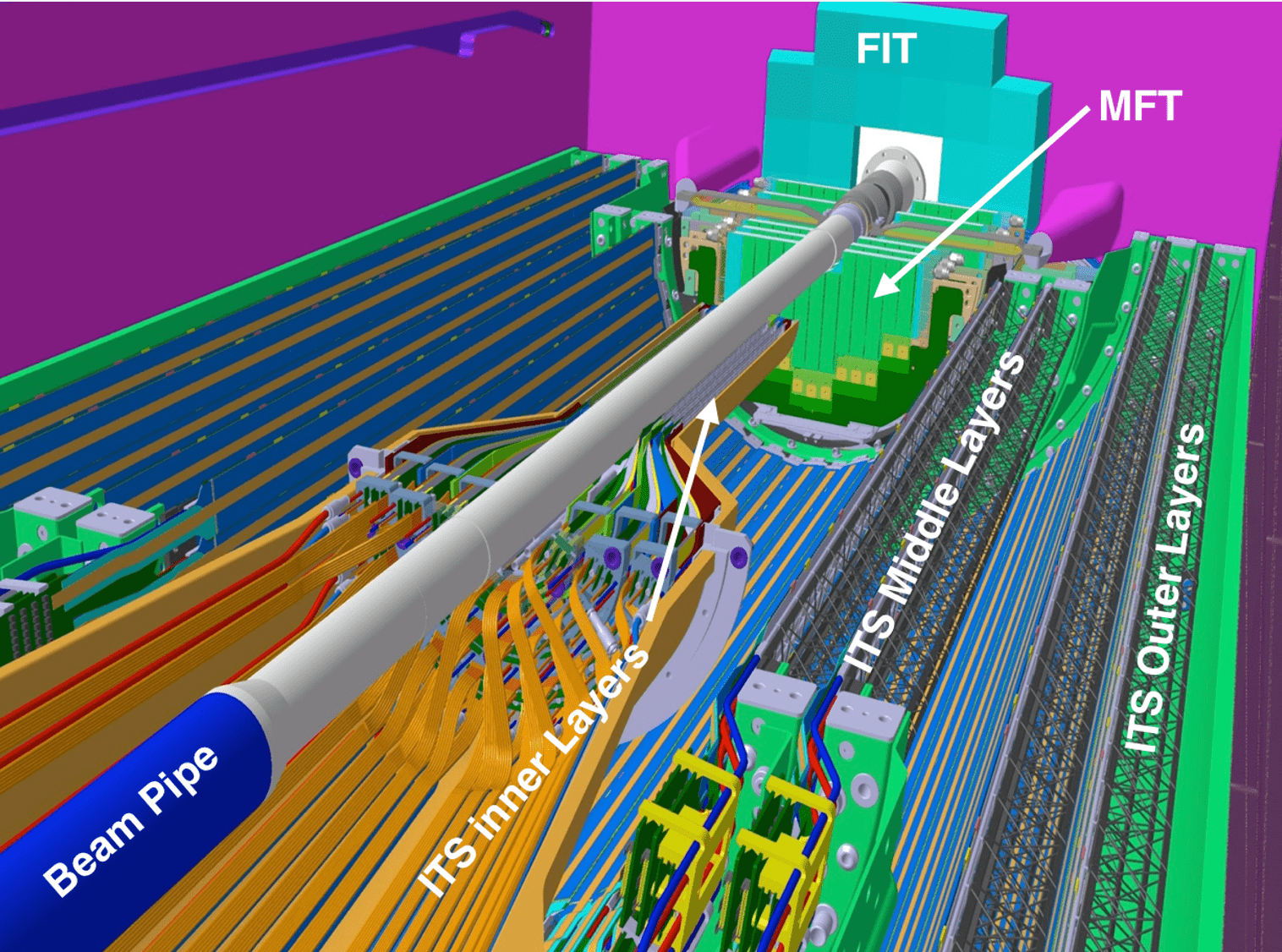}
\caption[Overview of Muon Forward Tracker]{Schematic view of the Muon Forward Tracker (left) and its integration with the central barrel (right).}
\label{mft_fig_environment}
\label{mft_fig_layout}
\end{figure}

The Muon Forward Tracker detector (MFT), see Fig.~\ref{mft_fig_environment}, is a high position resolution silicon detector, which has been designed to extend the physics program of the muon spectrometer (see Sec.~\ref{sec:muon}).
Its primary goal is to improve the pointing resolution of muons by matching the tracks reconstructed downstream of the hadron absorber to those reconstructed inside the MFT upstream of the absorber~\cite{mft:1981898}.
This approach allows the removal of multiple scattering effects in the hadron absorber and improves the pointing resolution of muon tracks down to about \SI{100}{\um}.
The MFT is located between the interaction point and the front absorber and surrounds the beam pipe at the closest possible distance.
It provides charged particle tracking in the pseudorapidity interval $-3.6<\eta<-2.45$, which covers most of the muon spectrometer acceptance.
The acceptance boundaries are defined on one side by the size of the beam pipe, and on the other side by the volume and position of the ITS2, the FIT-C and the beam pipe support, as shown in Fig.~\ref{mft_fig_environment}.

\subsubsection{Detector layout}

The MFT has a projective geometry (see Fig.~\ref{mft_fig_layout}) based on five disks, coaxial with the beam pipe and labelled D00 (innermost) to D04 (outermost), the first two (D00 and D01) being identical and the others (D02, D03 and D04) having their diameters increasing with the distance from the interaction point. To ease assembly and insertion, the detector is divided into two identical halves, labelled H0 for the bottom part and H1 for the top part.
The MFT is composed of a total of 936 ALPIDE silicon sensors (see Sec.~\ref{alpide:subsection:alpide}) distributed on both sides faces of the ten half-disks, and arranged in detection modules called ladders.
Each ladder is a hybrid integrated circuit with two to five sensors (depending on the position within the disk), which are glued and interconnected on a flexible printed circuit (FPC) board to provide the power and readout connections. In order to minimize the material budget, the silicon-pixel sensors constituting the MFT are thinned down to the same thickness of \SI{50}{\um} as the ITS2 inner barrel sensors (see Sec.~\ref{sec:its:intro}), and the FPC to which they are connected are made of polyamide with two layers of aluminum on either side. Each ladder is connected to a PCB that is located outside the acceptance, external to each half face.
The MFT contains 240 ladders whose positions were defined to ensure an 85\% overlap of the sensors between the two faces of each disk. The face of the half-disks is subdivided into four zones (each containing between three and five ladders) which yields a total of 80 zones for the full MFT.

\begin{figure}[h]
\centering
\includegraphics[width=.7\textwidth]{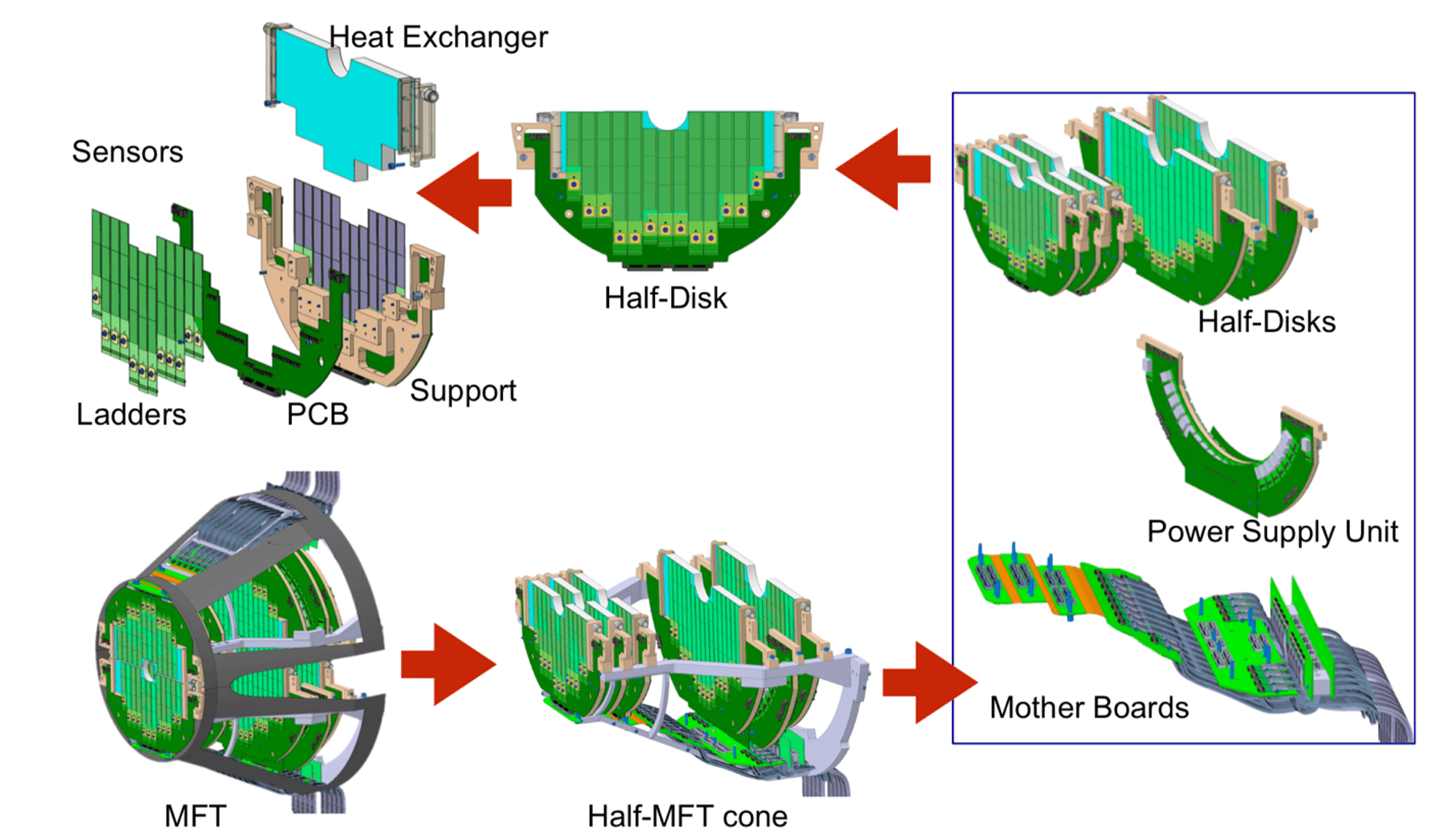}
\caption[MFT detector elements]{Detailed view of the elements composing the MFT detector.}
\label{mft_fig_exploded}
\end{figure}

\begin{figure}[h]
    \begin{minipage}[c]{.46\linewidth}
        \centering
        \includegraphics[width=\textwidth]{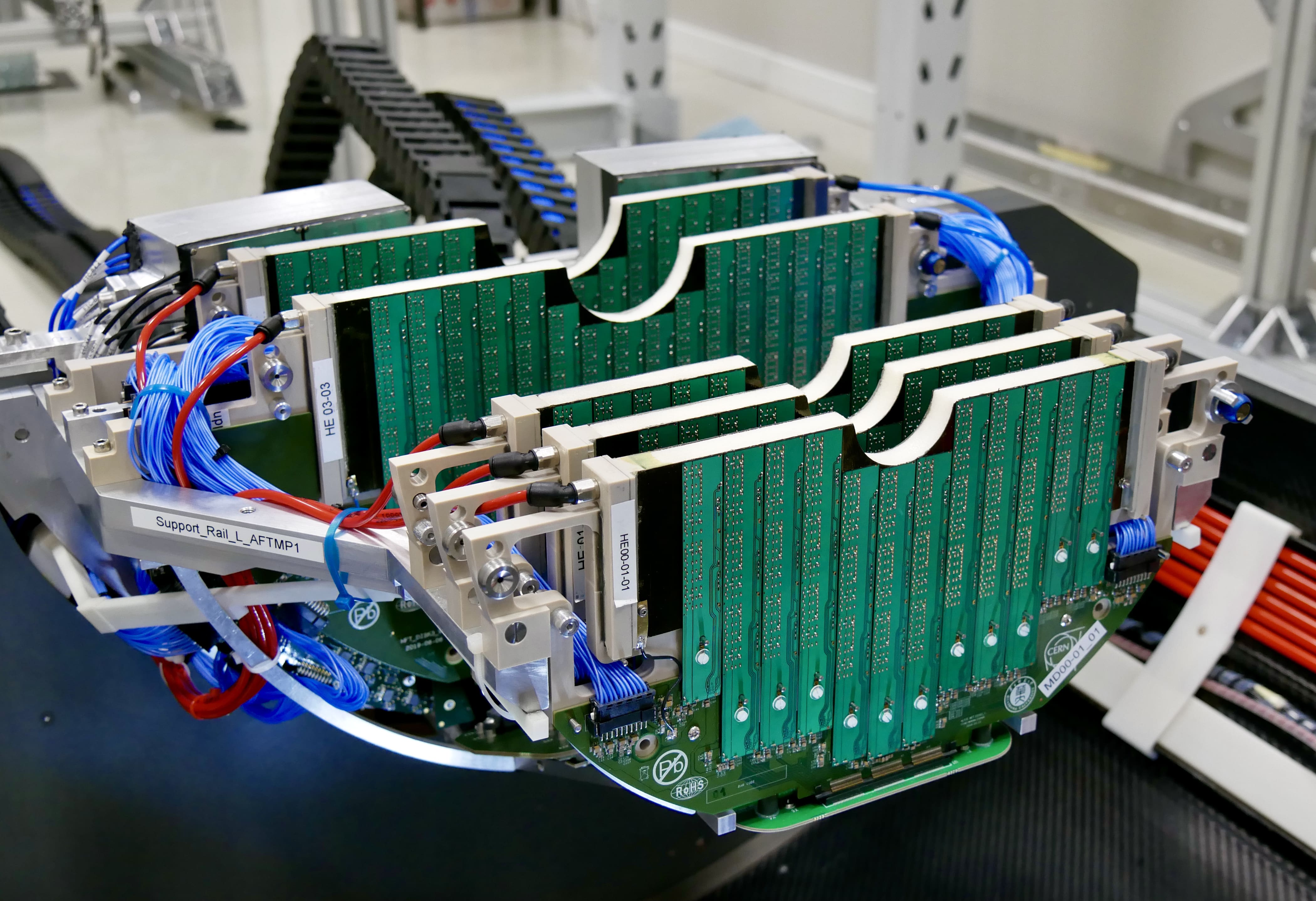}
    \end{minipage}
    \hfill%
    \begin{minipage}[c]{.46\linewidth}
        \centering
        \includegraphics[width=\textwidth]{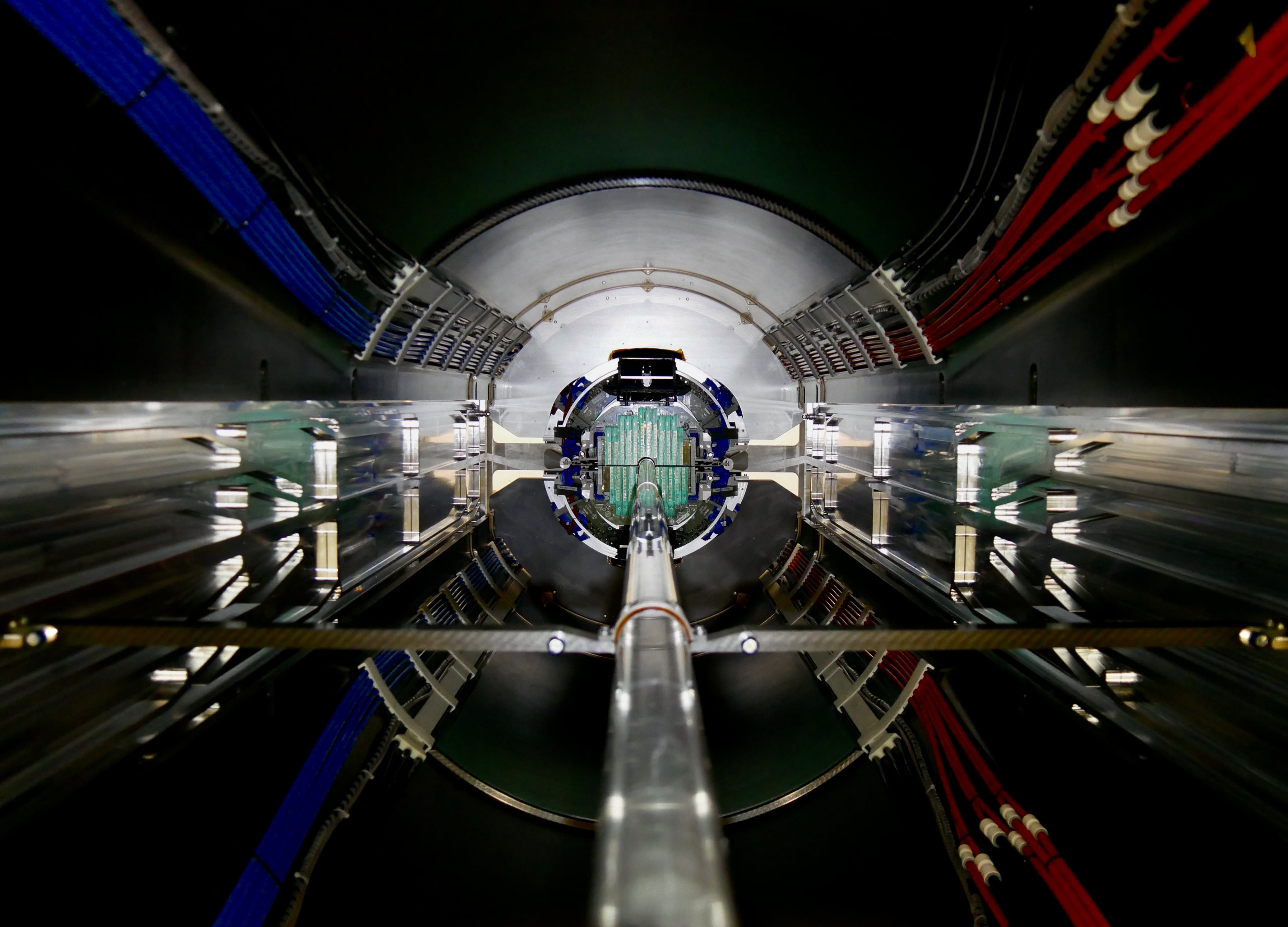}
    \end{minipage}
    \caption[FIT installation]{Left-hand panel: fully assembled half-cone of the MFT with patch-panel and the FIT-C detector. Right-hand panel: the MFT in its final position; the cooling and power services can be seen along the half-barrels.}
    \label{mft_installed}
\end{figure}

The ten half-disks are then assembled into half-cones.
The first three half-disks are connected to a set of motherboards that provide the connection of the readout lines with \SI{6.5}{\metre} long copper cables, which run alongside the whole absorber towards the front-end electronics boards.
For the two larger half-disks, the same type of copper cables is used, connected directly to the PCBs.
Each half-cone also houses a Power Supply Unit (PSU), which
controls and monitors the powering of the zones to guarantee the ladder safety
and is located outside the acceptance between the last two half-disks.
The disks and the PSU are water-cooled and air ventilation is used to ensure temperature homogeneity inside the confined space where the MFT is mounted. Figure~\ref{mft_fig_exploded} shows an exploded view of the different elements composing the detector.
The two half-cones are fixed to two end-cap patch-panels which in turn are fixed to large carbon fibre composite structures, called half-barrels, that are used to insert and position the MFT within the ALICE internal cage, see Sec.~\ref{sec:fit}. The services are routed along the half-barrels and through the patch-panels to reach the detector. The patch-panels are mechanical pieces used also to support the FIT-C detector and to interconnect the readout cables from the half-cones to the readout units, which are located \SI{6}{\meter} away beside the front absorber.
Figure~\ref{mft_installed} shows a fully assembled half-cone and the MFT in its final position.

\subsubsection{Ladder assembly and testing}

The basic element of the MFT detector, called ladder, is composed of an aluminum FPC on which silicon-pixels sensors are glued and interconnected.
The length of the ladders varies from 2 to 5 chips each to match the size of the half-layers.
Each FPC is equipped on one side with footprints for placing the sensors and a 70-pin connector for powering the chips and transmitting the high-speed readout signals, and on the other side with microelectronic components (resistors and capacitors) that decouple the power supply (analog and digital) of the sensors and adapt the impedance of the data lines.
Given their variable length, the FPC design was optimized in order to reduce the maximum voltage drop to 100~mV and to ensure the best transmission of the high-speed data lines.

\begin{figure}[h]
\centering
\includegraphics[width=.6\textwidth]{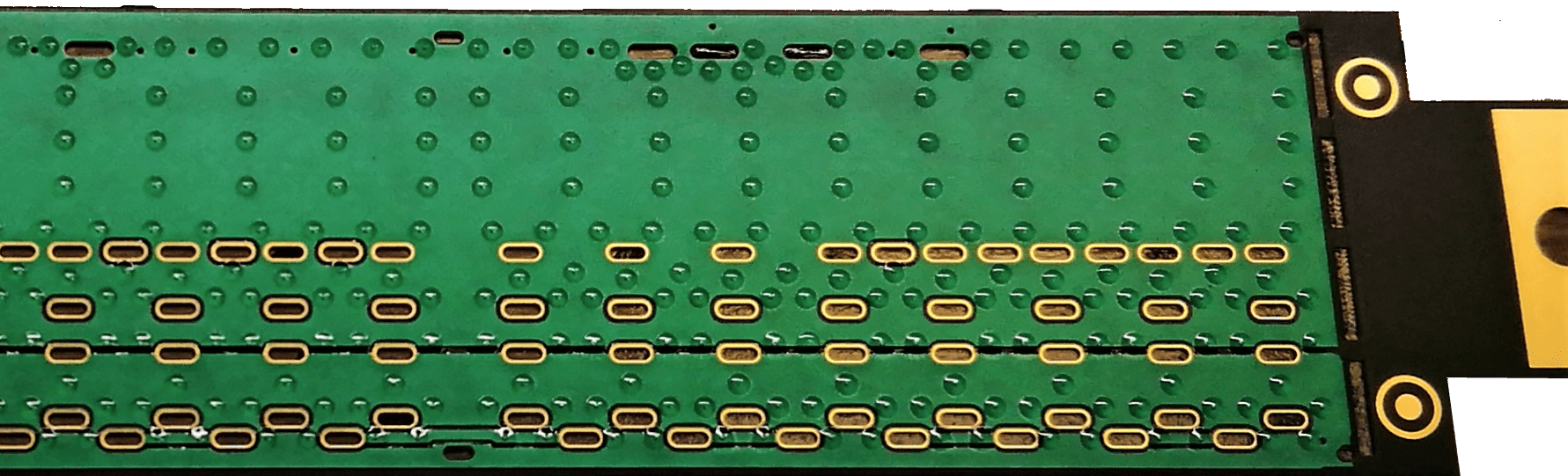}
\includegraphics[width=.6\textwidth]{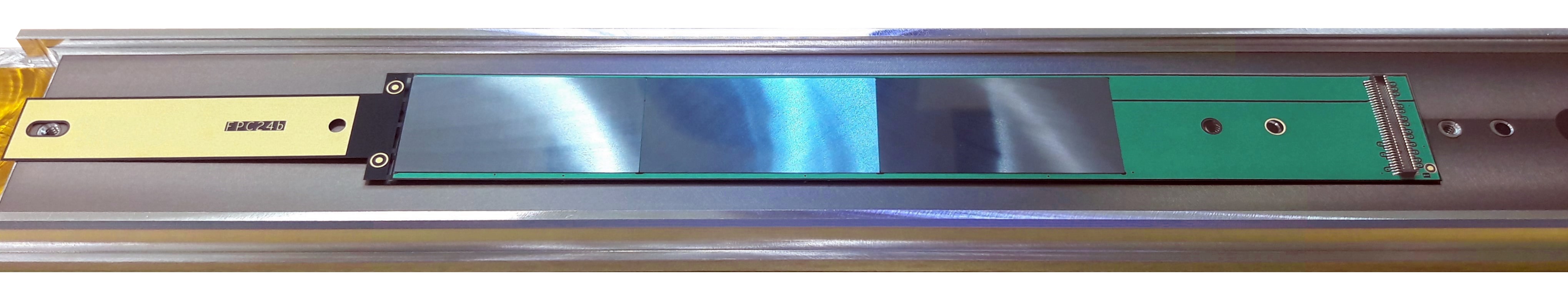}
\caption[MFT ladder]{Example of an assembled  MFT ladder. Upper picture: back side of the FPC with the glue spots on which the sensor are glued. Bottom picture: front view of assembled ladder.}
\label{fig:mft_ladder}
\end{figure}

The ladder assembly took place in the clean room of the CERN EP/DT Departmental Silicon Facility using a three-axis digitally controlled placement machine, called ALICIA (ALICE Integrated Circuit Inspection and Assembly machine), which places the chips with a precision of \SI{5}{\um} on a specially machined stainless steel support with a lattice of very small holes to hold the chips in position by suction. At the same time, the FPC is positioned on a suction support which also keeps it perfectly flat and positioned with a precision better than \SI{300}{\um}.
Small dots of Araldite-type two-component glue are applied to the FPC using a stainless-steel stencil with conical holes.
The FPC is flipped and positioned opposite to the chips with the help of precision centering pins. The weight of the FPC carrier provides sufficient pressure to spread out the glue and a spacer between the FPC and the chips ensures that the final thickness of the glue layer is around \SI{50}{\um}. After curing of the glue, the assembly is removed from ALICIA, visually inspected and brought to the CERN bond-lab, where the connection between the FPC and the chips is realized by ultrasonic micro-wire bonding on the metallized pads of the FPC.
Each micro-interconnection consists of 3 wires with \SI{25}{\um} diameter that pass through the vias of the FPC to be connected to the 74 pads on the surface of the sensors. The assembly of the ladder is then completed and a final visual inspection verifies the quality of the interconnections. Figure~\ref{fig:mft_ladder} shows a picture of an assembled ladder.

Once the ladder is assembled, it is transferred to the MFT laboratory where it undergoes a battery of tests to qualify its performance. In the laboratory, two test benches allow each ladder to be qualified from an electrical and functional point of view. First of all, the ladder is gradually powered with analog and digital voltage and its power consumption is checked. Then, the ladder is connected to an acquisition system developed specifically for the ITS2 and the MFT projects, which allows its functionality to be tested in terms of electronic noise, number of dead or defective pixels, and transmission speed of the digitized data. Each test is associated with a qualification grade which is a function of the performance measured against the expected specifications and automatically determined by the qualification software. At the end of these tests, the ladder is qualified according to four grades:
\begin{itemize}
\item ``Gold" for a ladder that works perfectly and whose pixels and circuitry respond exactly as expected.
\item ``Silver" when the number of pixels which do not respond correctly is between $0.1$ and $4$\textperthousand{} and the ladder can be used without any problem.
\item ``Bronze" when the number of defective pixels is between $0.4$ and $1$\% and the ladder, although functional, is used as spare rather than for equipping the MFT detector.
\item ``non-compliant" when a ladder does not pass the tests because of, for example, damaged chips, a defect of the FPC or improper handling. This ladder is discarded.
\end{itemize}

To fully equip the MFT (including one additional half detector and $20$\% spares modules), 500 ladders were manufactured, tested, qualified and mounted on the disks in one year with a rate of gold and silver qualified ladders of around 91\%.

\subsubsection{Half-Disks}
\begin{figure}
\centering
\includegraphics[width=.6\textwidth]{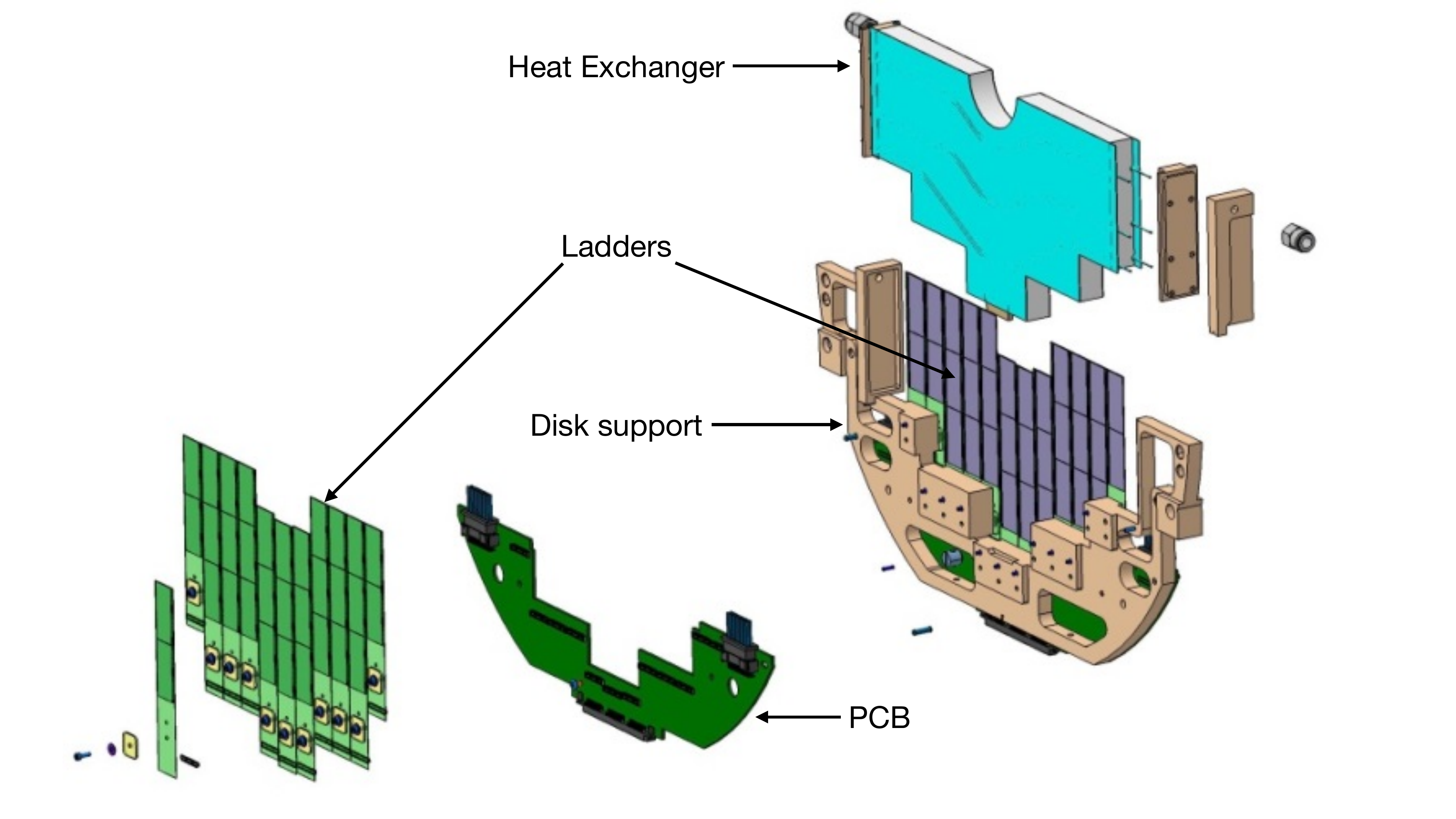}
\caption[Half-disk (exploded view)]{Exploded view of a half-disk (D00-01).}
\label{mfthalfdiskexploded}
\end{figure}

As shown in Fig.~\ref{mfthalfdiskexploded}, each half-disk is composed of a support to which a heat exchanger is glued and two PCBs are screwed. The ladders are glued onto each face of the heat exchanger, screwed to the support, and connected to the PCBs. Four different types of half-disks were designed since the first two disks (D00 and D01) are identical.

The half-disk supports, which form the mechanical interface between the elements of the half-disk and the cone structure, were also designed to ease integration of services (cooling pipes and power cables). To minimise the material budget, the support structures are made of PEEK (PolyEtherEtherKetone plastic).

The ladders are connected to a PCB to route the readout, slow control, and clock signals from the ladders to connectors located at the periphery of the half-disk.
The PCBs of half-disks D00, D01 and D02 are connected to  motherboards that relocate the connection to readout cables for integration reasons. For half-disks D03 and D04, the readout cables are directly connected to the PCBs.
The PCB also distributes the different voltages (analog, digital, and reverse bias) to the ladders. Two connectors are located on the left and right sides, one for zones 0 and 1, and the other for zones 2 and 3.
The PCB is equipped with decoupling capacitors located close to the ladder and power connectors. In addition, a temperature sensor (PT100) allows the measurement of the local temperature and the acquisition of a reference for the temperature information given by the ALPIDE chips. The temperature signal is sent to the PSU on a dedicated line of the power cables.

The heat exchanger was designed to keep the ALPIDE sensors at a temperature below 30$^\circ$C using water cooling and to have a total material budget per half-disk below 0.7\% of a radiation length.
It is composed of two K13D2U carbon fiber cold plates glued to each side of a 14~mm thick core made of Rohacell foam. 
To circulate cooling water under each sensor, 3 or 4 kapton tubes of 1~mm diameter are glued and covered by a carbon fleece. Two manifolds made of PEEK are glued on each side of each heat exchanger to distribute water through the kapton pipes.

The heat exchangers are qualified in several steps. First, the internal structure is inspected using X-ray tomography in order to check the quality of the gluing and the integrity of the cooling pipes. Before closing the second manifold, the water flow exiting each pipe is measured in order to verify that none of the pipes are pinched. Finally, cooling tests are performed by using resistive patches to simulate the heat generated by the sensors. The temperature is monitored with PT100 sensors located on the heat exchanger and a thermal camera. The goal is to check the homogeneity of the cooling.

\begin{figure}
\centering
\includegraphics[width=.6\textwidth]{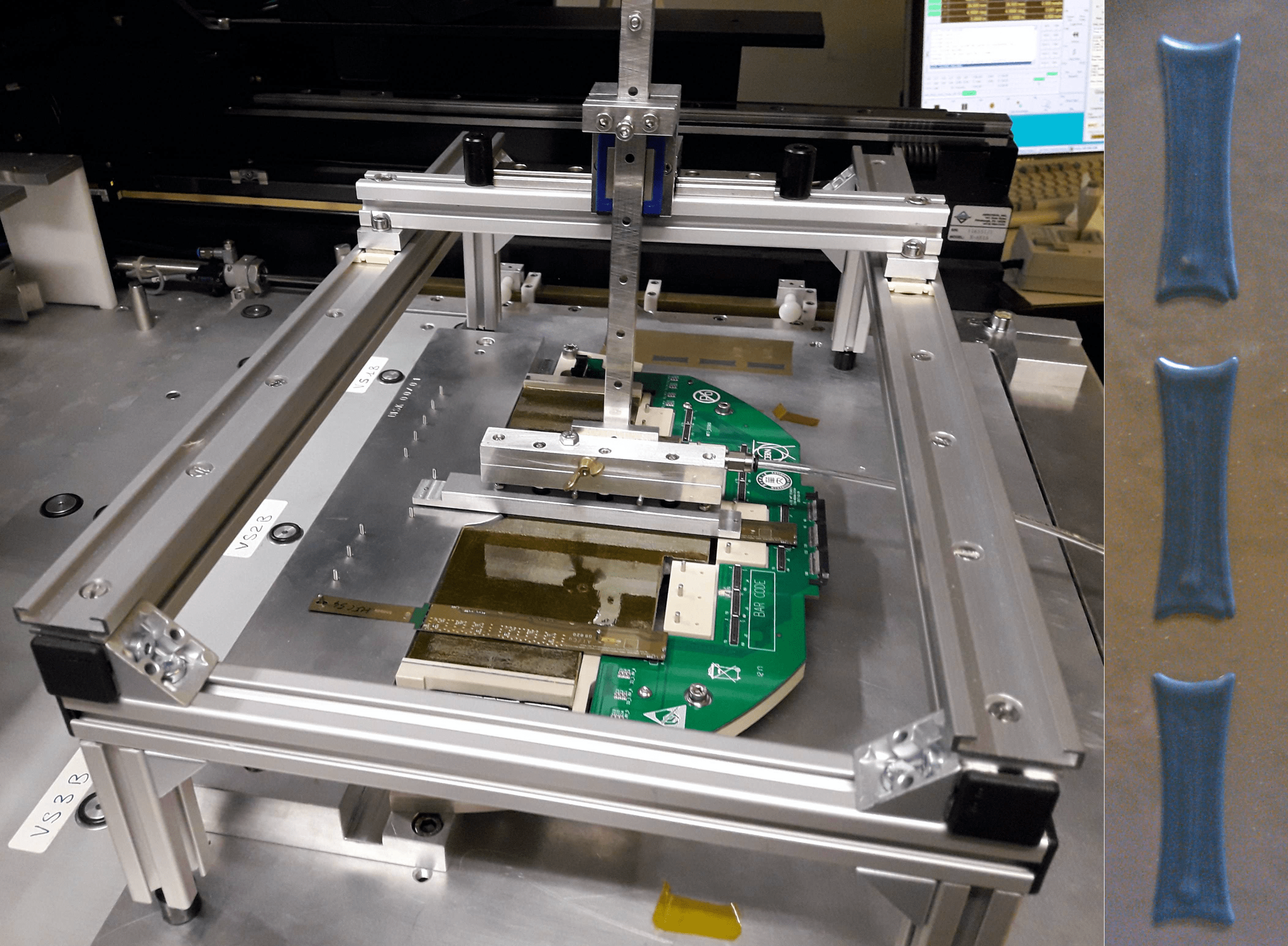}
\caption[MFT disk gluing]{Left: Half-disk during ladder gluing. Right: Glue deposition pattern.}
\label{mftglueondisk}
\end{figure}

The ladders are glued to half-disks using the Dow Corning SE4445-CV silicon glue. The pattern of glue deposition was studied to avoid any flow outside the area of each sensor (see Fig.~\ref{mftglueondisk}). The planarity of the heat exchanger is around \SI{50}{\um} and the final thickness of the glue is around \SI{50}{\um}. The ladders are positioned using a gantry and plugged into the disk PCB. Electrical and communication tests of the sensors are performed to check their proper functioning. In case of failure in the electrical and functional tests, the ladder can be replaced before the glue is fully cured. The remaining glue is removed and a new ladder can be glued.

\subsubsection{Cone and Barrel}
\begin{figure}
\centering
\includegraphics[width=.6\textwidth]{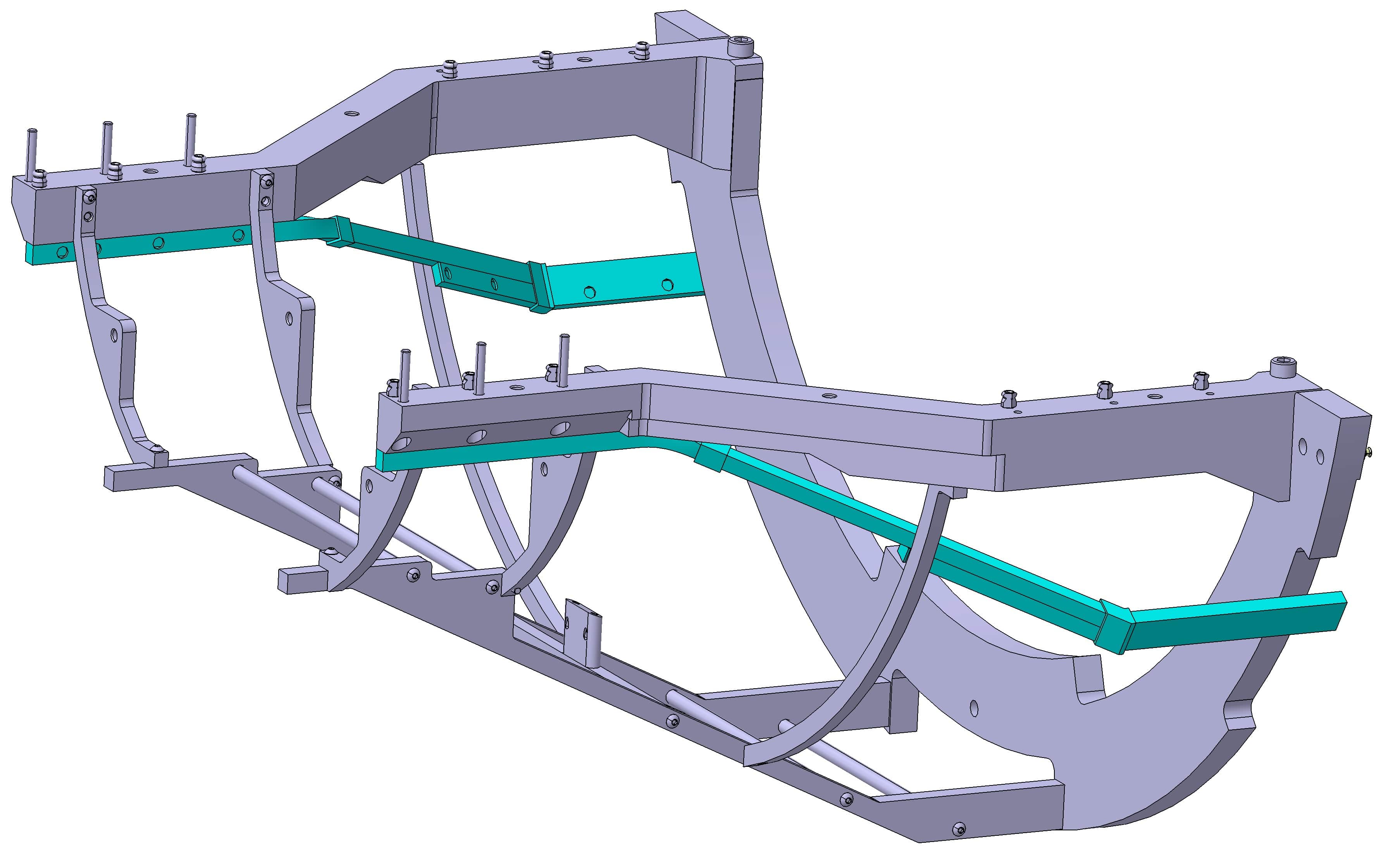}
\caption[Half-cone structure]{Half-cone structure with air ducts in light blue.}
\label{mftConeStructure}
\end{figure}

Each MFT half-cone is supported by a mechanical structure on which three motherboards for the readout are mounted (see Sec.~\ref{MFT-Readout}). The different half-disks and the PSUs (see Sec.~\ref{MFT-PSU}) with their support are mounted on this structure and the different services (readout cables, power cables, and cooling pipes) are routed inside this structure. The environment of the MFT, in particular the presence of the very fragile beam pipe, imposed several constraints on the design of the cone. As the MFT cone is supported from the side close to disk D04, the displacement due to gravity is the largest at disk D00. This displacement has to be kept below \SI{100}{\um} to avoid interference with the beam pipe support flanges which are positioned very close to the detector. The cone support structure was produced from aluminum. In order to homogenise the temperature inside the cone, air guides produced by a 3D-printer were added, see Fig.~\ref{mftConeStructure}. Reference targets can be fixed to the support structure for the purpose of geometry surveys.

Each half-cone is fixed onto a half-barrel which is built from composite material along which the services (power cables, water pipes, and air ducts) are run as shown in the right-hand panel of Fig.~\ref{mft_installed}. On the A-side, a patch-panel (PP2) is mounted to guide these services. On the C-side, the patch-panel has the following functions:
\begin{itemize}
\item position and support a half-cone
\item connect and guide the services from/to the half-cone (this is where the filter boards are fixed, see Sec.~\ref{MFT-PSU})
\item position and support the FIT-C detector.
\end{itemize}
Finally, the half-barrels are equipped with wheels that run on guide rails for the insertion of the half-cones into their final position.

\subsubsection{Services}
\label{MFT-PSU}
\begin{figure}
\centering
\includegraphics[width=.6\textwidth]{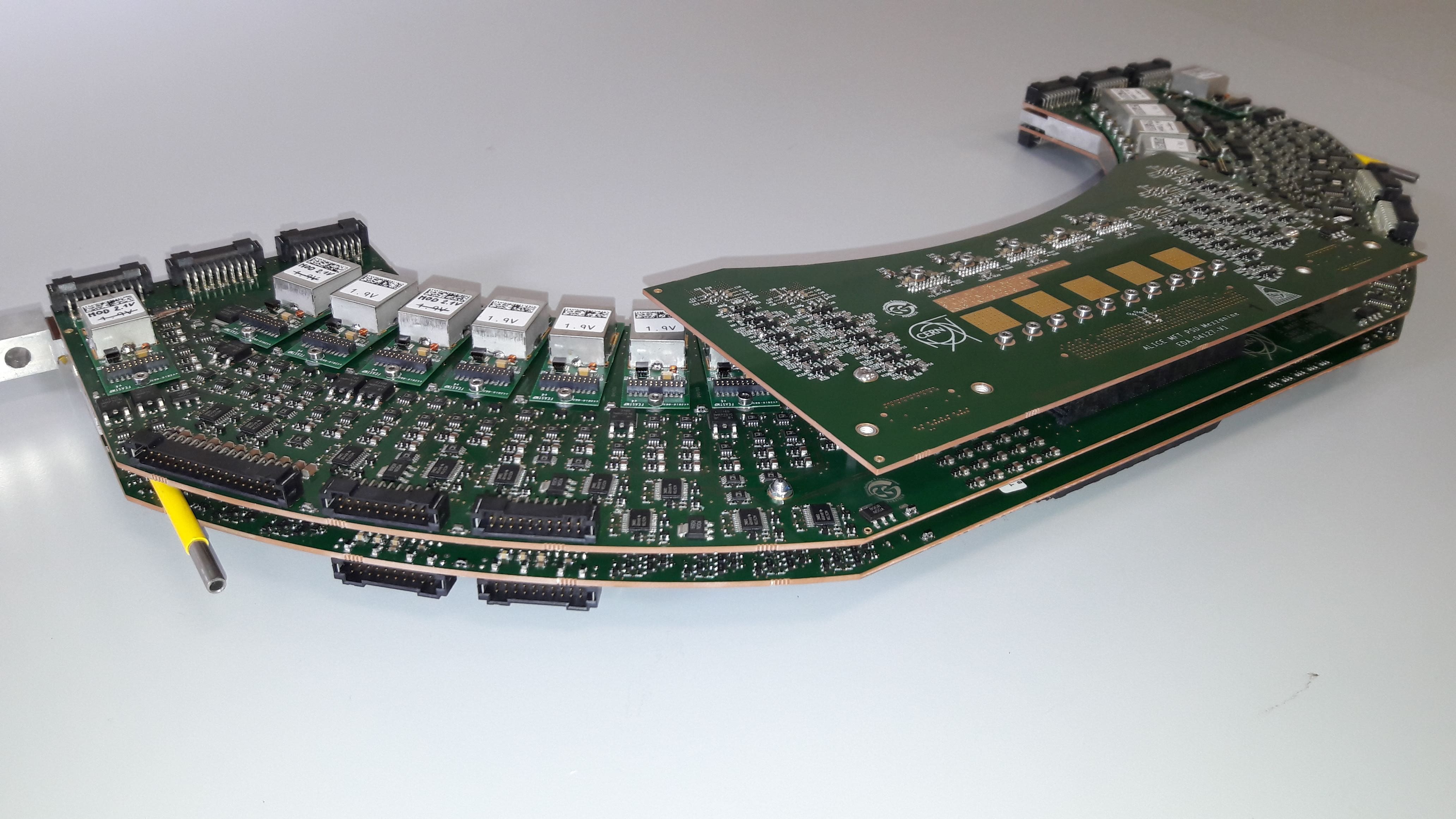}
\caption[MFT PSU boards]{MFT PSU boards with mezzanine assembled on their support.}
\label{mftPSU}
\end{figure}

The ALPIDE sensors require three different voltage supplies (analog, digital and reverse bias) which are locally generated via DC-DC converters in a PSU in order to minimize the material budget in the barrel by reducing the number of copper power lines and moving the fine power distribution as close as possible to the active detector area.
The MFT is equipped with four PSUs, each powering one face of the five half-disks of the same half-MFT.
Each PSU is composed of two boards (see Fig.~\ref{mftPSU}): a main board ensures power distribution from the DC-DC converters and a mezzanine board controls the main board and is equipped with one GBT-SCA to send measurements to the DCS (voltages, currents, temperatures, humidity, and status of zones). The different functionalities of the PSU are:
\begin{itemize}
\item Conversion of the eternal power supply for the analog and digital circuitry of the sensors by FEASTMP-CLP DC-DC converters \footnote{A DC-DC converter is an electronics circuit that converts a continuous current from one voltage level to another.}. One DC-DC converter is used to power two zones of the same half-disk face (0-1 or 2-3) with analog voltage. For digital voltage, one DC-DC converter is also used to power two zones for half-disks 00-01-02  and one zone for half-disks 03 and 04 since they have more sensors per zone and the output current is limited to \SI{4}{\ampere}. The voltage drop from the PSU to the sensors is taken into account and the output voltages of the DC-DC converters are adjusted accordingly.
\item Detection of latch-up events through the measurement of currents per zone (analog, digital, and reverse bias). In case of latch-up, all voltages of the zone are switched off and the information of the line that has generated this event is encoded and transmitted.
\item Communication with DCS is realised by the use of GBT-SCAs to control DC-DC converters, reset zones, adjust reverse bias voltage and thresholds on analog, digital and reverse bias currents. It is also used to send the measurements of voltages, currents, and temperatures of the half-disks, of the inlet and outlet of cooling of the PSU, of the ambient temperature (two sensors on the mezzanine board), and finally of the humidity (sensor located on the mezzanine board). The status of zones and the latch-up information are also sent to DCS.
\item Fail-safe procedure: in case of a loss of communication with the DCS for more than \SI{6}{\second}, all voltages are automatically switched off and the DC-DC converters are disabled.
\end{itemize}
All the PSUs are controlled through a dedicated CRU via four intermediate boards, called PSU-Interfaces (one per PSU), equipped with GBTx ASICs.

\subsubsection{Readout}
\label{MFT-Readout}
The MFT readout is based on the same general architecture used for the ITS2 detector (based on the Readout Unit front-end board, see Sec.~\ref{sec:its:readout}) arranged with an implementation specifically designed for the MFT geometry. In particular, the \SI{50}{\ohm} differential pairs connecting the sensors to the external readout are routed from the disk to the cone section, then to the patch panel, and finally along the sides of the ALICE front absorber to the readout crates. This requires a specific design of the routing elements and cables in order to provide a reliable connection quality, in particular for the high speed data link at \SI{1.2}{Gb/s} used for data transmission over a distance of several meters.
The design and development of the MFT readout architecture contains a larger number of connections and passive elements than the ITS2, due to constraints on the number of cables that can be routed through the barrel detectors. All signals are instead routed through the  limited space between the absorber and the TPC, requiring a sequence of passive dispatching elements, each one inducing specific constraints in terms of impedance adaption, which is a crucial parameter for the high-speed signal transmission quality.

\subsubsection{Detector commissioning}
\label{sec:mft_commissioning}
\begin{figure}
\centering
\includegraphics[width=.6\textwidth]{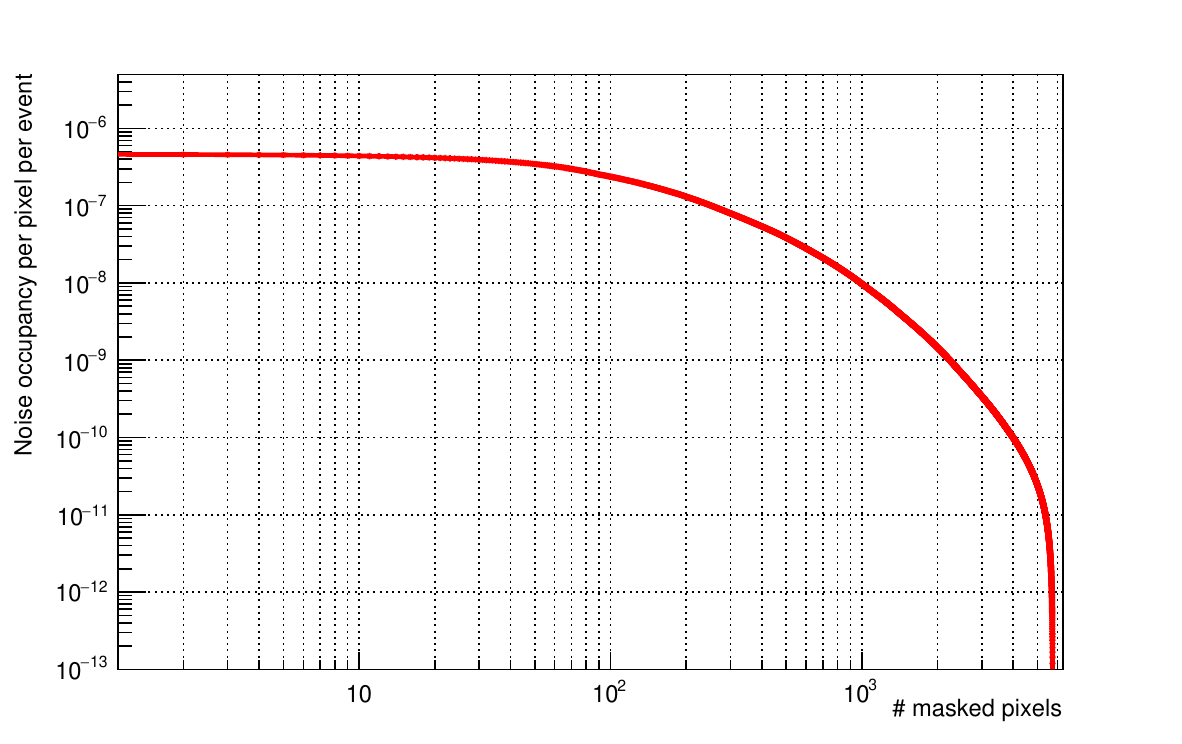}
\caption[MFT sensors noise rate]{Noise occupancy, measured in hit/pixel/event as a function of the number of masked pixels over the whole MFT detector.}
\label{mft_noise}
\end{figure}

The two halves of the MFT detector, together with a third spare half, were assembled by the end of 2019. Over almost one year, the detector was fully qualified and commissioned in the laboratory in order to assess and optimize its operation in terms of powering, cooling, and readout. A detailed study of the noise rates was performed and summarised in Fig.~\ref{mft_noise}, which shows the fake-hit rate as a function of the number of masked pixels. A noise occupancy below 10$^{-7}$ hits per pixel and per event is obtained by masking only 138 pixels out of a total of 490 millions pixels. This result is well within the specifications for the detector.

\begin{figure}
\centering
\includegraphics[width=.6\textwidth]{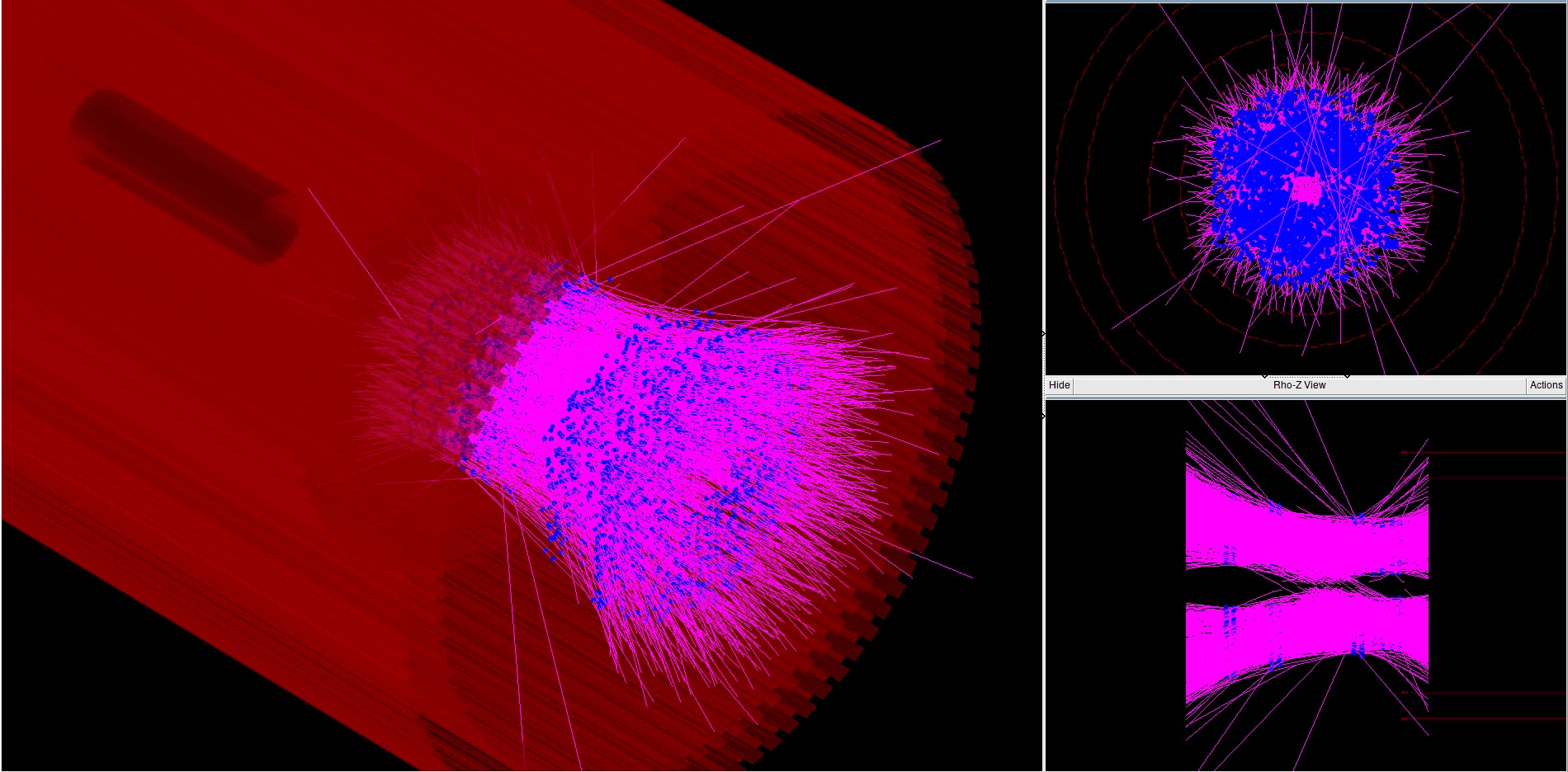}
\caption[Event display]{Display of the reconstructed muon tracks in the MFT from a TED shot event.}
\label{mft_ted}
\end{figure}

The installation of the MFT into the ALICE experiment took place in December 2020 and required several months of activity to route all the services and integrate this new detector within the ALICE central systems. A detailed commissioning phase confirmed that the noise levels measured in the laboratory were unchanged after the complex installation inside the ALICE cavern. Moreover, the MFT was ready to take data in October 2021 during the commissioning of the LHC injection lines, using proton beams injected into the transfer line, which are dumped on the Target Extraction Dump (TED). Interactions of the proton beam with the TED, which is located around \SI{30}{\metre} upstream from the ALICE cavern, produce a shower of muons traversing the ALICE detector. As shown in the event display in Fig.~\ref{mft_ted}, the MFT detector was able to detect and reconstruct these muon showers, proving its readiness for data taking in Run~3.

\clearpage
\subsection{Time Projection Chamber}
\label{sec:tpc}

This section summarizes the upgrade of the ALICE Time Projection Chamber
(TPC).
A more detailed description of the upgrade can be found
in~\cite{ALICETPC:2020ann,tpc:TDRtpcUpgrade}.

\subsubsection{Introduction}
\label{sec:tpc.intro}

\begin{figure}[t]
\centering
\includegraphics[width=0.7\linewidth]{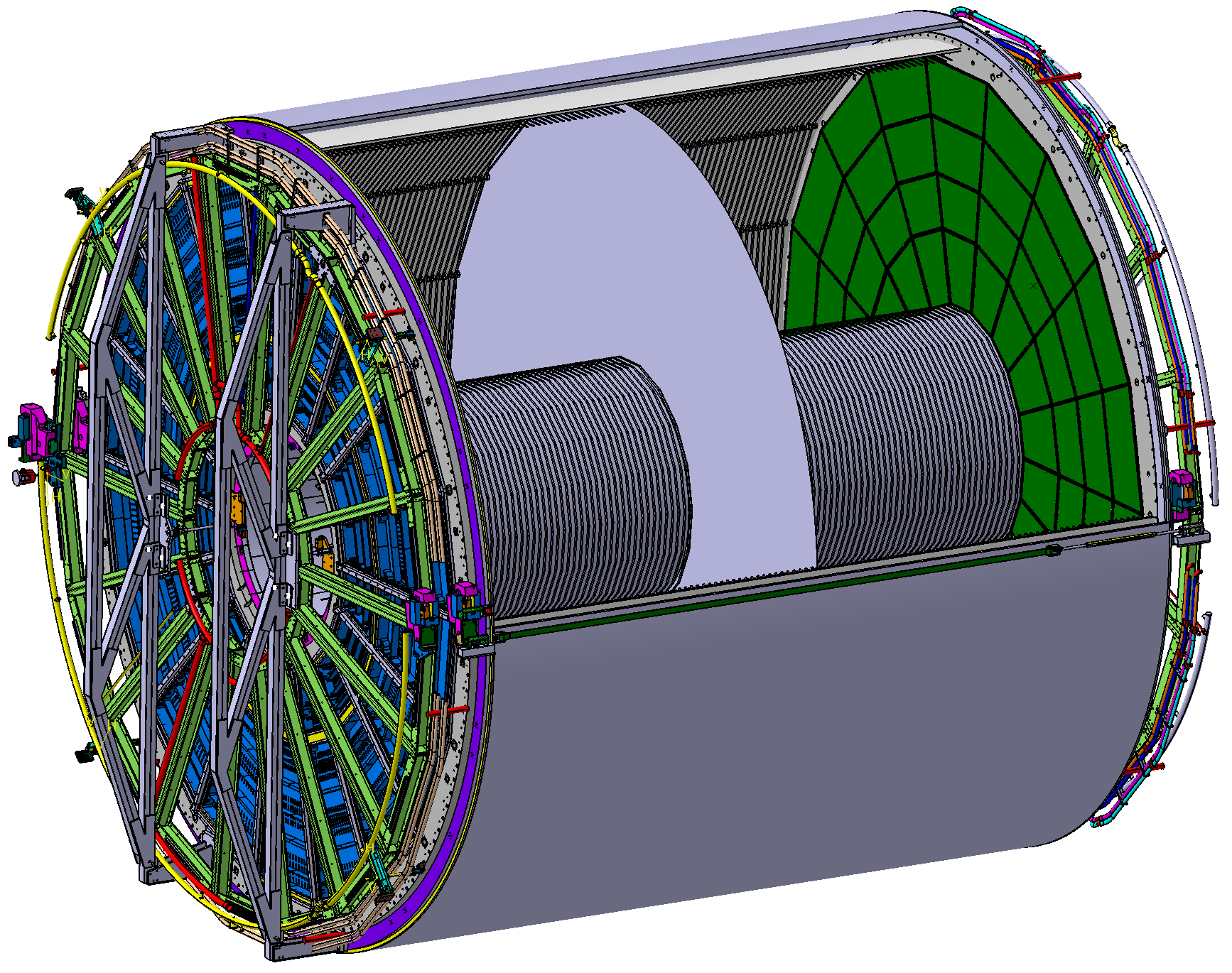}
\caption{Schematic view of the ALICE TPC.}
\label{fig:alitpc}
\end{figure}

The TPC was successfully operated in pp, p--Pb, Pb--Pb,
and Xe--Xe collisions at a variety of collision
energies~\cite{tpc:nim2010,Abelev:2014ffa} during LHC Runs 1 and 2
(2009 to 2018).
Its active volume has a cylindrical shape with a length and
outer diameter of about 5\,m, resulting in a total active volume of
88\,m$^3$ (see Fig.~\ref{fig:alitpc}).
It covers a symmetric pseudorapidity interval around midrapidity
($|\eta| < 0.9$) at full azimuth.
The field cage has a high-voltage electrode in its center, which
divides the active volume into halves.
The inner diameter of the central field cage drum is 114\,cm, which
provides the necessary space for the installation of the ITS.
Each of the two endplates is subdivided into 18 azimuthal sectors.
Each sector houses one inner (IROC) and one outer readout chamber
(OROC).

During Runs~1 and 2, the readout chambers were based on multiwire
proportional chamber (MWPC) technology.
MWPCs have to be operated with an active ion gating grid in order to
collect ions from the amplification region.
Otherwise, these ions would drift back into the drift volume, where
they would lead to substantial space-charge distortions.
However, triggered readout is not compatible with the goals of the
ALICE upgrade described in Chapter~\ref{chap:readout}.
Instead, the upgraded TPC must be read out continuously.
This implies that the previous readout system, including the
readout wire chambers and the front-end electronics needed to be
replaced.
At the same time, the excellent performance achieved in Runs~1 and 2
needed to be maintained in order to achieve the ambitious ALICE
physics program for Runs~3 and 4.
A d$E$/d$x$ resolution of 5\% (for isolated tracks) translates
into a requirement for the local energy resolution of better than
14\% at the $^{55}$Fe-peak.

On average, at a collision rate of 50\,kHz, tracks from five collisions
pile up within the TPC drift time window of about 100\,\si{\micro\second}.
Without continuous readout, not all of these interactions can be read out.
With continuous readout, however, novel gas amplification techniques are
required in order to provide sufficient ion blocking without an active
ion gate.
The requirement to keep the ion-induced space-charge distortions at a
tolerable level leads to an upper limit of 2\% for the fractional ion
backflow (defined as the ion escape probability per effective
electron-ion pair produced in the gas amplification stage) at the
operational gas gain of 2000.

Gas Electron Multipliers (GEMs)~\cite{tpc:sauli97} provide a viable solution to this challenge.
They can be arranged in stacks, creating layers of amplification
stages, which can be tuned accordingly.
With a careful optimization of the gain share among the GEMs, and by
efficiently blocking the path of back-drifting ions that emerge
from subsequent layers, the required low ion backflow can be achieved.

The gas mixture for the operation of the upgraded TPC is
Ne-CO$_2$-N$_2$ (90-10-5),~i.e.\ 90 parts of Ne, 10 parts of CO$_2$,
and 5 parts of N$_2$.
No changes to the existing gas hardware were necessary, since the TPC had
already been operated with this gas mixture in Runs~1 and 2.
A mixture based on Ne has the advantage of a higher ion mobility compared
to similar Ar-based mixtures, which directly reduces the magnitude of the
space-charge distortions by nearly a factor of two~\cite{tpc:deisting2017}.

Table~\ref{tab:tpc.sumgas} summarizes the most important TPC parameters.
The drift time for ions from the readout plane to the central electrode is
214\,ms at a drift field of 400\,V/cm.
Therefore, at 50\,kHz, around $10^4$ collisions partially contribute
to the space-charge distribution.

\begin{table}[t]
  \caption[TPC parameters]{Parameters of the upgraded TPC. Table taken from~\cite{ALICETPC:2020ann}.}
  \begin{center}
    \begin{tabular}{ll}
      \hline
      Detector gas& Ne-CO$_2$-N$_2$ (90-10-5) \\
      Gas volume & 88\,m$^3$ \\
      Drift voltage & -100\,kV \\
      Drift field & 400\,V/cm \\
      Maximal drift length & 250\,cm \\
      Electron drift velocity & \SI{2.58}{\cm\per\us}\\
      Maximum electron drift time & \SI{97}{\us} \\
      $\omega\tau$ ($B=0.5$\,T) & 0.32\\
      Electron diffusion coefficients & $D_{\rm T}=\SI{209}{\um}/\sqrt{\text{cm}}$\\
      & $D_{\rm L}=\SI{221}{\um}/\sqrt{\text{cm}}$\\
      Ion drift velocity & 1.168\,cm/ms\\
      Maximum ion drift time & 214\,ms\\
      \hline
    \end{tabular}
  \end{center}
  \label{tab:tpc.sumgas}
\end{table}

\subsubsection{Readout chamber design}
\label{sec:tpc.roc}

\begin{figure}[b]
  \centering
   \includegraphics[width=0.95\linewidth]{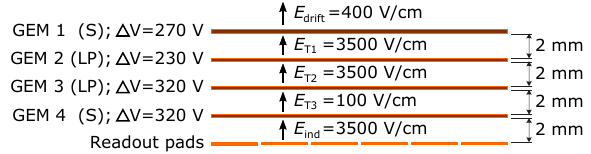}
  \caption[GEM stack]{Schematic view of a stack with four GEM foils. The baseline
  settings for the voltages across the four GEMs, the transfer fields between the
  GEMs, and the induction field between GEM\,4 and the pad plane are indicated as
  well.}
  \label{fig:4GEM-schematic}
\end{figure}

\begin{figure}[t]
  \centering
   \includegraphics[width=0.7\linewidth]{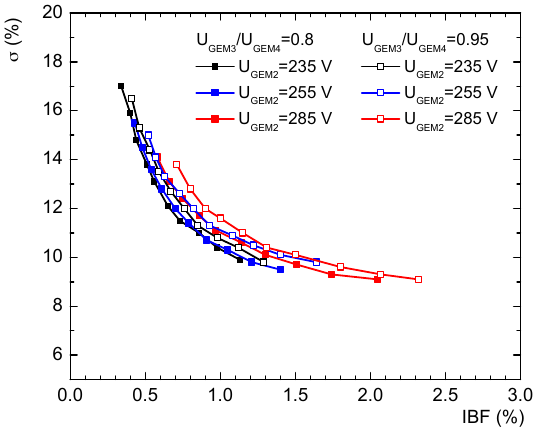}
  \caption[Energy resolution as function of ion backflow]{Energy resolution
    $\sigma(^{55}\rm{Fe})$ as a function of
    ion backflow (IBF) in a 4-GEM stack (S-LP-LP-S) in Ne-CO$_2$-N$_2$ (90-10-5).
    The gas gain is kept at 2000 in all measurements by adjusting the
    voltages on GEM\,3 and GEM\,4 at a fixed ratio of 0.8 or 0.95. Figure from
   ~\cite{ALICETPC:2020ann}.}
  \label{fig:tpc.sigmavsIBF}
\end{figure}

The new readout chambers of the upgraded TPC are based on stacks of four GEM foils.
Foils with standard (S, 140\,\si{\micro\meter}) and large (LP, 280\,\si{\micro\meter})
hole pitch are combined to an S-LP-LP-S configuration, which is shown in
Fig.~\ref{fig:4GEM-schematic}.
Most of the ions are produced in the last amplification step, i.e. GEM\,4.
Their drift path is efficiently blocked by the upper GEM layers by carefully
optimizing the GEM voltages and transfer fields, and by choosing GEM hole
patterns avoiding the accidental alignment of holes in subsequent layers.
As Fig.~\ref{fig:tpc.sigmavsIBF} shows, an extended operational region
satisfying the requirements (ion backflow below 2\% and
$\sigma(^{55}\text{Fe})$ below 14\%) can be found with an S-LP-LP-S
setup.
The data exhibit a characteristic anticorrelation between ion
backflow and the relative energy resolution $\sigma(^{55}\text{Fe})$.
This effect is largely due to the operational conditions at GEM\,1
since ions emerging from this layer have a large probability to escape
into the drift volume.
In order to minimize the number of ions produced at GEM\,1, the gas
amplification has to be reduced, which however also leads to an
effective loss of primary ionization, and therefore to a degradation of
the energy resolution.
 
\begin{figure}[htb]
  \centering
   \includegraphics[width=0.8\linewidth]{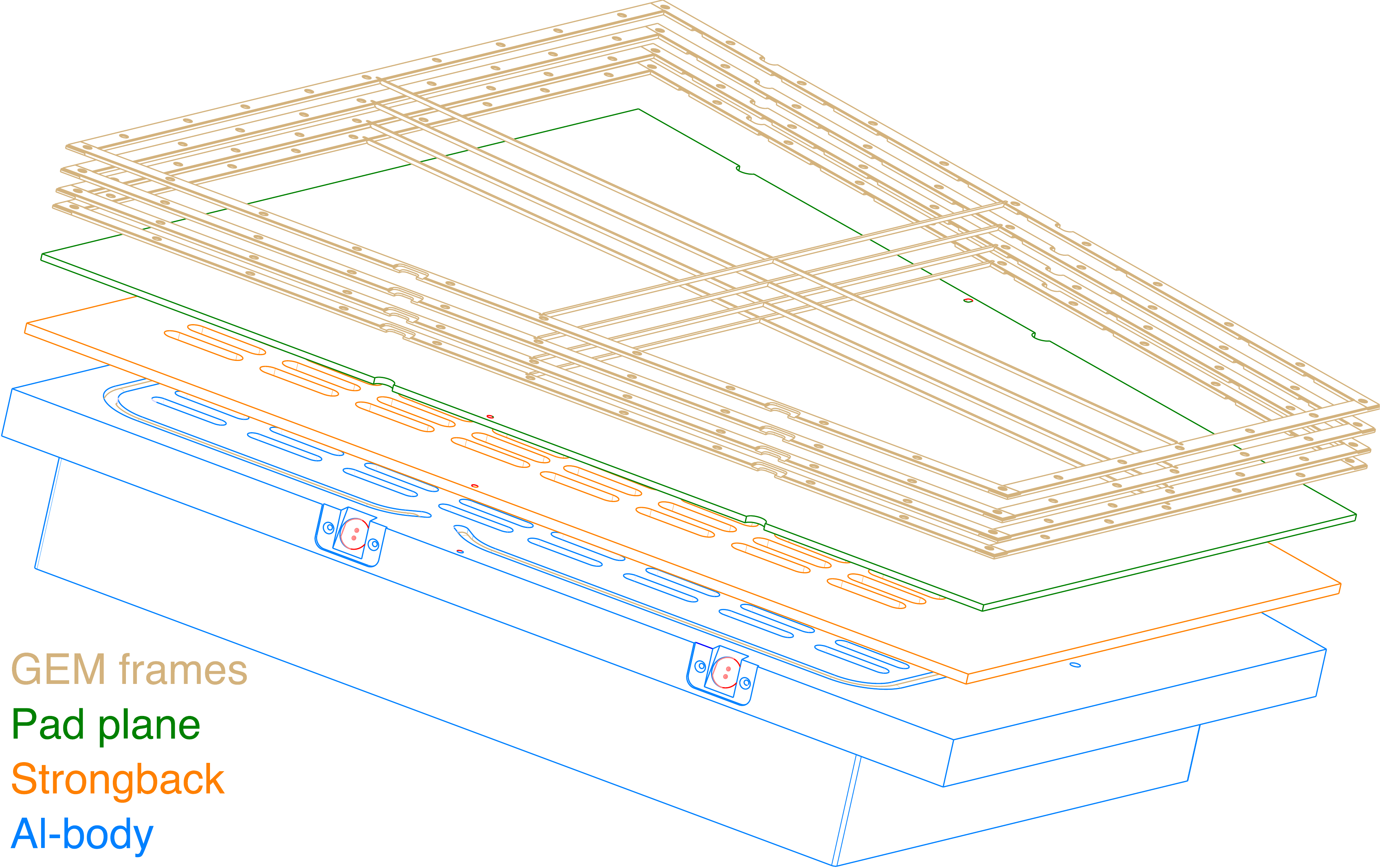}       
  \caption[Exploded view of IROC]{Exploded view of an IROC with chamber body components and
    GEM frames. Figure taken from~\cite{ALICETPC:2020ann}.}
  \label{fig:tpc.roc-exploded}
\end{figure}

A readout chamber consists of a trapezoidal aluminium frame
(\textit{Al-body}), a fiberglass plate (\textit{strongback}) and a
\textit{pad plane} made of a multilayer printed circuit board (PCB).
Figure~\ref{fig:tpc.roc-exploded} shows an IROC with the individual
components.
While an IROC is assembled from one Al-body, one strongback, one pad plane
and one GEM stack, an OROC is assembled from one Al-body, one
strongback, three pad planes and three GEM stacks labeled OROC\,1, OROC\,2
and OROC\,3.
The stacks consist of four GEM foils stretched and glued onto fiberglass epoxy
frames, each containing a spacer cross.
The geometrical parameters of the new readout chambers are summarized
in Table~\ref{tab:tpc.sumroc}.

\begin{table}[htb]\footnotesize
  \caption{Geometrical parameters of the new readout chambers.}
  \begin{center}
    \begin{tabular}{ll}
      \hline
      {\bf Readout chambers} & \\[0.25ex]
      Total number&$2\times2\times18 = 72$\\ 
      Readout technology&4-GEM stack, single mask,
                                  standard (S, 140\,\si{\micro\meter}) and large \\
                                  &(LP, 280\,\si{\micro\meter}) hole pitch GEMs in
                               \mbox{S-LP-LP-S} configuration\\
                               Effective gas gain&2000\\
      {\bf Inner  (IROC)} & \\[0.25ex]
      Total number&$2\times18=36$\\ 
      Active range&$848.5<r<1321$\,mm \\
      Number of HV segments per GEM foil & 18\\
      Pad rows &63\\
      Total pads (IROC)&5280\\
      $S$:$N$ (MIP)&20:1 \\[0.5ex]
      {\bf Outer  (OROC)} & \\[0.25ex]
      Total number & $2\times18=36$\\ 
      Active range & $1347<r<2464$\,mm \\
      Total pads (OROC) & 9280\\
      $S$:$N$(MIP)   &30:1\\
      Pad rows &89\\[0.5ex]
      {\bf OROC\,1} & \\[0.25ex]
      Active range & $1347<r<1687$\,mm\\
      Number of HV segments per GEM foil & 20\\
      Pad rows & 34\\
      Number of pads & 2880\\[0.5ex]
      {\bf OROC\,2} & \\[0.25ex]
      Active range & $1708<r<2068$\,mm\\
      Number of HV segments per GEM foil & 22\\
      Pad rows & 30\\
      Number of pads & 3200\\[0.5ex]
      {\bf OROC\,3} & \\[0.25ex]
      Active range & $2089<r<2464$\,mm\\
      Number of HV segments per GEM foil & 24\\
      Pad rows & 25\\
      Number of pads & 3200\\[0.5ex]
      \hline
    \end{tabular}
  \end{center}
  \label{tab:tpc.sumroc}
\end{table}

\begin{figure}[htb]
  \centering
   \includegraphics[width=0.8\linewidth]{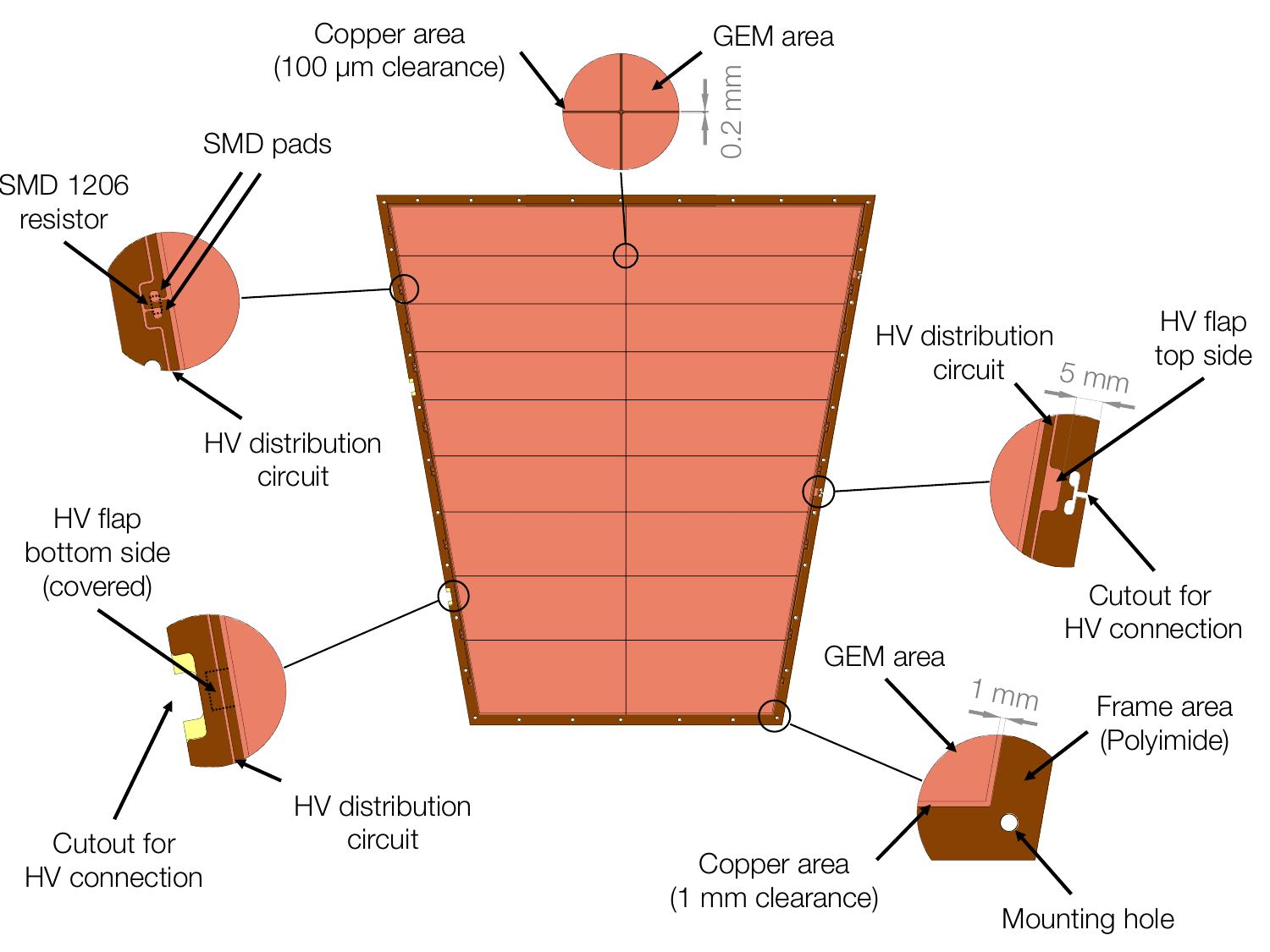}
   \caption[GEM design details]{GEM design details. The detail on top shows the 200\,$\mu$m
      separation between the copper sectors and an additional clearance
      of 100\,$\mu$m for the GEM holes from the copper edge.
      Figure taken from~\cite{ALICETPC:2020ann}.}
  \label{fig:tpc.gem-design} 
\end{figure}

Figure~\ref{fig:tpc.gem-design} shows the most important features of the GEM
design.
The design of the Al-bodies includes a copper pipe for temperature control
by water cooling.
The electric potential is provided to the GEM stacks via feedthroughs.
The strongback provides electrical insulation between pad plane and
Al-body and reduces the pad capacitance to ground.
The pad planes are made of a 3.2\,mm thick FR4 multilayer board.
The top PCB layer consists of copper readout pads, arranged in pad
rows.
The pad planes do not contain ground electrodes in order to minimize
the capacitance to ground at the preamplifier input.
The pads are connected to traces routed to vertical 40-pin female connectors
on the backside of the pad plane.
The routing of the traces is done using three additional PCB layers.
The signal routing was designed to minimize the trace length.
Four connectors in radial direction group 160~pads to connect to one
front-end card (see Sec.~\ref{sec:tpc.fee}). 

The GEM foils for the ALICE TPC upgrade were produced using the
single mask technique~\cite{tpc:singlemask2011}.
The top side of each foil is subdivided into high-voltage segments
with an area of about 100\,cm$^2$. The segmentation limits the currents in
case of electrical discharges and minimizes the affected area in case a
segment develops a short circuit.
The gap between the adjacent segments is 200\,\si{\micro\meter}.
An additional 100\,\si{\micro\meter} space is added between the segment
boundaries and the surface containing GEM holes.
Electric potentials are applied to a foil through wires soldered to HV
flaps placed on the top and bottom sides of the foil.
From here, the potential is distributed via a 1\,mm wide copper trace running
along three sides of the foil.
Each foil segment is connected to the HV trace via a 5\,M$\Omega$ loading
resistor ($R_{\text{load}}$).

\subsubsection{Foil production, chamber production and quality assurance}
\label{sec:tpc.prod}
  
In total, 45 IROCs and 40 OROCs were assembled over several years at several
production sites.
More than 300 shipments of material and subcomponents between the different
production sites were necessary.
Standardized transport and testing procedures and well defined assembly
and quality-assurance protocols were followed in order to ensure high
quality and reliability of the assembled ROCs.
All assembly activities involving GEM foils were performed in clean
rooms with class ISO 5 to 7, taking all precautions to avoid
contamination of the GEMs.

The GEM foils were extensively tested before and after each transport.
These tests consisted of optical inspection and a measurement of the
leakage current~\cite{tpc:ball2017}.
An excessive current of a segment points to a shorted GEM segment.
An increased current may come from a defect or contamination and was
considered a potential danger.
Shorted and contaminated GEMs were sent back to the production site
for cleaning.
An advanced quality assurance procedure, performed once for each GEM,
consisted in a long-term (at least 5 hours) leakage current measurement
and an optical
survey~\cite{tpc:brucken2018rej, tpc:brucken2018EPJ, tpc:hilden2018}.
During the optical survey, microscope photographs were taken of the
entire GEM.
The pictures were stitched together and analyzed for defects and
hole-size nonuniformities.
In total, 829 GEM foils were produced in the course of the project.
The final yield of good foils after all production and quality assurance
steps was about 91\%.

The framing of a GEM foil consisted of stretching the foil and positioning
it above a frame covered with a thin layer of epoxy glue.
The framed foils were subsequently mounted on the preassembled readout
chamber bodies, and the HV wires were soldered to the top and bottom
HV flaps.
The stacks are not glued, such that they can be disassembled and rebuilt
in case of problems.
After HV connection, resistance and capacitance across each foil
were measured in order to identify possible issues at the earliest
possible stage.
The assembled detectors were mounted in gas-tight test and transport
vessels and qualified, before being sent to CERN for acceptance tests,
storage and installation in the TPC field cage.

At CERN, the accepted ROCs underwent a final stability test under
irradiation.
For these tests, the nominal gas mixture and the final components
of the HV system were used.
These tests were performed in the ALICE cavern during LHC operation
or at the CERN Gamma Irradiation Facility
(GIF++)~\cite{tpc:capeans2009,tpc:jaekel2014}.

Those ROCs that were not accepted or replaced after initial
commissioning of the TPC at the surface have been refurbished
such that they are available as good spare chambers for an eventual
replacement campaign in the future.
The TPC could be brought to the surface again during the Long
Shutdown 3 of the LHC in the years 2026 to 2028.
The refurbishment of the ROCs includes in particular the assembly
and installation of new GEM stacks.

\subsubsection{Field cage}
\label{sec:tpc.fc}

The field cage of the TPC is described in detail in~\cite{tpc:nim2010}.
The central drift electrode is biased to a potential of about --100\,kV
and generates a uniform electric drift field with the help of
potential strips that are suspended close to the walls of the inner and
outer field cage vessels.
These strips are powered through a resistor chain housed in a water-cooled
rod.
In addition, aluminium strips are glued to the walls of the field cage
vessels at a certain distance.
These {\it guard rings} are powered through separate resistor chains.
Their purpose is to prevent local charging-up of the surfaces.
The potential of the last strip of each resistor chain (both resistor
rods and guard rings) is set to a value similar to that of the GEM\,1
top electrodes facing the drift volume, which is around --3.26\,kV
for the nominal configuration.
With respect to the original MWPC-based TPC, this potential is now higher,
and requires to be set at the last strips with additional power supplies.
The existing HV connections in flanges in the aluminium endplates of the
field cage, and the connections to the last strips had to be adapted for
the higher potential ratings.
Suitable last resistors to ground were chosen to allow for a small
current to ground.

\subsubsection{HV system}
\label{sec:tpc.hv}

The baseline settings for the voltages across the GEMs, for the transfer
fields between the GEMs, and for the induction field between GEM\,4 and
the pad plane were optimized with respect to operational stability under
the radiation load expected in Run\,3.
The settings are indicated in Fig.~\ref{fig:4GEM-schematic}.
The main feature is a very low transfer field $E_{\text{T3}}$ between
GEM\,3 and GEM\,4 of only 100\,V/cm.
The other two transfer fields and the induction field $E_{\text{ind}}$
are kept at typical values around 3500\,V/cm.
The highest gain is provided in GEM\,3 and GEM\,4, while the gain in
GEM\,1 is relatively low.
As a consequence, most ions are created around GEM\,4.
Their drift into the drift region of the TPC is hindered by the
large-pitch foils utilized for GEMs\,2 and 3, and by the very low
transfer field $E_{\text{T3}}$.

An equalization of the gain across all 144 GEM stacks on the TPC is
achieved by adjusting the voltages in GEM\,3 and GEM\,4.
The induction field is corrected correspondingly in order to ensure
that the potential on the GEM\,1 top electrode remains uniform over all
stacks, such that the uniformity of the drift field in the TPC drift
volume is not disturbed.

\begin{figure}[htb]
  \centering
  \includegraphics[width=1.0\textwidth]{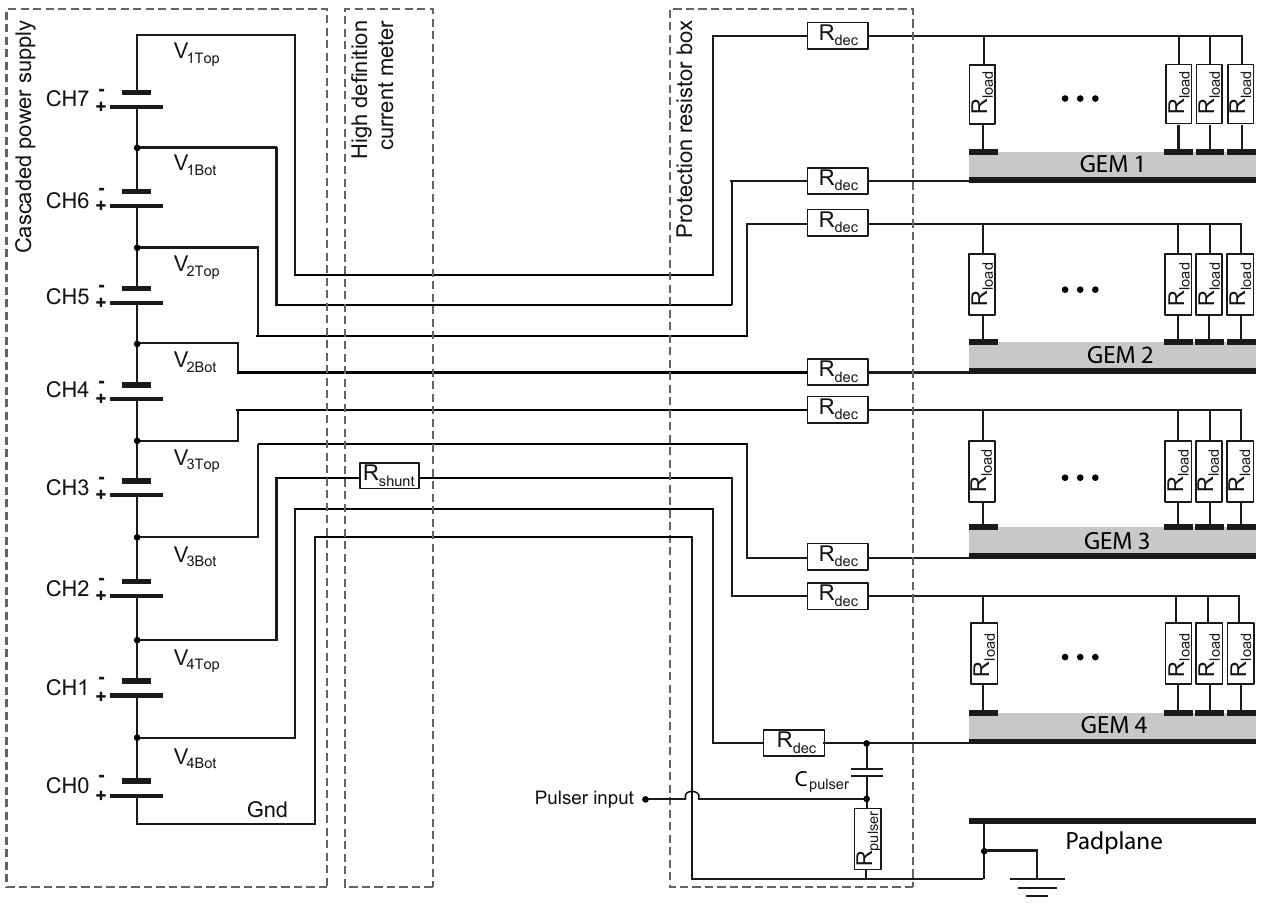}
  \caption[GEM stack powering]{Detailed powering scheme of a GEM stack.
    Each subsequent high-voltage channel is stacked on top of the
    lower-lying channel.
    The ground reference is defined by a separate line connected to the
    ground of the detector.
    The line for GEM\,4 top is shunted with a resistor ($R_{\text{shunt}}$)
    inside the high-definition current meter.
    Each line is connected to the detector through a decoupling resistor
    ($R_{\text{dec}}$).
    The signal from a calibration pulser is coupled via a capacitor
    ($C_{\text{pulser}}$) to the line for GEM\,4 bottom.
    Individual loading resistors ($R_{\text{load}}$) are mounted on all
    segments on the top sides of the GEMs.
    Figure taken from~\cite{ALICETPC:2020ann}.}
    \label{fig:tpc.hvscheme}
\end{figure}

A new HV system was designed for the operation of the GEM-based ROCs.
A detailed description can be found in~\cite{ALICETPC:2020ann}.
In order to maximize operational safety, while at the same time
providing the highest possible flexibility, a power supply system
with cascaded channels was chosen.
In this way, the potentials at the various electrodes can be easily
adjusted, and a safe operation of the GEM stacks can be guaranteed.
A schematic view of the high-voltage system including all loading and
protection resistors is shown in Fig.~\ref{fig:tpc.hvscheme}.
A shunt resistor in the voltage distribution for the GEM\,4 top electrode
allows to periodically read the currents for all GEM stacks.
During the operation of the TPC, current variations in the GEM stacks
will directly relate to variations of the local track density in the
TPC drift volume.
In a high-definition current meter the currents are digitized at a
rate of typically 1\,kHz (8\,kHz maximum) by 24-bit ADCs with a
resolution of 3\,nA.
From these data, three-dimensional maps of the space charge from
back-drifting ions can be extracted in order to calibrate drift-field
distortions (see Sec.~\ref{sec:tpc.calibration}).
Finally, the powering scheme also includes the possibility to inject a
pulser signal to the HV line of each GEM\,4 bottom electrode for
calibration purposes.

\subsubsection{Front-end electronics and readout}
\label{sec:tpc.fee}

\begin{figure}[htb]
  \begin{center}
    \includegraphics[width=0.9\textwidth]{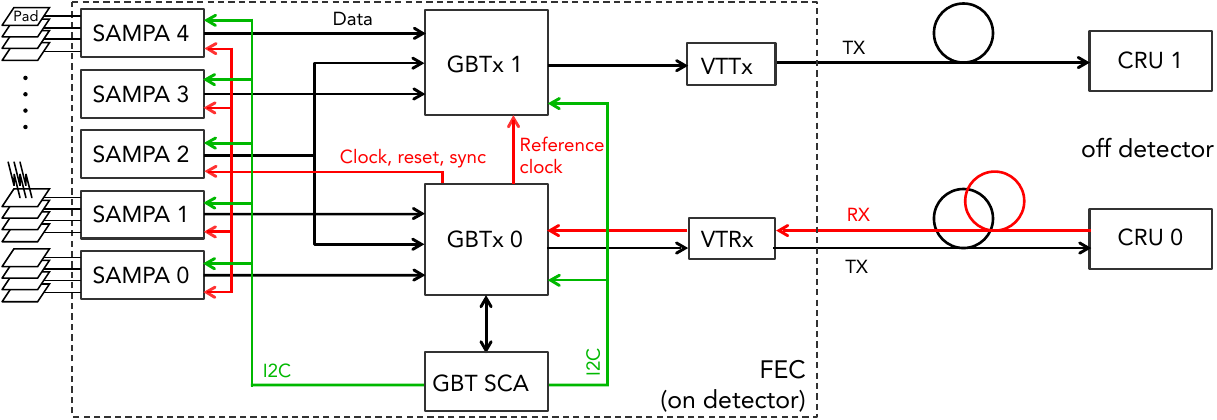}
  \end{center}
  \caption[TPC readout system]{Schematic view of the TPC readout system.
    Five SAMPA chips amplify, shape, and digitize the current
    signals picked up on the connected pads.
    Two GBTx ASICs multiplex the digitized data.
    GBTx\,0 forwards the data from two and a half SAMPA chips to a
    VTRx.
    GBTx\,1 forwards the data from the other two and a half SAMPAs to
    one VTTx module (two optical transmit channels).
    GBTx\,0 also receives configuration data and the reference clock
    through the VTRx.
    The reference clock is distributed to the other components.
    A GBT-SCA chip is used for monitoring and configuration.
    Figure taken from~\cite{ALICETPC:2020ann}.}
  \label{fig:tpc.fec-blockdiagram}
\end{figure}

A schematic view of the front-end electronics and readout system is
shown in Fig.~\ref{fig:tpc.fec-blockdiagram}.
A single Front-End Card (FEC) processes the signals from 160 input
channels.
The pulses are transformed into differential, semi-Gaussian voltage
signals and then digitized in five SAMPA ASICs (see
Sec.~\ref{sec:sampa}).
All ADC values are transmitted off-detector through two optical
links.
In this way, all data are available such that flexible filter
algorithms can be applied in the FPGA-based readout cards (CRU,
see Sec.~\ref{sec:cru}).
One down-link is needed for control and configuration of a FEC.
In total, 3276 FECs are needed to read out the TPC.
This leads to 6552 data links and a total data throughput of 3.28\,TB/s
to the 360 CRUs for TPC readout.
In the CRU FPGA the data are corrected for the common-mode effect
and ion tails are removed from the signals, before data reduction
(zero suppression) is applied.
Moreover, the signals are integrated for each channel over 1\,ms, and
these data are sent out separately as input to calculate three-dimensional
space-charge maps for distortion correction.
An overview over the correction of the TPC signals in the CRU and of
the calibration of the TPC data is given in Sec.~\ref{sec:tpc.calibration}.

\begin{table}[b]\footnotesize
  \caption[TPC front-end electronics parameters]{Parameters of the upgraded
    front-end electronics. Table taken from~\cite{ALICETPC:2020ann}.}
  \begin{center}
    \begin{tabular}{ll}
      \hline
      Readout mode & continuous\\
      Number of channels & 524160\\
      Number of FECs & 3276\\
      Signal polarity & negative \\
      Average system noise (ENC) & 670\,$e$\\
      Conversion gain & 20\,mV/fC\\
      Dynamic range & 100\,fC ($30\times\text{MIP}$)\\
      Peaking time & 160\,ns\\
      CSA saturation limit & 30\,nA\\
      ADC number of bits & 10\\
      ADC sampling rate & 5\,MSa/s\\
      Power consumption (total) & 56\,mW per channel\\
      \hline
    \end{tabular}
  \end{center}
  \label{tab:tpc.fec-specifications}
\end{table}

The parameters for the front-end electronics system are summarized in
Table~\ref{tab:tpc.fec-specifications}.
With respect to the readout system utilized in Runs~1 and
2~\cite{tpc:nim2010}, the new FEC has to meet two new requirements:
continuous readout and negative input signal polarity.
For the charge-sensitive amplifier (CSA), a saturation limit of 30\,nA
was required in order to accomodate the expected average rate of
primary ionization clusters (up to about 3\,nA per front-end channel)
and in addition fluctuations due to the local track multiplicity.
The conversion to digital values takes place with a gain of 20\,mV/fC,
at a sampling rate of 5\,MSa/s, and with a precision of 10 bit.

\begin{figure}[htb]
\begin{center}
\includegraphics[width=0.9\textwidth]{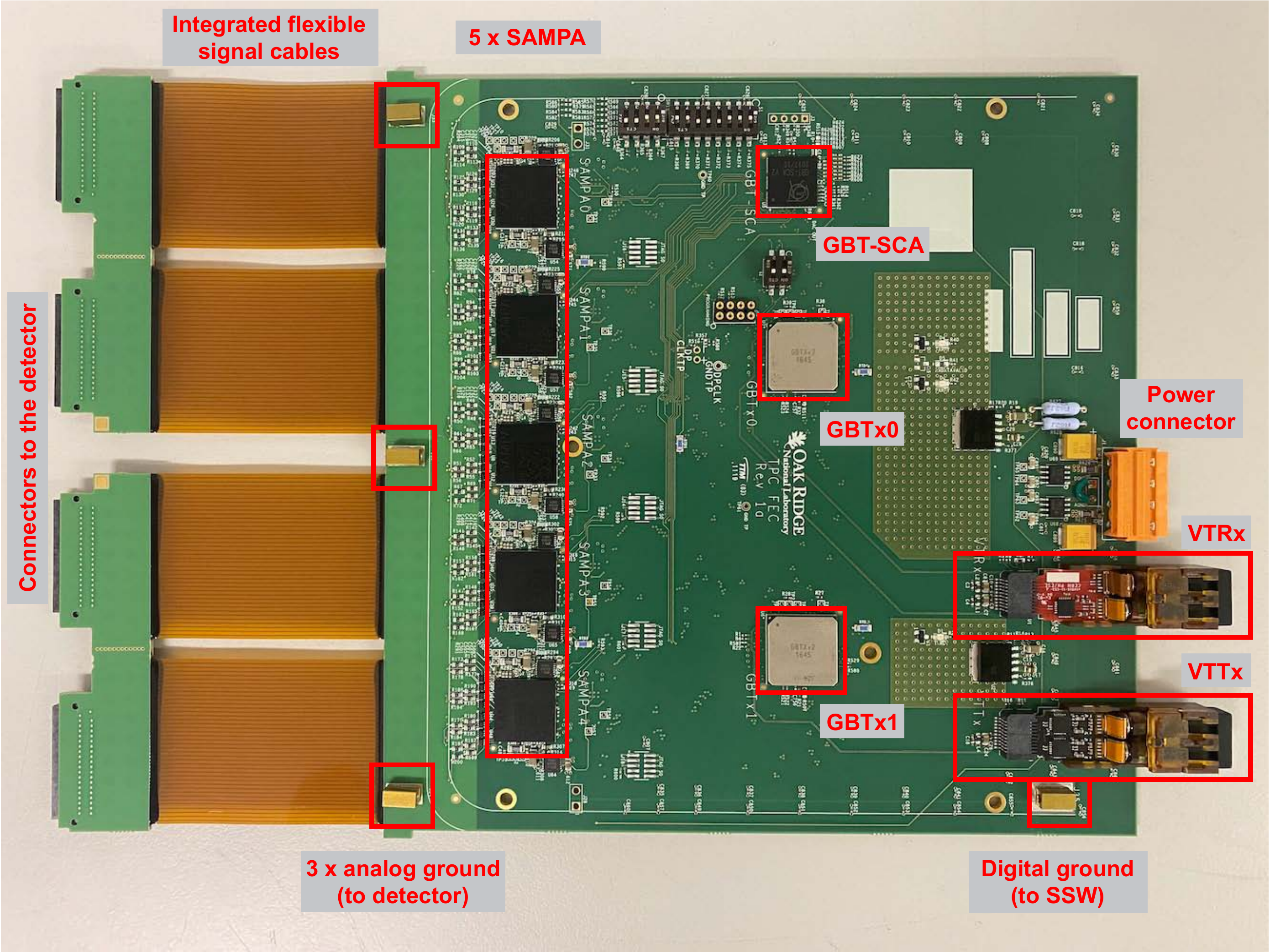}
\end{center}
\caption[TPC FEC layout]{Layout of the final TPC FEC PCB Rev.\,1a.
  The components are mounted on both sides of the board.
  The figure shows the top side with five SAMPAs, two GBTx, one
  GBT-SCA, one VTRx, one VTTx and some other components.
  On the bottom side a few additional small components and the
  connectors to the detector are placed. Figure taken from
 ~\cite{ALICETPC:2020ann}. }
\label{fig:tpc.fecrev1a}
\end{figure}

The TPC FEC is shown in Fig.~\ref{fig:tpc.fecrev1a}.
It has a similar form factor as the one used in Runs~1 and 2.
The 8-layer FEC PCB utilizes rigid-flex technology, where rigid and
flexible substrates are laminated together into a single structure.
The radiation-hard GBT link~\cite{tpc:moreira2009} system is used
(see Sec.~\ref{sec:cru}) for data transfer and control.
The clock for the digital circuitry is received from the CRU.
The electrical links between GBTx and SAMPA use the SLVS standard
and are operated at 160\,Mb/s.
The ADC sampling clock of 5\,MHz is derived by division from the
SLVS link clock speed inside the SAMPA chips.
The monitoring of FEC operational parameters is based on the GBT-SCA
12-bit ADC and includes 14 measurements per FEC.
FEC control is achieved via the GBT-SCA as well.

The SAMPA chips are operated in DAS mode, where the DSP is bypassed
(see Sec.~\ref{sec:sampa}).
In this mode, the power consumption is 9\,mW per channel, which adds
up to 1.5\,W for the whole FEC.
Additional power is needed for the GBT components and for the voltage
regulators.
The total power consumption for the full FEC is about 9\,W (56\,mW per
channel).
The power is supplied using the same Low Voltage (LV) system from
Runs~1 and 2~\cite{tpc:nim2010}.
In particular, two low voltage channels are used for each TPC sector
(91 FECs) to supply the analog (2.25\,V, 85\,A) and digital (3.25\,V,
185\,A) power.
Each FEC is surrounded by water-cooled copper envelopes.
The heat transfer from the hottest components (the two GBTx ASICs,
the five SAMPA ASICs and the voltage regulators) to the copper plates
is optimized by the addition of flexible heat-transfer pads.

\subsubsection{Installation}
\label{sec:tpc.installation}

The upgrade of the TPC was carried out in a 15-month period during
the Long Shutdown 2 of the LHC in a dedicated clean room on the
surface at the site of ALICE at LHC Point 2.
The procedure included the deinstallation of the old readout chambers
and the installation of the new GEM detectors, the modification of the
field cage, the installation of the new front-end electronics and
services, and a series of basic functionality and performance tests.

After the end of Run~2, the TPC was disconnected in December 2018 and
January 2019, and then moved from the experimental cavern to the surface
in February and March 2019.
It was installed in the clean room after removal of the old services
(cables, pipes and hoses), front-end electronics and Service Support
Wheels (SSW)~\cite{tpc:nim2010}, and after extensive cleaning.
Initially, all MWPC ROCs were uninstalled on one side of the field cage,
and the end plate was closed with aluminum panels.
In a second step, the new GEM ROCs were installed.
When all GEM chambers had been installed on the first side, the TPC was
lifted and rotated by 180 degrees for chamber installation on the
second side.

After completion of the ROC installation, the modified SSWs were
installed on the two sides of the TPC.
Each SSW supports the front-end electronics and related services (LV
and fibers), the HV infrastructure (protection resistor boxes), the
manifolds for the various circuits for cooling water, and the drift gas
manifolds.

In a next step, the front-end electronics (3276 FECs) were installed and
the corresponding services (power cables, fibers and cooling tubes) were
connected.
A first commissioning phase was then carried out for pairs of sectors in
order to verify all chambers and the electronics.
Various modes of data taking allowed for testing and improving the
readout and reconstruction workflow.
The acquired data were also used for calibration purposes and
validation of the detector simulation.
The data sample included pedestal, pulser and laser runs, as well as
samples containing cosmic tracks and charge clusters from irradiation
with an X-ray source.
In August 2020, the TPC was transported back to the experimental
cavern for installation into the ALICE magnet.

After installation in the central barrel of the ALICE experiment, the
TPC was connected to its service infrastructure in the winter of 2020/21.
The necessary connections include the hoses for the water cooling of the
electronics and of the auxilliary systems, the LV cables, the HV cables,
the fiber patch cords and a few additional cables (pulser and laser
control).

\subsubsection{Performance}
\label{sec:tpc.performance}

\begin{figure}[htb]
\begin{center}
\includegraphics[width=0.9\textwidth]{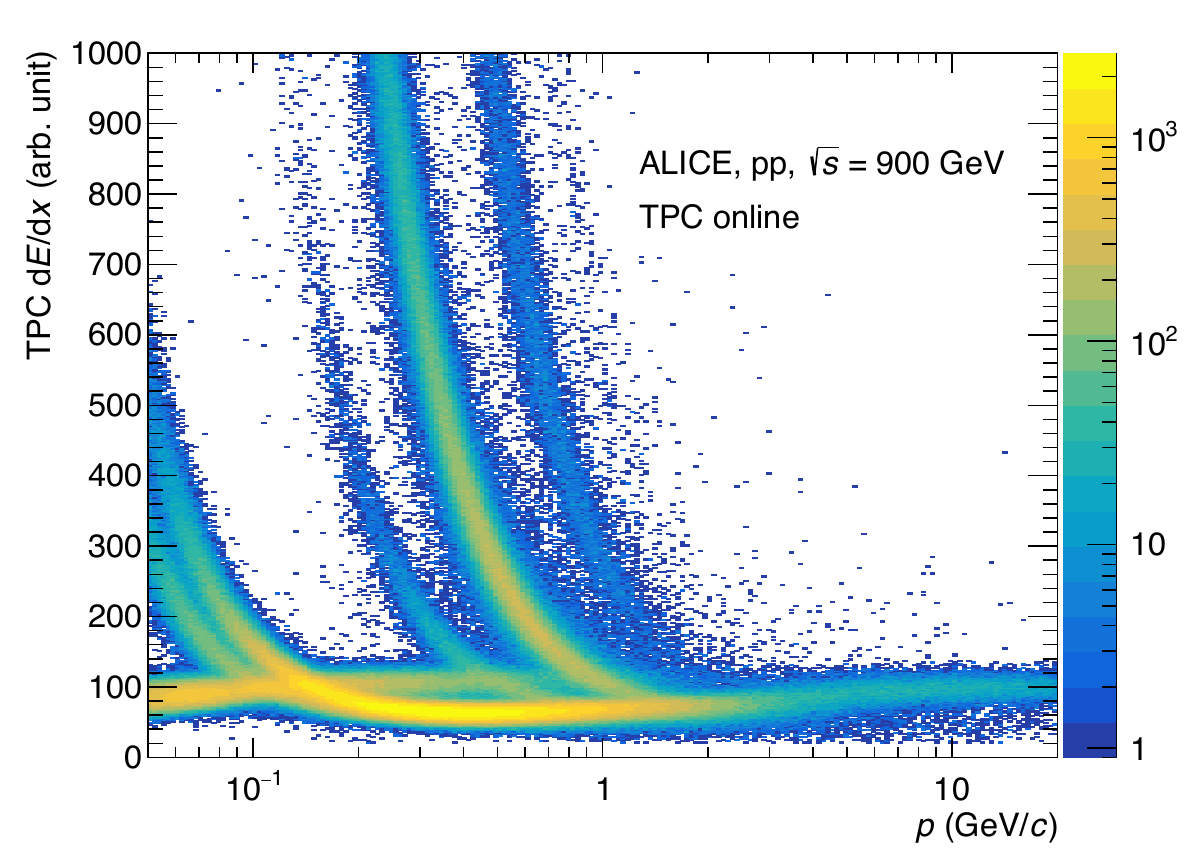}
\end{center}
\caption[TPC d$E$/d$x$ performance in pilot beam]{TPC d$E$/d$x$ as a function
  of momentum $p$. Online plot from quality control during pilot run with
  900\,GeV collisions in October 2021. No track selection criteria were applied.
  The magnetic field value was of 0.5\,T.}
\label{fig:tpc.dEdx}
\end{figure}

After connection and verification of the services, the commissioning of the data
processing chain, of the readout and reconstruction workflows and of
the calibration procedures started.
One first highlight were the pilot beams at low luminosity provided by the LHC
in October 2021.
Figure~\ref{fig:tpc.dEdx} shows an online plot from data recorded during this
period.
No track selection criteria and no calibration was applied in these data.
The data come from the online quality control system and demonstrate the
particle identification performance for tracks reconstructed online in the
graphics processing units installed in the O$^2$ EPN servers (see
Sec.~\ref{sec:epn}).

\begin{figure}[htb]
\begin{center}
\includegraphics[width=0.9\textwidth]{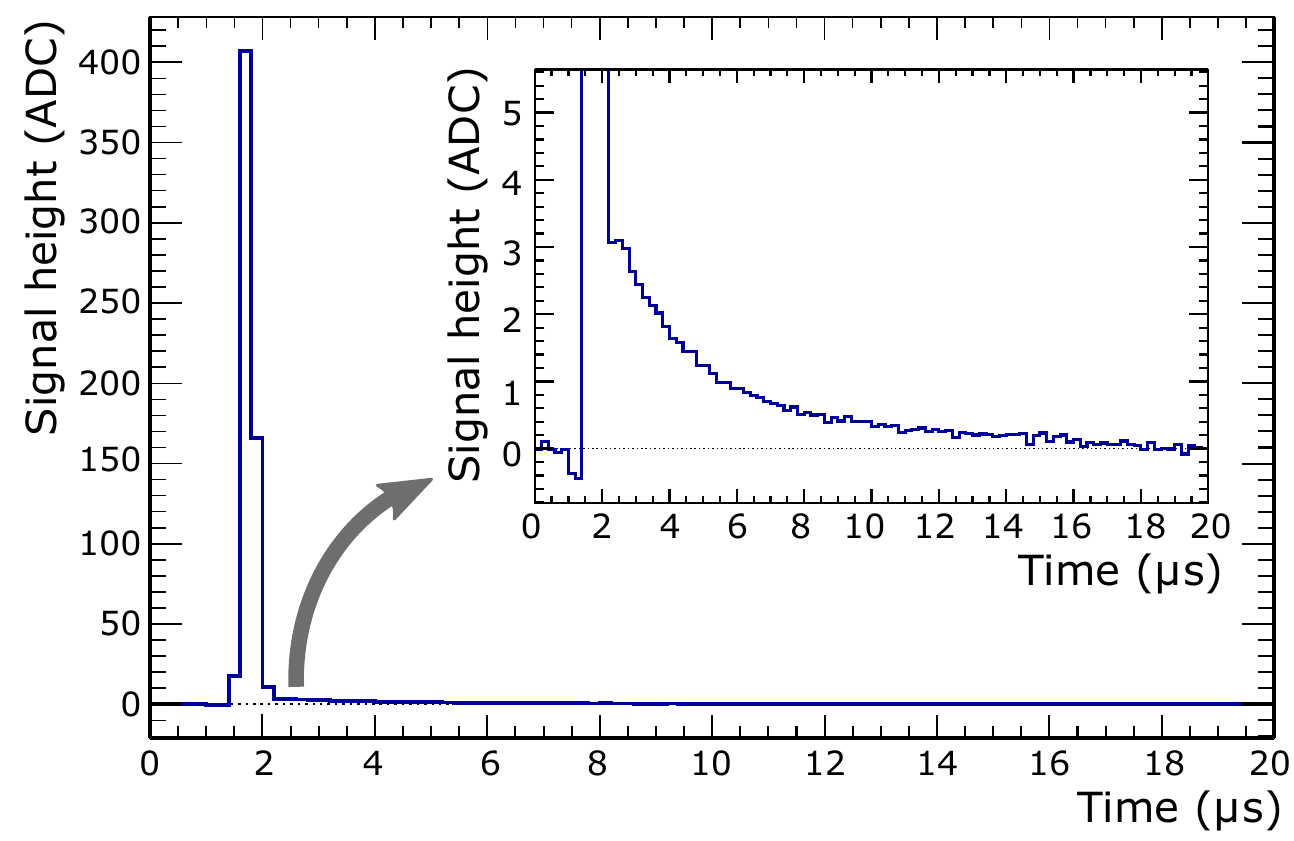}
\end{center}
\caption[Ion tail measured in TPC GEM]{Measured shape of the ion tail
in data recorded with laser tracks.}
\label{fig:tpc.it_meas}
\end{figure}

\begin{figure}[htb]
\begin{center}
\includegraphics[width=0.9\textwidth]{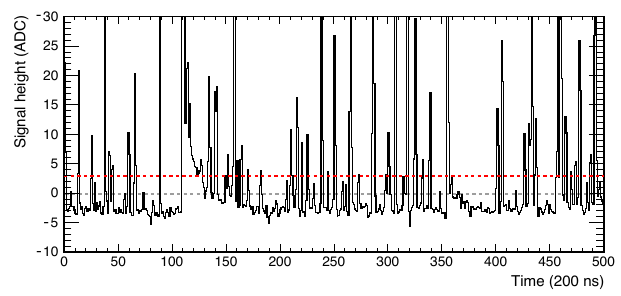}
\end{center}
\caption[Ion tail and common-mode effect in TPC GEM]{Visualization of the ion tail
and of the common-mode effect for one readout channel zoomed around the baseline region.
The data are from a simulation with 30\% pad occupancy and without noise from the
electronics.
The common-mode effect shifts the baseline to below zero. The ion tail is visible for
signals with large amplitude.
Both effects are corrected in the CRU FPGAs.}
\label{fig:tpc.it_cm_sim}
\end{figure}

From summer 2022 the TPC was recording data at higher luminosities routinely.
The data are affected by a baseline shift due to the common-mode effect.
This effect is well known, and occurs in the ROCs due to capacitive coupling of
the GEM\,4 bottom electrode to the readout pads.
In addition, during the analysis of the data collected during the commissioning
phase, a characteristic tail was identified, which appears in particular for
signals with large amplitudes.
Simulations showed that part of the tail is induced by ions that are created just
below the holes of GEM\,4.
Due to the local electric field configuration, these ions move fast enough to
induce a signal on the pad plane.
In addition, due to the rather high induction field applied for the TPC GEMs,
amplification occurs in the full induction gap.
The gain in the induction gap is very small, but nevertheless the produced ions
contribute to the ion tail.
A high-precision measurement of the ion tail (using overlay of many signals
from laser tracks in the TPC) is shown in Fig.~\ref{fig:tpc.it_meas}.
Fig.~\ref{fig:tpc.it_cm_sim} visualizes the common-mode effect together with the
effect of the ion tail on data collected by a single readout channel at high
occupancy.

\subsubsection{Calibration}
\label{sec:tpc.calibration}

In order to operate the detector, some basic calibration steps are necessary.
The baseline (pedestal value) and noise of each electronics channel are
extracted from the data collected in special pedestal calibration runs, where
all filtering on the CRU is inactive.
The mean of the baseline distribution defines the pedestal value, while
the sigma corresponds to the noise of a given channel.
The baseline values need to be uploaded to the logic in the CRU FPGAs in
order to subtract them from the input signals.
The thresholds for the zero suppression filter algorithm in the CRU are
derived from the noise values (typically 3\,$\sigma$).

The correction algorithm for the common-mode effect requires the upload
of configuration parameters (one per channel) into the logic in the CRU
FPGAs as well.
The parameter is extracted from the data collected in special calibration
pulser runs, where a pulser signal is injected into the HV line of each
GEM\,4 bottom electrode.
It describes the local geometry (distance between pad plane surface and
GEM\,4 bottom electrode), which is influenced by sagging of the GEM foils.

The correction algorithm for the ion tail filter requires two configuration
parameters for each channel, describing the shape of the tail and its amplitude
relative to the pulse amplitude.
These parameters are extracted for each pad from physics data aquired with beam.

On top of the basic calibrations, more complex calibration of the TPC data
is needed.
The drift velocity can be extracted from laser tracks that can be generated inside
the drift volume of the TPC in special laser calibration runs or during physics
runs.
The gain can be extracted for each pad in special calibration runs where the
radioactive decay of the $^{83}$Kr isotope is measured, or from analysing the
tracks generated by particles in the physics runs.
Different inputs may be used for the correction of the space-charge distortions.
\begin{itemize}
\item
An interpolation method using external track references in ITS, TRD and
TOF may be used.
This method was extensively tested during Run\,2, where some
distortions due to imperfections were already present in certain regions.
\item
A reference average distortion map may be scaled with the actual luminosity
of the interactions at the given moment for each time frame.
\item
The signals collected continuously in the TPC front-end electronics
are integrated (1\,ms integration time) for each channel inside the CRU and
may be used for building a three-dimensional map of the space-charge distribution.
\item
Finally, the information from the high-definition current meters that
sample the analog currents at all GEM\,4 top electrodes for all GEM stacks may
be used for building similar space-charge maps with lower granularity.
\end{itemize}
The latter two methods require as input the ion drift velocity at the given
electric field strength.

In addition, static distortions play an important role.
They are related to a small misalignment between readout chambers and central
electrode, and to misalignment of the magnetic field and the drift field
inside the field cage due to imperfections.
The static distortions are constant in time for a given detector configuration.

A two-stage process (see Chapter~\ref{chap:readout}) has been implemented for
the processing and calibration of physics data.
The first stage is performed synchronously with the collection of the data,
and focuses on cluster finding and the association of clusters to tracks.
For this purpose, the mean space-charge distortions scaled with the current
luminosity are used.
The reconstructed tracks have sufficient precision to allow matching to the
external detectors (ITS, TRD and TOF).
The compressed data are written to permanent storage. 
The second reconstruction stage is performed on the compressed data in
asynchronous mode.
It aims at a further improvement of the data quality, in particular in terms
of the space-charge distortion calibrations.
It may employ a combination of the described methods, as well as other more
refined calibration input.

\clearpage
\subsection{Fast Interaction Trigger}
\label{sec:fit}

\begin{figure}[htbp]
\begin{center}
\includegraphics[width=0.8\linewidth]{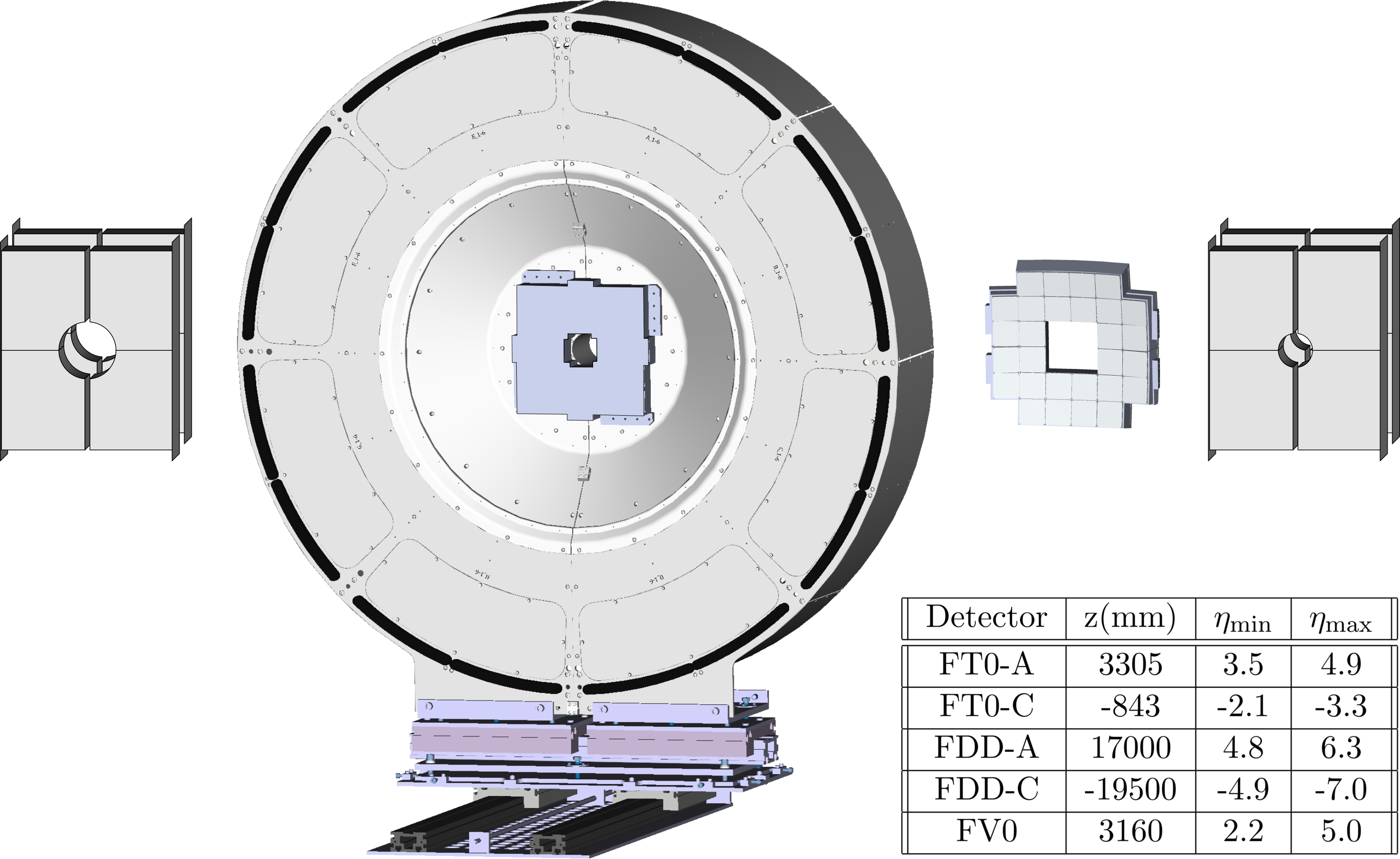}
\caption[Overview of FIT detectors]{View of the FIT detectors illustrating the relative sizes of each component.  From left to right FDD-A, FT0-A, FV0, FT0-C, and FDD-C are shown.  Note that FT0-A and FV0 have a common mechanical support. FT0-A is the small quadrangular structure in the centre of the large, circular FV0 support. Note that all detectors are planar with the exception of FT0-C, which has a concave shape centered on the IP. The inset table lists the distance from the interaction point and the pseudorapidity coverage for each component.}
\label{FITschematic}
\end{center}
\end{figure}

The Fast Interaction Trigger (FIT)\cite{Trzaska:2017reu} serves as an interaction trigger, online and offline luminometer, initial indicator of the vertex position, and forward multiplicity counter. Offline analysis of FIT data provides the precise collision time for TOF-based particle identification, yields the collision centrality and the event plane orientation, and provides the main input for the measurement of cross sections of diffractive processes. The FIT consists of five distinct detector stations, positioned at different locations along the beam line. Three different detector technologies, as detailed below, are used. An illustration of FIT is shown in Fig.~\ref{FITschematic}; the distance from the interaction point (IP) and pseudorapidity coverage of the different arrays are displayed in the inset table.
The naming convention relates to the similar ALICE detectors used during Run~2. FT0 is the successor of T0~\cite{Bondila:2005}, which owes its name to the fact that it was used to provide a start time. FV0 is the successor of V0~\cite{V0performance:2014}, which provided the vertex location. Finally, FDD (Forward Diffractive Detector) is the successor of ALICE diffractive detector, AD, which detects charged particles at large pseudorapidity for the selection of diffractive and ultra-peripheral events.~\cite{broz:2020mb}.

A new, fast electronics and readout system~\cite{Finogeev:2020qkf} that can handle the larger interaction rates in Runs~3 and 4 has been designed and implemented for all FIT subdetectors ~\cite{Trzaska:2020zzl}.

\subsubsection{FT0}
 
The FT0 consists of two arrays of quartz Cherenkov radiators, FT0-A and FT0-C, which are optically coupled to MicroChannel
Plate-based photomultipliers (MCP).
The FT0-A is located at 3.3\,m from the IP and comprises 24 MCPs and 96 quartz radiators.
Due to the close proximity to the IP, the FT0-C support has a convex shape (as seen from the IP), positioning all 28 MCPs such that each of the 112 quartz radiators is at a distance of 84\,cm from the nominal IP.
The Photonis XP85002/FIT-Q MCPs are factory-customized versions of the Planacon XP85012. The customization is a new back-plane design for FT0 which groups the usual 64 anodes into four outputs, one for each of the four optically isolated quartz radiators, each with a thickness of 2\,cm and an area of $2.65\times 2.65$\,cm$^2$. This segmentation provides the granularity for measurements of multiplicity in central \PbPb{} collisions, while minimizing the dead areas due to MCP edges and the optical isolation of the radiators. In order to obtain the best possible timing resolution, the signal path from each MCP anode to the front-end electronics has the same length. The intrinsic time resolution of each quadrant is $\sigma_{\rm t} \approx 13$\,ps~\cite{Melikyan:2020owp}. Accounting for signal deterioration along the 30\,m long signal cables and processing by the 
front-end electronics, the achieved time resolution of FT0 is about 25\,ps for a single minimum-ionising particle.
Simulation studies of FT0 with the PYTHIA event generator~\cite{Sjostrand:2019zhc} and the GEANT detector response simulation~\cite{Brun:1987ma} indicate that the efficiency of the minimum bias trigger for pp collisions is  $\geq 98\%$ for the OR of the two sides and $\geq 77\%$ for coincidences between FT0-A and FT0-C.

\subsubsection{FV0}
\begin{figure}[htbp]
\begin{center}
\includegraphics[width=0.9\linewidth]{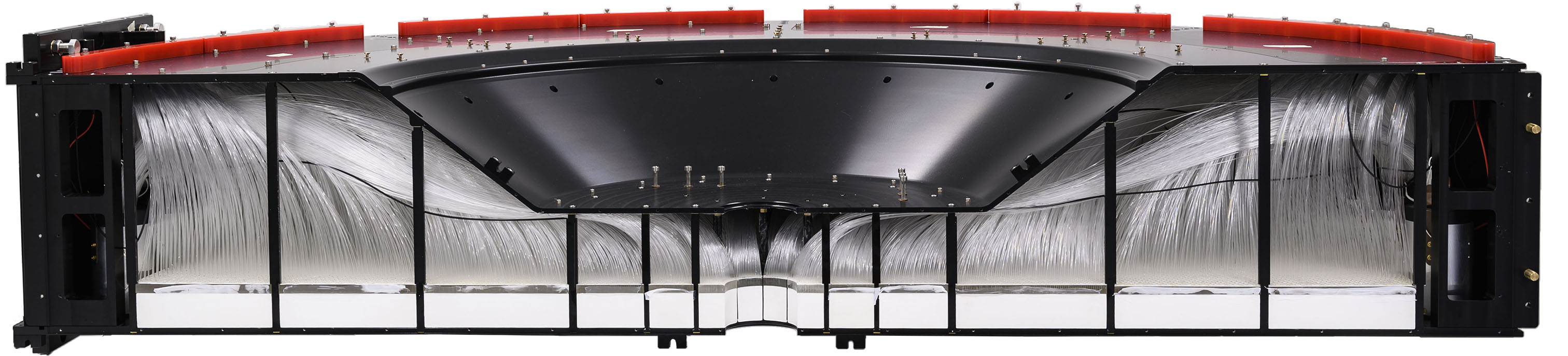}
\caption[Picture of FV0]{Photograph of one half of the FV0. The optical fibers connect the scintillators to the PMTs on the rim of the support structure, the black structure seen here. The center wall has been removed to show the scintillator, the surface matrix structure, and the optical fibers.}
\label{FV0fig}
\end{center}
\end{figure}

FV0 is a large, segmented scintillator disk with a novel light collection scheme~\cite{grabski2019new}, assuring short pulses, to achieve a single MIP time resolution of about 200\,ps, and a very uniform response across the entire detection surface. The active element of FV0 is a 4\,cm thick EJ-204 plastic scintillator divided into five concentric rings of equal pseudorapidity coverage. The outer diameter of the largest ring is 144\,cm and the inner diameter of the smallest is 8\,cm. The four inner rings are subdivided into eight sectors of 45 degrees each, while the outermost ring, due to its large area, has 16 sectors. A grid of equal-length, clear Ashai fibers is attached to the back side (as viewed from the IP) of the scintillator as can be seen in Fig.~\ref{FV0fig}. At the other end, the fibers from each sector are bundled and optically coupled to Hamamatsu R5924-70 PMTs. This way, the 48 sectors of FV0 are mapped to 48~independent readout channels. This segmentation, combined with the information from the other forward detectors, is sufficient to yield the required centrality and event plane resolution. Together with FT0, FV0 provides the needed input to generate minimum bias and multiplicity triggers at the 'minus one' trigger level (LM). With a total latency below \SI{425}{\ns}, this is the fastest trigger in ALICE. In addition, the FV0 monitors the LHC background conditions and luminosity. 

\subsubsection{FDD}
The FDD~\cite{Rojastorres} comprises two nearly identical arrays, FDD-A and FDD-C, surrounding the beam pipe on opposite sides of the IP. Each array consists of eight rectangular scintillator pads with a size of $21.6 \times 18.1 \times 2.5\,$cm$^3$. These eight pads are assembled in two overlapping layers of four sectors each. To make clearance for the beam pipe, a quadrant was removed from the innermost corner of each scintillator plate. The radius of the removed quadrant is \SI{6.2}{\cm} on the FDD-A scintillators and \SI{3.7}{\cm} on the FDD-C, as illustrated in Fig.~\ref{FITschematic}. Each pad has two wavelength shifting (WLS)
bars attached to the opposite sides of the scintillator. Clear optical fibers carry the light
from the WLS to H8409-70 PMTs. There are eight independent FDD channels on each side of the IP.

The FDD covers a large pseudorapidity interval (see table in Fig.~\ref{FITschematic}) and is sensitive to the presence of even a single MIP. As such, it is an ideal system to tag interactions characterised by large rapidity gaps as those from photon-induced ultra-peripheral collisions or diffractive processes. The main physics goals to which the FDD contributes to the pp program are the studies of centrally produced exclusive states, measurements of cross sections for single and double diffraction, and inelastic processes. Regarding the physics objectives in \PbPb{} and \pPb{} collisions, the FDD provides an independent measurement of centrality based on the charged-particle multiplicity in an intermediate pseudorapidity range between the ITS and the ZDC and contributes to the selection of ultra-peripheral collisions as well as their classification into exclusive or dissociated channels.

\subsubsection{Electronics and readout scheme}

\begin{figure}[htbp]
\begin{center}
\includegraphics[width=0.9\linewidth]{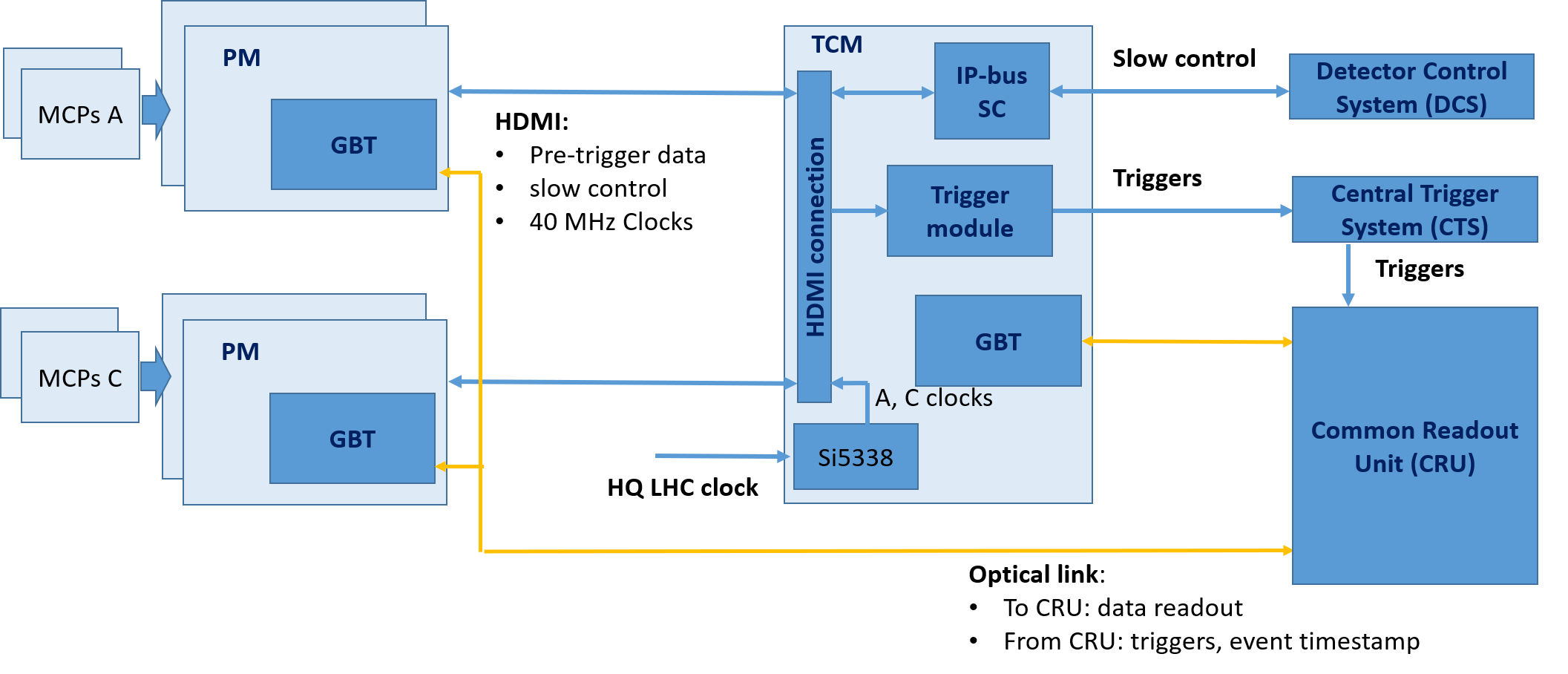}
\caption[Block diagram FIT readout]{Schematic diagram of the FIT readout electronics. The MicroChannel Plates (MCP) are described in the FT0 section.}
\label{fig:ft0_elec}
\end{center}
\end{figure}

All three subsystems of FIT use the same front-end and readout electronics based on just two custom-designed modules: a Processing Module (PM) and a Trigger and Clock Module (TCM). One PM provides twelve independent inputs. Each subdetector has only one TCM while the number of PMs is determined by the number of channels. Each PM is connected to the dedicated TCM via an HDMI cable to transmit "pre-trigger" data, slow-control data, and the LHC clock. The commands, configuration data, and status data are sent from the detector control system to the TCMs via a 1\,Gb Ethernet optical link using an IPbus (UDP-based protocol)~\cite{Finogeev:2020qkf}. A schematic diagram of the FIT electronics using FT0 as an example, is shown in Fig.~\ref{fig:ft0_elec}. The triggers and the measured event rates for the luminosity
measurements are transmitted from the TCMs via the same connection. The PMs are
configured from TCMs via an HDMI SPI connection. The PMs and TCMs are connected
to the ALICE DAQ with GBT links. The FIT delivers the produced trigger signals to the central trigger system.
Custom-made laser calibration systems provide pulses used for time and amplitude calibration, as well as monitoring of ageing and radiation damage of the FIT detector components.

The FIT detectors were installed in ALICE in 2021. Initial commissioning of the detectors was performed in October 2021 with low-intensity proton beams in the LHC at a collision energy of 450\,GeV. These pp collisions were used to check both the integration of the FIT detectors in the ALICE data processing chain, and to get a first, preliminary look at FIT performance.  We note that such low multiplicity collisions give only a lower bound on the performance of the FIT detectors, particularly on the time resolution of FT0. Using this data set, the time resolution of FT0 was found to be 26\,ps.  Further checks on this first data set are being performed, and full integration with the online systems and software framework is being completed at the time of writing.

\clearpage
\subsection{Muon System}
\label{sec:muon}

The forward muon arm was described in ~\cite{Aamodt:2008zz}. It consists of a
composite absorber of about 10 interaction lengths, made from layers of
both high- and low-Z materials located at a distance of 90\,cm from the nominal
interaction point, a large dipole magnet with a 3\,Tm integrated field
placed outside the L3 barrel magnet, ten planes of very thin, high
granularity, cathode pad tracking stations. The muon arm is completed by a second muon filter (seven interaction lengths of iron) located after the last tracking
station and upstream from four planes of resistive plate chambers which are used for muon identification.
The spectrometer is shielded by a dense conical
absorber tube, of about 60\,cm outer diameter, which protects the
chambers from secondary particles created in the beam pipe.

The increased luminosity of the LHC at the ALICE interaction point after LS2 required an upgrade of
the front-end and readout electronics on both the muon
tracking and muon identifier subsystems.

\subsubsection{Muon Tracking}
\label{sec:mch}

\begin{figure}[h]
  \centering
  \includegraphics[width=.45\textwidth]{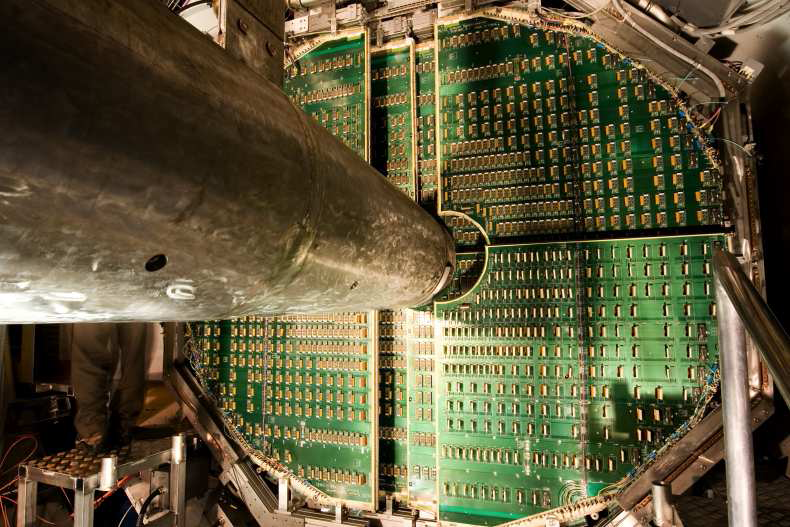}
  \hspace{0.05\textwidth}
  \includegraphics[width=.45\textwidth]{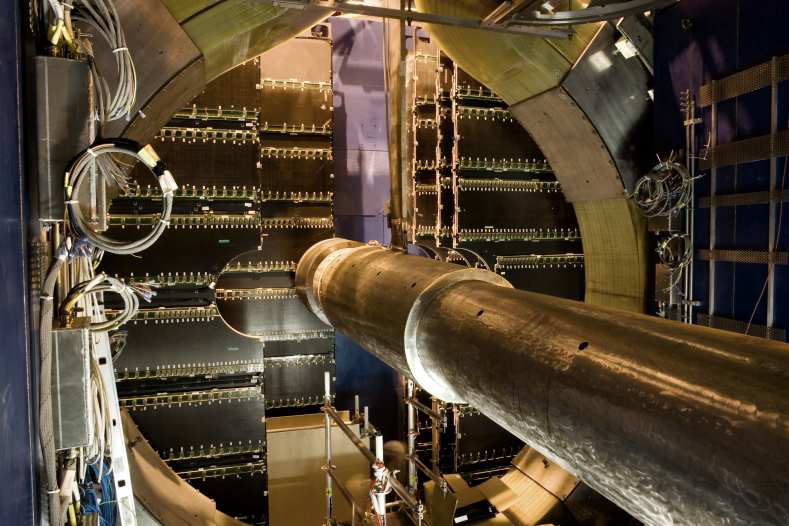}
   \caption[Tracking system]{Left: Station 2 of the tracking system; the readout electronics are distributed on the surface of a quadrant. Right: Stations 4 and 5 of the tracking system; the readout electronics are mounted along the top and bottom edges of the slats.}
  \label{quad+slat}
\end{figure}

The Muon Tracking Chambers (MCH)~\cite{Aamodt:2008zz} consist of 156 multiwire proportional
chambers with cathode pad readout (cathode pad chambers) with more than
one million electronic channels.
The system has five tracking stations, each of which is composed of two
chambers. Because of the different sizes of the stations, ranging from a few square meters for station 1 to more than 30\,m$^{2}$  for station 5, two different designs were adopted. The first two stations are based on a quadrant structure, with the readout
  electronics distributed on their surface (left panel of Fig.~\ref{quad+slat} ). Four independent quadrants form one chamber. For the larger stations (3 to 5), a slat architecture was chosen (right panel of Fig.~\ref{quad+slat} ). The largest slat size is 40 $\times$ 240\,cm$^{2}$ and the electronics are mounted on the top and bottom parts of
  each slat. Slats are mounted on a support to form one
  half-chamber. One half-chamber consists of 9 slats for station 3,
  and 13 slats for stations 4 and 5. The tracking system
  covers a total area of about 100\,m$^{2}$.

The detector chambers are unchanged from Runs~1 and 2, while the front-end and the readout electronics were upgraded to accommodate the larger interaction rates for Runs~3 and 4.

\begin{figure}[t]
  \centering
  \includegraphics[width=1\textwidth]{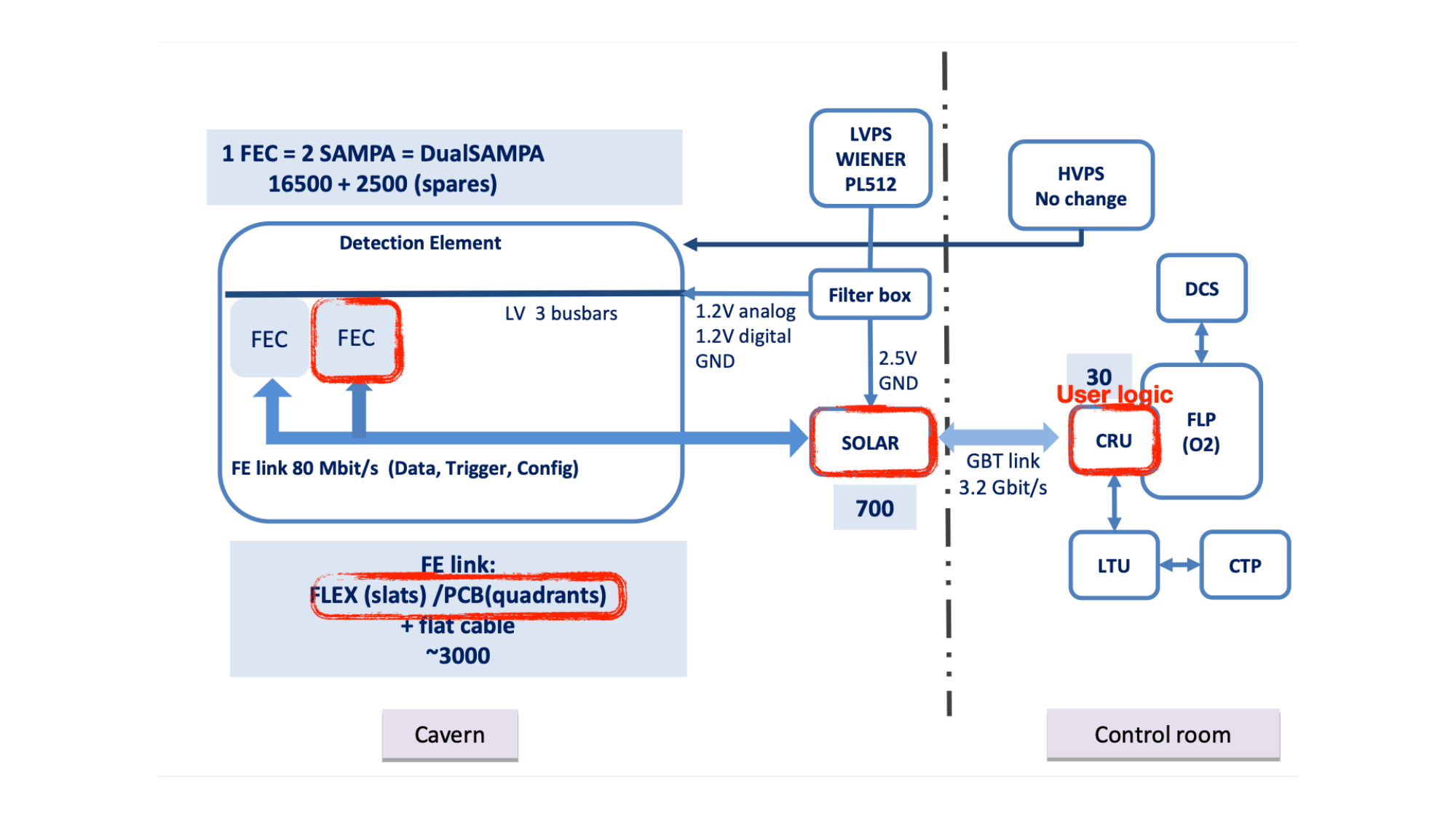}
     \caption[Tracking readout scheme]{The Muon Tracking readout scheme}
  \label{readoutscheme}
\end{figure}

The electronics can run either in the default dead-time-free continuous
mode or in triggered mode. The readout data flow is schematically shown in
Fig.~\ref{readoutscheme}. The Front-End Cards (FEC) continuously send
data at 80\,Mbit/s through an electrical link to the SOLAR (Sampa to
Optical Link for Alice Readout) readout boards which connect to the CRU (see Sec.~\ref{sec:cru})
through GBT optical links at 3.2\,Gbit/s.

\paragraph{The DualSAMPA front-end electronic cards\\}

\begin{figure}[t]
  \centering
  \includegraphics[width=1\textwidth,height=8cm]{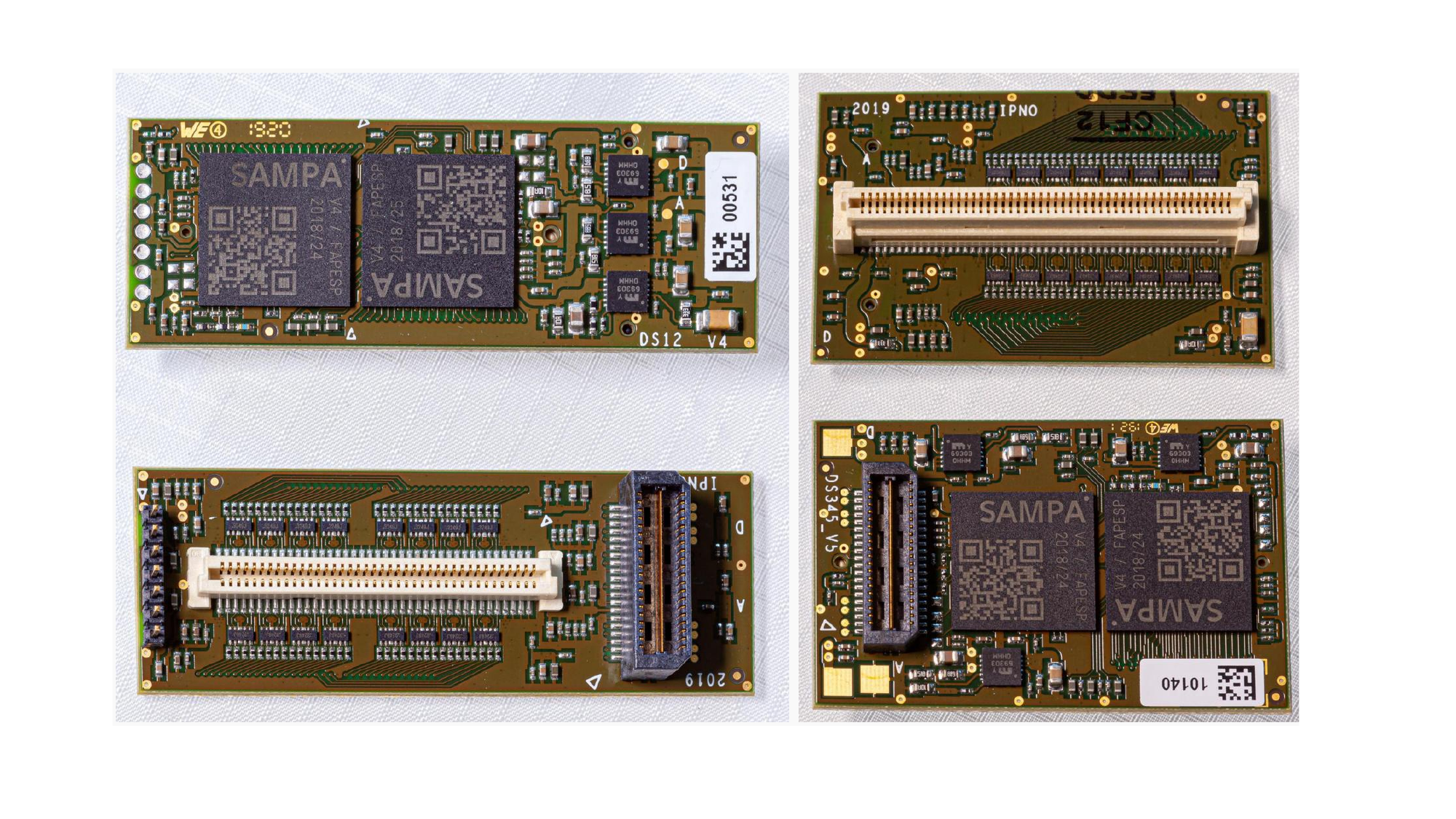}
     \caption[DualSAMPA]{The two geometries of the DualSAMPA boards (DS12 on the left and DS345 on the right),
       with the white connector plug socket on PCB and on the other side the black connector connecting to SOLAR boards.}
  \label{dualsampa}
\end{figure}

The front-end electronic cards, called DualSAMPA, host two chained SAMPA chips (see Section~\ref{sec:sampa}) of 32
channels each and three low voltage regulators. Since the detectors are the same
ones used in Runs 1 and 2, the dimensions and the layout of the connectors for the DualSAMPA cards on the
electronic PCBs remain the same as for the previous FEC. Moreover, two
types of cards were produced, each with the same functionalities but with
different dimensions to suit the quadrant and
slat detector layouts. The DualSAMPA board is shown in Fig.~\ref{dualsampa}).

Out of the 19300 DualSAMPA produced (11000 DS345 for slats of stations 3, 4, and
5, and 8300 DS12 for quadrants of stations 1 and 2), 16900 are
installed in the cavern (9700 DS345, 7200 DS12).

\paragraph{The readout electronic FLEX links and large electronic PCBs\\}

\begin{figure}[t]
  \centering
  \includegraphics[width=.45\textwidth,height=5cm]{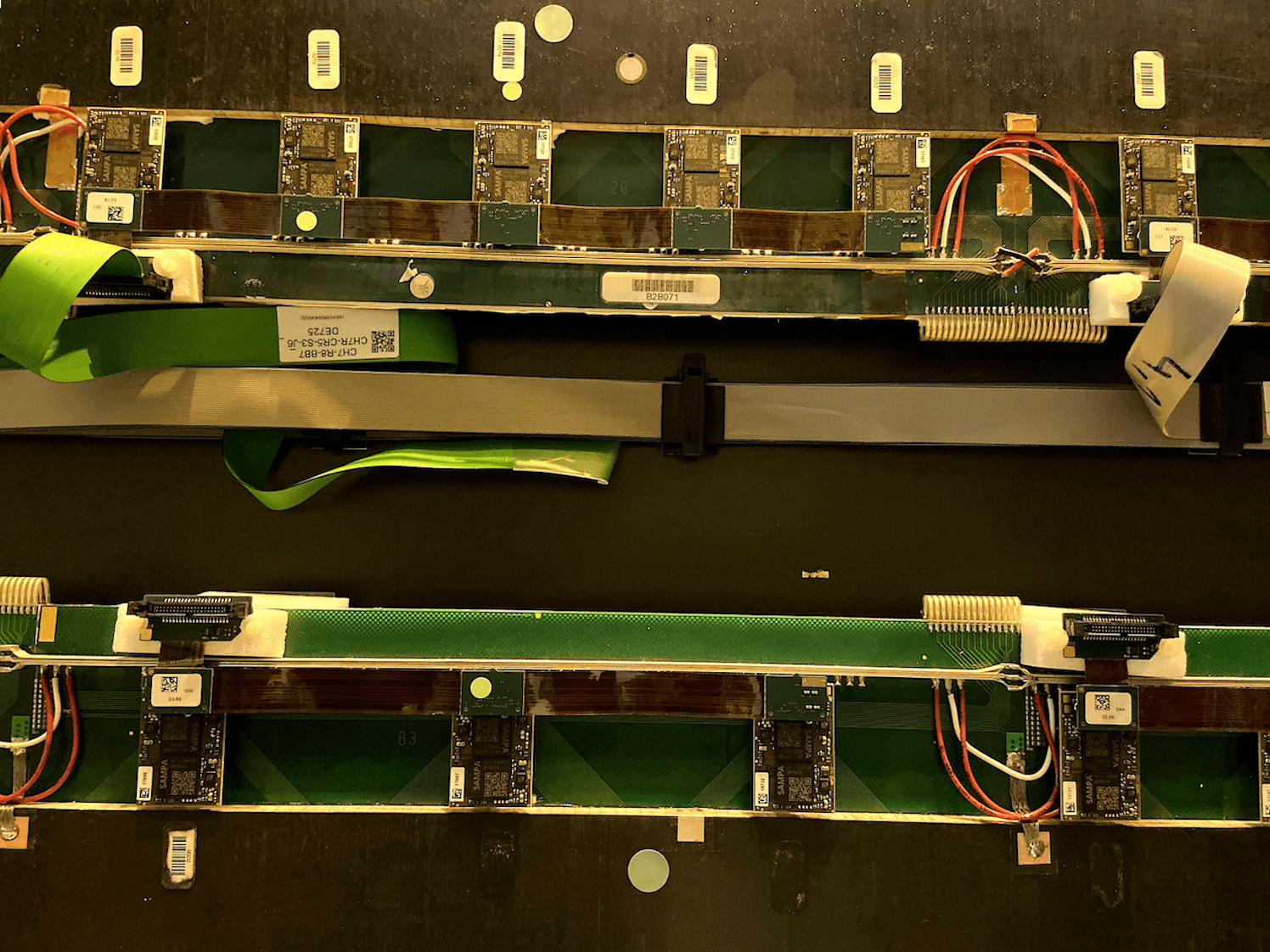}
  \hspace{0.05\textwidth}
  \includegraphics[width=.45\textwidth,height=5cm]{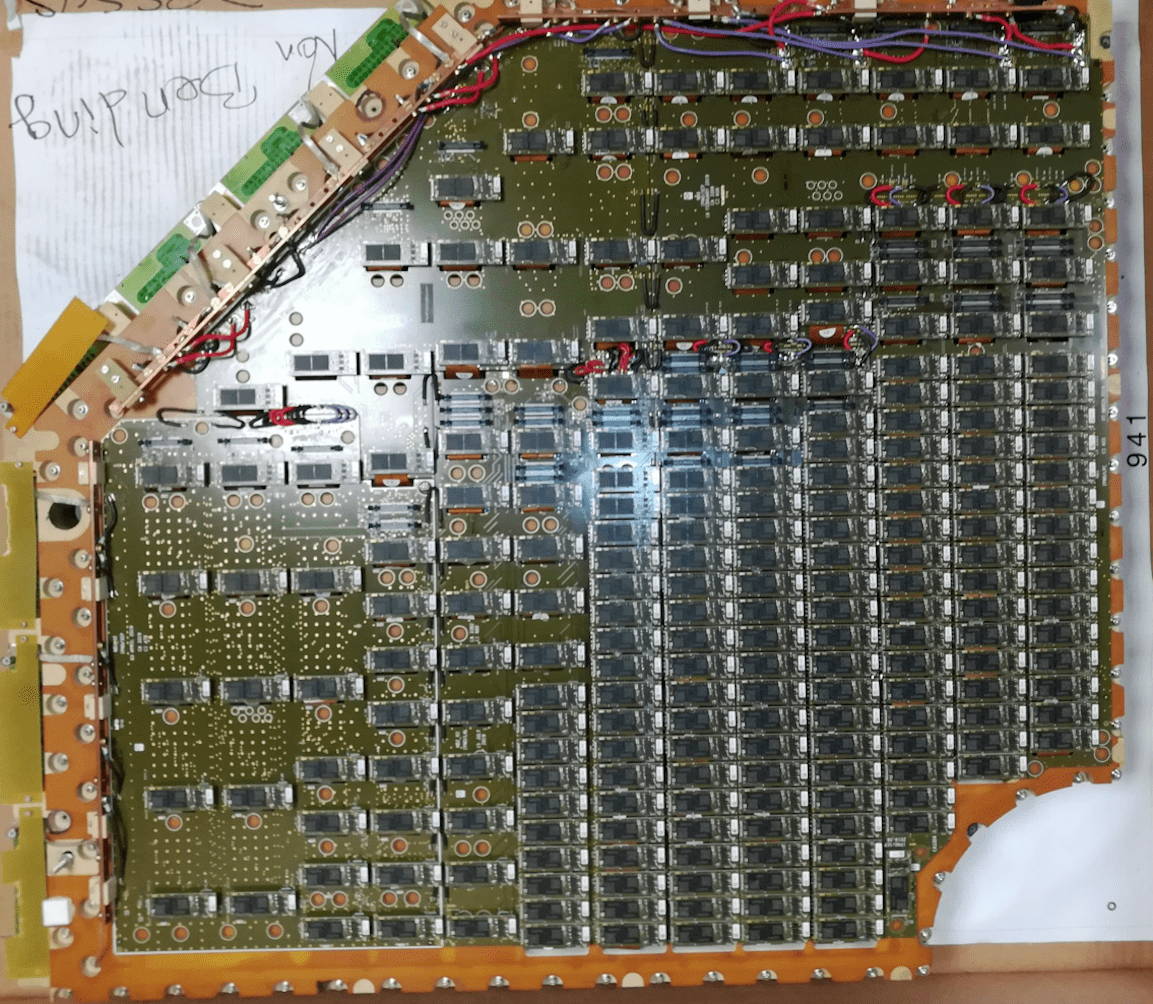}
   \caption[Electronic links]{Left: Flex mounted on a slat connecting five
     DualSAMPA cards linking through a green flat ribbon cable to the readout
     board. Right: Large electronic PCBs covering the surface of a quadrant.}
  \label{flex+pcb}
\end{figure}

The link between the DualSAMPA and the readout cards consists of a
flexible circuit (FLEX) and a flat ribbon cable for the slats, while a
large electronic PCB and a flat ribbon cable are used for the
quadrants (see left and right panels of 
Fig.~\ref{flex+pcb}).

\begin{figure}[t]
  \centering
  \includegraphics[width=0.8\textwidth, height= 4cm]{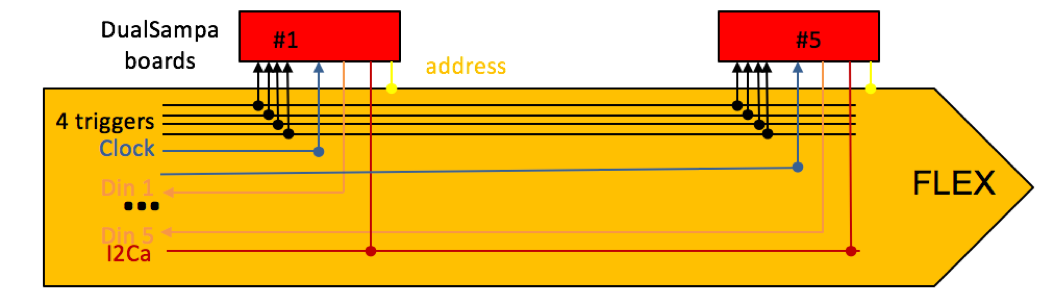}
     \caption[FLEX scheme]{ The FLEX scheme.}
  \label{flexscheme}
\end{figure}

Each DualSAMPA has dedicated data and clock lines while the
trigger lines are daisy chained to feed up to 5 DualSAMPA (see
Fig.~\ref{flexscheme}) using an I2C line. An active buffer was added to the I2C line to ensure a
good signal integrity.

More than 3000 FLEX PCBs  of 24 different types were produced depending on the number of DualSAMPA to address, the geometry, and the pad density; 2760 of these were
installed, the remainder serving as spares.
 
\begin{figure}[t]
  \centering
\includegraphics[width=.45\textwidth,height = 6 cm]{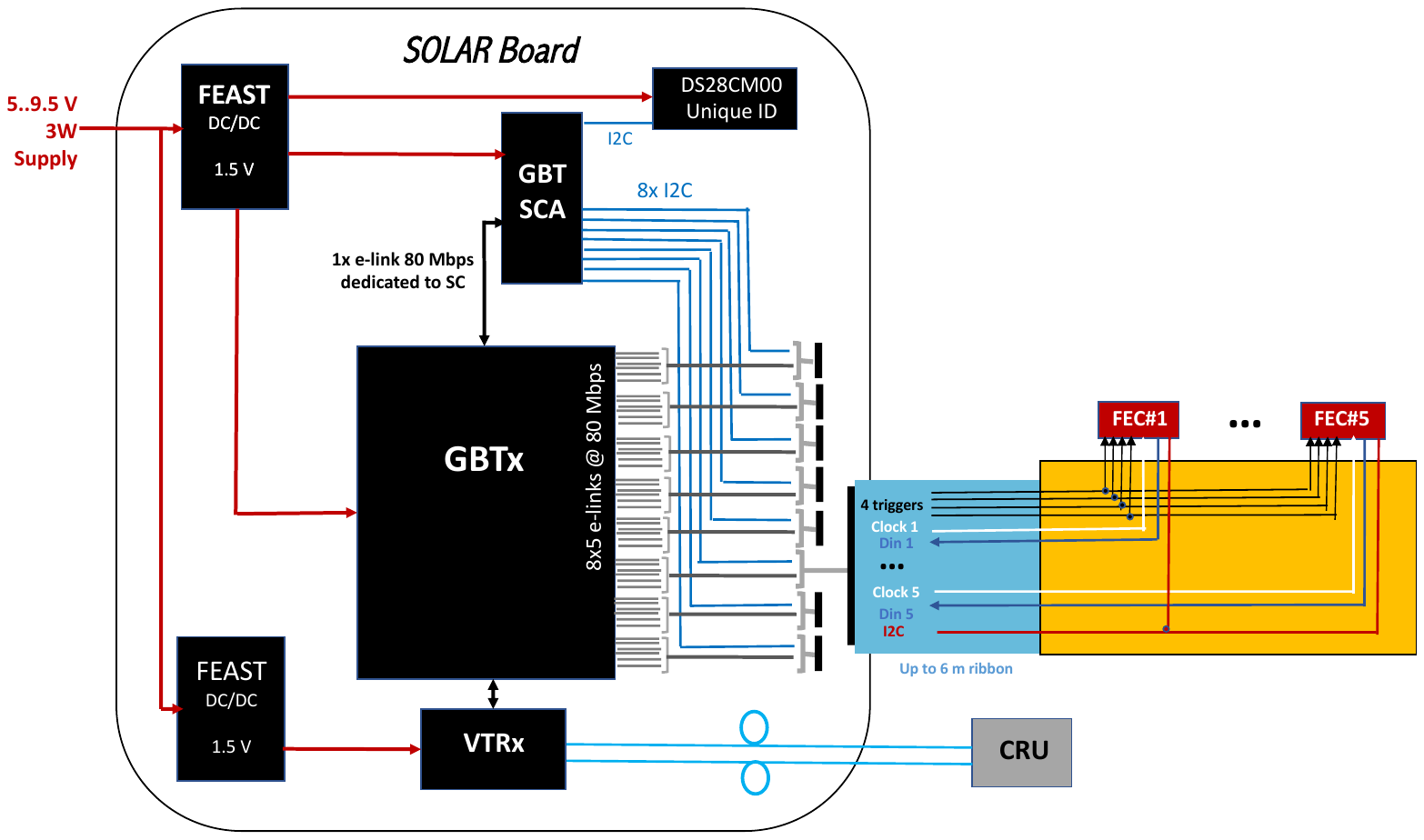}
  \hspace{0.05\textwidth}
  \includegraphics[width=.45\textwidth,height = 6 cm]{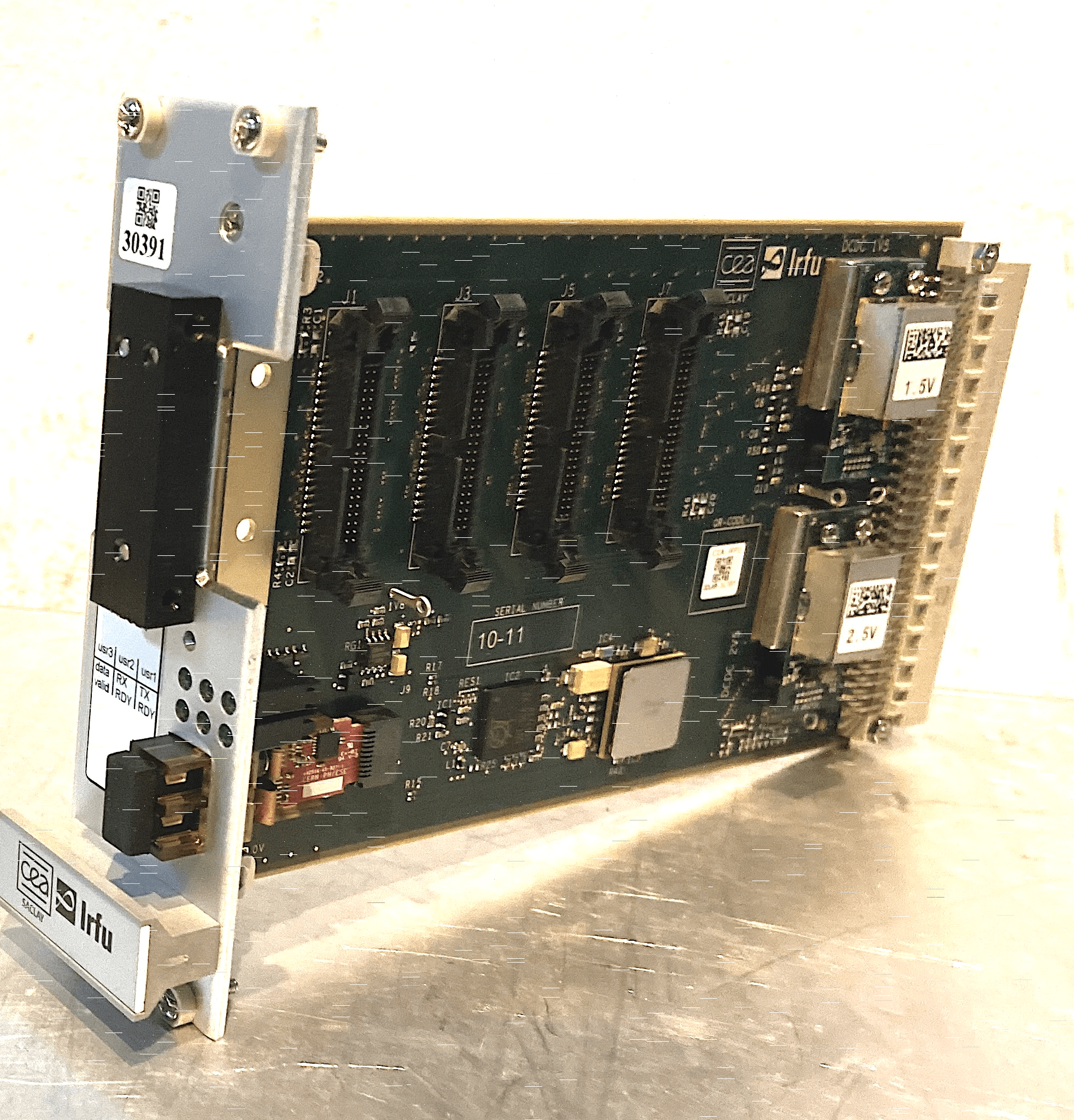}
     \caption[SOLAR scheme]{The SOLAR board: functional diagram (left panel) and photo of the board itself (right).}
  \label{solarscheme}
\end{figure}

\paragraph{The SOLAR readout cards\\}
Each FLEX/ribbon cable is plugged into one of the eight ports of a SOLAR readout
board, allowing this latter to read out up to 40 DualSAMPA boards (see
Fig.~\ref{solarscheme}). The GBTx chip of the SOLAR board acts as a serializer to send the
signals from the different DualSAMPA to the CRU through GBT optical links. The SOLAR board
also hosts a GBT-SCA chip handling the eight I2C
command and control lines, one optical transmitter/receiver VTRx and two
DC/DC FEAST converters. 
 
A total of 624 SOLAR boards are hosted in 112 custom SOLAR crates, with up to six boards each.

\paragraph{The data flow from SAMPA to the CRU User Logic\\}
In the SAMPA chip, the signal of each electronic channel is amplified with a gain of 4\,mV/fC,
waveformed with a shaping time of 300\,ns, then sampled and digitized at 10\,MHz in
a 10-bit ADC, and is eventually digitally processed with a baseline
correction and a zero-suppression before being formatted. The SAMPA format
consists of data samples from a signal waveform with its time stamp and
size together with a header containing mainly the bunch crossing number, the SAMPA
address and the channel address of the SAMPA chip.

The signals of the 64 channels of the two chained SAMPA chips of a FEC
are serialized at 80\,~Mbit/s (2 bits at 40\,~MHz). The first port of the
SOLAR board handles the first 2 bits of the first DualSAMPA while the second port takes care of the 2 bits of
the second DualSAMPA and so on, combining all 40 ports, which results
in a 3.2\,~Gbit/s data optical transmission to one input of a CRU. The
electrical and optical links are always transmitting data, independent
of the type of information (physics data, synchronisation, etc.).

\begin{figure}[t]
  \centering
  \includegraphics[width=0.7\textwidth,height=3cm]{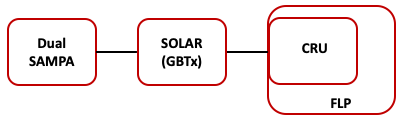}
     \caption[Data flow]{Data flow scheme.}
  \label{dataflow}
\end{figure}

\begin{figure}[h]
  \centering
  \includegraphics[width=1\textwidth]{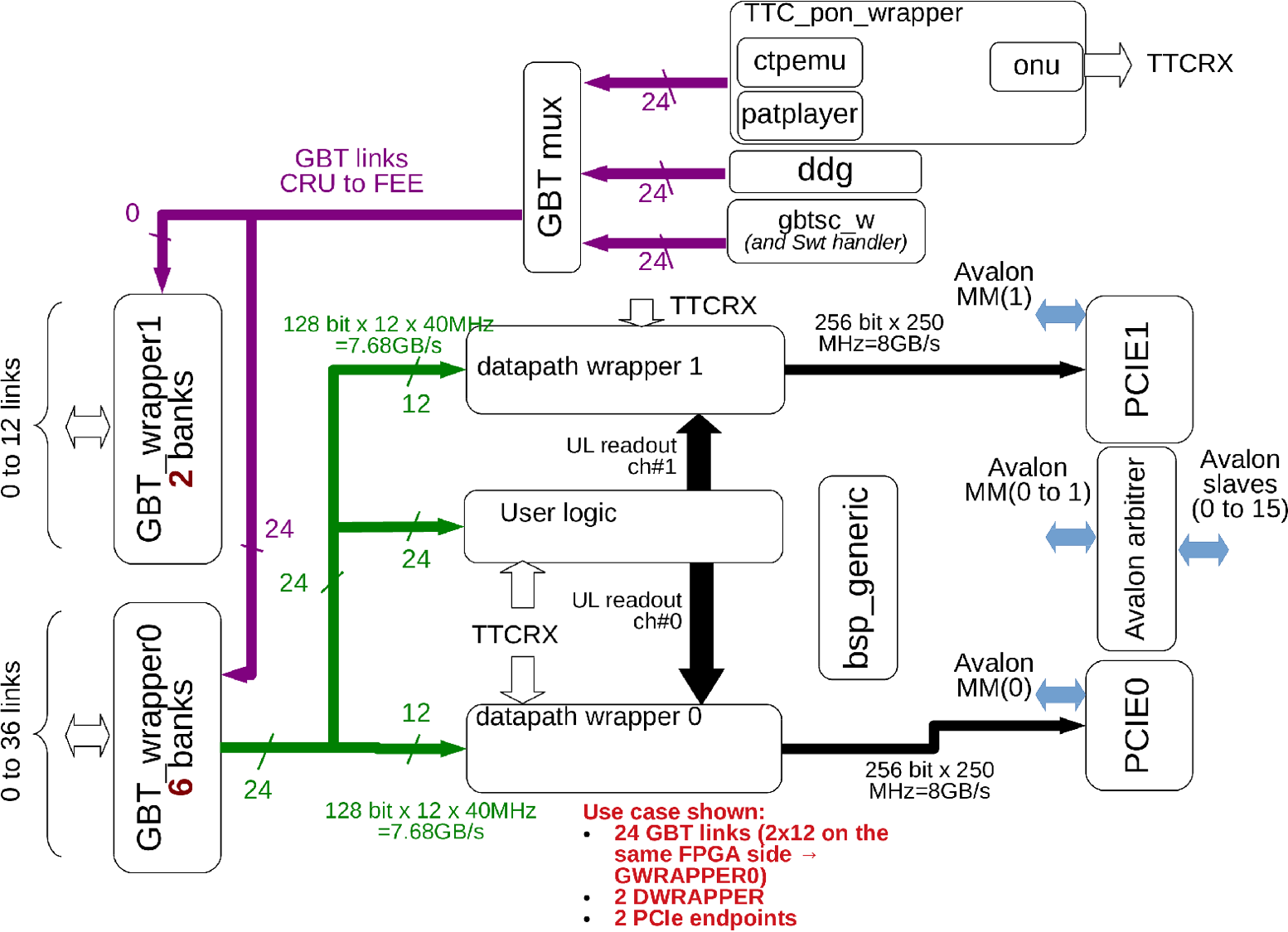}
     \caption[CRU Scheme]{CRU scheme.}
  \label{cru-ul}
\end{figure}

The MCH CRU user logic receives data from the 24 GBT links (see Sec.~\ref{sec:cru}), each one handling 40 DualSAMPA channels. For each GBT
link, the user logic deserializes the 80 bits, forms the SAMPA
words, removes the SYNCH words and inserts error checks and
configuration conditions. These 64-bit SAMPA words contain the payload, the GBT
link identifier, the DualSAMPA channel identifier, and error bits. The
user logic then embeds the TTC signals into this stream, constructs
the RDH (Readout Data Header) and transmits words of 256 bits to the Front Level
Processor (FLP, see Sec.~\ref{sec:flp})  (see Figs.~\ref{dataflow} and~\ref{cru-ul}).

No specific processing is performed in the FLPs. The cluster finding, the cluster fitting and the track finding are done on the EPNs (see Section
\ref{sec:epn}).

Quality Control (QC) processes have two steps: the QC error check task
is perfomed on the entire raw data at the FLP level, while the
detector occupancy and pseudo-efficiency are monitored from decoded
data samples in QC tasks on EPNs. While the detector occupancy QC uses
digits (signal from each front-end electronic channel),  the
pseudo-efficiency task uses the pre-clusters (groups of pad hits that
are close in time and space). The charge of the pre-clusters is also
verified. These tasks will allow the monitoring of the detector performance.

\subsubsection{Muon Identifier}

The Muon Identifier (MID) is the present designation of the Muon Trigger system~\cite{Aamodt:2008zz}, which was operational in ALICE during LHC Runs~1 and 2. 

\begin{figure}
\centering 
\includegraphics[width=.6\textwidth]{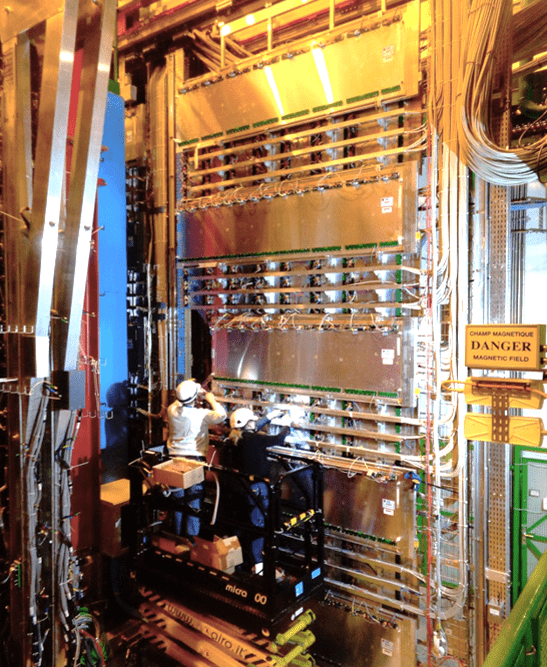}
\caption{Overview of one MID half-plane in open position.}
\label{midhalfplaneopen}
\end{figure}

The detector is composed of 72 single gap Resistive Plate Chamber (RPC) detectors, organised in two stations of two planes each, located at a distance of 16\,m and 17\,m from 
the interaction point, respectively. In both stations the two planes are 17\,cm apart. The total detection area is about 150\,m$^{\rm 2}$. An overview picture of one half-plane of the MID, in open (maintenance) position, taken during the FEERIC card 
installation (see next section) in 2019, is shown in Fig.~\ref{midhalfplaneopen}.

The RPC signals are collected by means of a total of 20992 readout strips, each of them connected to Front-End Electronics (FEE).
The output signals from the FEE, in LVDS standard with a width of 25\,ns,
are propagated via multiwire copper cables to the local cards, which are part of the readout electronics.

The FEE cards, which are located on the RPC detectors, were replaced during LS2. The main motivation is to reduce the ageing of the RPCs and improve the rate capability during the upcoming data taking periods. 
The ASIC of the past FEE, called ADULT~\cite{mid:ADULT}, was upgraded to a new type, called FEERIC~\cite{mid:FEERICref1,mid:FEERICref2}.
Unlike ADULT, FEERIC performs amplification of the RPC analog signals. Thanks to this upgrade, the ALICE RPCs can be operated at lower gain, with a reduction by a factor 3--5 of the charge produced in the gas gap,  
hence limiting ageing effects.

The readout electronics, composed of 234 local cards and 16 regional cards, was also completely replaced to sustain the larger data flow associated with the higher collision rate in Runs~3 and 4.

Although all RPC detectors were still operational at the end of Run~2, a few of them were drawing a relatively large current after having accumulated a charge up to \SI{20}{\milli\coulomb\per\square\cm} in the gas.
It was therefore decided to replace those RPCs with completely new ones. For the longer term, a crucial R\&D on new environment-friendly 
gas mixtures~\cite{mid:RPCgasmix} for RPCs, based on tetrafluoropropene, which is characterised by a very low Global Warming Potential (GWP), has been launched.

\paragraph{FEERIC electronics\\}
 
\begin{figure}
\centering 
\includegraphics[width=.8\textwidth]{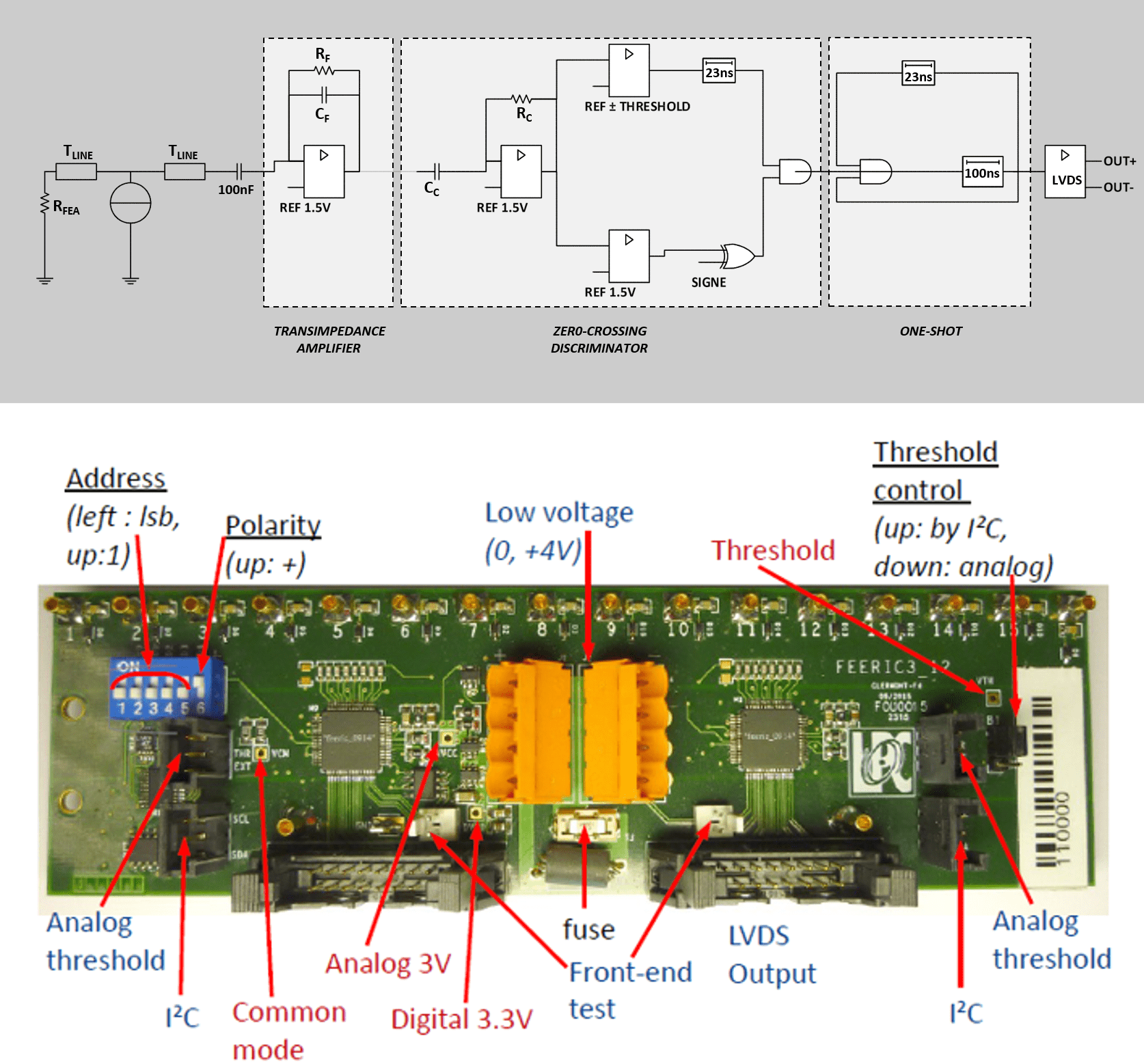}
\caption[FEERIC architecture and picture]{FEERIC ASIC architecture (top) and FEERIC card picture (bottom).}
\label{midFEERIC}
\end{figure}

The FEERIC 8-channel ASIC is designed in the AMS \SI{0.35}{\micro\meter} CMOS technology. Its main components are (see Fig.~\ref{midFEERIC}, top scheme) a transimpedance amplifier, a zero-crossing discriminator, and a one-shot circuit
which inhibits retriggering during \SI{100}{\nano\second}. The operating threshold is typically \SI{70}{\milli\volt} corresponding to a charge of approximately \SI{130}{\femto\coulomb} at the readout strip level. Details of the performance of the FEERIC electronics 
are given in~\cite{mid:FEERIC-PRR}. Figure~\ref{midFEERIC}, bottom panel, shows a picture of a FEERIC card. A total of 2720 cards (spare included) were produced in the second half of 2017.
The installation of the FEERIC cards on the RPCs in the ALICE cavern was completed in July 2019. 

During Run~2, one of the 72 ALICE RPCs was equipped with FEERIC electronics and showed satisfactory performance and stability~\cite{mid:FEERICref2}. The charge released in the gas gap, around 30\,pC per charged particle 
crossing the RPC at nominal high voltage, was typically four times smaller as compared to the one with ADULT.

\begin{figure}
\centering 
\includegraphics[width=.8\textwidth]{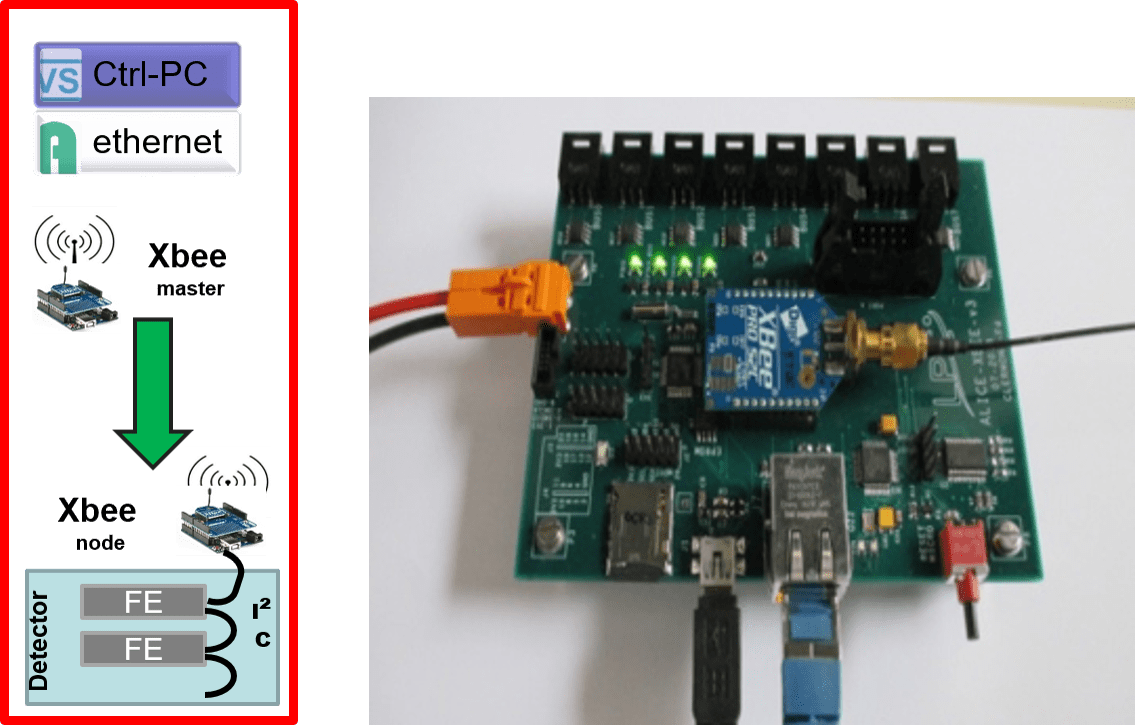}
\caption[Wireless threshold distribution]{Wireless threshold distribution scheme (left) and master or node electronics card (right).}
\label{midTHRdistri}
\end{figure}

A new wireless threshold distribution for the FEERIC cards was developed. 
A total of two masters (one on each side of the cavern) and 24 nodes close to the RPCs (see Fig.~\ref{midTHRdistri}, right panel) were installed in the ALICE cavern in 2019 to remotely control the threshold of each of the 2384 installed 
FEERIC cards. The masters are controlled via ethernet and communicate via the high level ZIGBEE wireless protocol with the nodes, which are I2C chained to the FEERIC cards on the RPCs (see Fig.~\ref{midTHRdistri}, 
left panel). A further upgrade of this system to a more powerful WiFi Ethernet-based system is planned for the winter shutdown of 2022. Both master and node share the same hardware and firmware. A card acts either as master or node, depending on the configuration stored in its EEPROM which also retains the memory of the last requested threshold values. The latter are restored at power on.

\paragraph{Readout electronics\\}

The LVDS digital signals from the FEERIC electronics, so-called strip-patterns of 16\,bit length (one bit corresponding to one RPC readout strip), are received by the local cards.
Each local card receives the strip patterns corresponding to 16 horizontal and 16 (or 8 in some cases) vertical readout strips, on both sides of the same RPC, from each of the four detection planes. The vertical readout strips (maximum length 73\,cm) 
cover the full RPC width while the horizontal strips (maximum length 55\,cm) are segmented all along the RPC length. The details of each local card inputs are given in~\cite{mid:LOC-MTR}. The entire setup consists of 234 local cards, housed in 16 VME-9U 
crates used as mechanical support and for power supply.
The signals from up to 16 local cards are collected by a regional card via the e-links on the J2 bus in each crate. Each regional card is interfaced to a Common Readout Unit (CRU) by means of two GBT links at 3.2\,Gb/s. 

\begin{figure}
\centering 
\includegraphics[width=1.\textwidth]{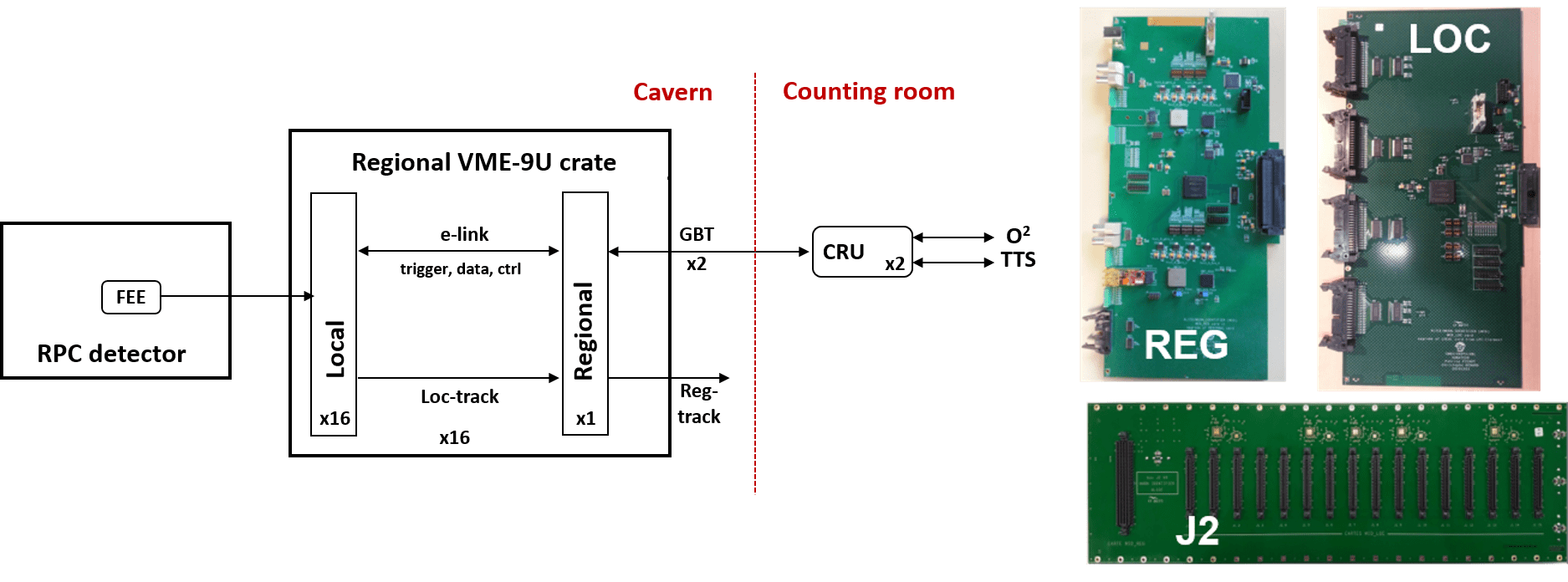}
\caption[MID readout architecture]{MID readout architecture (left) and readout cards (right picture) with local (LOC) (top right), regional (REG) (top left) and J2 bus (bottom) between local and regional.}
\label{midRO}
\end{figure}

In total this project has two CRUs, housed in one single FLP desktop computer. The MID readout architecture is shown in the left panel of Fig.~\ref{midRO}, while a picture of the three types of readout cards 
is shown in the right panel. Simulations of the expected bandwidth for Pb--Pb collisions at 50\,kHz rate, based on Run~2 data, indicate that 
this design includes a safety factor of more than one order of magnitude. 

Data corresponding to MID self-triggered physics events~\cite{mid:ROweb,mid:DataNote} are transmitted from the local and regional cards to the CRU using the GBT up-links. Colliding beam particles in the LHC are of course the main source of such events. 
Every 25\,ns (40\,MHz) a new self-triggered event is potentially stored in the local and regional FIFOs. Only non-empty events are stored in these FIFOs. In the standard configuration a non-empty event corresponds to, at least, a non-zero strip pattern.

It is important to note that it takes five clock cycles at 40\,MHz, per self-triggered event, for the transfer of the regional FIFO data and 9--21 clock cycles for the local ones, depending on the
number of non-empty detector planes in this last case. This means that the data from different local and regional cards, corresponding to the same bunch crossing, arrive asynchronously in the CRU. This also means that the 
local and regional FIFOs saturate in case they are filled at the full clock frequency of 40\,MHz.
For instance, the FIFO saturation could happen in case of a very high level of noise at the FEE level. It should be noted that a busy bit would be set in such a case.

At the first stage of data processing in the CRU, the user logic performs zero-suppression and raw data header construction using the central trigger (CTP) orbit information. 
The output of the user logic is transmitted by words of 256 bits to the FLP. At this level, the data coming from the different GBT links are assembled in C++ structures and synchronized to provide the information 
corresponding to a given bunch crossing. The local and regional cards always operate in continuous readout mode. However, the system can also run in triggered mode, with the CRU transmitting only data corresponding to a time window centred 
on a bunch crossing which coincides with a trigger from the CTP.

The local and regional cards respond also to all types of triggers delivered by the CTP and received via the GBT down-link~\cite{mid:ROweb}. 

The commissioning of the complete MID detection chain has started in the autumn of 2020. First muon tracks, in coincidence with MFT and MCH, have been registered in October 2021 by dumping proton beams in the TED, as explained in Sec.~\ref{sec:mft_commissioning}.

\clearpage
\subsection{Transition Radiation Detector}

The construction, operation and performance of the TRD is presented in~\cite{Acharya:2017lco}.
The TRD contributes to the overall momentum determination of charged particles, as it provides up to six track segments (tracklets, each of a geometric length of about \SI{3.5}{\cm}) for each charged particle in the acceptance ($|\eta|<0.9$). In addition, the TRD allows to identify electrons via the detection of transition radiation. Using a likelihood method enhanced with machine learning techniques, it is possible to suppress pions by a factor of more than 100 while retaining an
 electron (positron) efficiency of 90\%.

The TRD has the capability to trigger on events based on the charged particle content, incl. the identification of electrons, within about \SI{8}{\us} after a collision. This feature was used in LHC Run~2 to select collisions with charmonia, jets or atomic nuclei.
Here we focus on modifications implemented for the high rate running in LHC Runs~3 and 4.

\subsubsection{High-voltage distribution and common mode}

During TRD operation in Runs~1 and 2 a number of anode and drift channels of individual chambers developed high currents and were eventually not operational any more. Based on similar experience on the TPC, and on experience from the repair of one TRD supermodule (SM) during LS1, the built-in decoupling capacitors in the on-detector high-voltage distribution system were suspected to cause the observed behaviour. At that time, the construction of the TRD was still ongoing, and the last four SMs were built without certain capacitors (4.7\,nF) in the high-voltage distribution system. In total, until the end of Run~2, 70 anode channels and 20 drift channels were taken out of operation from a total of 522 chambers installed in 18 SMs.

\begin{figure}
  \centering
  \includegraphics[width=0.5\textwidth]{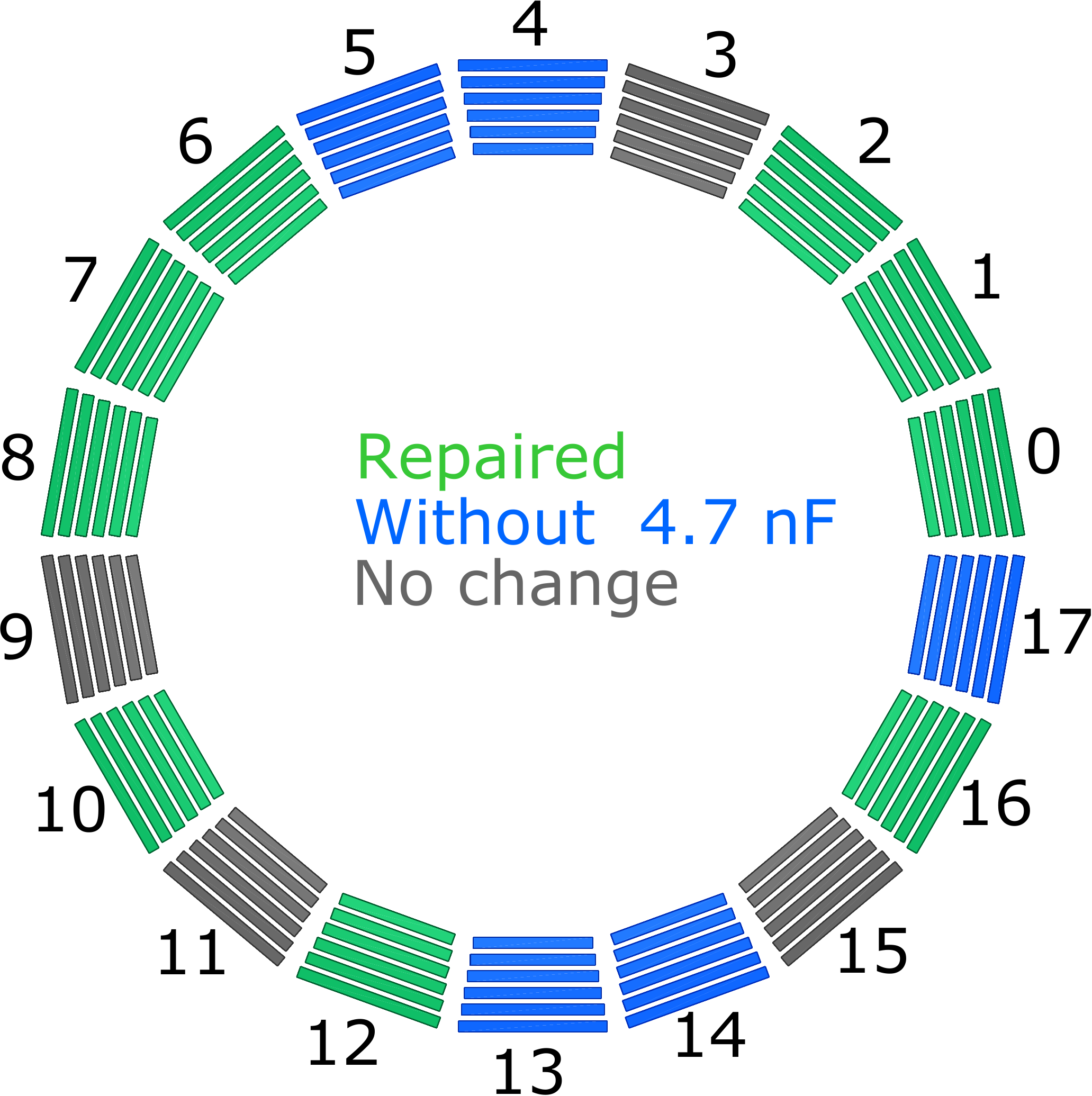}
  \caption[High-voltage status TRD supermodules]{The status of TRD supermodules concerning capacitors in the high-voltage distribution.}
  \label{fig:TRD_status}
\end{figure}

The high-voltage distribution system with the decoupling capacitors on filter boards is mounted directly on each chamber and therefore encased in the hull of the SMs. Via milling cut-outs into the casing and by removing the top cover, it was possible to access the filter boards of all 30 chambers in a SM. 
Each anode and each drift channel hold a 4.7\,nF capacitor; the anode wire plane is segmented in eight or six sectors of two pad rows each, decoupled from each other by a 2.2\,nF capacitor. Measurements of capacitors that were removed confirmed the reason for the high-voltage failures, explaining the observed issues. Most of the problems could be traced to failing 4.7\,nF capacitors, but it was found that also a small fraction (in the percent range) of the 2.2\,nF capacitors had failed. Therefore, all capacitors on the filter boards were removed from a total of 9 TRD SMs. This number was determined by the turnaround time of SM deinstallation from the space frame of ALICE, repair, test and reinstallation during the first year of LS2. Before reinstallation of each SM, long-term high voltage, low voltage, cooling and readout tests were performed to ensure proper detector operation.
It turned out that 96\,\% (80 out of 83 not operational chambers) in the nine SMs could be restored. 
Figure~\ref{fig:TRD_status} displays the configuration of the individual SMs in terms of installed decoupling capacitors.
Based on experience, the expected failures of remaining capacitors until end of Run~4 is estimated to be low enough such that good tracking capability in all sectors is ensured for the entire period of operation.

\begin{figure}
    \centering
    \includegraphics[width=0.6\textwidth]{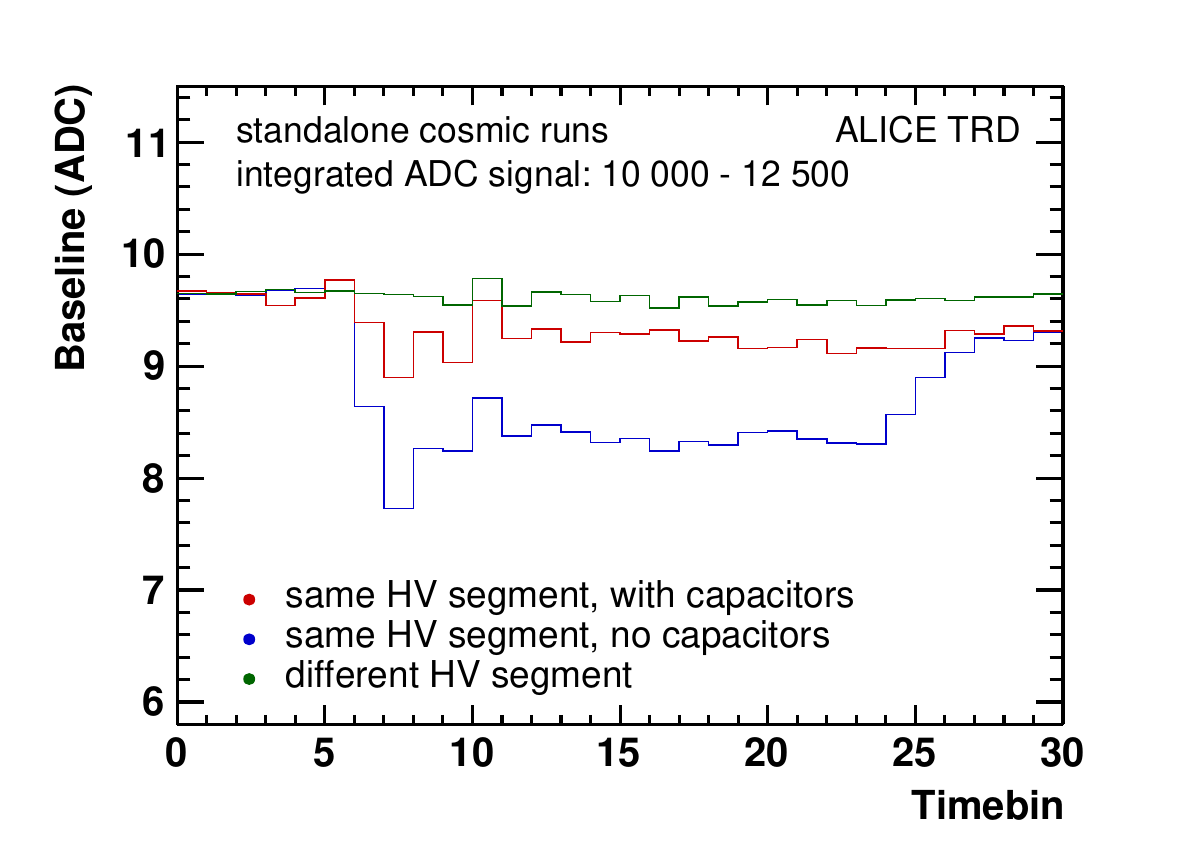}
    \caption[TRD common-mode signal]{Induced common-mode signal with and without capacitors for the anode high-voltage. The baseline of pads in the same high-voltage segment as a cosmic-ray particle with an integrated signal between 10\,000 and 12\,500 ADC counts is shown before (red) and after (blue) the removal of the capacitors. For comparision, the baseline from pads in a high-voltage segment without hits is shown in green.}
    \label{fig:common_mode_ph}
\end{figure}

As the capacitors were meant to buffer high charge deposits in the chambers, their removal results in larger induced common-mode signals on readout pads in the same high-voltage segment. The measured common-mode signal is shown in Fig.~\ref{fig:common_mode_ph} and is about three times larger than with the capacitors in place, consistent with the expectation from the remaining capacitance of the readout chamber.
This effect will be corrected at the software level based on the measured local charge deposit.

\subsubsection{Readout}

The readout chain has been optimised in the past for a high event inspection rate at Level~-1 (LM) with a fast calculation of the L1 trigger contribution (LM tracklet data readout time $< \SI{8}{\micro \second}$, L1 decision time $< \SI{6}{\micro \second}$), while transferring large, high resolution raw data for events accepted beyond the L1 level (L1 raw data readout time $\approx$ \SI{300}{\micro \second})~\cite{Acharya:2017lco}. In Run~3, the L1 trigger functionality is no longer required and the detector must provide readout rates as high as feasible while writing all events to permanent storage. No data shall be discarded in the readout chain and the fraction of recorded events in \SI{1}{\mega \hertz} interaction rate pp collisions or \SI{50}{\kilo \hertz} Pb--Pb collisions shall be maximised.

The applied solution is presented in the following sections. Simulations confirm that it enables collecting more than 70\% of the events in a \SI{50}{\kilo \hertz} interaction rate \PbPb{} running scenario.

\paragraph{Optimisation of the existing FEE.}

In order to achieve a high event-readout rate in Run~3, only tracklets are read out, a mode which has been used to find fast L1 trigger contributions in Run~2.
The maximum data volume per LM trigger and per Multi Chip Module (MCM, processing signals from 18 readout pads) is four words of 32 bits each. The usage of the available bits is no longer optimised for triggering, but for physics analysis.

Previously, each MCM processed and transmitted up to four tracklets, where each tracklet was transmitted as a 32-bit word.
However, even in the most central \PbPb{} collisions, a track density of four tracklets per MCM has been rarely reached. Therefore, only three tracklets per MCM are allowed in the Run~3 data format. The estimated fraction of tracklets lost by this measure in central \PbPb{} events is below 1\,\%.
The freed-up 32 bit word is used as a header to store position information about the MCM and eight bits per tracklet are reserved for PID information. It is followed by one to three 32-bit tracklet words that store the position within the MCM, slope and additional twelve bits of PID information of the tracklet. The details are shown in Table~\ref{tbl:tracklet:format}.

\begin{table}[!ht]
  \caption[TRD tracklet data format]{TRD tracklet data format. Each MCM that has reconstructed at least one tracklet will send a header with shared coordinate information and eight bits of PID information per tracklet. For each reconstructed tracklet, one additional payload word with additional position and PID information, as well as the reconstructed tracklet angle (slope), will be stored.}
    \label{tab:trackletformat}
    \begin{tabular}{lc}
         Header & 
        \bitpattern[startBit=31]
        {1}[1] {padrow}[4] {col}[2] {HPID2}[8] {HPID1}[8] {HPID0}[8] {1}[1]/
         \\
        Payload (1-3x)
        &
        \bitpattern[startBit=31] {position}[11] {LPID}[12] {slope}[8] 0[1]/
    \end{tabular}
  \label{tbl:tracklet:format}
\end{table}

The PID information per reconstructed tracklet will increase from eight to 20 bits, which will be used to store charge information from three time slices with six or seven bit dynamic range each. Simulations have shown that the expected performance with this data format is similar to an offline analysis with the same number of time slices. The tracklet position and slope will also be stored with higher precision than in previous runs. 
         
In Run 3, the TRD uses a physics trigger sent by the CTP at LM latency (\SI{575}{\ns}, see Sec.~\ref{sec:cts}). In addition, the TRD supports a new trigger type, called calibration trigger. The calibration trigger, also sent at LM latency, enables the shipping of tracklet data and, additionally, the full raw data. This allows to trigger a full readout for a small fraction of events, facilitating detector calibration. Apart from that, the calibration trigger is interpreted by the FEE as a command to reload its configuration parameters from hamming protected memory areas. This is a precaution measure and mitigates the impact of Single Event Upsets (SEUs) on data taking, sporadically observed on some isolated half chambers as Link Monitor Errors (LMEs). An LME of a particular half chamber occurs whenever the data sent by the half chamber cannot be correlated with the corresponding triggers.

\paragraph{Common Readout Unit (CRU).}

For Run~3, the Global Tracking Units used previously~\cite{Acharya:2017lco} are replaced by CRUs (see Sec.~\ref{sec:cru}).
The CRUs receive the data directly from the FEE via 1044 custom optical links based on 8-bit/10-bit encoding.
Every CRU provides 30 link inputs, implying that in total 36 CRUs are in use (two per TRD SM). They are housed in twelve First Level Processors (FLPs).

The FPGA firmware on the CRU is composed of a common logic, and a TRD-specific user logic. It controls the readout process of the detector and receives, buffers and formats the data for the O$^2$ system. All CRUs are connected to the LTU to receive trigger information and to signal a detector busy status to the CTP (see Sec.~\ref{sec:system}). Each CRU determines an individual busy status contribution depending on the status of the readout of the connected FEE links. The CTP combines the busy status contributions from all CRUs in order to determine a global busy status of the detector.

Before being written to permanent storage, the data are reformatted and compressed, optionally either on the FLPs or EPNs (see Sect.~\ref{sec:flp} and~\ref{sec:epn}). 
The following points describe sequentially the process of acquiring an event, explaining the role of the CRU and the interactions with other readout components:

\begin{enumerate}
\item The Central Trigger Processor (CTP) sends a trigger at LM latency (physics or calibration) via the Local Trigger Unit (LTU) to the FEE and to the CRUs in parallel. The trigger to the FEE is shipped via a legacy Timing, Trigger and Control (TTC)~\cite{Taylor:592719} network.
The trigger to the CRU is sent via trigger distribution networks, using the new Trigger and Timing Control via TTC-PON technology. Nine networks are necessary to achieve the minimum latency (see Sec.~\ref{sec:CTP}). The CRUs store all information from the received trigger message (e.g. orbit and bunch crossing id) in internal buffers.
    
    \item Upon the arrival of the trigger, the FEE begins recording the data, while primary charge drifts towards the anode region. Each CRU receives the trigger at approximately the same time and internally opens a time window to wait for the input links to send all the acquired data. The timeout is programmable. In addition, each CRU generates its busy status contribution and sends it to the CTP. The TTC-PON upstream communication feature is used to transmit the busy status signal. This prevents the CTP from sending any other trigger as long as any CRU contributes an active busy signal in order to avoid confusion of the FEE state machines.
    
    \item When the FEE has acquired and processed the data, it starts shipping them via the optical links. At the end of the transmission, the FEE appends specific end markers. The CRUs record the data received on all input links.
    In case no data end marker is recognised by the CRU within the programmable timeout or data words are received outside the data expectation window, the CRU marks the concerned link as erroneous (LME) and excludes it from data taking until a manual or automated recovery takes place. The CRU stores all received data in large internal data buffers with size sufficient to hold entire calibration data events at maximum multiplicity. When the CRU has confirmed the reception of end makers on all active links, or the timeouts have been reached, the CRU releases its busy status contribution. The CTP considers the detector as busy until all 36 CRUs have released their busy contribution. 
    
    \item Once the detector side of the event acquisition is finalised, and the data are stored in internal buffers, the CRU is ready to acquire the next event. The buffered data are reformatted and shipped to the readout system in parallel. The CRU packs the data into packets of a maximum size of \SI{8}{\kilo B} and equips these packets with Raw Data Headers (RDHs). In addition, TRD-specific headers are inserted into the data stream. The headers contain various information, in particular the trigger timestamp information needed in order to link the acquired data to other detector data during the reconstruction.
    
    \end{enumerate}

\subsubsection{Detector control}

A special feature of the upgraded TRD DCS system is that the readout chain status of all half chambers is made available to the DCS system by the CRU. The CRU firmware contains a dedicated error state machine for all half chambers. If a connected half chamber shows a misbehavior that can be detected at the CRU level, the corresponding state machine enters an error state. This error state is stored in a dedicated CRU register, which is read by the DCS system via the ALFRED~\cite{Jadlovsky:ICALEPCS2017-THPHA208}  system (Sec.~\ref{sec:dcs_software}). The obtained status of all half chambers is displayed on a dedicated DCS panel in order to monitor LMEs.

\subsubsection{Standalone tracking}
\label{sec:TRD_stand_alone_tracking}
A standalone tracking algorithm for the TRD was implemented using a Kalman filter approach. The seeding uses the direction and position information of all pairs of TRD tracklets. The track reconstruction efficiency and transverse momentum resolution was determined by matching tracklets to tracks reconstructed by the TPC.
For the TRD standalone tracking, a momentum resolution of about 9\% for 500\,MeV/$c$ particles was achieved for the case of six tracklets in the fit. By including the primary vertex information as an additional constraint, a momentum resolution better than 4\% was achieved. The TRD standalone tracking algorithm was used to identify and study photon conversions and nuclear interactions in front of and within the TRD. It will also be used for the TRD drift velocity calibration in Run~3.

\subsubsection{Calibration}

\begin{figure}[b]
  \centering
    \includegraphics[width=0.42\linewidth]{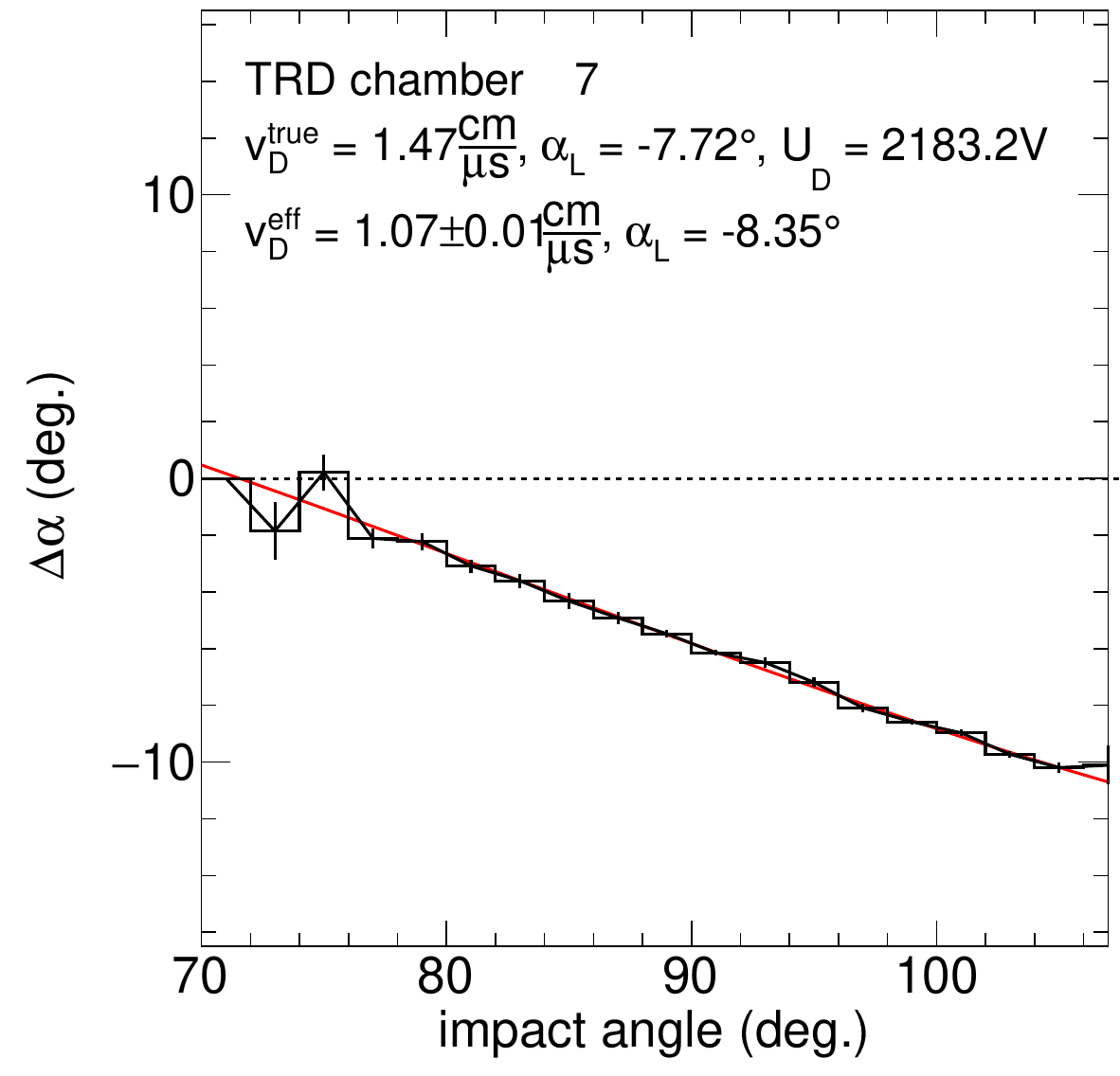}
    \includegraphics[width=0.45\linewidth]{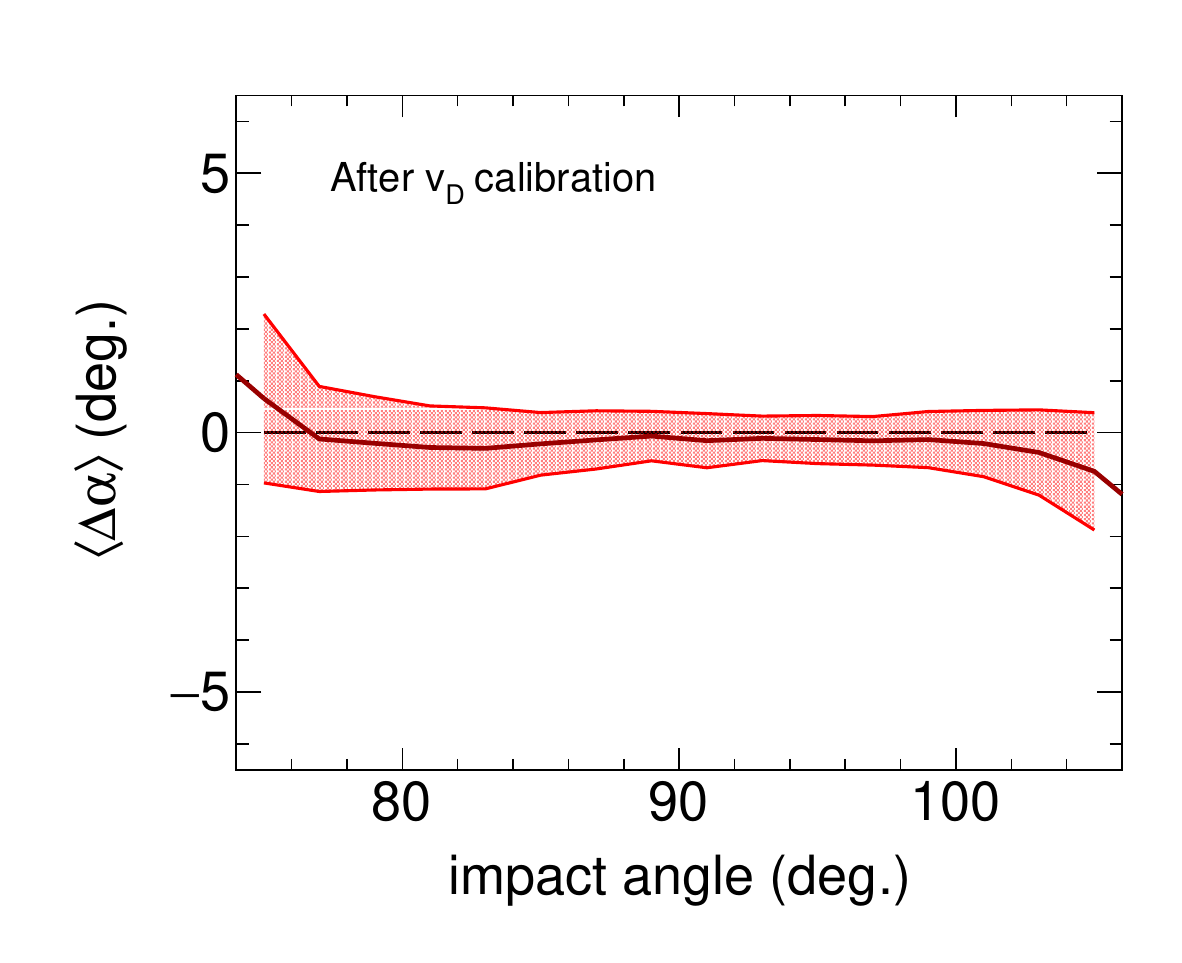}
  \caption[TRD calibration]{Left: $\Delta\alpha$ versus impact angle for a typical TRD chamber in Run~2, having a fixed uncalibrated drift velocity. The quoted values refer to the Run~2 calibration procedure (upper row) and to the new calibration scheme (lower row). Right: Average $\Delta\alpha$ versus impact angle for all TRD chambers after the calibration was applied. The red band shows the RMS of the distribution.}
  \label{fig:calibrate_5}
  \end{figure}

The Run 3 TRD calibration procedure is similar to the one employed before, except for the drift velocity calibration, which is based on a new development. The angle between a TRD tracklet and the corresponding TRD track, $\Delta \alpha$, is measured as a function of the track impact angle for each chamber. 
A model with two free paramaters, the effective drift velocity $v_{\rm D}^{\rm eff}$ and the Lorentz angle $\alpha_{\rm L}$ (the angle between the velocity of drifting electrons and the drift field), is used to fit the distributions. A typical example
is depicted in Fig.~\ref{fig:calibrate_5} (left). The effective drift velocity is compensating the ion tail effect which is systematically changing the tracklet angle. The physical true drift velocity $v^{\rm true}_{\rm D}$ is about 35\% larger than $v_{\rm D}^{\rm eff}$. A closure test with \SI{5e4}{} events using Run~2 data demonstrates that the average angular difference between tracklets and TRD track is zero, as shown in Fig.~\ref{fig:calibrate_5} (right).

About \SI{4e5}{} minimum bias pp-equivalent events are needed for an update of the calibration parameters. This is similar to what was used in Run~2 with about 600 to 3000  tracklets per chamber. The seeding and Kalman filter procedures need on average 10\,ms per p--Pb event. In total, not more than 20 minutes for one update of the calibration parameters is needed on a single CPU core. 

\subsubsection{Quality Control}

The QC system consists of tasks that are running in various parts of the O$^2$ system and produce QC objects, mostly in the form of histograms.
The following items are controlled:

\begin{itemize}

    \item The data arriving from the FEE via the CRU are validated, allowing to detect disabled or malfunctioning parts in the readout tree or SEUs in the FEE. 
    
    \item Zero-suppressed ADC data from all calibration events are analyzed to reconstruct the average, time-dependent signal shape for each of the 522 readout chambers. These histograms are versatile low-level monitoring tools for many aspects of the operation of the TRD, including trigger timing, drift velocity and gas gain.
    
    \item Tracklets from a small fraction of events are used to monitor the local reconstruction of track segments in the FEE of each chamber.
    
    \item The tracking QC monitors the efficiency of the synchronous and asynchronous reconstruction algorithms at the tracklet and track level.
    
    \item Residuals between reconstructed tracks and tracklets are analyzed in the asynchronous stage to monitor the impact of alignment and calibration on the detector performance.
\end{itemize}

The data from these QC tasks are further processed by checker algorithms to provide automated notifications and trending. 

In Run~3, the upgraded TRD system successfully recorded calibration data (for gain uniformity correction) with a $^{83}$Kr source in standalone mode, and collision data with the whole ALICE setup.

\clearpage
\subsection{Time-of-Flight detector}

The ALICE Time-Of-Flight (TOF) detector~\cite{TOF1,TOF2} is a large
array of  Multi-gap Resistive-Plate Chamber (MRPC) strip detectors,
where each strip is read out by 96 pads each with $2.5 \times \SI{3.5}{\centi\meter}^2$ area. Groups of 91 strips are organized in supermodules, covering the 18 sectors of the ALICE spaceframe. Each of the supermodules is read out by four custom VME crates, each hosting nine or ten Time-to-digital-converter Readout Module (TRM) boards, one Data Readout Module (DRM) card and one Local Trigger Module (LTM). While the DRM acts as master and has interfaces with the central systems, the LTM elaborates trigger information and sets the threshold on the NINO ASIC chips hosted on the front-end cards.

The TOF upgrade for Run 3 mostly concerns part of its readout electronics, to accomplish continuous readout, aligning with the ITS and the TPC, and with the aim to exploit at maximum its particle identification discriminating power in the intermediate momentum range. The intervention needed to adapt to a continuous readout was relatively limited thanks to the very small intrinsic dead time (\SI{\sim 10}{\nano\second}) of the MRPC detector and its front-end electronics and the fact that the High Performance TDC (HPTDC) has on-board buffering resources for digitized data, as detailed in the next sub-section. 

\subsubsection{Implementation of continuous readout}
\label{sec:tof_cont_readout}

The TRM cards are equipped with 30 HPTDCs operated in very high resolution mode, with \SI{24.4}{\pico\second} bin width. The specifications of the HPTDC (and its performance once integrated in the TRM cards) are detailed elsewhere~\cite{Akindinov:2004gf}, but it is important to recall here its trigger matching function. Based on time tags, the HPTDC allows the trigger latency to be programmable over a large dynamic range and also supports overlapping triggers, where individual hits may be assigned to multiple events. Once a trigger is received, only stored hits starting from a given time and for a limited matching window are moved to the readout FIFO and made ready for further stages of readout. During Runs~1 and 2, with a limited high-rate capacity in the barrel detectors of ALICE, the trigger was limited to a few kHz. In this situation, the internal HPTDC buffers for the TOF were configured with a latency window of \SI{6500}{\nano\second} (corresponding to the latency of the triggers reaching the TOF crates) and with a matching window of \SI{600}{\nano\second}, to comfortably collect all hits registered in the TOF detector associated to the triggered collision.

\begin{figure}[b!]
\centering
\includegraphics[width=1.0\textwidth]{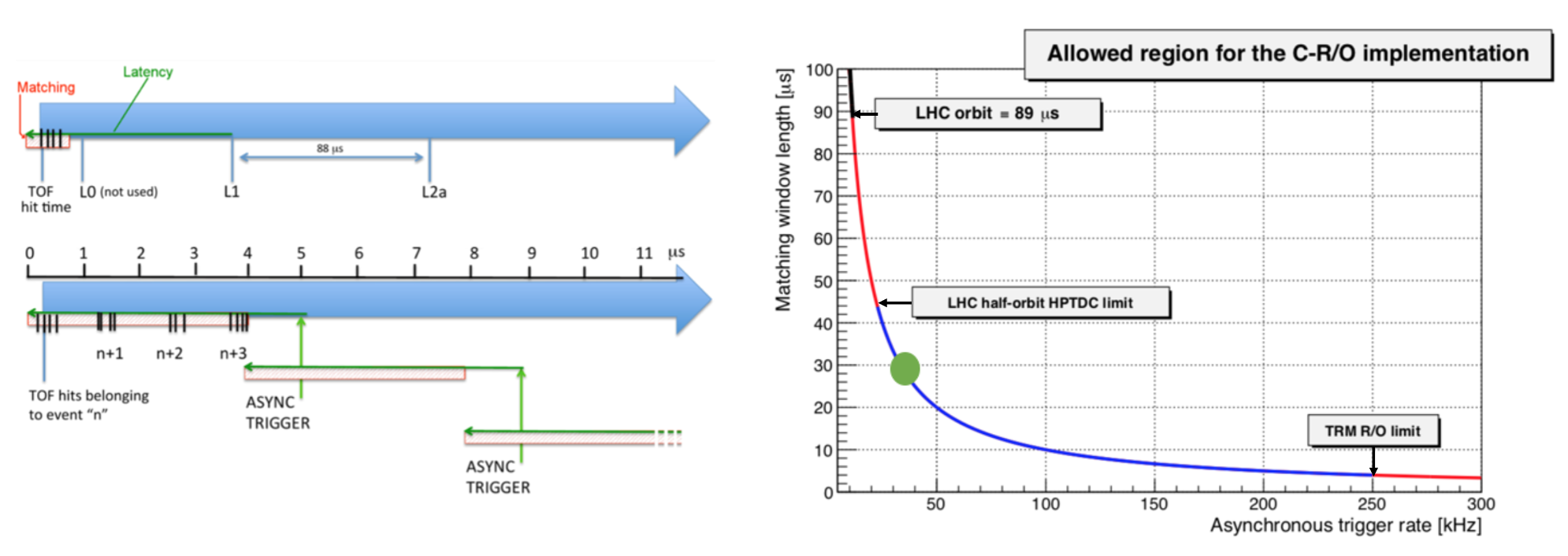} 
\caption[TOF continuous readout implementation]{(left) HPTDC programming in Run 1 and 2 operations (top arrow) and in Run 3 (bottom arrow). The three trigger levels L0, L1 and L2a are replaced by a periodic trigger with a given frequency, mimicking a continuous readout. All hits (black lines) are read out and can be associated to physical events at a later stage. (right) Possible selection of parameters (fixed trigger frequency $f_\mathrm{T}$ and matching window width $m_\mathrm{w}$) to realize a continuous readout. The green circle corresponds to the chosen point of operations.}
\label{fig:cro}
\end{figure}

\begin{figure}[t!]
\centering
\includegraphics[width=0.6\textwidth]{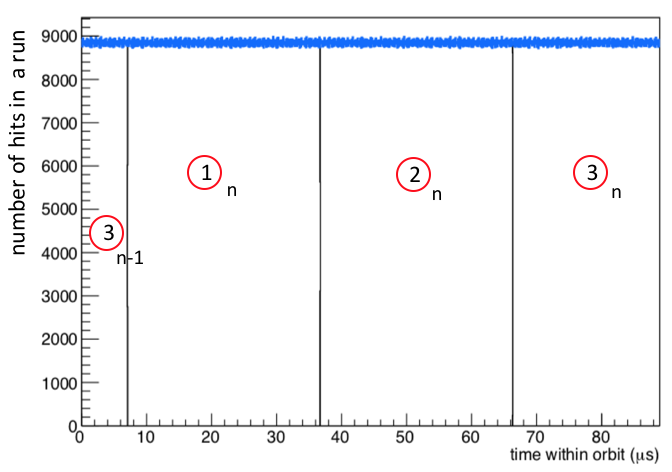} 
\caption[TOF hit time distribution readout in continuous mode]{Hit time within the orbit of randomized hits sent at fixed rate to HPTDC inside a TOF crate operated in continuous readout mode. No holes are seen in the distribution, which means that random hits are received in all 3564 bunch crossings through the whole LHC orbit.}
\label{fig:hits}
\end{figure}

In this configuration, continuos readout may be achieved by applying a strictly periodic trigger
with frequency $f_{\rm T}$ and matching window $m_{\rm w} = 1/f_{\rm T}$. 
Figure~\ref{fig:cro} (left) illustrates the underlying idea. Delivering a trigger with a constant \SI{50}{\kilo\hertz} frequency,
and setting latency and matching windows of \SI{20}{\micro\second}, all hits are
readout. Figure~\ref{fig:cro} (right) shows the curve of allowed values, together with the limitations of the system. On the one hand,
the latency window cannot be set at a value larger than half of an LHC
orbit. On the other hand, as discussed in the ALICE Readout Upgrade
TDR~\cite{Antonioli:2013ppp}, the trigger frequency cannot be too
high, given the time spent reading the HPTDC chains in the TRM cards
(two HPTDC chains of 15 chips) with a fixed readout deadtime of \SI{3.2}{\micro\second} for token-passing operations among chips alone. 
More generally, the readout time on average has to be less than $1/f_{\rm T}$. Considering also the readout time over the VME backplane (up to 10 TRMs per crate have to be read) an optimal operation point was found with $f_{\rm T}=\SI{33}{\kilo\hertz}$.  The procedure was verified in a test system, sending random hits to several TRM cards, programmed with appropriate latency and matching windows (\SI{29800}{\nano\second}). The periodic triggers, hereafter labelled as TOF special triggers (TT), were delivered at fixed bunch crossing. The orbit is split in three parts with TT occurring at BC numbers 51, 1177, and 2673. Given the flat distribution hits over the LHC orbit, Fig.~\ref{fig:hits} shows that no hits were lost.

\subsubsection{The new Data Readout Module (DRM2)}
In order to keep up with the planned increase of luminosity and of
the interaction rate (up to 1~MHz in pp collisions and 50
kHz in Pb--Pb collisions), a new Digital Readout Module 2 (DRM2), was
designed~\cite{Falchieri:2018fqw}. With respect to the existing DRM module (hereafter DRM1) it has a more modern FPGA (Microsemi IGLOO2). Overall,
it replaces the connections to the central readout and from the
central trigger in the DRM1, which were based respectively on the Detector Data
Link (DDL)~\cite{ref_DDL} and the TTC system~\cite{Taylor:592719}, with just one bidirectional GBT link (see Sec.~\ref{sec:cru}).
The GBTx ASIC is hosted on the DRM2 and the GBT protocol is implemented in the FPGA on the CRU at the receiving
end. The new system for each DRM2 has a user bandwidth to the central
readout system (CRU) of 3.2\,Gb/s, corresponding to the bandwidth available on a single GBT link.

As mentioned, the readout is implemented with special TOF triggers at
fixed bunch crossing values with $f_{\rm T}=\SI{33}{\kilo\hertz}$, setting a matching window of \SI{30}{\micro\second} in the HPTDC installed in the TRMs to achieve continuous readout. The same link is also used for receiving triggers and a low-jitter clock, which is distributed to the front-end electronics as primary clock. For the TOF detector, the quality of this clock is crucial, and a campaign of measurements on the clock received from the common readout unit has been carried out. A clock jitter (RMS) as low as \SI{\sim 10}{\pico\second} was measured in the laboratory, which is compatible with the requirements. Nevertheless, a dedicated line of clock distribution is available (same as during Runs~1 and 2) with a similar jitter.

\begin{figure}[t!]
\centering
\includegraphics[width=0.6\textwidth]{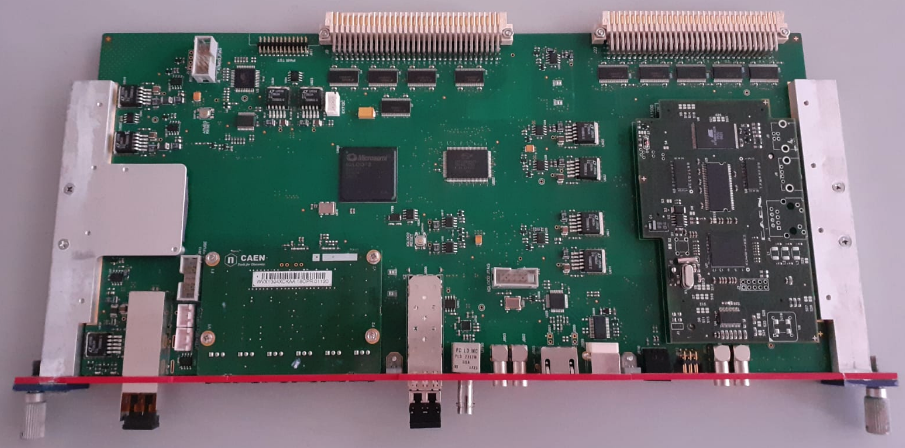} 
\caption[TOF DRM2 card components]{The DRM2 card: on the left the VTRX transceiver and the GBTx ASIC (covered by a heat dissipating panel). On the right the ARM piggy-back card is visible. The additional optical receivers for the SCL and the LHC clock are in the middle of the front panel.}
\label{fig:drm2}
\end{figure}

A picture of the DRM2 card is shown in Fig.~\ref{fig:drm2}. It is a narrow 9U VME card ($\SI{16}{\centi\meter}\times \SI{33}{\centi\meter}$) with the same form factor as the DRM1 and the TRM boards. The heart of the board is a Microsemi Flash-based IGLOO2 FPGA (M2GL090-FG676 with silicon revision 3), which drives the trigger and data flows inside the crate. This device has been chosen since the expected TID (Total Ionizing Dose) for the board (placed at $\approx$4 meters from the beam pipe) is 0.13 krads in 10 years, which is acceptable for such a device. The advantage is that the FPGA configuration memory is immune to single event upsets, so that scrubbing is not needed. Results of irradiation tests on several components of the DRM2, including the IGLOO2 FPGA, commercial optical transceivers, and staging RAM have been reported in~\cite{Falchieri:2019ulw}.

The FPGA's GBTx connection consists of a single 40-bit large parallel lane of 80 MHz differential signals. The same configuration was
previously tested on a GBTx test board developed before designing the DRM2, where a bit-error ratio lower than $10^{-14}$ and a total jitter
on the received clock around \SI{50}{\pico\second} had been measured~\cite{Falchieri:2017ofi}. As on the DRM1, an additional optical Slow Control Link (SCL) is
implemented. The SCL has a dual role: configuration and monitoring. It provides a firmware implementation of the CONET2 protocol developed by CAEN~\cite{CONET2}. Via this link, all DRM2 are
connected to commercial A3818 PCIe cards, housed in Linux machines hosted in the DCS network. The SCL is used for configuration of the
front-end electronics and programming of all VME cards. In addition, while the data collected are immediately sent to the CRU via the GBT link, the firmware also stores them in the staging RAM ($1\text{M}\times 36$ bit SSRAM from Cypress: CY7C1460KV33) for transfer (1 MB buffers) to the DCS machines. From these data (only a portion of all data are inspected via the SCL), some values such as temperatures are stripped and reported in the DCS via DIM servers. In addition, Quality Control programs run on these data. 

A block of hardware inherited by DRM1 is the ARM microprocessor
mounted on the A1500 piggy-back card provided by CAEN. This CPU
implements, via JTAG over the VME backplane, the programming of the Actel
APA750 and APA600 installed in the TRM and LTM cards, respectively. The
connection on the front panel to the console port of this CPU was
improved, with respect to the DRM1, by using a commercial RS232-USB
interface to provide a more modern USB interface to connect a laptop
computer. The ARM CPU is also able to program the firmware of the
IGLOO2 FPGA. As for the A1500 mounted on the DRM1, thanks to the
modified Ethernet interface (which has been validated to operate in
magnetic field), all the firmware updates of the VME cards (DRM2, TRM
and LTM) can be remotely executed.

Finally, the DRM2 distributes the clock to all VME cards inside the crate. This is an entirely new functionality with respect to the DRM1. Previously, only every second TOF crate had a clock distribution module (Clock and Pulse Distribution Module, CPDM). This complicated the power-up and configuration sequences, and created single points of failure (e.g.\ via the dependence on the crate power supply). The DRM2 distributes a local clock to all cards in the VME crate in a user-selectable way, using either the clock received via GBTx or the clock received directly from the LHC interface. For the latter, an optical receiver from PD-LD/NECSEL is used, with ST plug-type, with pinout compatible with the Truelight TRR-1B43-000, which was previously widely used for TTC applications. This configuration minimizes the jitter of the clock distributed to the TRM cards.

All the DRM1s were removed and disassembled during the first months of
2019. The A1500 ARM piggy-back cards were tested and prepared for
installation on the DRM2 cards. The procedure for validation and test of the
DRM2 production (completed during 2019) is described
in~\cite{Falchieri:2019mpp}. The installation of all DRM2 cards,
partially delayed by the Covid-19 pandemic, was ended in June 2020 with full commissioning starting in September 2020. The MRPC high voltages were ramped to the nominal value ($V=\SI{6500}{\volt}$ on each stack) in the same month. The TOF detector actively participated to the data taking with the LHC pilot beams in October 2021.

The front-end software needed for configuration of the electronics was upgraded and deployed in a staged way during spring 2020.

\subsubsection{Additional upgrades in low voltage and quality control systems}

During LS2, several other TOF systems linked to the readout upgrade were subject to key improvements and maintenance to prepare for the intense data taking foreseen in Run~3. Among many interventions, we highlight in particular:

\begin{itemize}
\item The DC/DC systems (CAEN modules A1395 and A1396): these modules supply power to the four crates of each TOF supermodule. They receive a DC \SI{48}{\volt} power supply via bus-bars from outside the L3 Magnet and provide LV power supply for the VME boards and the front-end cards on the MRPC modules. A solid state fuse that was subject to frequent breakdown was replaced. Additionally, a study via proton irradiation at the Centro di Protonterapia in Trento in 2019 investigated the cause of SEU events registered in 2018 at high irradiation-rates that produced sudden loss of communication with the module. The addition of a filter capacitor on the reset line of the microprocessor on the A1396 fixed the problem. A full refurbishment of all modules (entailing dismounting 216 modules from the detector) was completed in 2019-2020.

\item All DRM2s are equipped with an ARM microprocessor (AT91RM9200 from Atmel) running Linux. During Runs~1 and 2 these CPUs were used exclusively to perform firmware upgrades on the VME boards via the JTAG interface on the VME backplane using Actel software for APA FPGAs. Using the cross-platform development tools provided, a full slow control DIM server has been deployed on these CPUs. This provides an additional channel to monitor voltages and temperature on the cards (even if the SCL is not connected). More importantly, thanks to a different hardware implementation on the DRM2 with respect to the DRM1, via the server running on ARM CPUs it will be possible to reset the CONET link and to reset the Microsemi FPGA of the DRM2. These two emergency resets may be used in case of loss of communication with the DRM2 (on the SCL), without the need of executing a power cycle. This may be useful especially because the DRM2 provides the primary clock to all TRM cards and a DRM2 power cycle would cause the loss of the clock and therefore the need of a power cycle in all VME slots in the crate.

\item The procedure for the control and validation of the recorded data has been integrated into the O$^{2}$ framework under the project of the Quality Control (QC). 
For example, counters reporting errors detected in some HPTDC on the TRM are provided, as well as information on the hit rate on all 150\,000 channels of the TOF detector. Noisy channels, identified by a hit rate greater than \SI{1}{\kilo\hertz}, can be disabled individually.
Quality control tasks monitor the reconstruction and calibration process at various levels in order to provide a detailed insight into the various steps of the data processing. The QC code runs on dedicated computing nodes (FLPs or EPNs) to monitor the TOF data stream, and in the DCS system, sampling data through the SCL.
For the SCL part, the QC task is primarily intended to monitor temperatures of front-end electronics while readout is ongoing, as well as to monitor the hit rate, turning off quickly very noisy channels that could prevent the continuous readout (on average each event has to be read out in \SI{30}{\micro\second}, see Sec.~\ref{sec:tof_cont_readout}). While a certain degree of duplication exists, the FLP and EPN QC provides more aggregated information for the whole TOF, like the average multiplicity per matching window.

\item The data flow from the CRUs is processed by the CPUs to perform the first level of data decoding and some preprocessing, which produces a second level of raw data, providing a zero-suppressed data stream where the relevant information is stored in a compact format. This effectively reduces the output bandwidth from the FLP to the EPNs by a factor of four for very high multiplicity events, and by a much larger factor for low multiplicity events. This allows the framework to make the best use of the available computing resources on the FLPs by performing low-level data monitoring. The QC system is able to access the preprocessed data directly on the FLPs for monitoring of the raw data stream as early as possible.
\end{itemize}

\clearpage
\subsection{High-momentum particle identification}

\subsubsection{Introduction}
The ALICE High Momentum Particle IDentification (HMPID) detector~\cite{Beole:1998yq, Aamodt:2008zz} is designed to identify hadrons at $p_{\rm{T}} >1$\,GeV/$c$. 
It is based on Ring Imaging Cherenkov detectors (RICH).
Seven MWPCs, equipped with CsI segmented photocathodes, detect Cherenkov patterns.
Together with the momentum measured by the TPC, they allow the determination of the particle mass.
During Run~3, the main goal of the detector is to identify light nuclei and corresponding anti-nuclei at high transverse momenta in the central rapidity region, up to 12\,GeV/$c$ for the deuteron and triton, and up to 10 GeV/$c$ for $^3$He.
During LS2, the readout firmware was upgraded to allow an increase of the event readout rate. 
In addition, to measure the inelastic cross section of (anti-)deuterons in the momentum range 0.2 to 2.2\,GeV/$c$, two aluminum absorbers of \SI{8}{\cm} thickness, were installed in front of two RICH modules. In one module, two out of three Cherenkov radiator gas vessels were leaking, whereas the second RICH was located in a favourable position for the installation of the second absorber, needed for the required statistical abundance of the measurement. The consequent loss of acceptance of the PID measurement is largely compensated with the remaining modules by an event readout rate ten times higher than in Run 2.
The detector stayed in place during the long shutdown, since moving it to a laboratory for repairing and upgrades was considered too risky. In turn a re-design of the readout firmware was carried out as explained in the next section.

\subsubsection{Upgrading of readout firmware and trigger}

\begin{figure}
  \centering
  \includegraphics[width=.9\textwidth]{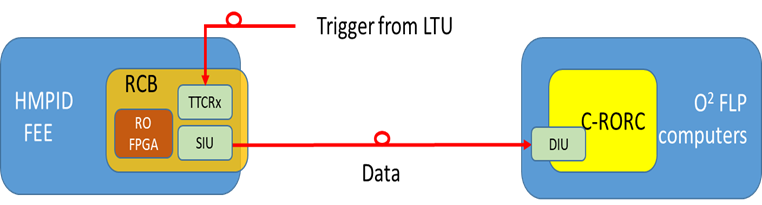}
  \caption[HMPID front-end electronics]{From left to right: HMPID Front-End Electronics (FEE), Readout Control Board (RCB) with the readout FPGA, the TTCRx and the Source Interface Unit (SIU). On the right, the C-RORC cards installed on O$^{2}$ FLP computers.}
  \label{fig:hmp:fee_overview}
\end{figure}

Figure~\ref{fig:hmp:fee_overview} shows the block diagram of the HMPID readout chain.
The Readout Control Board (RCB) houses the readout FPGA, the TTCRx mezzanine card and the Source Interface Unit (SIU) interfacing the trigger and DAQ systems. 
The FPGAs firmware (FW) synchronises the trigger and data readout and is a key element of the readout performance. 
The C-RORC cards are installed on O$^{2}$ First Level Processor (FLP) computers and connected via optical links to the HMPID readout electronics.

\begin{figure}[htb]
  \centering
  \includegraphics[width=0.6\columnwidth]{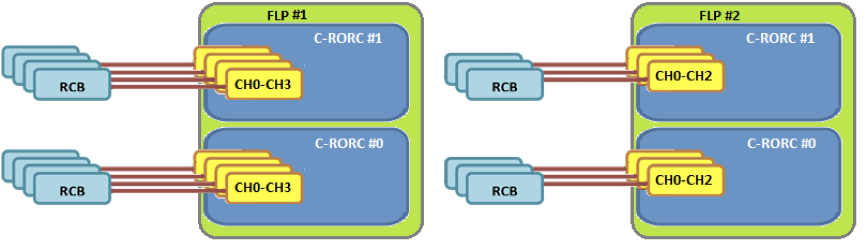}
  \caption[HMPID data acquisition]{HMPID full DAQ structure. Four C-RORC boards are installed on two First Level Processor (FLP) computers of the O$^2$ data acquisition environment. Fourteen optical links connect the 14 RCBs, two per RICH module.}
  \label{fig:hmp:ro}
\end{figure}

The block diagram for the HMPID DAQ system is shown in Fig.~\ref{fig:hmp:ro}. 
Fourteen optical links, two per RICH  module, are connected to four C-RORC cards, on two FLPs .

\subsubsection{New readout firmware and readout rate}

In Run~2, the detector was operated at lower readout rates, in line with the TPC (2\,kHz and 800\,Hz, respectively, in pp and \PbPb collisions). As a result, the effective increase in data sample size between Run~2 and 3 can be up to a factor 14 and 10, respectively, in pp and \PbPb collisions. 
In order to improve the readout speed in Run~3, a new readout firmware was designed, tested in the laboratory, and finally deployed on the RICH modules. In laboratory tests, the HMPID has reached a readout rate up to \SI{28}{\kilo\hertz} for \pp collisions (readout data headers only, a factor five improvement with respect to Run~2) and \SI{9}{\kilo\hertz} in \PbPb collisions (about a factor three improvement with respect to Run~2).

The readout firmware runs on the ALTERA Stratix II EP2S15F484C5 FPGA housed on the Readout and Control Board (RCB). The firmware improvements listed below resulted in higher and more stable data acquisition rates. The improvements are:
\begin{itemize}
  \item shortening of the event readout time by \SI{90}{\micro\second} due to the omission in Run 3 of the L2a trigger level with its long latency,
  \item skipping of empty readout columns (in pp collisions only one out of seven events has a track in the HMPID acceptance, which results in an occupancy of ~0.17\%),
  \item the digitization and data transfer from FEE cards to the column memory buffer, which is now carried out in parallel to the decoding of the L1 trigger message in the TTCRx card. On average, it completes after \SI{15.6}{\us} from the arrival of the LM trigger,
  \item masking of failing electronic columns, which is applied during the configuration of the FEE.
\end{itemize}

\begin{figure}[htb]
  \centering
  \includegraphics[width=0.7\columnwidth]{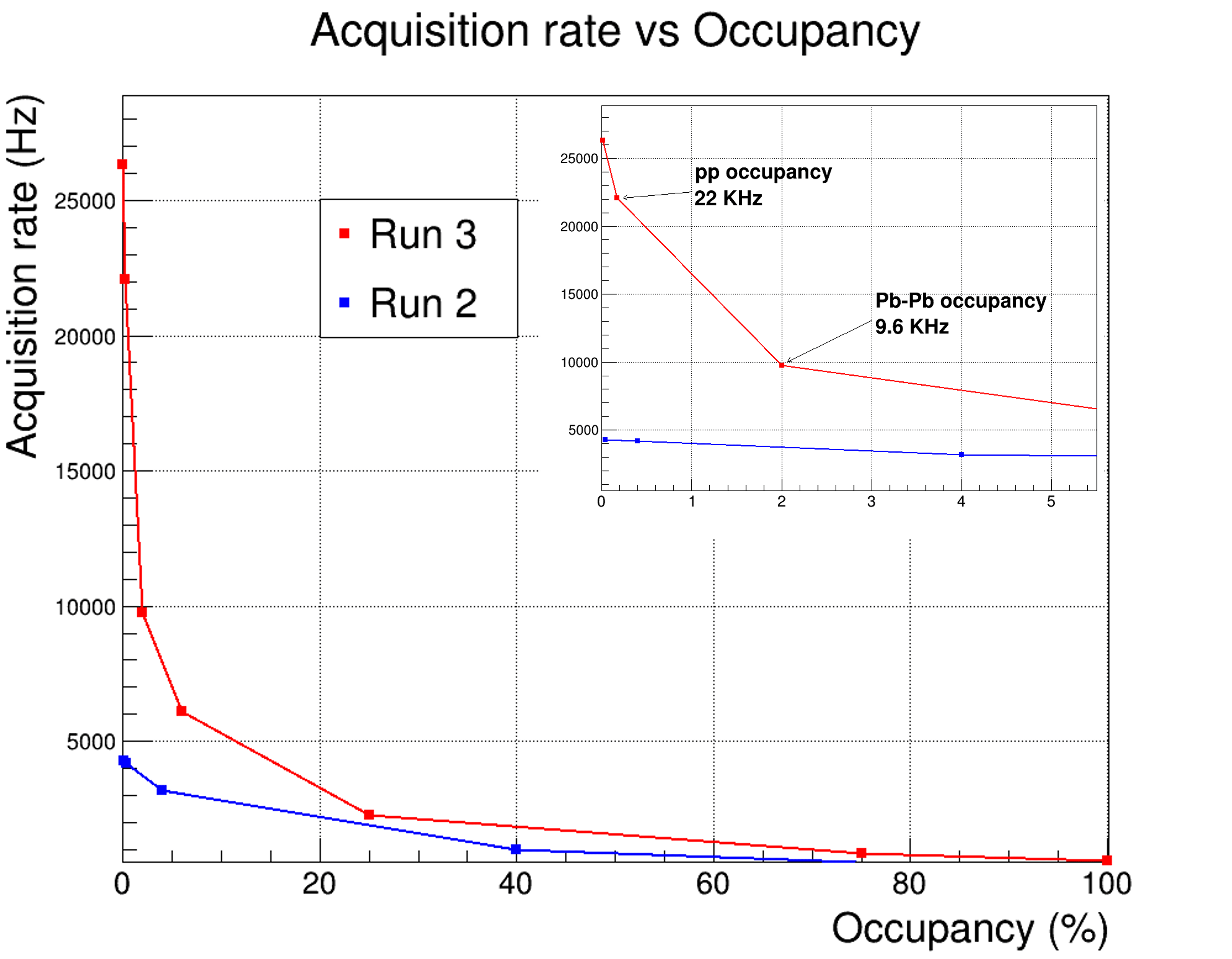}
  \caption[HMPID event rate as a function of occupancy]{Event rate as a function of the detector occupancy, which is on average about 0.17\% and 2\% in \pp and \PbPb collisions, respectively. %
}
  \label{fig:hmp:rate}
\end{figure}

The final readout performance, as measured in the ALICE experiment with simulated occupancy, is shown in Fig.~\ref{fig:hmp:rate}. 
With 0.17\% occupancy (pp collisions), the measured acquisition rate is \SI{22}{\kilo\hertz}, whereas with 2\% occupancy (in Pb--Pb collisions) it is \SI{9.6}{\kilo\hertz}. 
During the LHC pilot beam campaign in October 2021, the HMPID recorded events at 11.2\,kHz with the zero bias trigger, coherently with the measured readout performance. In fact, the LHC filling schema had two colliding bunches with a \SI{\sim 20}{\us} separation. On average, only the first one was accepted in the HMPID, whereas the second colliding bunch was usually rejected by the busy time of \SI{45}{\us}. Differently spaced colliding bunches, as expected during the normal operation of the LHC, will allow the full exploitation of the readout performance.

\subsubsection{Detector calibration formalism}

The detector-calibration method has changed with respect to Run~2. 
During a standalone run without zero suppression, a dedicated workflow at the EPN level computes the average values of the pedestal distributions and the corresponding standard deviation.
These values are archived in the Calibration and Constants Data Base (CCDB, see Sec.~\ref{chap:readout}).
The DCS retrieves these data from the CCDB and uploads the pedestal values and the standard deviations in the readout electronics, via the ALFRED mechanism (see Sec.~\ref{sec:dcs}). 
The granularity of this calibration mechanism is at the level of a single readout column, with a single calibration file per column. Each C-RORC link configures 24 columns, which correspond to the right or left half of a RICH module.

\subsubsection{Other subsystems}

\paragraph{Detector Control System}

The DCS for the HMPID was upgraded providing the following new features:
\begin{itemize}
\item uploading the pedestal and sigma values in the readout electronics via the new ALFRED formalism; 
\item monitoring the busy time of a single RO link;
\item automatically ramp up tripped HV channels.
\end{itemize}

\paragraph{Absorbers for anti-deuteron inelastic cross section measurement} 

Another important achievement during LS2 is the installation of two aluminium absorbers of \SI{8}{\cm} thickness, corresponding to half an interaction length for the anti-deuteron inelastic cross section measurement in the momentum range of 0.2 to 2.2\,GeV/$c$. 
The (anti-)deuterons impinging on the two absorbers will be identified using the d$E/$d$x$ measured by the TPC and the time-of-flight measured by the TOF detector. The detection of secondary particles produced in the hadronic interaction with the target nuclei will be carried out by the pad-segmented cathodes of the HMPID-MWPCs installed right behind the absorbers.

\begin{figure}[htb]
  \centering
  \includegraphics[width=0.7\columnwidth]{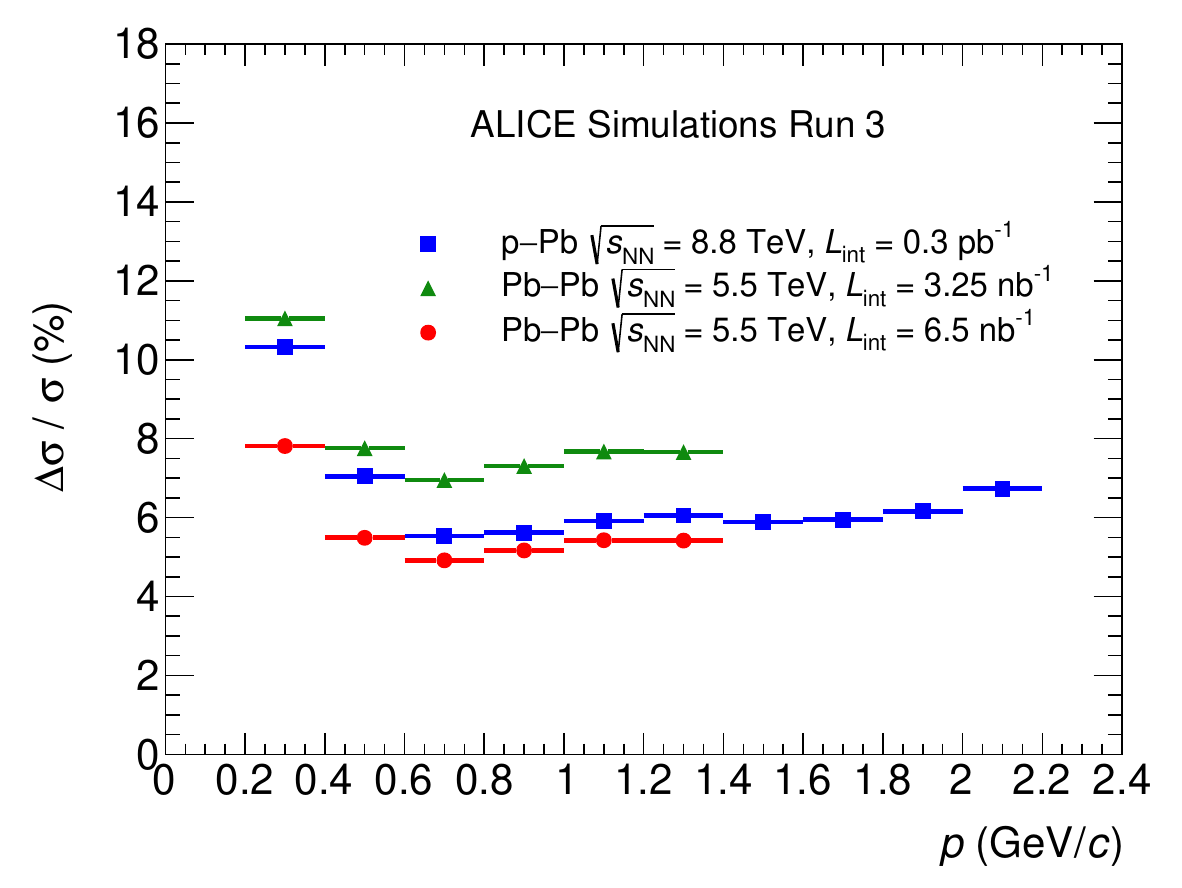}
  \caption[HMPID event rate as a function of occupancy]{Expected relative statistical uncertainty on the anti-deuteron absorption cross section in \pPb and \PbPb collisions for the full Run 3 data sample and for one half of the Pb--Pb sample.}
  \label{fig:hmp:crosssection}
\end{figure}

During Run 3 the expected statistical precision of this measurement is expected to be in the range 5--10\% in the momentum interval $0.2 < p < 2.2$ GeV/$c$ for \pPb collisions at $\snn = 8.8$ TeV and 5--8\% in the momentum interval $0.2 < p < 1.4$ GeV/$c$ for Pb-–Pb collisions at $\snn = 5.5$ TeV (see Fig.~\ref{fig:hmp:crosssection}). A systematic uncertainty of maximum 5.5\% is expected based on a conservative estimate.

In ALICE an effort is ongoing to measure the (anti-)deuteron inelastic cross section, also at low momentum, using the TRD as an absorber. This detector has an average mass number $\langle A \rangle$ $\approx$ 8 considering the gas mixture, the active detector materials, and the support structure. The measurement using aluminum ($A$ = 27) as a target will provide complementary information to other existing approaches and these measurements will allow the study of the mass-number dependence of the (anti-)deuteron inelastic cross section at low momentum, which is currently unknown.

\clearpage
\subsection{Electromagnetic Calorimeter}

The Electromagnetic Calorimeter (EMCal)~\cite{Cortese:2008zza, Allen:2010stl} was designed for the measurements of electrons from heavy-flavor hadron decays, of the electromagnetic component of jets, and of direct photons and neutral mesons. The calibration procedures and achieved performance during Runs 1 and 2 are described in detail in~\cite{emcPerfPaper}.

The calorimeter remains a trigger detector during Run~3 and will continue to provide L0, L1-$\gamma$ and L1-jet triggers (described in Sec.~\ref{sec:cts}). It is also possible to read out the calorimeter with any other trigger provided by the CTP (e.g. FIT minimum bias trigger).
The hardware did not need any modifications to comply with the Run~3 requirements. However, in order to be able to operate during Run~3, the following actions were taken: spare hardware was produced in order to ensure operation during Runs~3 and 4, and the front-end electronics firmware was upgraded in order to satisfy the specifications for Run~3. Details will be given below.

The EMCal is a shashlik-type lead-scintillator sampling calorimeter comprising 4416 individual modules that are grouped into 
20~Super Modules (SM). Each of the modules is composed of four optically isolated towers, resulting in 17\,664 individual towers in total.  
The optical readout of each tower is provided using wavelength shifting fibers coupled to an Avalanche Photo Diode (APD).

The front face dimensions of the towers are ~6$\times$6\,cm$^{2}$ resulting in individual tower acceptance of $\Delta  \eta \times \Delta  \varphi \simeq 0.0143 \times 0.0143$ at $\eta = 0$. The towers are arranged within the SMs such that each tower is approximately 
projective to the interaction vertex in $\eta$ and $\varphi$. The towers are operated at $\sim 25^{\circ}$C ambient temperature with a nominal APD gain of $\simeq 30$, to achieve a 14-bit effective dynamic energy range from $\sim 16$\,MeV to $\sim 250$\,GeV per tower.

The overall design of the calorimeter is heavily influenced by its integration within the ALICE setup~\cite{Aamodt:2008zz}. SMs of 3 different sizes are used: full-size, 2/3-size and 1/3-size. 
Each full-size SM consists of $12\times 24=288$ modules arranged in 24 strip modules of $21\times 1$ modules each. The 1/3 and 2/3 size SMs consist of $4\times24=96$ and $12\times16=192$ modules, respectively.

\begin{figure}[tbh!]
\begin{center}
\includegraphics[width=0.5\textwidth]{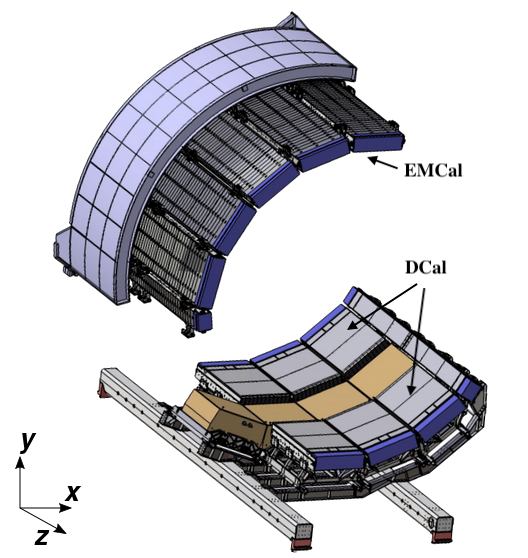}
\caption[Schematic view of EMCal]{Schematic view of the EMCal, consisting of two disjunct detector segments, in the top-left hemisphere and the bottom-right hemisphere (DCal), covering approximately opposite locations in azimuth. The PHOS calorimeter inside the DCal segment is indicated in brown.}
\label{fig:EMCAL_DCAL_3D}
\end{center}
\end{figure}

The detector consists of two parts, that cover two different regions in azimuth, as illustrated in Fig.~\ref{fig:EMCAL_DCAL_3D} (see also Fig.~\ref{fig:intro:alice_run3}). The main segment of the EMCal consists of ten full-size SMs and two 1/3-size SMs covering  $|\eta| < 0.7$ in azimuth and $80^{\circ} < \varphi < 187^{\circ}$ in azimuth, while six 2/3-size SMs and two 1/3-size SMs are installed around the PHOS detector, covering $0.22 < |\eta| < 0.7, 260^{\circ} < \varphi < 320^{\circ} $ and $|\eta| < 0.7, 320^{\circ} < \varphi < 327^{\circ}$. The latter part of the detector is some times referred to as Di-Jet Calorimeter (DCal). In the following, we will use the term EMCal to refer to the full system, and DCal only when the distinction between the two segments is useful.

The SMs are located at $R\simeq 4.3$\,m in radial distance from the beamline, inserted into support frames situated between the time-of-flight  detector and the ALICE L3 magnet. The weight of a single full-size SM is $\simeq$ 7.7 tons, and the total weight of all 20 SMs is $\simeq$ 120 tons. 
More details regarding the mechanical structure and Front-End Electronics (FEE) can be found in~\cite{Cortese:2008zza, Allen:2010stl}.

An individual EMCal tower is read out with an avalanche photodiode and preamplifier mounted on the tower. The preamplifier signal is split into energy and trigger shaper channels on the FEE boards~\cite{Muller:2006jr}. The energy shaper signals are sampled at 10\,MHz with 10-bit resolution using ALTRO 
chips~\cite{EsteveBosch:2003bj}. Prior to digitisation, each energy signal is split into high and low gain channels, each shaped separately, with a gain ratio of 16 to provide an effective dynamic range of 14 bit. Each FEE board provides readout of the high and low gain channels from 32 towers.

The ALTRO chips are configured
to record 15 10-bit time (pre-)samples per readout channel per event to cover the 1.5\,$\mu$s integration window. The data are compressed by discarding samples close to the reference level (pedestal) that contain no useful information ("zero suppression"), reducing substantially
the data volume. The pedestals are obtained from special runs with no pre-programmed pedestal or signal present.

\subsubsection{The readout system}
Each SM is equipped with a readout concentrator, the so called Scalable Readout Unit (SRU)~\cite{Zhang:2014ioa}.
The SRU interconnects with each FEE board through a custom daughter card which was designed for the EMCal FEE board. It provides interface compatibility between the SRU and the EMCal FEE board to provide the Data, Trigger, Clock and Control (DTC) links. 
The maximum bandwidth of a DTC link on the SRU is 2\,Gb/s. In the EMCal application, the bandwidth of the DTC link is conservatively limited to 20\,MB/s due to the hardware capability of the rather outdated FEE board FPGA (Altera ACEX 1K Family EP1k100QC208-3). However, the DTC link does not limit the EMCal data throughput.

Each SRU has a total of 40 point-to-point links to connect to 37 FEE and three TRU boards for the full size EMCal SMs, and sends the data to an FLP through two Detector Data Links (DDL), see Sec.~\ref{sec:flp}. 
The SRU board integrates a TTCrx (LHC Trigger, Timing, and Control receiver)~\cite{Christiansen:1996dg}, which can receive trigger and timing information from the ALICE Trigger system. It also has three SFP+ ports directly connected to the FPGA’s high speed serial transceivers for serial data transport at up to 5\,Gb/s and an additional SFP+ port that provides a 10\,Gb Ethernet link. For the EMCal application, one of these transceivers is used for the Ethernet connection to the ALICE detector control system, while the other two transceivers are used for the two DDL links to transmit the data to an FLP.

\subsubsection{Trigger} 
The EMCal provides inputs to the L0 and L1 trigger decisions in ALICE (Sec.~\ref{sec:CTP}). The trigger subsystem resides in specific hardware boards.
The analog signals of $2\times2$ adjacent towers are summed in the FEE boards and transmitted to a Trigger Region Unit (TRU) board, where the 
$2\times2$ tower sums from twelve FEE cards ($2\times2$ sums from 96 channels) are digitized at the LHC clock frequency of 40\,MHz~\cite{Kral:2012ae}. 
The digitized $2\times2$ tower sums are summed over time samples 
with pre-sample pedestal subtraction to provide an integral energy measurement, referred to as time sum. 
Finally, overlapping $4\times 4$ tower digital sums are formed within each TRU and a peak finding algorithm is used to find a signal peak.  
Each $4\times 4$ sum signal peak amplitude is then compared against a threshold to provide a L0 trigger output that indicates the presence of a high energy shower in the TRU region (1 TRU covers 1/3 of the area for a full-size SM). 
The L0 trigger decision from each TRU is passed to a Summary Trigger Unit (STU), which performs the logical OR of the L0 outputs from all TRUs to provide a single L0 input to the ALICE CTP.

Upon reception of an accepted L0 trigger from the CTP, the digitized time-summed $2\times2$ tower sums from each TRU are passed to the STU. In the STU the $4\times4$ overlapping tower sums are 
formed again, but across TRU boundaries over the full acceptance to provide an improved L1 high energy shower trigger referred to as L1-$\gamma$ trigger~\cite{Bourrion:2012vn}. 
At the same time, tower sums over a large $8\times 8$ trigger channel window ($16\times 16$ towers) and a
$16\times 16$ trigger channel window ($32\times 32$ towers)  are also formed to provide a L1 jet trigger. Both L1 triggers allow to define two thresholds for the event selection.

In order to reduce the bias due to multiplicity fluctuations in heavy-ion collisions, there is a direct communication between the STUs of the main EMCal segment and the DCal to consider the underlying event background in the online L1 trigger decision.
The background is estimated  based on the median of the energies deposited in $8\times 8$ trigger channel ($16\times 16$ towers) windows in the opposing segment of the detector.
The background is subtracted from the signal amplitude and then compared against a threshold to provide L1 triggers. 

\subsubsection{Spare production}
In order to guarantee a smooth operation of the detector through Run~3, additional FEE boards were produced. They are identical to the ones used during Runs~1 and 2. A total of 100 front-end cards and six TRUs, amounting to 15\,\% of the units used in the experimental cavern, were produced. In addition, 2 STUs were produced as spares.

\subsubsection{Front-end electronics firmware upgrade}
\begin{figure}[t]
    \centering
    \includegraphics[scale=0.8]{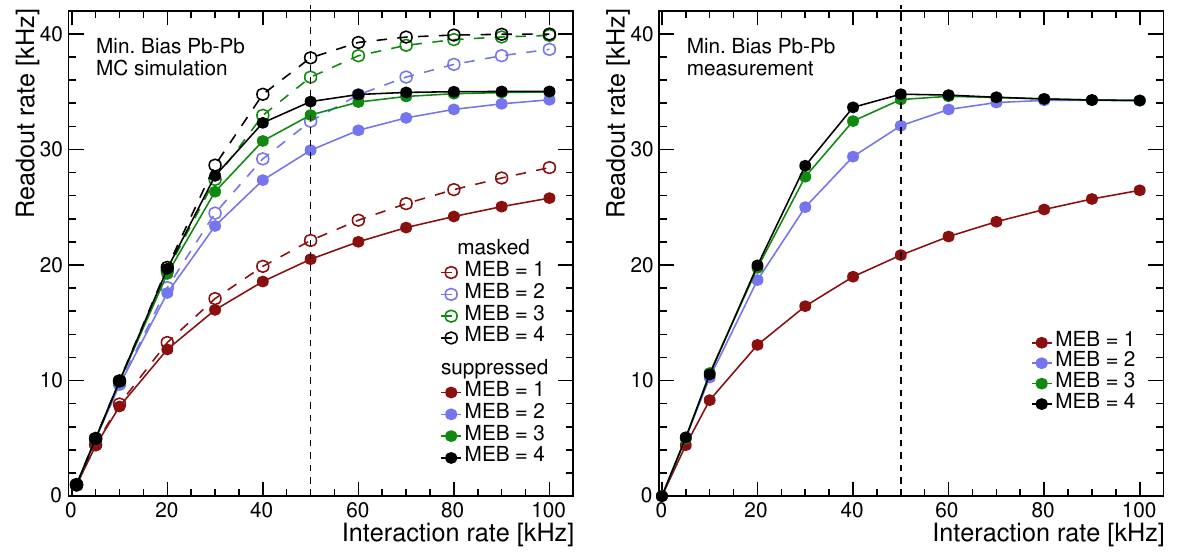}
    \caption[SRU readout rate]{SRU readout rate as a function of interaction rate for different Multi-Event Buffering (MEB) schemes. The left panel shows simulated data, while the right panel shows measured performance.}
    \label{fig::emcal::SRUrate}
\end{figure}

The firmwares of SRU and of STU had to be upgraded in order to satisfy the requirements concerning the readout rate for Run~3.
In particular, in order to increase the readout rate for the anticipated 50\,kHz minimum bias \PbPb\ interaction rate for Run~3, a multi-event buffering (MEB) logic was implemented in the SRU firmware, allowing to accept a trigger while the data from the previous trigger is being processed. 
The importance of multi-event buffering for the data recording rate as a function of interaction rate is shown in Fig.~\ref{fig::emcal::SRUrate}. 
The left-hand plot shows the predictions expected from Monte Carlo simulations, and the results from measurements with black events are shown on the right-hand plot. 
To estimate the SRU readout rate, which depends on detector occupancy, some \PbPb\ data from Run~2 were used. 
Pedestal data are used to create the load expected for minimum bias \PbPb{} collisions. For early readout rate estimates~\cite{Antonioli:2013ppp}, the detector occupancy was emulated by masking ALTRO channels in the FEE configuration (open markers on the left-hand side plot).
This was improved later by applying a relatively high value for the baseline and by suppressing the data in the ALTRO channels (full markers). This procedure yields results that are closer to real data. 
The readout rate decreased by $\sim10\%$ compared to preliminary expectations for both single- and muti-event buffering configurations.
With four event buffers, a readout rate of $\sim35$\,kHz is expected for minimum bias \PbPb\ collisions at 50\,kHz, and the results from Monte Carlo simulations are in a good agreement with the results from measurements at Point 2.

In addition, further improvements were implemented in the SRU firmware for increasing the readout stability and the physics performance. 
In particular, a synchronization between the LHC 40\,MHz clock and the ALTRO 10\,MHz clock was implemented in order to perform online time calibration during Run~3.

The firmware of the STU was upgraded to conform to the Run~3 trigger and DAQ protocols. The STU readout time highly depends on the data size to be sent from STU to DAQ. It is $\sim$ \SI{50}{\us} when sending the data from all trigger channels needed for the full QC. However, for physics analysis only the channels contributing in the trigger decision are selected by the STU FPGA, resulting in a readout time of 
$\sim$ \SI{20}{\us}. During normal operation for physics data taking, the data from all channels will be recorded for only $\lesssim 1$\% of events, and the average readout time is expected to be close to \SI{20}{\us}.

\subsubsection{Data compression}

In order to reduce the amount of data written to tape, a fit of the time series of raw ADC signals is performed per tower, extracting amplitude and peak time. This fit is performed during the synchronous reconstruction. Amplitude and time, as well as tower index and gain type are encoded in a 48-bit word per tower. Further compression can be achieved by removing low tower energy signals which are rejected in the clusterization process and will therefore not contribute to physics measurements.  

\subsubsection{Calibration}

The calibration procedure is based on existing calibration procedures used during Run~2. At the beginning of the data taking process of Run~3, a sample of events will be used to determine the absolute energy scale for each tower, based on the comparison of the reconstructed $\pi^{0}$ mass to the nominal mass. Identification of bad channels, which need to be removed in the analysis process, and calibration of the time measurement with respect to the collision time are based on the event-by-event tower energy and time measurement and are performed for all data blocks.
Calibration histograms are filled and calibration parameters are determined in an automatized procedure during the synchronous and asynchronous reconstruction using the O$^2$ calibration software framework. The remaining time-dependence of the energy calibration resulting mainly from the sensitivity to the temperature are calibrated using the light-emitting diode (LED) system, by generating an ultra-bright blue light triggered by the CTP~\cite{Cortese:2008zza}. 

\subsubsection{Quality Control}

Monitoring of the data quality is based on the Quality Control framework within the ALICE O$^2$ computing framework. The EMCal Quality Control is designed to provide sufficient information in order to identify problems during the data taking process, and to decide on the usability of data blocks for physics measurements. Quality Control for EMCal consists of the following tasks:

\begin{itemize}
    \item Raw Data level: A fraction of the raw data is inspected in order to find problematic parts of the detector, and to check for errors in the raw stream
    \item Digit level: Tower-based quantities (energy, position, time) after the fit to the raw data are monitored in order to find inactive or noisy regions of the detector.
    \item Cluster: %
    A calorimeter cluster, an aggregate of adjacent calorimeter cells with energy above the noise threshold, is the main object
    delivered by the reconstruction software. Basic EMCal cluster observables, such as energy, position, number of contributing towers, and others, are monitored at the level of the synchronous and asynchronous reconstruction stages for a fraction of the data.
    \item Trigger level: Trigger-level digits are monitored for a small fraction of the data in order to identify noisy regions in the trigger system.
\end{itemize}

The Quality Control is histogram-based. The histograms are further processed to produce derived observables which are monitored continuously to follow the time evolution of the detector state.

\clearpage
\subsection{Photon Spectrometer}

The Photon Spectrometer (PHOS)~\cite{ALICE:1999ozr} is a precise electromagnetic
calorimeter which specialises in the detection of photons with high energy and spatial
resolutions. PHOS covers a limited acceptance at mid rapidity
$|y|<0.13$ and azimuthal angle $250^\circ < \varphi <
320^\circ$. The layout of the PHOS detector surrounded by the DCal 
is depicted in Fig.~\ref{fig:EMCAL_DCAL_3D}. The physics objectives of the PHOS detector are the
measurement of direct photon yields in the energy range from $\approx
0.1$ to 100\,GeV, azimuthal anisotropy of photon emission, photon-hadron correlations, as well as
measurements of light neutral mesons $\pi^0$, $\eta$, $\omega$ with transverse momenta $p_{\rm T}$ above
$\approx 0.6$\,GeV/c, with the upper $p_{\rm T}$ limit being driven
mainly by available statistics.

\subsubsection{Detector layout}

\begin{figure}[htb]
  \includegraphics[width=0.45\hsize]{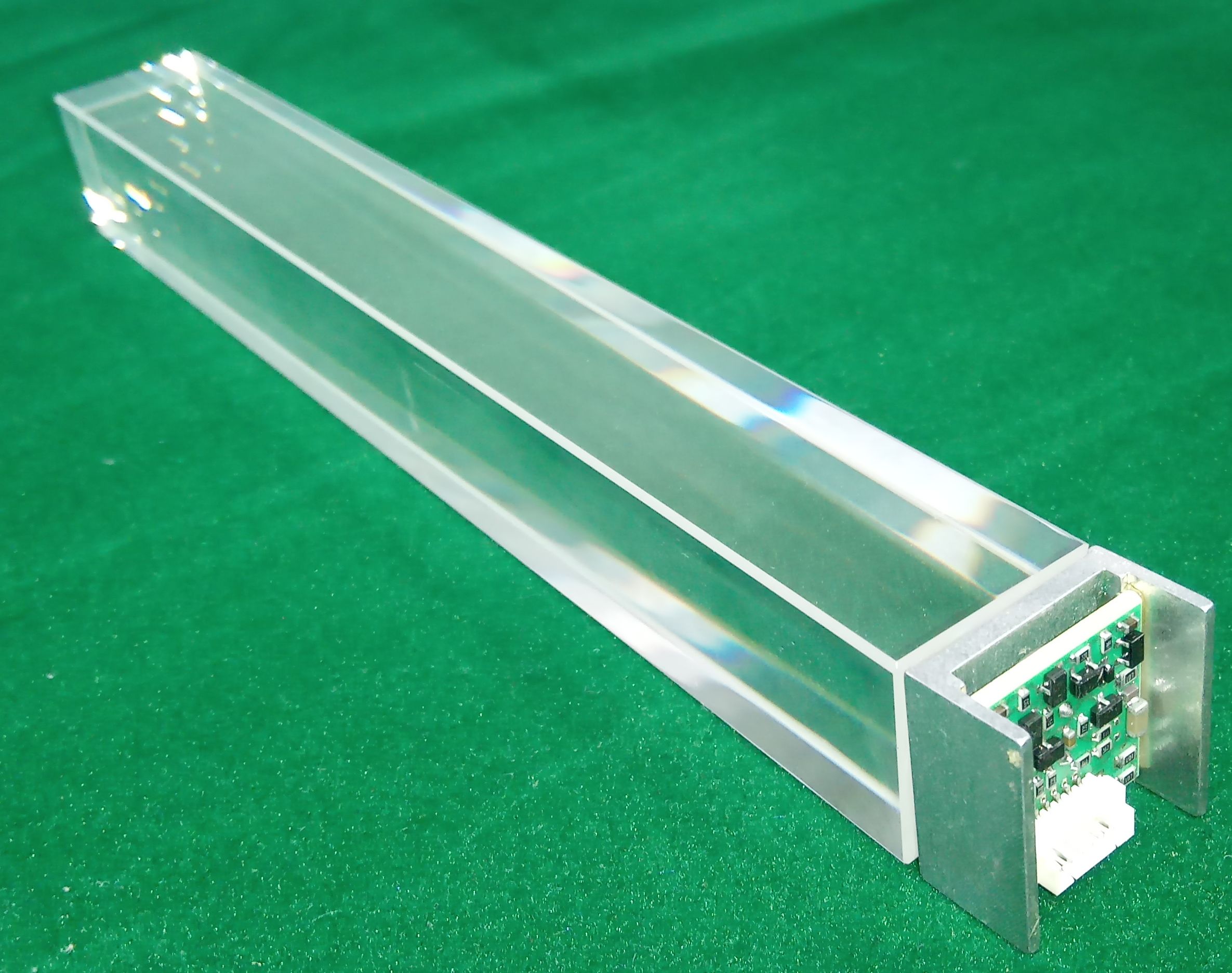}
  \hfil
  \includegraphics[width=0.45\hsize]{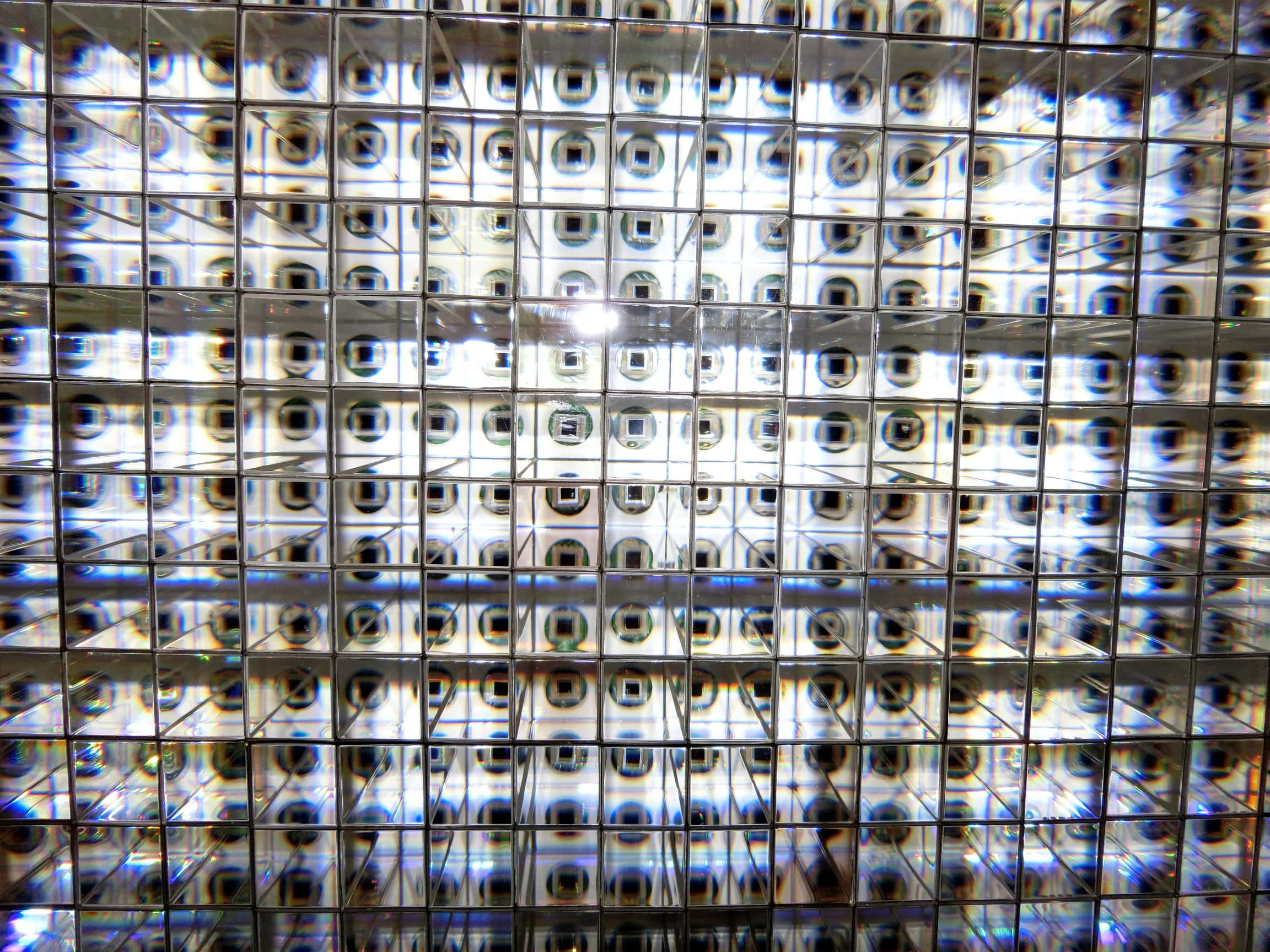}
  \caption[PHOS module]{PHOS detection element consisting of a PbWO$_4$ crystal and
    APD with preamplifier (left) and a fraction of a cell matrix of one
    PHOS module (right).}
  \label{fig:phs:crystals}
\end{figure}

PHOS consists of 4 modules assembled from the active detection
elements consisting of lead tungstate (PbWO$_4$) crystals with 
avalanche photodiode (APD) photodetectors and preamplifiers. These detection
elements, called cells, compose rectangular matrices which are called modules. 
Three PHOS modules covering the azimuthal angle range $260^\circ <
\varphi < 320^\circ$ consist of $64\times 56$ cells,
and one module at the angles $250^\circ < \varphi < 260^\circ$ is a
matrix of $32\times 56$ cells. Figure~\ref{fig:phs:crystals} depicts
one cell and cells stacked into a module matrix.
The front surface of each crystal is positioned at a distance of
460\,cm from the beam axis.

The active detection elements of the cells are made of lead tungstate,
PbWO$_4$, an inorganic crystalline scintillator which has a high density $\rho=8.3$\,g/cm$^3$, a radiation length $X_0=0.89$\,cm, and Moli\`ere radius
$R_M=2.0$\,cm. The light yield of the
crystal is about $0.3\%$ of the light yield of NaI. The luminescence
of PbWO$_4$ has a wide spread in the region of visible photons with 
a maximum at $\lambda_{\rm max} = \SI{420}{\nano\meter}$.
The light yield of the crystals changes by 
$-2.5\%$ for every Kelvin temperature change around the operating temperature.

The PHOS modules are operated at a temperature of $-25^\circ$C to achieve an increase of the light yield by a factor of three compared to room
temperature. In order to ensure these working conditions and to
provide thermal stabilization of $0.1\%$, the crystal matrices of the
PHOS modules are housed in a thermoinsulated volume cooled down by
C$_6$F$_{14}$ for which the flow is provided by the
cooling plant installed outside the ALICE solenoid magnet at about
10\,m from the PHOS modules. The PHOS modules also contain a so-called
``warm volume'' with front-end electronics.
The warm volume and the cold crystal volume are contained in air-tight boxes,
through which dry nitrogen is blown in order to maintain a low humidity.
The environment inside the modules is monitored by a set of temperature and humidity sensors.

\subsubsection{Readout}

For LHC Run~3, like in the previous Runs~1 and 2, PHOS remains a triggered detector,
i.e. acquiring data upon receiving a trigger from the ALICE trigger
system.
The new ALICE trigger protocol provides 2-level
triggers with the level-0 trigger generated \SI{0.8}{\micro\second}
after the collision, followed by the level-1 trigger with a
latency of \SI{6.5}{\micro\second}. The PHOS readout system receives the L0
trigger generated by one of the ALICE trigger detectors (FIT, EMCal, TOF,
PHOS, etc.). The choice of the L0 trigger source is configured by
the central trigger processor (CTP). After receiving the L0 trigger,
the PHOS readout raises the busy signal to block further reception of
triggers until the accepted trigger is processed and shipped to the
FLP upon receiving the L1 trigger. The busy signal is lowered after
sending the whole data payload to the FLP. The dead time depends
on the payload size and varies from 20 to \SI{55}{\micro\second}.
Shipping the L1 trigger is performed by the CTP, and the absence of the L1 trigger
within the time window corresponding to the L0--L1 latency is
considered as a rejection of the triggered event. 

\begin{figure}[htb]
  \centering
  \includegraphics[width=0.5\hsize]{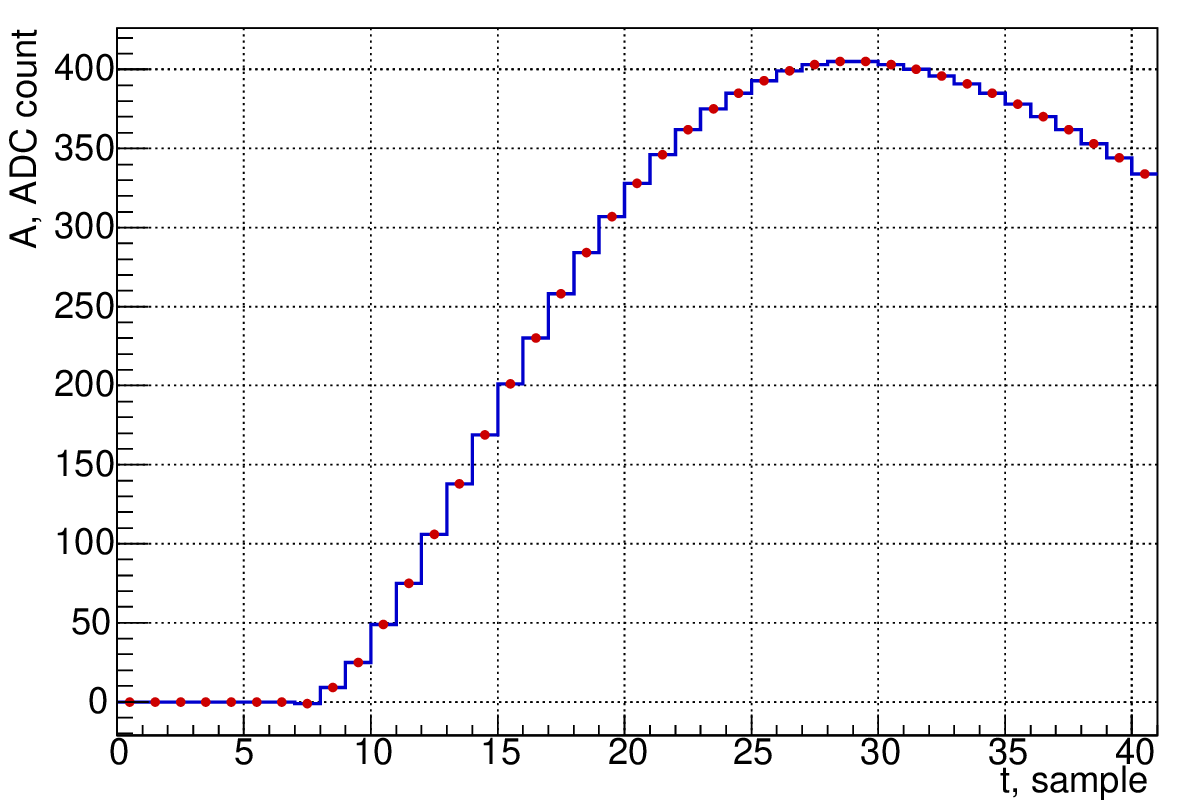}
  \caption[Signal waveform of PHOS FEE channel]{Digitized signal waveform from one channel of the PHOS FEE board. Sampling time is 100\,ns, digitization of sampled amplitude is 10 bits. }
  \label{fig:phs:PHOS_signalShape}
\end{figure}

Energy deposited in each PHOS cell by high-energy particles is
detected by an APD Hamamatsu S8664-55 with a
sensitive area of $5\times5$\,mm$^2$. The APD gain is adjusted to the
nominal value of 50 by setting the bias voltage with an accuracy of
$10^{-3}$ in the range from 210 to 400\,V. The APD signals are passed to
charge-sensitive preamplifiers with an output signal
proportional to the APD charge conversion. Dual-gain shapers with
an average gain ratio of 16.7 generate semi-Gaussian signals with
a rise time of \SI{2.1}{\micro\second}, which are further digitized by a 10-bit
sampling ADC (ALTRO~\cite{EsteveBosch:2003bj}) at a sampling rate of
10\,MHz. The dynamic range of photons detected in PHOS spans from 5\,MeV
to 5\,GeV in the high-gain channels and from 80\,MeV to 80\,GeV in the
low-gain channels. The number of samples is configurable via ALTRO
registers and is chosen to be 37 in order to cover the rising edge
of the signal and its maximum. The sampled digitized waveform of a
signal in one channel is shown in Fig.\ref{fig:phs:PHOS_signalShape}.
One front-end board processes 64 signals generated by high-gain
channels and low-gain channels from 32 PHOS cells. Data collection
from ALTRO, FEE board configuration and data shipping to the readout
units is provided by the Altera ACEX 1 K Family EP1k100 FPGA. The design
of the PHOS front-end board is described in~\cite{Muller:2006bt}.

The PHOS generates triggers at the L0 and
L1 levels to select events with high-energy deposition in the PHOS
cells. The input to the trigger decision starts from the analog sum of
the amplitudes of the group of $2\times 2$ cells implemented in the
FEE boards as ``fast-OR'' signals. Each FEE board produces 8 such
``fast-OR'' signals. The L0 trigger is produced by the Trigger Region
Units (TRU)~\cite{Kral:2012ae} covering an area of $16\times 28$~cells. The
TRU measures the energy deposits in a sliding window of $2\times 2$
``fast-OR'' channels, or $4\times 4$~cells. If the energy in at least
one window exceeds the configurable threshold, the TRU generates the
L0 trigger. The whole detector is inspected by 28 TRUs. All 28
L0 triggers generated by the TRUs are collected by the Summary Trigger
Unit~\cite{Bourrion:2010js} (STU), which performs a logical OR
operation of the inputs and generates the common L0 trigger if at
least one TRU generated a trigger. The TRU boards deployed by PHOS
are similar to those designed for the EMCal detector, with the only
difference being the number of channels: PHOS TRU has 112 channels,
whereas the EMCal one has 96 channels. While the STU boards are
electronically identical for PHOS and EMCal, different firmware is used in the two cases.

FEE boards are read out by a point-to-point protocol via the
designated DTC links (described earlier in the EMCal section) to the
readout concentrator, the Scalable Readout Unit (SRU)~\cite{Zhang:2014ioa}. One
SRU can serve up to 40 front-end boards connected to its 40 DTC
ports. The PHOS readout topology is closely related to the geometry of the PHOS
modules, using 28 FEE boards and 2 TRU per SRU. The whole
PHOS detector is read out by 14~SRUs. Triggering and synchronization
of the SRU is provided by the TTC signal distributed by the ALICE
central trigger. The TTC clock and trigger is propagated by the SRU to
each FEE board or TRU via the DTC links. The SRU raises the busy
signal upon receiving the L0 trigger via TTC and releases the busy
signal when all FEE boards and TRU are read out. Data collected by the
SRU are shipped to the FLP via the DDL link with a bandwidth of
2.125\,Gb/s. All electronic modules (FEE, TRU, SRU)
remain unchanged compared to their hardware state during Run~2. However,
the upgrade concerned the SRU firmware which was adapted to comply with
the new trigger protocol and was modified from the 3-level trigger
sequence in Run~2 to the L0-L1 trigger sequence described
above. Shipping data from SRU to FLP immediately after receiving the L1 trigger
results in a significant reduction of the busy time, allowing an increase of the readout rate.

\subsubsection{Performance}

While the PHOS active detection elements, photodetectors and front-end
electronics in Run~3 remain the same as they were during Run~2, the
physics performance of the detector achieved during Run~2
remains valid for the upcoming Run~3.

The high light yield, short radiation length and small Moli\`ere radius of lead
tungstate, enable high energy
and spatial resolutions for the PHOS. The energy resolution measured in a beam test
is parameterized by the equation~\cite{Aleksandrov:2005yu}
\begin{equation}
  \frac{\sigma_E}{E} = \sqrt{\left(\frac{a}{E}\right)^2 +
    \frac{b^2}{E} + c^2}
  \label{eq:PHOS:eresolution}
\end{equation}
with $a=0.013$\,GeV, $b=0.036$\,GeV$^{1/2}$, $c=0.011$ and the photon
energy $E$ is expressed in GeV. Spatial resolution was evaluated
in Monte Carlo simulations and indirectly confirmed by the mass
resolution of $\pi^0$ mesons in data collected by PHOS during LHC
Runs~1 and 2. The value of the noise parameter $a$ of the energy resolution
(\ref{eq:PHOS:eresolution}) is determined by the APD intrinsic noise
and design of the APD preamplifiers and FEE boards. The stochastic
term $b$ is driven by the light yield of the PbWO$_4$ scintillator at
the nominal working temperature and by the light
collection efficiency defined by the surface area of the APD. Since
the PHOS electronics did not change since Run~1 and 2, the energy
resolution parametrized by (\ref{eq:PHOS:eresolution}) remains valid
for Run~3.

The spatial resolution is parametrised as~\cite{Alessandro:2006yt}
\begin{equation}
  \sigma_x = \sqrt{A^2 + \frac{B^2}{E}},
\end{equation}
where parameters $A$ and $B$ depend on the photon incident angle;
averaging over all angles yields $A=0.96$\,mm, $B=2.29$\,mm$\cdot \mbox{GeV}^{1/2}$.
One of the key performance parameters of PHOS is the two-photon separation distance, i.e. the distance at which individual photons striking the PHOS surface can be identified. As discussed in
\cite{Alessandro:2006yt}, PHOS can distinguish photon showers split by at
least one PHOS cell, $\delta_r = 2.2$\,cm. This feature is especially
important in the high-multiplicity environment of heavy-ion collisions,
as well as for resolving single photons from $\pi^0$ decay at high \pt{}.

The mass resolution for $\pi^0$ and $\eta$ mesons in pp collisions at
$\sqrt{s}=13$\,TeV in Run~2 is discussed
in~\cite{Acharya:2019rum}. The mass resolution is affected by large
incident angles of photons at low transverse momenta and by the
splitting procedure of overlapping showers at high $p_{\rm T}$. Photon
showers from $\pi^0$ decay start to overlap in PHOS at $p_{\rm
  T}>25$\,GeV/$c$, and shower splitting efficiency is reduced with the
growth of the $\pi^0$'s $p_{\rm T}$. At $p_{\rm T}\approx 50$\,GeV/$c$
the most probable distance between decay photons from the $\pi^0$
becomes to be one PHOS cell, then the photons cannot be resolved
anymore. At such high $p_{\rm T}$, the reconstructed $\pi^0$ mass is
distorted by overlapping showers, and the mass resolution becomes
rather large. The widths of the $\pi^0$ and $\eta$ mesons measured
with PHOS in Run~3 will remain the same, $\sigma_{\pi^0} =
4.5~\mbox{MeV}/c^2$ and $\sigma_{\eta} = 15~\mbox{MeV}/c^2$.

\begin{figure}[htb]
  \centering{
    \includegraphics[width=0.48\hsize]{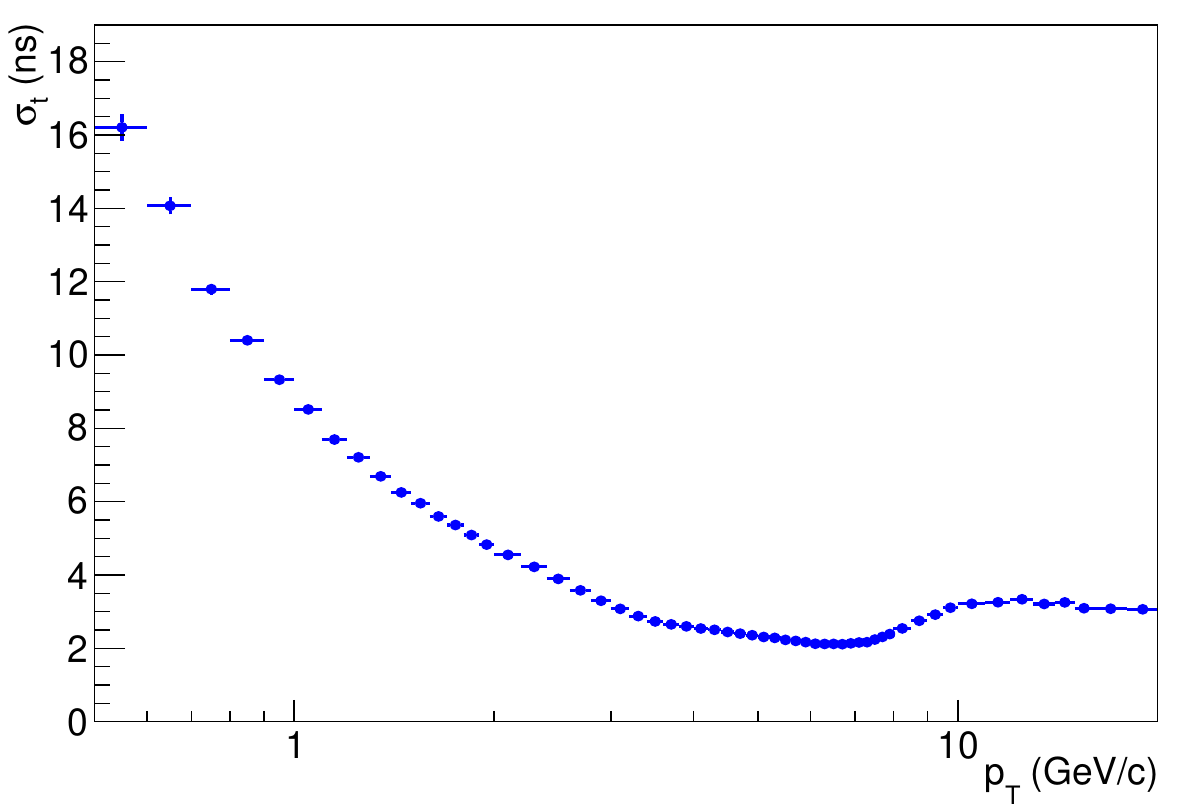}
  }
  \caption[PHOS time resolution]{PHOS time resolution dependence
    on photon transverse momentum during the Run~2 data taking.}
  \label{fig:phs:PHOS_timing_resolution}
\end{figure}

Photon identification in PHOS is based on charged-particle background rejection using matching between charged-particle tracks and clusters, as well as selections of cluster-shape parameters~\cite{Alessandro:2006yt} to discriminate between
electromagnetic and hadronic showers. Measurements of the arrival time could also be
a strong criterion to identify fast neutral clusters produced by
photons and slow clusters produced by heavy neutral hadrons such as
neutrons and antineutrons. The intrinsic time resolution of PbWO$_4$ is
rather good and can reach $\sigma_t=0.15$\,ns at photon energy
$E=1$\,GeV. However, the front-end electronics deployed by PHOS is not
designed for precise time measurement. Timing-resolution dependence on
photon energy, measured using physics data collected by PHOS during
Run~2, is shown in Fig.~\ref{fig:phs:PHOS_timing_resolution}.  At energies below a few hundred MeV,
the time resolution $\sigma_t$ rises above 10~ns. The best time resolution of $2$\,ns is achieved
for an energy $E \approx 5$\,GeV, but then deteriorates because
the low-gain channel is used for larger signals. The achieved time resolution is
sufficient to separate photons of energies $E>1$~GeV produced in different bunch crossings
with 25-ns intervals.

As mentioned above, the PHOS performance will not change in Run~3. Hence, the
PHOS energy and time resolutions and the related systematic uncertainties
on photon and neutral meson measurements achieved in Run~2 will not
improve. However, the ability to collect larger data samples with the new readout
strategy will allow to improve the statistical uncertainties by a factor of two to
five with respect to Run~2.

\clearpage
\subsection{Zero-Degree Calorimeter}

The aim of the ZDC upgrade is to cope with the high collision rate foreseen for Runs~3 and 4. Since the zero-degree calorimeters behaved well with the irradiations during the Run~1 and 2 operations, the calorimeter stacks are unchanged\cite{Aamodt:2008zz}. However, the infrastructure had to be consolidated and the readout system upgraded.

Concerning the infrastructure, two actions were performed. Firstly, the control electronics of the movable platform was upgraded. This platform is used to move the ZDC calorimeters in a garage position where it is shielded from potential beam losses during beam injection or adjustment operations. Moreover, it allows to align them with the neutron (proton) average impact position during data taking. Secondly, additional power supplies for the voltage dividers of the ZDC photomutipliers were installed, in order to stabilize the gain in the high event rate conditions that are foreseen.

The main upgrade activity concerned a new readout system based on faster electronics. In fact the Run~1 and 2 readout electronics were based on VME charge-to-digital converters with a conversion time of $\sim\SI{10}{\micro\second}$ that cannot sustain an event rate of 50 to 100\,kHz without dead time (taking also into account a possible luminosity increase beyond the LS3 baseline). Moreover, in order to fully exploit the ALICE physics potential in ultra-peripheral heavy-ion collisions, the ZDC aims to take data in continuous (autotrigger) readout mode. This operating condition is particularly challenging since the acceptance of the ZDC not only covers nucleon emission from hadronic interactions but also the ones resulting from electromagnetic dissociation~\cite{Pshenichnov2001,Pshenichnov2011,ALICE:2012aa} that have $\sim 50$ times higher cross sections for Pb--Pb collisions at LHC energies. The designed Pb--Pb readout rate of $\SI{100}{\kilo\hertz}$ will be accompanied by an additional $\sim\SI{5}{\mega\hertz}$ event rate, mostly uncorrelated among the two neutron ZDCs (ZNA on side ``A'' and ZNC on side ``C'' of the experiment, at positive and negative pseudorapidities, respectively), resulting from electromagnetic interactions that do not involve barrel detectors.

Because of the low number of channels to be instrumented, the new readout system is based on commercial digitizers, in particular ALICE will use VITA 57 FPGA Mezzanine Card (FMC) digitizers, that allow a continuous sampling of the signal waveform followed by a real time analysis on an FPGA. 
The adequate bandwidth available through the FMC connection from the digitizer to the FPGA allows for the full waveform to be analyzed. Fast trigger and selection algorithms are executed on the FPGA and the relevant portions of the waveform (see below) are transferred to the acquisition and reconstruction system via optical GBT links (see Sec.~\ref{sec:cru}).

To preserve the time and charge resolution 
and to match the bandwidth of the ZDC signals, the digitizers should have about $\SI{12}{\bit}$ resolution (with an effective number of bits of $\sim\SI{10}{\bit}$) with a sampling frequency of
$\SIrange[range-units=single,range-phrase=\div]{0.5}{1}{\giga\hertz}$. Since the photomultiplier signal is unipolar the digitizer has to be DC coupled. After evaluating a few modules, the ADC3112 FMC~\cite{IOXOSADC3112} mounting digitizer ADS5409~\cite{TIADS5409} was chosen. Thanks to the shielded location of the readout electronics there is no requirement of radiation hardness. The FMC is hosted on the carrier ``Intelligent FPGA Controller'' IFC\_1211~\cite{IOXOSIFC1211} with a Kintex UltraScale XCKU40 FPGA. The ADC3112 on-board oscillator is locked to the LHC revolution frequency recovered from a GBT link and dispatched through the FMC connector. Since the ADC will acquire 24 samples per bunch crossing, it will run at a frequency of
$\sim\SI{960}{\mega\hertz}$. Internally each ADC channel is acquired by two digitizers which work in interleaved mode. In order to reduce the data size the low pass filtering with digital downsampling is enabled on the ADC. This has the benefit of improving the measurement accuracy by averaging over the even and odd samples removing the need to correct for the slightly different gains and offsets between the two circuits. The data throughput to the FPGA will therefore be reduced to 12 samples per bunch crossing at $\sim\SI{480}{\mega\sps}$, simplifying the firmware design.

A critical aspect of the ZDC operation in Run~3 is triggering at high rates in \PbPb with the bunch spacing reduced to $\si{50}$ or $\SI{25}{\nano\second}$ since the duration of the photomultiplier signals will be comparable or longer than the bunch spacing. This is complicated by the large signal dynamics (from 1 to $\sim60$ neutrons in the acceptance of the neutron calorimeters). In order to identify the presence of a signal, a differential trigger algorithm was developed. Samples at different times are compared (sample $y_{\textrm{i}}$ with sample $y_{\textrm{i+shift}}$ where $\textrm{shift}$ is a tunable parameter from 3 to 5 samples). If two consecutive differences are above threshold, 
the trigger condition is satisfied, effectively rejecting fake triggers due to electronic noise, and the bunch is flagged for acquisition. This autotrigger condition drives the acquisition in continuous readout mode while in triggered mode the readout system acquires data regardless of the autotrigger flag. The same flags are used also to measure the interaction rate that is used to estimate the instantaneous luminosity.

The measurements of signal arrival time and amplitude need to take into account the baseline (pedestal) oscillations and the possible presence of a signal in an earlier bunch crossing (pile-up).

Two methods for pedestal evaluation were implemented. Given the bunch structure of LHC that alternates ``trains'' of colliding bunches to ``gaps'' where no collisions can occur, it is possible to measure the pedestal considering portions of the digitized data where no collision can occur. These are prescribed by a filling map uploaded on the front-end at each fill. Using this information the pedestal average for each LHC orbit is computed and then transmitted on GBT. This allows taking into account a possible low frequency drift of the baseline and obtaining an accurate reference. A second method allows to effectively subtract the pedestal in presence of noise at higher frequencies. For each trigger (or autotrigger), in addition to the bunch where the signal peaks ($BC_0$), the 12 samples of the preceding bunch crossing will be transferred in order to evaluate and correctly subtract the pedestal in case of a significant discrepancy in the orbit average computed with the first method.

For what concerns the pile-up from a signal in an earlier bunch crossing, in autotrigger mode all ZDC signals are transmitted and reconstructed, allowing to identify and correct for pile-up. On the other hand, in triggered mode, the firmware ensures that the information on the signal inducing pile-up is not lost due to trigger selectivity. Consequently, for each triggered bunch crossing, up to four bunch crossings will be transferred: the triggered and the preceding one (pedestal evaluation) and additionally $BC_{-2}$ and $BC_{-3}$ in case a pile-up signal is detected.

\begin{figure}
    \includegraphics[viewport=0bp 0bp 569.4bp 736.111bp,clip,width=0.5\columnwidth]{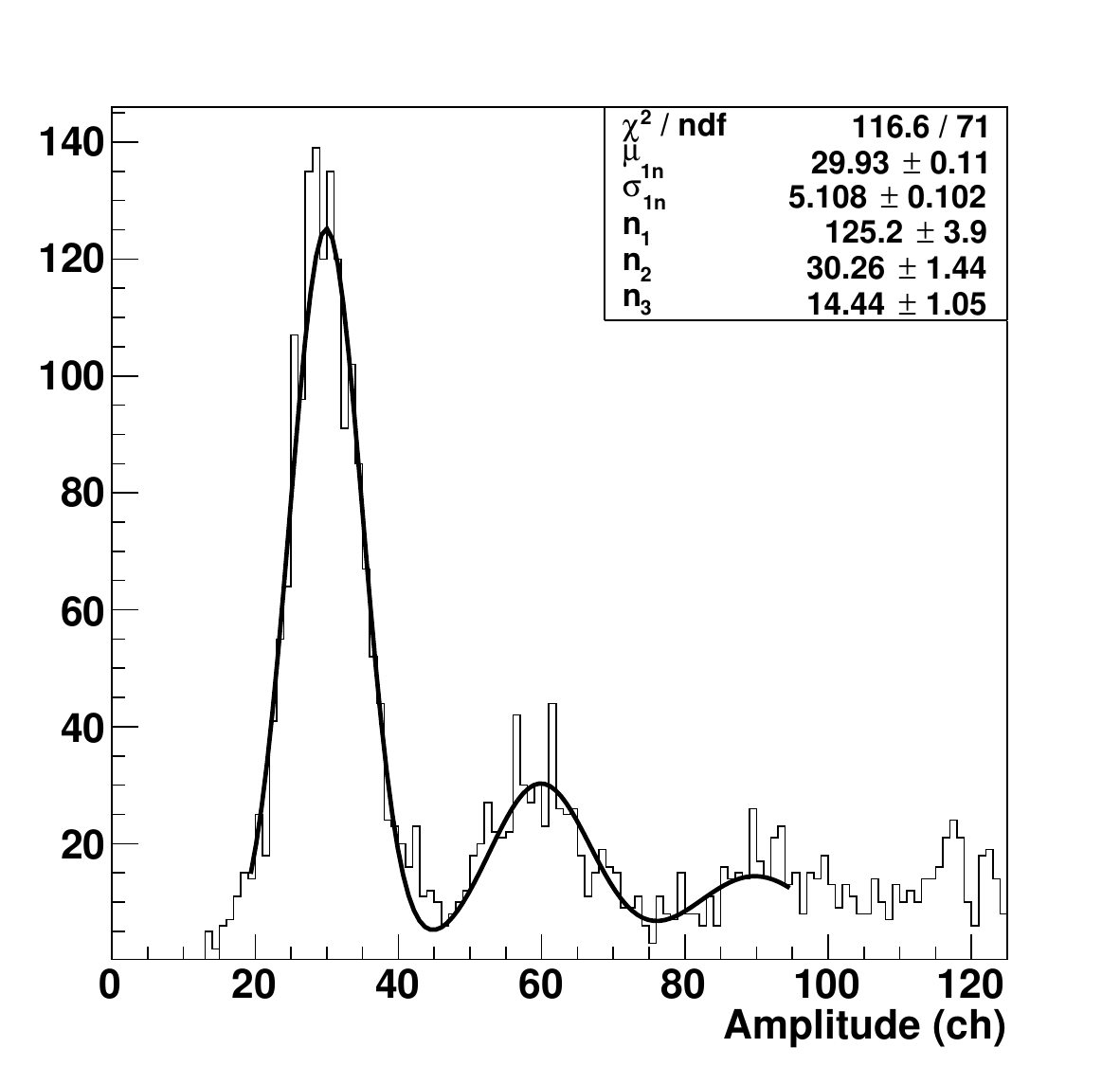}\includegraphics[viewport=0bp 0bp 569.4bp 736.111bp,clip,width=0.5\columnwidth]{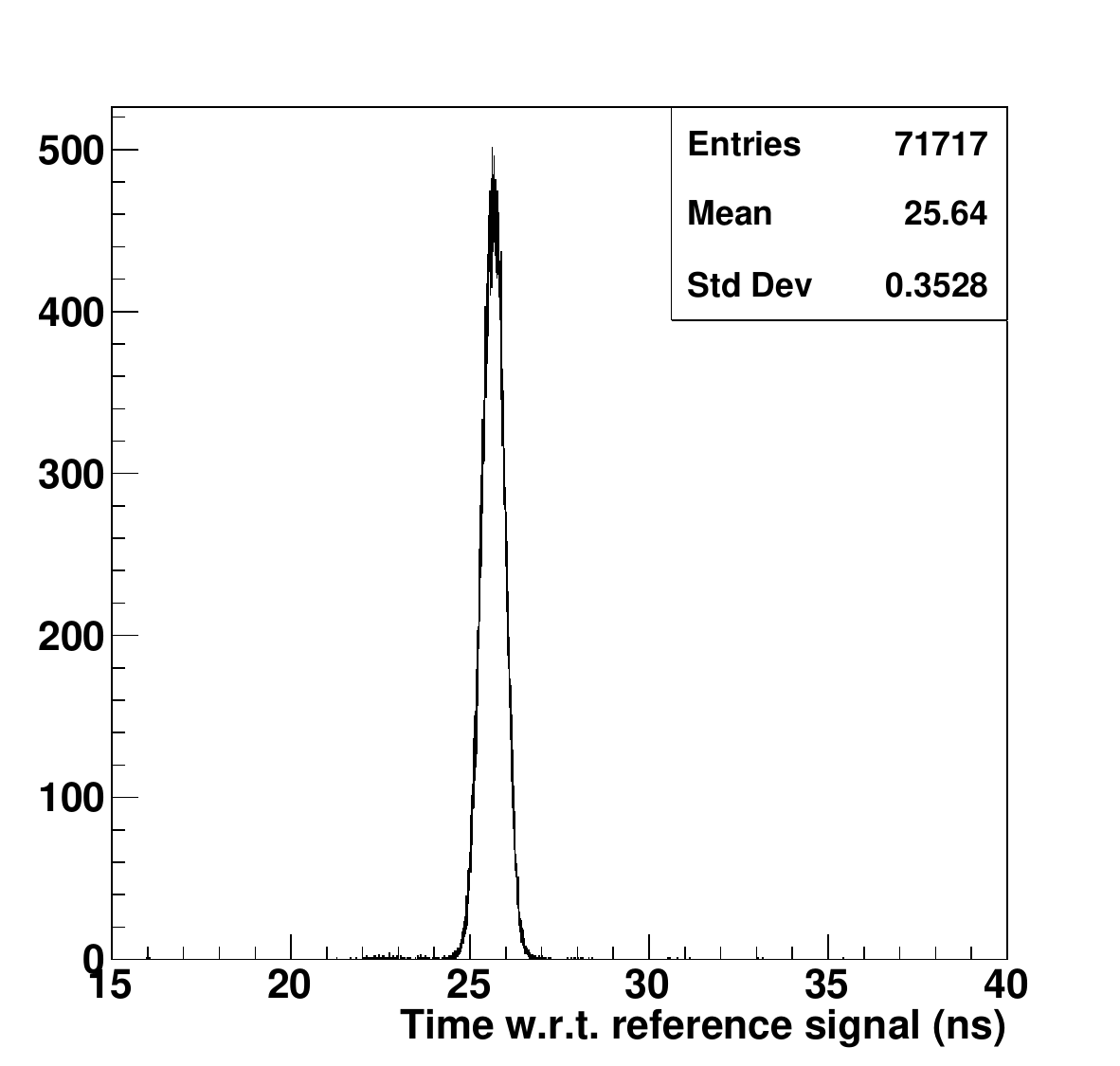}

    \caption[Performance of ZDC digitizer]{Performance of the digitizer during Pb--Pb 2018 data taking in the operating conditions chosen for Run~3. On the left plot: the lower part of the triggered spectrum of ZNC common photomultiplier in Pb--Pb collisions where the emission of a single $\SI{2.76}{\tera\electronvolt}$ neutron and multiples are visible. The spectrum is fitted to a superposition of gaussian functions whose peak positions $\mu_i$ are related to the neutron multiplicity by the relation $\mu_i=\mu_{1n}\times i$ and their widths by the relation $\sigma_{i}=\sigma_{1n}\sqrt{i}$, where $i$ is the neutron multiplicity and $\mu_{1n}$ and $\sigma_{1n}$ are the mean and the r.m.s. of the single neutron peak, respectively. The autotrigger algoritm effectively rejects pedestal events. On the right plot: the arrival time of ZNC common photomultiplier
    signals w.r.t. the reference ALICE L0 trigger signal.}
    \label{fig:zdc-signals}
\end{figure}

During \PbPb data taking in 2018 a prototype of the ZDC system was tested in parallel to the ALICE data acquisition by using a custom system based on Labview reading the ADC\_3112 mounted on a Xilinx evaluation board or using the IOxOS IFC\_1210 carrier. An example of the achieved performances is shown in Fig.~\ref{fig:zdc-signals}. The resolution on $\SI{2.76}{\tera\electronvolt}$ single neutron emission detected by ZNC is $\sim17\%$, resulting in an improvement w.r.t. the $\sim20\%$ of the previous electronics. The time resolution w.r.t. the ALICE L0 trigger is $\sim\SI{0.35}{\nano\second}$, a value that is comparable with the performance of the previous system.

\section{Mechanics and integration}
\label{chap:integration}

The principal layout and infrastructure of the original ALICE detector is described in Ref.~\cite{Aamodt:2008zz}. During LS1 (2013 to 2014), the DCal detector had been added as an extension of the electromagnetic calorimeter on a 60 degree azimuthal acceptance opposite of the EMCal detector.
For this purpose, new support rails and a new support structure holding the PHOS and the DCal modules were installed in the bottom part of the L3 magnet. These new support rails are also used for injecting 10000\,m$^3$/h of cold air into the L3 magnet volume for stabilising the air temperature around the ALICE detector.

For the LS2 upgrade, the global mechanical structures of ALICE remained unchanged. The most important modification was related to the support of the beam pipe and the ITS2 detector. In the original ALICE setup, the TPC had to be moved to the parking position in order to carry out maintenance of the ITS2 detector. This required the disconnection of about 30\% of all ALICE services and would therefore only have been possible in a long shutdown of more than one year. In addition, the beam pipe, ITS2, and TPC were connected in a way that did not allow relative adjustment, so alignment of the beam pipe with the nominal LHC beamline required the adjustment of the TPC or even of the entire ALICE experiment. Such an operation had been carried out in 2008.

\begin{figure}[ht]
 \begin{center}
  \includegraphics[width=8cm]{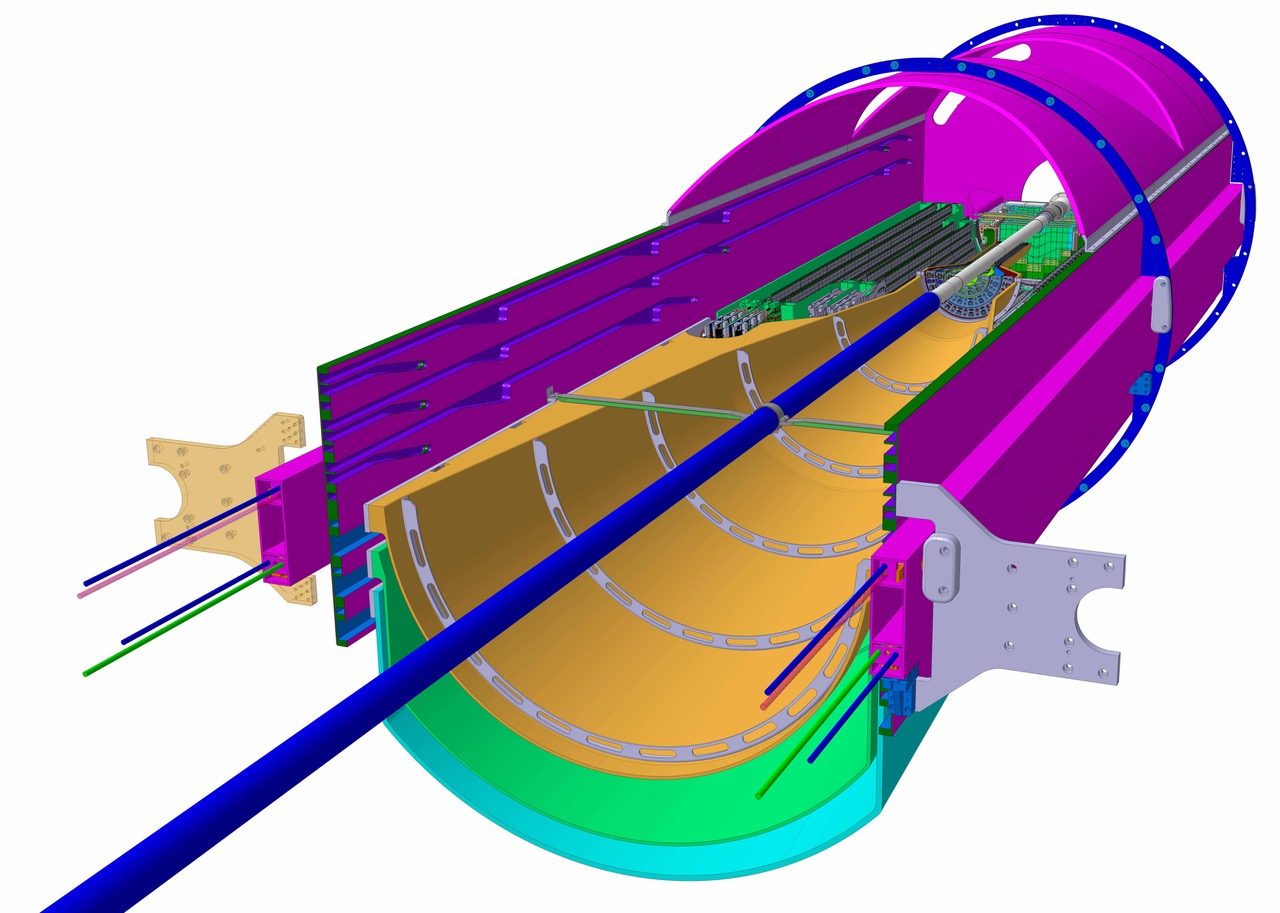}
   \caption[Support of beampipe, ITS2, and MFT]{The cage: a support structure for beam pipe, ITS2, and MFT. Shown is the cage (magenta) with the bottom half of ITS and MFT as well as the beam pipe already installed.}
  \label{cage}
  \end{center}
\end{figure}

For the ALICE~2 detector, the support structures of the ITS2 and the beam pipe inside the TPC were therefore completely re-designed. The cage, a  support structure made from carbon fiber material, was installed inside the TPC as shown in Fig.~\ref{cage}. The cage holds the beam pipe  and has a rail system that allows the installation of the ITS2 and MFT detectors with the TPC in place. This makes it possible to perform maintenance of the ITS2 detector during a year end technical stop of about three months. 
In addition, it allows the alignment of the beam pipe with the nominal beamline within a range of $\pm$\SI{4}{\mm} without the need to move the TPC.
 
\begin{figure}[ht]
 \begin{center}
  \includegraphics[width=12cm]{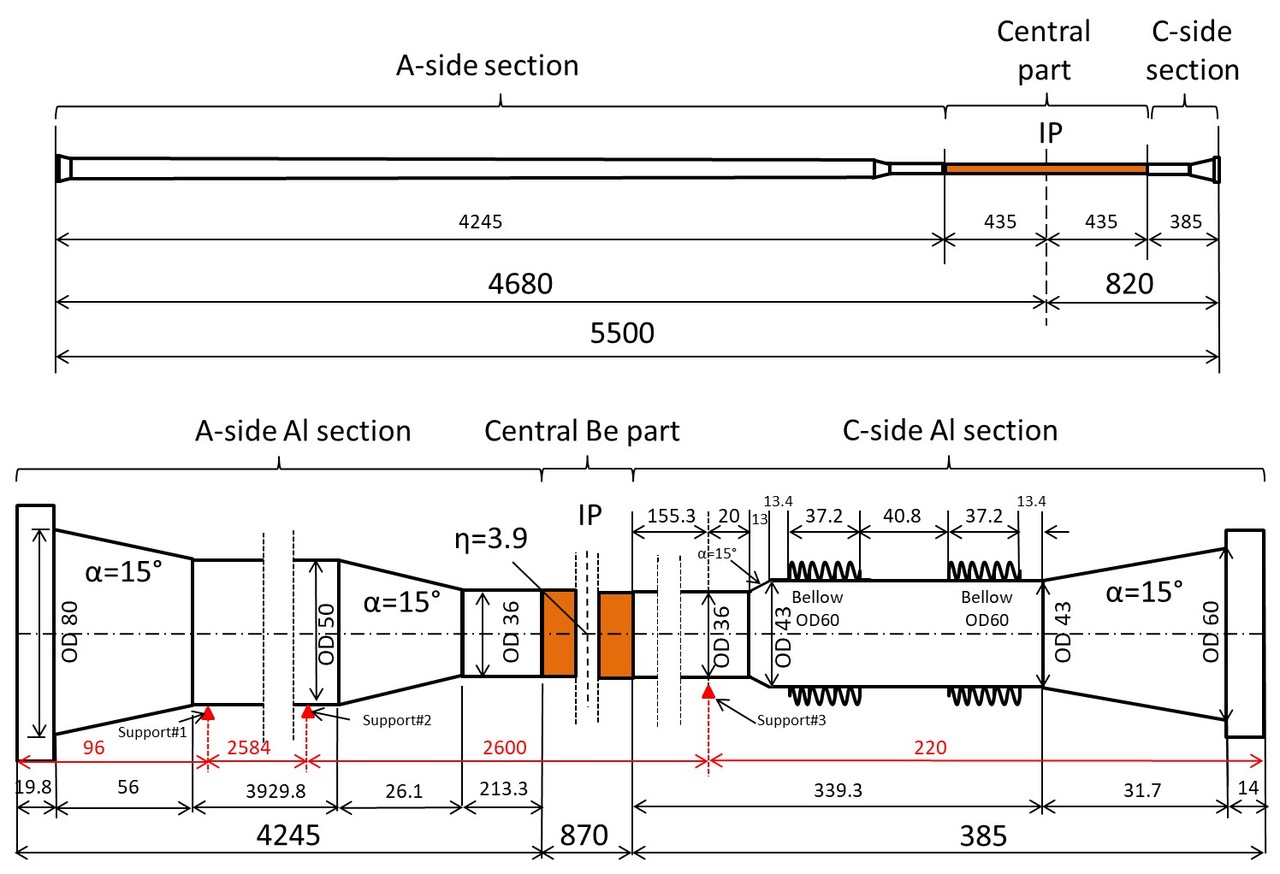}
   \caption[ALICE 2 beampipe]{Beam pipe installed for the ALICE~2 detector with an outer diameter of \SI{36}{\mm}.}
  \label{beampipe}
  \end{center}
\end{figure}

The new ALICE beam pipe has a central beryllium section with a  length of \SI{888}{\mm}, an outer diameter of \SI{36}{\mm}, and wall thickness of \SI{0.8}{\mm} (Fig.~\ref{beampipe}).

\section{Readout and data processing}
\label{chap:readout}

In this section, an overview of the readout concepts and the data flow is given.
Subsequently, the individual systems for experiment and detector control, triggering, data acquisition, synchronous and asynchronous event reconstruction, and the processing of analysis object data are discussed.

\subsection{Readout data flow}
\begin{figure*}[hbtp]
  \begin{center}
    \includegraphics[width=1.\textwidth]{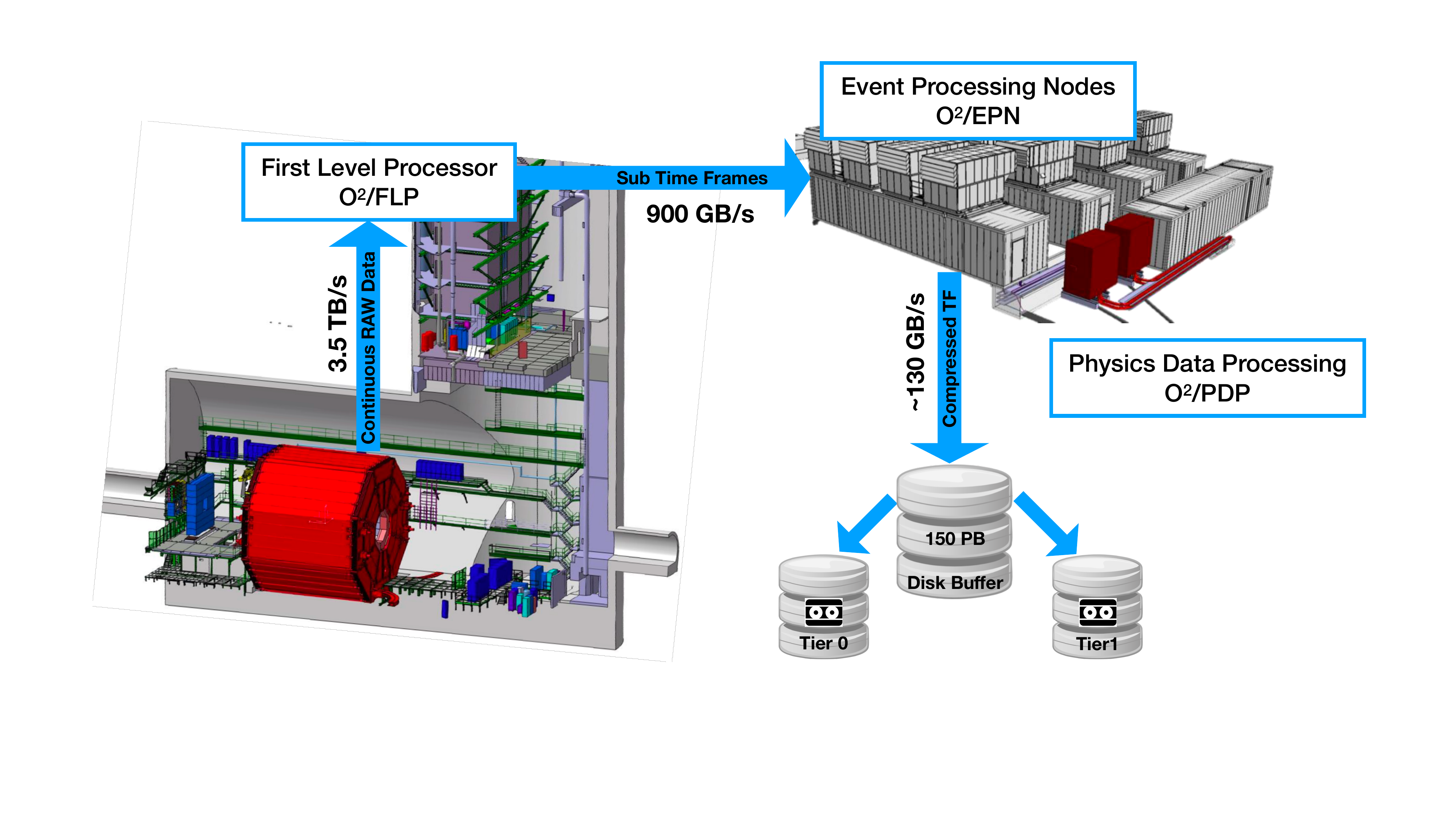}
  \end{center}
  \caption[Readout and processing overview]{Overview of the components of the O$^2$ data read out and processing systems and the main data flows.}
  \label{fig:o2dp}
\end{figure*}

In order to minimise the costs and requirements for data processing and storage, 
the ALICE computing model for Runs~3 and 4 is designed for a maximum compression of the data volume read out from the detectors synchronously with data taking~\cite{Buncic:2011297}. In order to compress the large data flow from the TPC, tracks are reconstructed online. Moreover, data for detector calibration are extracted during online processing avoiding additional offline calibration passes over the full data set.
Online data processing is performed in two steps on the ALICE online/offline facility (${\rm O}^2$) located at Point~2.
The facility consists of two types of computing nodes:  the First Level Processor (FLP) located in the experiment access shaft (CR1), and the Event Processing Nodes (EPN) in dedicated computing containers (CR0), see Fig.~\ref{fig:o2dp}.
The facility provides also the network for data distribution, large disk storage capacity as well as interfaces with the GRID and the permanent data store at the Tier~0 computing center. 

The upgraded online system supports both continuous and triggered readout. Legacy sub-systems not upgraded to continuous readout are not capable of reading the full event rate and thus require a hardware trigger signal. These detectors are
therefore read out whenever they are not busy. Triggered readout for all detectors is also used for commissioning and calibration runs. Data produced by the detectors are transferred to the Common Readout Units (CRU) (see Sec.~\ref{sec:cru}) where they are compressed, multiplexed, and then transferred to the memory of the FLPs.

During the revolution period of the LHC (\SI{\sim 88.92}{\micro\second} = "LHC orbit") there is an LHC filling scheme dependent number of bunch crossings (BC) at which collisions can occur. The ALICE data stream is divided into so called heartbeat frames (HBF) which have a duration of one LHC orbit and are synchronized with the LHC clock. A configurable number of HBFs form a time frame (TF), which represents the data container for data processing and replaces the traditional event entity. The nominal TF length is 128 orbits (\SI{\sim 11.4}{\milli \second}). At 50\,kHz interaction rate, it contains on average 569~\PbPb{} collisions. Continuous and triggered data are tagged by HBF and BC identifiers.

The FLPs perform a
first level of data compression to 900\,GB/s by zero suppression. In addition, they have the possibility to
perform calibration tasks based on local information from the part of the detector they serve.
One example is the TPC for which a first calibration step is already performed on the CRU. The signals from the GEM
readout detectors feature an ion tail and at high occupancy a common baseline shift, that is best removed as early as possible.
A Sub Time Frame (STF) comprises all HBFs belonging to a TF from one FLP. After all FLPs have built their STFs of an individual TF, an available EPN is selected and all STFs are sent there and the full TF is built.

A dedicated FLP is used to collect and process data from the Detector Control System (DCS) in two workflows. 
The first one processes DCS data shipped via the ALICE datapoint server and stores detector conditions like voltages, temperature, and pressure in compact objects (see also Sec.~\ref{sec:DCS_conditions_data}). The second one processes configuration files sent by detectors as well as LHC information. The calibration objects are stored in the condition and calibration database (CCDB)
and from there they are read by the following processing stages.
Another dedicated FLP is used to collect all trigger signals sent by the CTP to the detectors.

The EPN farm consists of 280 servers hosting 8 GPUs and
64 CPU cores each. The capacity has been dimensioned such that it can achieve
a first reconstruction pass (referred to as synchronous reconstruction), extraction of calibration
objects for subsequent asynchronous reconstruction passes, and data compression.
The compressed data are aggregated into so-called compressed time frames (CTF) replacing the original raw data and
written to a
disk buffer at an output rate of about 130\,GB/s. The disk buffer has a raw capacity of 150\,PB and is managed by the EOS system~\cite{Peters_2015}. The erasure coding configuration used for storage protection reduces the usable capacity to about 120\,PB.
 
Calibration data from EPNs are aggregated on dedicated nodes,
processed, and stored in the CCDB. 
CCDB objects are distributed back to the whole ${\rm O}^2$ farm through
multi-casting and migrated to the offline CCDB as well as to GRID storage elements, for usage by the ongoing synchronous reconstruction steps and for the later asynchronous processing and simulation, respectively.

The CTFs are transferred to the GRID for archiving. 
After data taking and full detector calibration, two or more asynchronous
reconstruction passes are performed on the GRID as well as on the EPN farm. 
The output of these reconstruction passes is stored as Analysis  Object Data (AOD), the input for physics analysis. For specific physics signals, a further data size
reduction and speed-up of the corresponding analyses is achieved by filtering out events of interest and writing out only the minimum event information needed.
The processing of pp data follows the same chain with an additional step of selection of interesting collisions during an asynchronous reconstruction pass and reduction of the CTFs by keeping only the
data
associated to these collisions. Reconstruction passes are followed by Monte Carlo production cycles taking into account the time dependent detector conditions.

Besides the computing infrastructure, a common software framework has
been developed within which all 
online and offline components are operated~\cite{Eulisse2019}. It
consists of three main layers.
The Transport Layer has been developed in collaboration with GSI (FAIR) and it uses the FairMQ message
passing toolkit~\cite{Rybalchenko:2019biz} with FairMQDevices as its main building blocks.
It enables efficient parallelism by providing
abstraction of network and inter-process communication as well as
by supporting shared memory backed message passing for devices on the same node. The data model
provides language agnostic and extensible descriptions of messages that are passed
between devices~\cite{ref_O2_data_model}.
It  provides support for various
back-ends such as a so called zero-copy format (a format that optimises performance by allowing to efficiently map files or portions thereof to memory and to share buffers between processes), 
serialisation based on the ROOT data analysis framework~\cite{Brun:1997pa}, and Apache Arrow~\cite{apacheArrow}
for analysis and 
integration with external tools.
Finally, the Data Processing Layer
(DPL) abstracts computation as a set of data processors
organized in a logical data flow specifying how data are transformed.
Depending on the deployment environment, the data
flow is mapped to a concrete topology and from there to a set of processes
running FairMQ devices. %

\begin{figure*}[hbtp]
  \begin{center}
    \includegraphics[width=.99\textwidth]{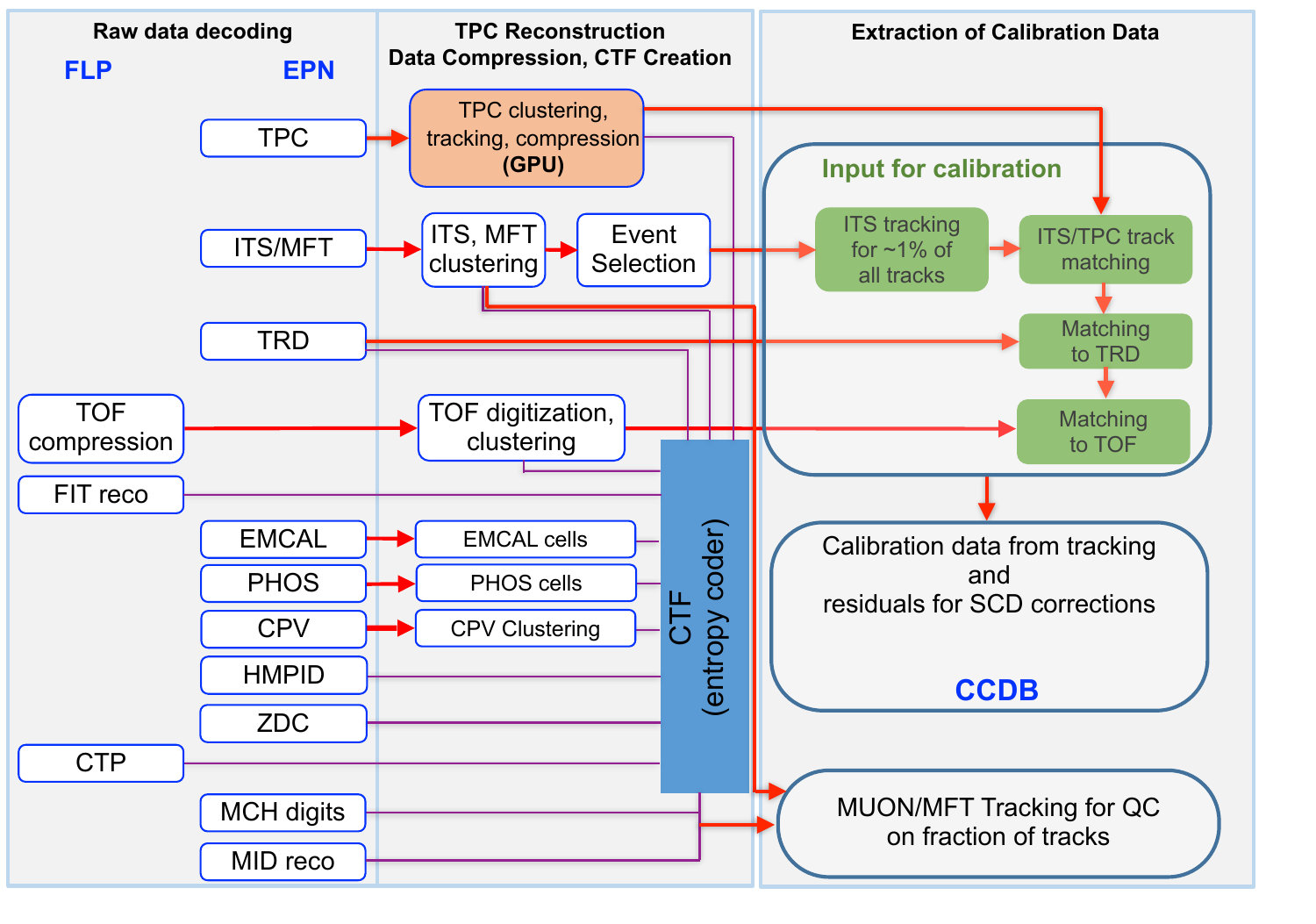}
  \end{center}
  \caption{Synchronous reconstruction workflow.}
  \label{fig:rwf}
\end{figure*}

\subsubsection{Synchronous reconstruction}
A schematic representation of the synchronous reconstruction workflow is shown in Fig.~\ref{fig:rwf}.
The main objectives of
the synchronous processing are the reduction of the data rate from the TPC,
which accounts for most
of the
raw data volume and the extraction of data for calibration. 
This is achieved by performing clustering and full track reconstruction in the
TPC and removing background hits from the data. Moreover, cluster space point
coordinates are stored as relative coordinates, thus reducing the entropy and allowing for efficient ANS entropy
encoding~\cite{ref_rANS} of the data.
The TPC space charge distortion calibration uses the information of fully reconstructed barrel tracks including ITS, TOF, and
TRD information. However, only a small fraction of all tracks needs to be fully reconstructed to gain
sufficient data. Hence, full TPC reconstruction needed for data compression
is the most demanding step in terms of computing time. The online processing makes extensive use of
Graphic Processing Units (GPU), which provide a significant speed-up
by about a factor of 50~\cite{Rohr_2021} compared to one CPU core in an EPN server, without compromising the physics performance.

The TPC reconstruction code has been developed starting from the existing Run~2 High Level Trigger (HLT) algorithms.
It starts with the cluster finding and is followed by tracking comprising the track
finding, track merging, fitting, and compression steps.
The presence of Space Charge Distortions (SCD) of up to 10\,cm
represents a particular challenge for the reconstruction of
continuous data. In absence of triggers, which provide reference for the drift time estimate, the $z$-positions of clusters
are unknown. However, this information is needed for $z$-dependent quantities used during track reconstruction:
the corrections of the SCDs,
the magnetic field strength,
and the cluster error parameterisation.
Therefore, TPC tracking is first performed without these corrections. Since the distortion effects are
smooth, the track finding is not strongly affected. Track seeds are extrapolated to the beam line
and the most probable $z$-coordinate is calculated under the assumption that the track is from a primary particle and the vertex
is at the interaction point. If the track turns out to be a secondary, an average pseudorapidity is assumed.
The track is refitted with the corresponding corrections. The average SCD corrections require a first order correction map obtained from simulation or a previous reference run which is scaled by the instantaneous luminosity.
In addition, the 1D integrated digital currents containing information about the fluctuations of the number of ion pile-up events and of the track multiplicity are used to achieve a partial correction of SCD fluctuations. 
During synchronous reconstruction, cluster positions can be corrected with a precision of $\cal{O}(\mathrm{mm})$ which is sufficient for correct cluster associations to tracks. The full correction with a precision of $\cal{O}(\SI{100}{\micro\meter})$ will be performed during 
asynchronous reconstruction.

Two options for TPC data compression are supported by the software. In the first
option (A) clusters from background (for example from noisy pads or charge clouds
related to low momentum protons) and clusters that are associated to or in the proximity of
background tracks are rejected. Background tracks include those from very low momentum
particles spiralling around the magnetic field lines,
track segments with large inclination with respect to the TPC pad rows, and clusters from secondary legs of looping low-momentum tracks used for physics. 
However, clusters in a tube around good tracks are protected.
For the second option (B), only clusters that are attached to, or
in the proximity of identified good tracks that may be used for physics analysis are kept.
The estimated rejection fractions for options A and B are $12.5-39.1\%$ and $37-53\%$, respectively.
While option B yields lower data size it bears the risk that in case the SCD corrections are not precise enough
track merging and partially also track following might lose good tracks or parts thereof.
Optimal performance of option A requires identification of hits from particles with momenta below
10\,MeV/$c$, since they contribute about 15\% of all TPC hits. Tracking in this momentum region is challenging and is currently under development.  

Further data size compression is achieved by converting the cluster properties from the  single-precision floating point format used in reconstruction to custom integer and floating point formats with exactly as many bits as  needed  for the intrinsic TPC resolution.
The entropy is reduced before the ANS encoding for further data compression. This includes the following steps.
Coordinates of hits that are not assigned to tracks
are sorted by geometrical coordinates and the difference to the previous
hit is stored. Raw coordinates (row, pad, time) of hits assigned to tracks
are stored relative to the extrapolated track (Track Model Compression).
Cluster properties, maximum charge, total charge, and cluster size, are encoded together in order to profit from their correlation.

Synchronous data processing of the remaining detectors is performed on CPU cores in parallel with the GPU processing.
For the ITS and the muon spectrometer system (MFT, MCH, MID), processing starts with space point reconstruction
(clustering). For the barrel calorimeters EMCal, DCal, and PHOS, the cell
properties (time, amplitude) are determined by
fitting the raw time distributions. Clusterization is performed in
order to select cells to write to the CTF, while final clustering
is performed during analysis. Data for time calibration and
dead-channel maps are extracted. 
For FT0, the reconstruction of collision times is performed for the needs of barrel global tracking and vertexing. FT0, FV0, and FDD digits converted from raw data are stored in the CTF.

For a subsample of
tracks selected from peripheral collisions (about 1\% of all tracks),
full tracking including all barrel detectors is performed, i.e. ITS tracking after clustering,
matching of ITS tracks to TPC tracks, and finally track matching to TRD and TOF. As in Run 2, residuals between
global tracks and TPC clusters are used to create 3-dimensional space charge distortion maps with a granularity of  1-2~minutes when running with Pb beams and 10~minutes in pp collisions. These maps together with the TPC integrated digital currents recorded during 
synchronous processing become part of the calibration used in asynchronous processing.

Global barrel tracks are also used to obtain
fast TPC drift time and TRD calibration (gain, $t_0$, $E \times B$, and drift
velocity). Moreover, the drift of the LHC clock with time (due to temperature
changes that impact the fiber refractive index and the distribution of the LHC clock time to the experiments), which affects the reference for the time of flight measurement as a global offset,
is calibrated using global tracks matched to TOF. At the same
time, the TOF channel-level offset in the measured times related to
the cable lengths and electronics is determined. In addition, other calibration algorithms are running during online reconstruction, particularly, for the determination of the interaction region, calorimeter bad channels, and gain parameters. 
The general way to perform these online calibrations is to extract for every TF compact data related to the parameters being calibrated and send them to dedicated aggregator servers. The workflows running on these servers attribute the incoming calibration data to time slots with a granularity characteristic for each calibration type and automatically create a CCDB object for every slot once they have accumulated enough data for processing. 
During synchronous processing, also input data are accumulated
for those calibration constants that need a large amount of data  or are too demanding to be determined synchronously. One example is the TOF channel time slewing. The corresponding calibration information is extracted before the asynchronous
reconstruction takes place, and the CCDB is updated.

The final processing step consists in compressing all data stored in the CTF using the rANS algorithm, a variant of
Asymmetric Numeral System coders, which allows to reach the entropy limit~\cite{ref_rANS, DBLP:journals/corr/Duda13}.

\subsection{First-Level Processors}
\label{sec:flp}

The O$^2$/FLP
subsystem includes the First-Level Processors (FLPs) detector readout farm, the data quality control system, and the services for control, configuration, monitoring, logging, and bookkeeping. 

\subsubsection{The FLP detector readout farm}

\begin{table}[h!]
\centering
\caption{FLP readout farm used to transfer the data from the detectors to the O$^2$ system.} 
\label{tab:flp_farm_links}
\begin{tabular} { l l r r r r r r}
\hline
Detector &  Link 	&  \multicolumn{2}{c}{Readout links}  &  \multicolumn{2}{c}{Readout boards} & Readout nodes\tabularnewline
         &  type 	& DDL   	& GBT       & C-RORC& CRU	& FLPs \tabularnewline
\hline
CPV     & GBT           &               & 16         &       & 1         &   1 \tabularnewline
CTP     & GBT           &               & 14        &       & 1         &   1 \tabularnewline
EMC     & DDL           & 40            &           & 8     &           &   2 \tabularnewline
FIT     & GBT           &               & 34        &       & 3         &   3 \tabularnewline
HMP     & DDL           & 14            &           & 4     &           &   2 \tabularnewline
ITS     & GBT           &               & 432       &       & 22        &  11 \tabularnewline
MCH     & GBT           &               & 550       &       & 30        &  11 \tabularnewline
MFT     & GBT           &               & 304       &       & 11        &   5 \tabularnewline
MID     & GBT           &               & 32        &       & 2         &   1 \tabularnewline
PHS     & DDL           & 16            &           & 4     &           &   2 \tabularnewline
TOF     & GBT           &               & 72        &       & 4         &   2 \tabularnewline
TPC     & GBT           &               & 5832      &       & 361       & 145 \tabularnewline
TRD     & Custom        &               & 1044      &       & 36        &  12 \tabularnewline
ZDC     & GBT           &               & 1         &       & 1         &   1 \tabularnewline
\hline
Total   &               & 69           & 9291      & 16    & 472       & 199

\tabularnewline
\end{tabular}
\end{table}

The readout farm consists of 199 nodes, with 488 readout cards (472 new CRUs and 16 legacy C-RORC) to transfer the data from each detector to the O$^2$ system. 
The number of FLP nodes and readout cards associated with each detector is given in Table~\ref{tab:flp_farm_links}.
The total nominal readout bandwidth amounts to 3.5\,TB/s from the detector electronics to the readout cards where it is compressed before the transfer to the memory of the FLP servers. Most of the detectors use the GBT link and the CRU (see Sec.~\ref{sec:cru}) adopted for this upgrade.
The system is also backward compatible with the Detector Data Link (DDL)~\cite{ref_DDL} and the Common ReadOut Receiver Card (C-RORC)~\cite{ref_CRORC} used during the LHC Runs~1 and 2. 

The server selected for the FLPs is the Dell Poweredge R740. The selection has been done after numerous hardware and software tests\,\cite{ref_FLP} and a competitive tender. Each FLP is equipped with 96\,GB of DDR memory and two CPUs. The CPUs are of two different flavours of the Intel Cascade Lake generation (the Silver 4210 or the Gold 6230 with 10 or 20 hardware cores, respectively) depending on the processing needs of the detector. Each FLP hosts up to three CRUs, up to four C-RORCs, and one Infiniband network interface, each using one PCIe Gen3 x16 slot. The readout software performance allows data to be transferred from three CRUs simultaneously to the FLP memory for a total input throughput of 330\,Gb/s corresponding  to 85\% of the maximum PCIe Gen3 bandwidth. The maximum output bandwidth available to the Infiniband network is 100\,Gb/s.

\begin{figure}[!h]
\centering
\includegraphics [width=100mm] {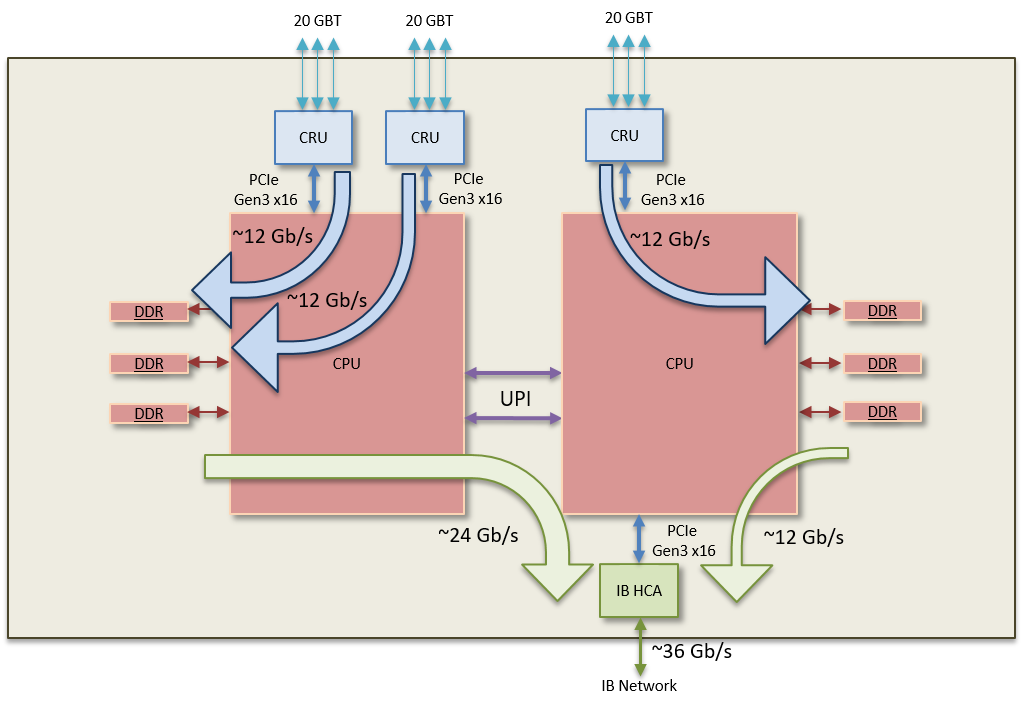}
\caption[FLP dataflow]{Simultaneous dataflows inside the FLP from the CRUs to the DDR memories and from the memory to the Infiniband network to the EPN farm.}
\label{fig_RO}
\end{figure}

The first layer on top of
the PCIe interface of the cards is the PDA (Portable Driver Architecture) UIO (Userspace IO) kernel module\,\cite{ref_PDA}. PDA also provides a user space library in C\,\cite{ref_PDA_lib} which supports PCIe device enumeration and provides a handle to PCI devices.
The readout software includes the readout program and the readoutCard library\,\cite{Alexopoulos:2020nhk} which orchestrate the simultaneous data transfers from the GBT links to the FLP memory as shown in Fig.~\ref{fig_RO}. The transfer of data to the EPN farm is handled by the O$^2$ data distribution (see Sec.~\ref{sec:epn}). 
 
\subsubsection{Data quality control}

The online execution of the calibration and the reconstruction and the replacement of the raw input data by compressed data make reliable data Quality Control (QC) mandatory. Its main purposes are to quickly identify and overcome problems during data taking and to provide good quality data for physics analyses. It is also crucial to ensure that the data processing behaves as expected, especially when running synchronously with the data taking.

\begin{figure}[!h]
\centering
\includegraphics [width=100mm] {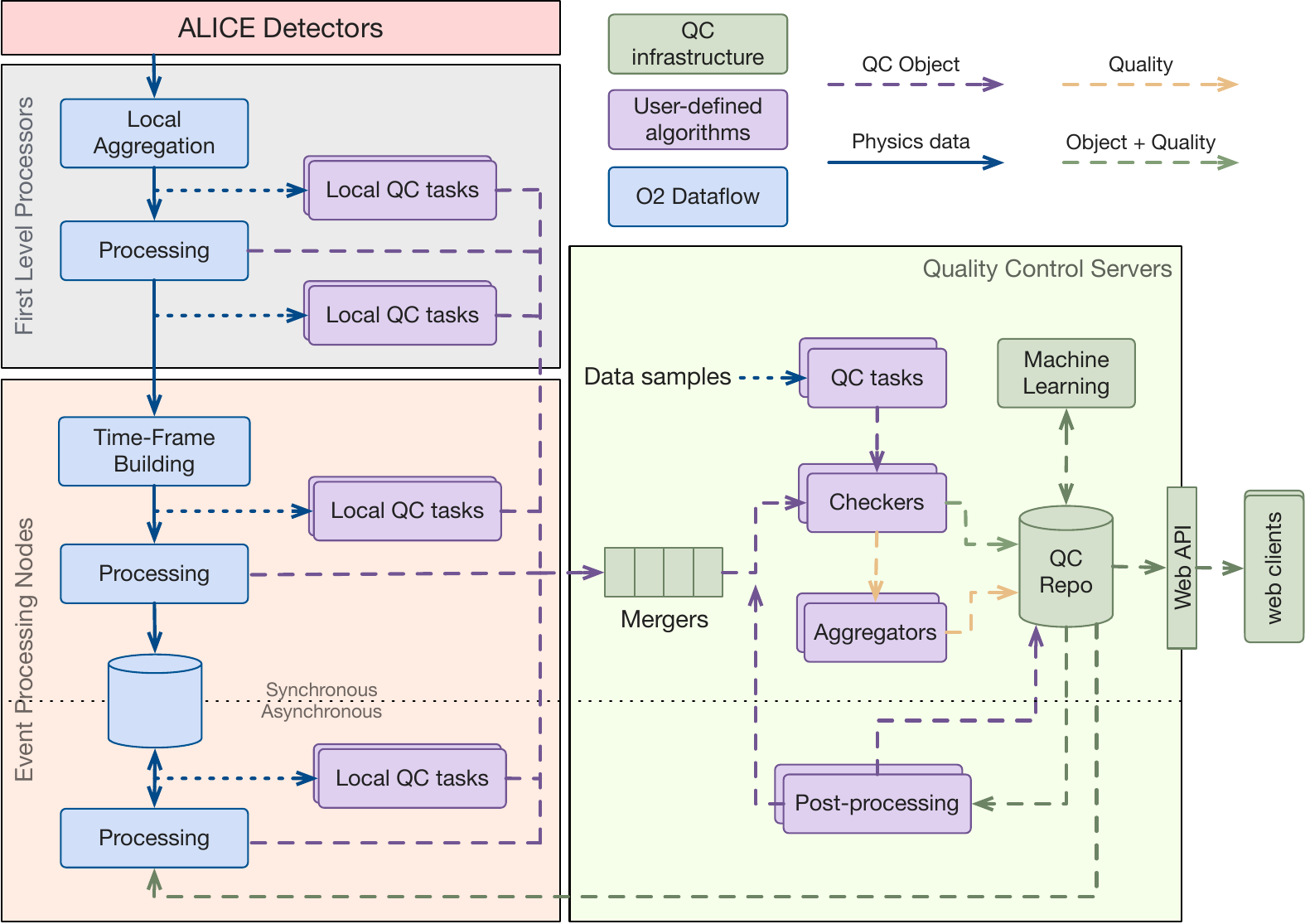}
\caption{O$^2$ Quality Control design.}
\label{fig_QC}
\end{figure}

The O$^2$  QC system\,\cite{ref_QC} includes a distributed software framework as shown in Fig.\,\ref{fig_QC}.

    Data samples are selected following a pseudo-random sampling and configurable policies at key points in the dataflow and are dispatched to local (on the FLPs and the EPNs) or remote (on QC servers) QC tasks executing detector-specific algorithms. Their results are published as QC objects, for example hit distributions in sub-detectors, which are typically represented as ROOT\,\cite{ref_ROOT} histograms. The results of the QC tasks running in parallel on many nodes are assembled by the mergers.  
    Checkers evaluate the quality of the objects, resulting in QC qualities, that summarise e.g.\ whether the hit distributions are good or bad. The QC qualities can optionally be aggregated and are stored together with the QC objects
    in the QC repository. 
    This database has reused the software developed for the CCDB
    of ALICE O$^2$.
    The post-processing component encompasses asynchronous tasks such as correlation and trending of data derived from QC objects and qualities. It is triggered periodically, manually, or on certain events (e.g. start of run or end of fill). The Machine Learning component will be a particular type of post-processing.
    QC and quality objects are accessible to shifters and experts through a web-based QC GUI.
\subsubsection{Services}
\paragraph {Web User Interface framework}
The Web User Interface (Web UI) framework provides the core functionalities and building blocks to easily create rich web applications.
The server side features REST and WebSocket API, authentication via CERN Single Sign-On and authorisation using CERN e-groups.
The client-side features Cascading Style Sheet building blocks for the user interface, 
asynchronous data fetching (Ajax), and bidirectional sockets (WebSockets).
Several O$^2$/FLP GUIs are based on this web interface: AliECS, InfoLogger, QC, and Bookkeeping. 

\paragraph{Control and configuration}
The ALICE Experiment Control System (AliECS)\,\cite{ref_aliecs} integrates the experiment control and configuration, the FLP farm control, and a high-level control interface to the O$^2$/EPN cluster. It implements a distributed state machine to represent the aggregated state of the constituent O$^2$ processes of a data-driven workflow. 
Furthermore, it allows reconfiguration of running processes and simultaneous operation of multiple worflows, with easy reallocation of resources among workflows. Finally, it reacts to inputs, handling events from the user, the LHC, the trigger system, the DCS, and the FLP cluster
itself with a high degree of autonomy. 

\begin{figure}[!h]
\centering
\includegraphics [width=100mm] {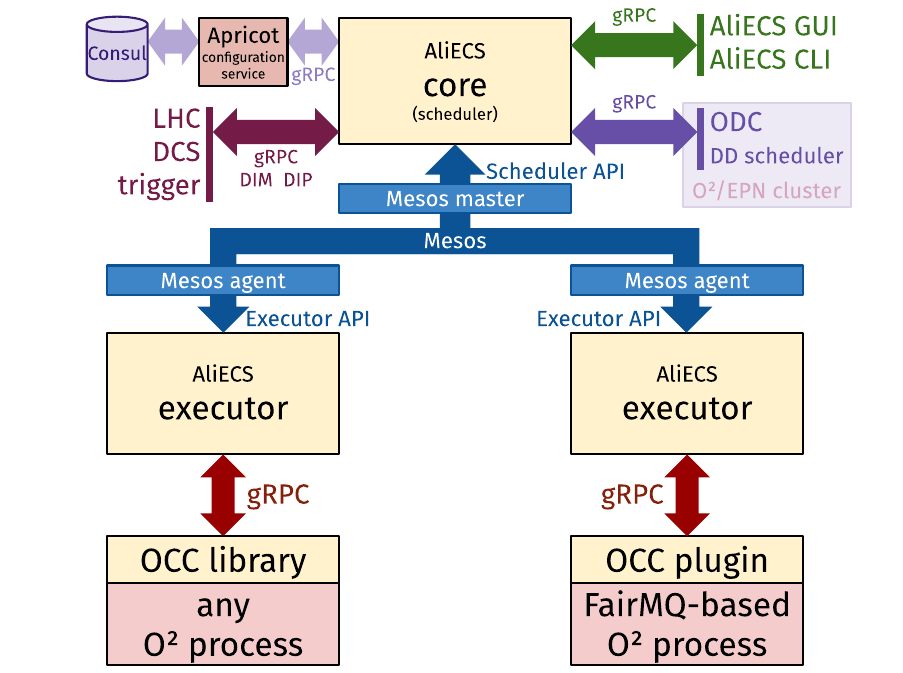}
\caption{AliECS design.}
\label{fig_aliecs}
\end{figure}

Figure~\ref{fig_aliecs} shows the architecture of the system. The AliECS core is the control scheduler implementing the distributed state machine communicating over the google Remote Procedure Call (gRPC) protocol with the operator using the GUI and other interactive applications based on the AliECS Command Line Interfaces (CLI). The AliECS also uses a variety of communication protocols for 
the exchanges with other systems: DIP with the LHC, gRPC with the trigger system, and DIM for the communication with the DCS. 
Apache Mesos~\cite{ref_mesos} is used by AliECS as cluster resource management system for the management of O$^2$/FLP components, resources, and tasks inside the O$^2$/FLP facility, effectively enabling the developer to program against the datacenter (i.e., the O$^2$/FLP facility at LHC Point 2) as if it was a single pool of resources. 
AliECS supports two O$^2$ Configuration and Control (OCC) interfaces to Mesos agents: either through the OCC library or through an OCC plugin for all the processes based on FairMQ, part of ALFA~\cite{ref_alfa}, which is the common O$^2$ transport layer for physics data.

AliECS interfaces with Consul~\cite{ref_consul}, a key-value store which acts as the configuration repository of the system. Once acquired by the AliECS core, configuration information is processed into an in-memory hierarchical key-value store, and from there it is fed into a template system in order to generate task deployment and configuration structures.

Most components of AliECS are written in Go~\cite{ref_go}, a statically typed general purpose programming language in the tradition of C, which is particularly suitable for distributed system development because of its advanced synchronization and threading facilities.

\begin{figure}[!h]
\centering
\includegraphics [width=100mm] {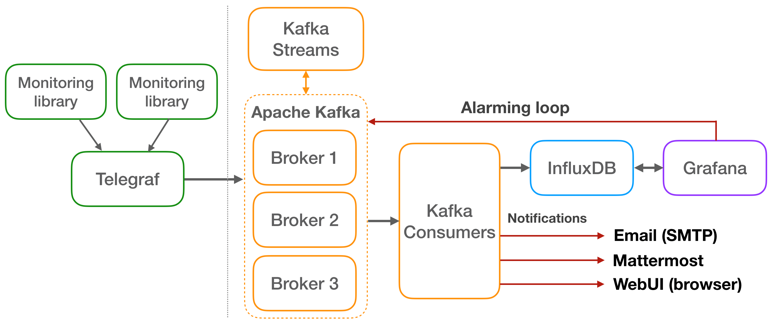}
\caption{The O$^2$ computing system monitoring design.}
\label{fig_monitoring}
\end{figure}

\paragraph{Monitoring}
The monitoring subsystem~\cite{ref_monitor1, ref_monitor2} provides a complete overview of the overall system health and detects performance degradation and component failures by collecting, processing, storing, and visualising values from hardware and software sensors and probes. As presented in Fig.~\ref{fig_monitoring}, metrics are sent to the system from both Telegraf~\cite{ref_telegraf} (for system metrics) and the C++ monitoring library (via Telegraf, for application metrics). These metrics are processed in an Apache Kafka~\cite{ref_Kafka} cluster and later written to an InfluxDB~\cite{ref_influxdb} time-series database for permanent storage.

The InfluxDB time-series database supports downsampling,
which decreases the value resolution over time
reducing the total database size. It is planned to keep high resolution metrics for several days. After that, time metrics will be downsampled in order to decrease the number of points and store them until the end of the calendar year. 

The system includes a data visualisation interface based on Grafana~\cite{ref_grafana} and channels
for alarms and reporting.

\paragraph{Logging}
The logging system has been adapted from the ALICE Run~2 DAQ software~\cite{ref_logging}. A new web-based user interface has been developed in addition to the existing GUIs.
\paragraph{Bookkeeping}
A new bookkeeping system~\cite{ref_bookkeeping} has been developed. It unifies two functionalities: gathering, storing, and presenting metadata associated with the operations of the ALICE detector and tracking the asynchronous processing of the physics data. The front-end and back-end are based on the WebUI framework like the other applications and are adaptive to various clients such as tablets, mobile devices, and other screens. 
The back-end includes a relational database and a REST API specified in the OpenAPI standard that allows to easily build bindings in various languages (as C++ and Go).

\subsubsection{Installation and commissioning}
The O$^2$/FLP system has replaced the former DAQ system used during the LHC Run 1 and 2 in the Counting Room 1 (CR1) located in the access shaft of the ALICE experimental cavern at the LHC Point 2. 
All the optical fibers transferring the data from the detectors to the CRUs have been installed in four campaigns: February--June 2019, October--November 2019, February--March 2020, and November--December 2020.

The FLP specification was reviewed in April and May 2019 and the purchase order made in August 2019. The FLPs were delivered in several batches from September to November 2019. The FLPs have then been prepared to house the CRUs which required a mechanical modification of the chassis from September 2019 to February 2020. The connection of fibers to the CRUs (readout and trigger) was performed from April to August 2020 and the cabling to the network from July to September 2020.

The FLP software has been released as a coherent set of packages monthly since July 2019 and weekly from August 2021.

The test of the FLP system with detector electronics started first in the laboratories in June 2018. Its commissioning with (large pieces of) individual detectors started on the surface in June 2019, and in the experimental cavern at the LHC Point 2 in March 2020. The global tests with several detectors began in July 2021 and the first realistic experience with beams was collected during the LHC pilot beam in October 2021.

\subsection{Event Processing Nodes}
\label{sec:epn}

The EPN farm is designed to perform a first online data reconstruction pass, extract detector calibration objects, and reduce the data volume in order to fit into the available storage buffer space of about 80\, PB.
The compression algorithms rely on data reconstruction properties, and the subsequent asynchronous reconstruction passes rely on the calibration objects.
During periods when data is not being collected, the EPN farm, in addition to other resources, will be used for the asynchronous reprocessing of the data and will contribute computing resources to the physics analysis of the previously recorded data.

\subsubsection{EPN farm}

\begin{figure}[ht]
\centering
\includegraphics[width=0.70\textwidth]{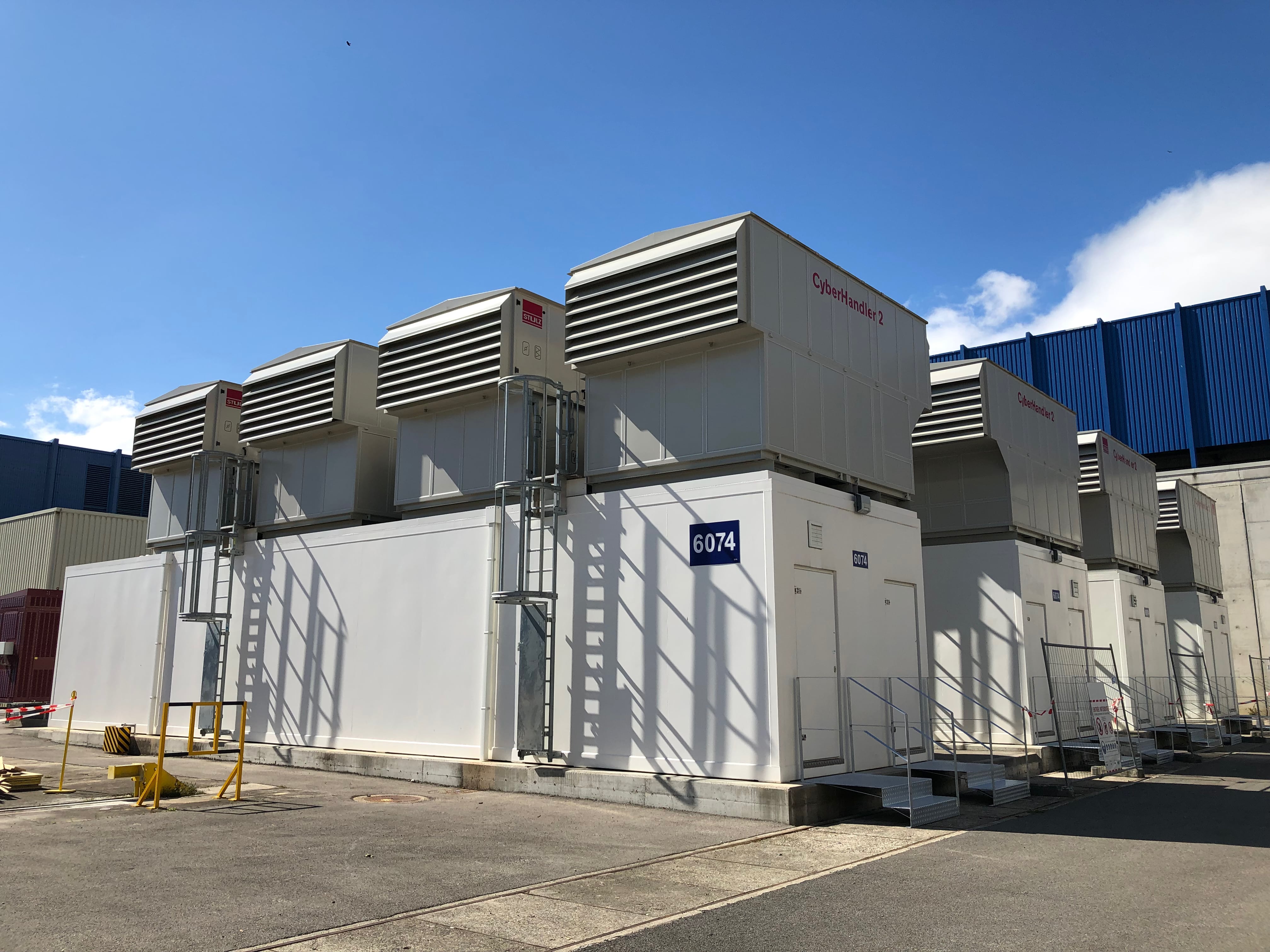}
\caption{The ALICE CR0 data centre which houses the EPN farm.} 
\label{fig:containers}
\end{figure}

Due to the increased Run~3 computing needs and the resulting space and cooling requirements~\cite{Buncic:2011297}, a new data centre for the EPN farm was built on the surface at Point 2 of the LHC, close to the ALICE detector. 
The new EPN data centre shown in Fig.~\ref{fig:containers} consists of four modules for standard Information Technology (IT) equipment and one infrastructure module. 
Each IT module has a cooling capacity of 525\,kW and allows for power densities of up to 1\,kW computing load per rack height unit (\SI{4.5}{\cm}). 
This design allows the use of highly integrated servers, with the maximum number of supported GPUs per server. 

During data taking, the EPN farm will receive up to $\sim$900\,GB/s from the FLP farm.
This data rate will then need to be reduced to $\sim$130 GB/s in real time, in order to write the data to the final storage space, see Fig.~\ref{fig:o2dp} .
The EPN farm is connected to the FLPs via a fast InfiniBand HDR network with a total throughput of 14.4\,Tbit/s. 
Connectivity to the disk buffer in the CERN IT data centre is realized via Ethernet with 100\,Gbit/s link speed and a total bandwidth of 2.4\,Tbit/s, in a high availability setup.%

The EPN farm consists of 280 EPN servers, which provide the necessary computing power 
for the synchronous Run~3 data processing, required by the O$^2$ software. 
The servers were dimensioned benchmarking the compute performance with simulated data via the O$^2$ software. 
This determined the required CPU cores, number of GPUs, and size of the memory per server. 
In the current state of the computing and software infrastructure, 230 EPNs are needed to process the 50\,kHz Pb--Pb data, replaying simulated data. 
A 4U Supermicro GPU server was chosen for its capability to house eight double-width GPUs as well as an InfiniBand HDR host adapter. 
The servers are equipped with two 32 core AMD Rome CPUs (PSE-ROM7452-0057) with 512\,GB DDR4-3200 RAM, eight AMD MI50 GPUs with 32\,GB Memory, a 1\,TB NVMe disk, along with an InfiniBand high data rate (HDR) host channel adaptor (HCA) operated at 100\,Gbit/s.

\subsubsection{EPN installation}

The first containers for the data centre were delivered at the end of September 2018, the last two containers at the end of July 2019. 
Extensive load tests were performed to commission the control of the cooling system, before installing the IT equipment. 
The first usage of the data centre was via a test system using old Run~2 servers, called vertical slice, in October 2019. 
The first batch of the final network was installed and enabled the commissioning of the infrastructure. 
The vertical slice commissioning allowed the testing of the complete chain, from FLP to EPN to the EOS Open Storage system hosted in the computing center at CERN, in a reduced capacity. 
In June 2020, the final network for the EPN farm was installed in preparation for the arrival of the EPN servers, and following the Production Readiness Review for the EPN servers in August 2020, the order was prepared. 
The installation of the servers into the final rack positions was performed in January 2021. 
Another round of tuning for the cooling system was done with the production servers at the beginning of 2021, to optimize settings to the final Run~3 system. 
The EPNs were used for commissioning at Point 2 since beginning of 2021. 
The final part of the network, the gateways from the InfiniBand network to the Ethernet network of the storage facility, was finalised by June 2022 due to delays in the availability of the gateways and issues with operating multiple gateways in a high-availability cluster. 

\subsubsection{O$^2$ data distribution}

The data flow in the O$^2$ system starts with the CRU performing direct memory access (DMA) transfers of the detector data into the memory of the FLP. 
The CRU DMA region is mapped as a shared memory region, which allows efficient intra-node communication between readout, data distribution, and local synchronous processing tasks. 
The zero-copy network transfers are implemented using remote direct memory access (RDMA) protocols of the InfiniBand network interface.
Detector data corresponding to the configured number of HeartBeat Frames (HBF), nominal 128 and up to 256, are aggregated 
from all FLPs on a single EPN node, forming the Time Frame (TF), which is the input for synchronous reconstruction.

\paragraph{O$^2$ data distribution network}

By far the largest data bandwidth requirement in the O$^2$ network comes from the readout data stream from FLPs to EPNs. 
As the data moves from all FLPs to a single EPN node, as mandated by synchronous processing, the flow must be actively regulated to avoid any points of congestion.

\begin{figure}[ht]
\centering
\includegraphics[width=0.95\textwidth]{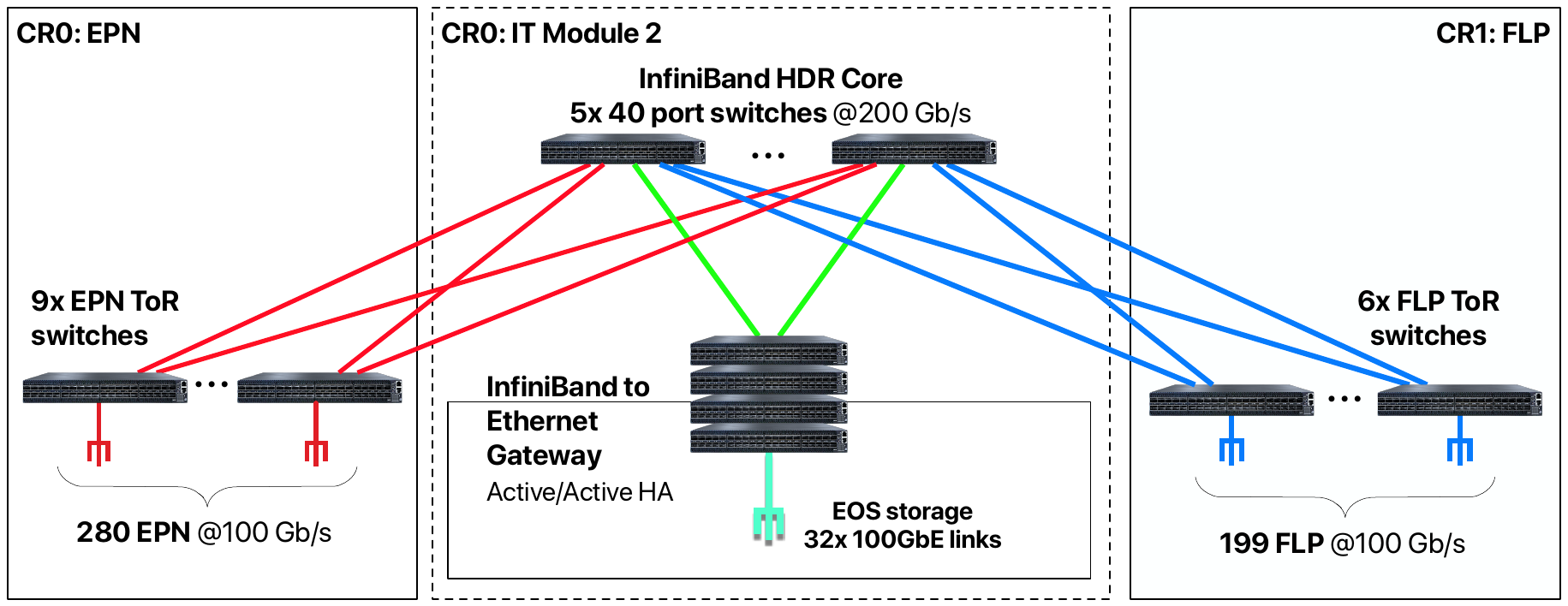}
\caption{Network diagram of the Run~3 O$^2$ facility. }
\answered {should the figure show the data path from left to right instead of right to left, number of FLPs needs to be unified with FLP chapter and readout. Reply: No. of FLPs set to 199 (number from Vasco) and we prefer to keep the figure as is, as it's not necessarily showing a flow from right to left (EOS is in the centre)}
\label{fig:alice_network}
\end{figure}

The network diagram of the entire O$^2$ facility is shown in Fig.~\ref{fig:alice_network}.
To satisfy individual FLP data rates,  which vary from FLP to FLP depending on, e.g.\,the connected subdetector and number of installed CRUs, a 100\,Gb/s InfiniBand network interface was chosen. 
The same interface is used for EPN nodes. 
The architecture of the InfiniBand network is implemented using a two-level folded-Clos topology network, often referred to as a fat tree. 
The network is built using 40 port switches for both core and top-of-the-rack (ToR) level switches. 
Two nodes interface at 100\,Gb/s using a single switch port utilizing copper splitter cables. 
The core of the network is implemented using fiber optic cables operating at 200\,Gb/s. 
Following the requirements, the network fabric features non-blocking communication from FLPs to EPNs, but implements a high blocking ratio within individual EPN and FLP sub-segments, where the bandwidth is not required. 
Additionally, three InfiniBand to Ethernet gateways, with 8 links at 100\,Gb/s each, provide the required throughput for mass storage and other services for offline processing.

\paragraph{O$^2$ data distribution software}

The ALICE detector implements a readout scheme with both continuous and triggered readout
where the full online reconstruction is performed during the data acquisition. 
The data stream from the detectors is grouped into time frames, where one particular TF is processed on a single EPN. 
The detector data arrive into the cluster of 199
input nodes (FLP). 

Given the large number of
FLPs, such a scheme has to implement a deep pipeline, where in the worst case scenario the number of TFs in-building approaches the number of FLP senders. 
However, as the transfers use RDMA read primitives, the EPNs are able to pull Sub-Time Frames (STFs) close to the line rate of network interfaces, without creating congestion in the network or the receivers. This allows for optimal transfers of the STFs and reduces the TF building pipeline.
The data rates across FLPs from different detectors range from 10\,kB/s up to 8\,GB/s, where the size of STFs depends on number of HBFs in a STF and can fluctuate in presence of changing conditions or equipment faults. 
Therefore, the size
of time slices can vary strongly in time and the processing time of time slices is also quite variable. 
Therefore, the data distribution scheduling framework has to take such fluctuations into account. 
There is a trade-off between the implementation of larger buffers on the processing nodes, allowing to average some of the fluctuations, and the overall latency of the time-slice processing. 
Further, the system has to be stable against failures, such as the crash of a processing job or failure of a processing node. 
In such cases, the data loss is localized only to the TFs currently being processed by the affected process or node. 
The data distribution system is designed to accommodate changing processing requirements by allowing the addition or removal of EPNs into the ongoing data taking run.
This enables the use of processing nodes for offline processing during times when there is a low load on the online system. 

\begin{figure}[ht]
\centering
\includegraphics[width=0.95\textwidth]{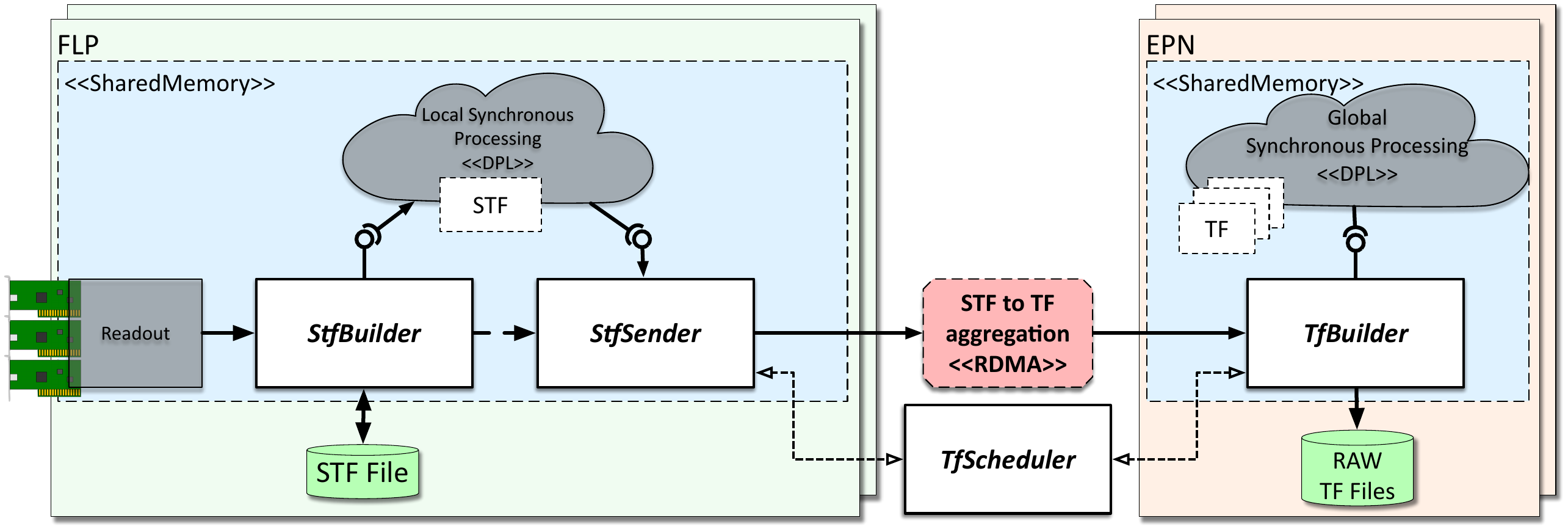}
\caption{Data distribution software framework.}
\label{fig:dd-software}
\end{figure}

The high-bandwidth many-to-one data flow that is needed to assemble full TFs on the EPNs is managed by a data distribution scheduling framework.
Figure~\ref{fig:dd-software} shows the main components of the developed data distribution system: sub-time-frame builder (StfBuilder) and sender (StfSender) on FLP nodes, time frame builder (TfBuilder) on EPN nodes, and time frame scheduler (TfScheduler) on an EPN infrastructure node which orchestrates the data distribution components and regulates the data flow. 
Detector data (HBFs) are transmitted to the StfBuilder, which publishes STF objects for the local processing and the quality control. 
The StfSenders report the availability and information about STFs to the scheduler, which selects a suitable EPN node where the TF can be built, once all STF components are available. %
Therefore, the scheduler keeps track of the utilization of all EPNs and buffer states of sending nodes. 
As several TFs are being built at the same time, due to the many-to-one traffic nature of TF aggregation, the scheduler keeps track of network fabric utilization to avoid creating congestion hotspots. 
Once TfBuilders are given information about all the STFs, they fetch the data from FLPs using remote DMA (RDMA), decreasing the CPU utilization on both the sender and receiver side. 
In the case when the data cannot be scheduled to any EPN, the StfSenders are instructed to drop the data in order to avoid creating back pressure. %
The dropping of complete TFs will occur only when there is not enough processing power available, e.g. insufficient number of EPNs in the partition, or in the presence of failures, e.g. frontend or readout misconfiguration or network issues.

The TF scheduling task keeps track of the utilization of TF destination buffers on EPNs and utilization of the shared memory (readout) buffer on FLPs. In case EPN or FLP buffer utilization is reaching the high watermark (configurable, and typically 32\,GiB for FLP and 112\,GiB for EPN), the TF scheduler throttles the transfers. 
This ensures the network transfers are not propagating back pressure to the FLP readout processes. 
Overflowing FLP buffers could be a result of misconfiguration of the readout card or frontend, where an FLP would generate more than 100\,Gb/s of raw data or an unstable network link resulting in reduced network bandwidth. 
Sufficient EPN buffers might not be available if the number of allocated EPNs is not sufficient to process the incoming TFs.
In the nominal case, when there is no back pressure in the whole data flow and processing chain, the TF scheduler assigns an EPN for each individual TF, maintaining even EPN and network utilization.

\subsection{Physics data processing}

\subsubsection{Asynchronous reconstruction}
Pb--Pb and pp data taking with synchronous reconstruction is followed by an about four to six week period during which the final calibration constants are evaluated. For some detectors, this requires also a short calibration pass over CTF data before the full asynchronous production passes can start.

The final calibration is performed during reconstruction passes, in particular targeting full correction of TPC space-charge distortions and nominal resolution. At this stage, all detectors are included in the reconstruction. The TPC tracks are matched to ITS
tracks and propagated to the outer detectors. The global tracks are established
by combining information from
multiple detectors, and improved track fits are performed.
Primary vertices are reconstructed and secondary vertices are identified in order to
reconstruct V0 and cascade candidates. For long-lived particles decaying at large
radius and producing  TPC tracks unconstrained by other detectors, continuous readout poses an additional challenge. Every pair of unconstrained TPC tracks needs to be
tested for multiple hypotheses of V0s from different primary vertices compatible with
the allowed time (or $z$) range of the tracks. Since the TPC track corrections depend on their $z$ position, this may even require on-the-fly re-calibration and refit of
TPC tracks.  In a final step, the particle identification hypothesis is assigned, based on combined information from all detectors. For the muon spectrometer system, stand-alone tracking is performed for MFT and
MCH, followed by matching of MFT-MCH track-segments and track selection to form global muon tracks. At least two full passes of asynchronous reconstruction are planned to achieve the full performance.

For pp data at full energy, the first reconstruction pass includes an event selection procedure in order to reduce the
overall data size. In addition to physics events of interest, such as heavy-flavor,
high-multiplicity, or diffractive events, events needed for the TPC distortion calibration are
selected. The CTF size is reduced by only keeping the clusters associated to tracks that point to the
primary vertex of a selected collision within $\pm\SI{30}{\centi\meter}$ in $z$. The goal is an event rejection factor of 1000 leading to a CTF reduction
to 1.2\% of the original size. Data from reference runs with pp collisions at the same centre-of-mass energy as the \PbPb{} data taking will not be pre-selected, but fully transfered to mass storage.

When the EPN farm is not (fully) used for synchronous processing, e.g.\,outside data taking periods, it will be used for asynchonous reconstruction.
During asynchronous reconstruction, the number of processing steps is larger than in synchronous reconstruction and without further code running on
the GPU processing is CPU bound. To make optimal use of the GPU resources on the EPN farm also during asynchronous reconstruction, ALICE aims to offload more processing steps onto the GPU with the final goal to run the complete barrel tracking on GPUs. The reconstruction code is written using generic C++ code and can run on different GPU hardware. This opens also the possibility to run reconstruction efficiently on heterogeneous computing platforms that become available on the GRID.

Reconstruction passes are followed by Monte Carlo (MC) simulation productions.
Physics analysis will be performed using GRID computing resources and on dedicated analysis facilities, using AODs from collision data and from MC productions as input
and produces additional physics objects like fully reconstructed charmed hadrons and jets.

\subsubsection{Simulation}
Physics simulation comprises primary event simulation, the transport of particles through the detector
geometry, detector response simulation, and digitisation of the detector signals.
The O$^2$ software framework for simulation~\cite{Wenzel2019} has been developed within the ALFA project, an ALICE/FAIR
collaborative effort based on common components such as FairRoot~\cite{Al-Turany:2012zfk} and FairMQ~\cite{Rybalchenko:2019biz}.

GEANT4~\cite{ALLISON2016186} is employed as the main transport engine.
As for AliRoot in Runs 1 and 2~\cite{fca2003}, the O$^2$ simulation
framework uses external transport codes through the Virtual Monte Carlo (VMC) layer~\cite{Hrivnacova:619573}. This
allows also the use of GEANT3~\cite{Brun:1987ma} and FLUKA~\cite{Battistoni:2015epi} with the same user code, for example for studies of systematic uncertainties or radiation calculations. The detector geometry is described using ROOT/TGeo and
the detector response using the VMC API and callbacks. As part of new developments for O$^2$
the VMC interface has been extended to interfacing fast detector
simulation components which can replace detailed simulation in parts of the detector or
for certain particle types~\cite{Volkel2020}.
Preserving the VMC interface has allowed efficient porting of detector code from
AliRoot into O$^2$.

The O$^2$ simulation framework has been developed with two main objectives
in mind: the possibility to leverage opportunistic resources, in
particular High Performance Computing (HPC) facilities, which frequently offer only very short processing time
windows, and performance optimization through parallelism on top of the capability of individual parts, going beyond standard event multi-threading of GEANT4.
To this end, the simulation process is broken up into individual subprocesses:
primary particle generation (event server), detector simulation, and I/O
processes. These subprocesses run in parallel and interact with each
other via sending and receiving messages. Parallelism is improved by
further dividing the event simulation task into the processing of
sub-events. Multiple independent detector simulation worker devices are
instantiated at the same time. Each of them asks the event
server for work chunks to process, where a chunk is either a full event
or a sub-event. Hence, the system is able to process multiple events in
parallel or collaborate on the simulation of a single event
concurrently. A strategy based on so called late forking makes optimal use of common memory between the different processes. Processing speed-up as a function of simulation workers shows ideal strong
scaling up to the physical number of cores. By reducing the processing time for a unit of work, the framework naturally supports the usage of opportunistic resources providing short processing time windows. In addition, there is the possibility to combine the results of various smaller transport simulations during digitization, so that a large and costly timeframe simulation can be split across multiple smaller jobs if necessary.

The output of detector response simulations are hits typically containing space-point
and energy loss information of particles passing sensitive detectors. They serve as the
input to the digitization step in which also the Time Frames are created. Since at the peak Pb--Pb luminosity on average five events can
overlap within the drift-time of the TPC, a simulated time frame cannot be assembled
from independently digitized events. Hence, the digitisation workflow takes into
account the contributions from different collisions to the same digits. 

Since the full simulation of Pb--Pb collisions is very time consuming, an optimized
simulation strategy, named embedding, has been developed for AliRoot and used in
production during Run~2. The background events are reused multiple times and overlaid with rare signal events
Owing to the overlapping features
mentioned above, the O$^2$ simulation framework supports this strategy naturally. The maximum time gain by embedding is limited by the time spent in digitization, in particular for the TPC.
For this reason, effort has been put into reducing digitization time to a minimum. The fraction of digitization time of the total simulation 
time is $\approx$ 10\% without embedding and reaches 40\% with embedding.

In addition to strategies such as embedding, an efficient MC workload execution engine based on a directed-acyclic graph scheduler was developed. This engine performs dynamic scheduling of tasks in the MC processing chain with the goal to make optimal use of multi-core GRID resources. Moreover, it naturally brings novel features, such as checkpointing or start-stop-continue possibilities to the processing. This is important for debugging or to split the processing over multiple GRID jobs.

\subsubsection{Analysis}
In Run~3, about 4\,PB of AODs will be produced per Pb--Pb running period and
a total of about 50\,PB will be accumulated in Runs~3 and 4. Considering a typical analysis
turnaround cycle of a few days for the full data set and assuming that all data is read only once, a data throughput of the order of up to
100\,GB/s is required. In order to meet this requirement, an optimized analysis model as well as a new analysis framework have been designed.

In order to achieve a fast turnaround cycle for analysis code validation and cut
optimization, 10\% of each data set (including simulated data) are copied to 
dedicated analysis facilities with exclusive access for ALICE (at the time of writing GSI, Darmstadt and Wigner, Budapest). 
Each of these consists of 20000 CPU cores, equipped with fast local storage and an internal network capable of sustaining
high rates of data transfer from the storage to the computing nodes. The
facility only analyzes local data, to reduce problems due to slow external network
connections or remote storage instabilities. The fast internal network allows for
data to be moved quickly from the storage to the nodes.
Analysis task validation on the analysis facility before running over full data sets on the GRID
avoids inefficiencies in the most costly stage of processing.
Moreover, a large reduction of processing time is expected from the systematic
usage of so called derived data sets of reduced size. This is in particular the
case for analysis of rare processes.
Derived data sets can be obtained through event selection (filtering) and/or event
data reduction (selection of only those quantities strictly needed for a specific analysis).

The new data analysis framework~\cite{Alkin:2021mfo} fully leverages the DPL and is built on top of it  offering an even higher level of abstraction for the benefit of analysis code writers.
As in Run~2, analysis is organized in trains consisting of wagons, the individual analysis tasks~\cite{Quishpe:2021pls}. In the new framework wagons correspond to a group of DPL devices allowing to process the tasks in parallel and to remove crashing tasks from the train.
Data are represented in memory as flat tables similar to a relational database and stored as flat
ROOT trees. This saves the processing time for de-serialization needed for the nested C++ objects used in the old framework.
In order to keep the size on disk small, a number of quantities are recomputed automatically when the data are read from disk. Significant development has been done to perform these operations transparent to the user (building on C++17 extensions).
The in-memory tables are implemented using Apache Arrow, an Open Source cross-language
development platform~\cite{apacheArrow}. It provides interoperability between external tools like Python Pandas~\cite{mckinney-proc-scipy-2010},
Apache Spark~\cite{Spark}, and many others. The compatibility with ROOT is guaranteed by using the TArrowDS data
source which allows using Arrow with RDataFrame. Besides I/O efficiency, the new data format
naturally allows for optimized vectorized processing and declarative analysis.
The frameworks API isolates many of the advanced features from the user and data access methods are similar to the ones used in the analysis framework in Runs~1 and 2. This facilitates porting of user analysis code into the new framework.

\subsection{Central Trigger System}
\label{sec:CTP}
\label{sec:cts}
ALICE operates at an interaction rate of 50\,kHz for \PbPb collisions and up to 1\,MHz for pp and \pPb collisions. The majority of ALICE detectors are read out continuously. A minimum bias trigger signal is recorded along with the continuous data in order to flag collisions. For legacy detectors that are not upgraded to continuous readout, the minimum bias trigger initiates the readout if their readout electronics are available and not busy from a previous readout operation.
For pp running, event filtering based on fully reconstructed events will run on the EPN farm.
The upgraded central trigger system (CTS)~\cite{ref_CTS} provides clock, timing, and trigger signals.

\subsubsection{Requirements of the Central Trigger System}
The CTS supports continuous and triggered readout. Detectors upgraded to continuous readout use triggered mode for commissioning and dedicated runs only. 
The CTS governs the continuous readout by sending regular heartbeat (HB) triggers to the front-end of upgraded detectors to synchronise data streams of the detectors and to adjust the data-taking bandwidth by either sending a HB accept (HBa) or a HB reject (HBr) trigger message. The CTS also provides minimum bias triggers
 at three different latencies, referred to as LM, L0, and L1, depending on the timing requirements of each detector. The CTS operates without dead-time by processing trigger inputs and distributing the corresponding trigger output signal for each bunch crossing. The CTS is connected to the ALICE readout via its own dedicated CRU, such that trigger decisions are recorded together with the detector data. 
In addition, the system may be used to monitor the status of all CRUs and is able to throttle the readout rate depending on the status of the CRU buffers. %

\subsubsection{Trigger hardware and interfaces}
\label{sec:ctp_interfaces}

\begin{figure}[t]
\centering
\includegraphics[width=0.7\textwidth]{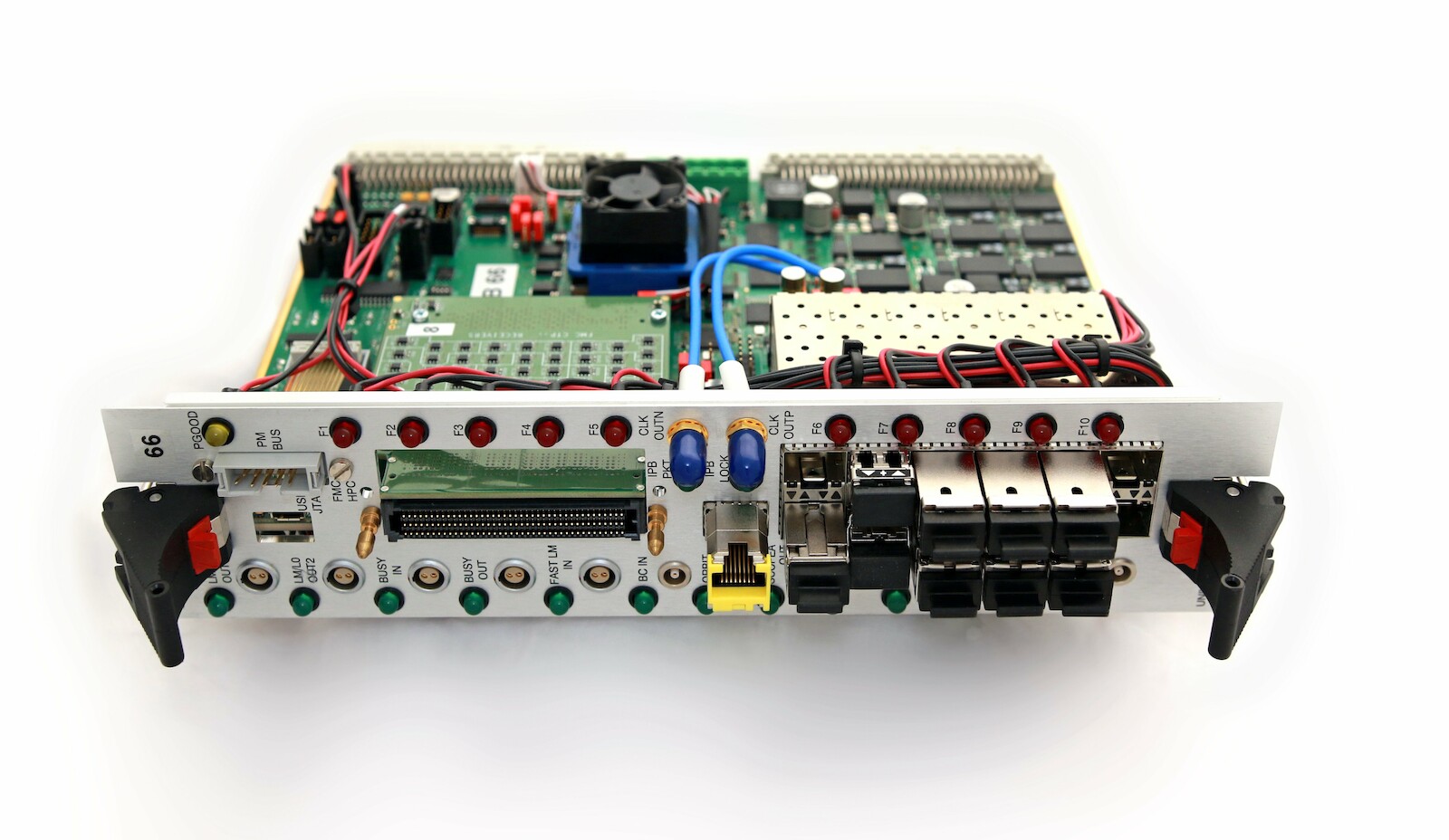}
\caption{\label{ctp_photo} Photograph of a CTP module.}
\end{figure}

Just like the trigger system for Runs~1 and 2~\cite{Aamodt:2008zz}, the CTS system is located in the experimental cavern. It employs a two-stage distribution system, which includes a Central Trigger Processor (CTP, see Fig.~\ref{ctp_photo}) and Local Trigger Units (LTU).   The CTP receives the LHC timing signals and the trigger input signals, and is connected via bidirectional TTC-PON optical links~\cite{Mendes:2017aok} to up to 18 Local Trigger Units (LTU), one for each detector. 
The standard CTS timing and trigger signal distribution path is from the CTS via the detector specific CRUs to the detector front-ends via bidirectional, radiation-tolerant GBT links~\cite{tpc:moreira2009}.
Detectors that require latency-critical trigger signals receive these trigger signals additionally on a direct path from the CTS to the detector front-ends on GBT links.
Legacy detectors not supporting continuous readout are read out via C-RORC readout cards~\cite{ref_CRORC,Alexopoulos:2020nhk} and require a hardware trigger signal to initiate the readout. They receive the clock and trigger signals via the legacy TTC system~\cite{Taylor:592719}. 
In commissioning runs, the LTUs may be decoupled from the CTP and used to emulate CTP signals for testing purposes. 

The CTP and LTU are based on identical PCBs. Each board contains one Xilinx Kintex Ultrascale FPGA, two 1\,GB DDR4 memories, two Si5345 PLLs, one FME-HPC connector, two six-fold SFP+ cages, a single SFP+ cage, and two UCD90120A power controllers. The CTP and LTU boards only differ by the installed FPGA: the CTP board is equipped with the more performant XCKU060-2FFVA1156E; the LTU uses the pin-compatible XCKU040-2FFVA1156E FPGAs.
The boards feature a triple-width 6U VME format using the VME backplane for power supply only, with all data interfaces being available via the front panel.

The CTP board utilises an FPGA mezzanine card  with a total of 72 LVDS I/O connections: two differential links are used for clock signals, 48 for trigger inputs (12 LM, 12 L0, and 24 L1), four for BUSY inputs from legacy detectors, and two for direct LM trigger outputs for detectors requiring a minimum latency trigger. Some of the LTUs are equipped with a commercial FMC S-18 card to extend the number of optical connections to the detectors from 12 to 19.
The LHC clock and orbit signals are connected via Lemo connections.
The cards are remotely programmable via their JTAG ports connected to an Ethernet adapter and are controlled via an IPbus interface~\cite{Larrea:2015wra}.
The CTS allows monitoring of internal counters, including trigger inputs, subdetector BUSY signals, and an internal snapshot memory.

\subsubsection{Trigger protocol and data format}
The minimum bias trigger input signals are delivered to the CTP by the FIT detector. The TOF, EMCal, DCal, and PHOS detectors also deliver trigger inputs for dedicated run scenarios.
The CTS aligns the trigger inputs and synchronises them to the BC clock. The trigger algorithm is applied using a lookup table and produces the trigger output signals. 
The latencies for the trigger input signals to reach the CTS are \SI{425}{\nano\second}, \SI{1200}{\ns}, and \SI{6100}{\ns} for LM, L0, and L1, respectively. The CTS processing and signal propagation time is about \SI{150}{\nano\second}. 
The total latency from interaction to output trigger signal is \SI{575}{\nano\second}.
The CTS can generate internal triggers controlled by software which are used for debugging and detector calibration.

The CTS allows grouping of detectors into up to 18 clusters, forming a data acquisition partition independent from other clusters. Naturally, it also foresees the inclusion and exclusion of individual detectors depending on run conditions. Similar to Runs~1 and 2, the trigger signal distribution for the triggered legacy detectors is protected by a BUSY signal, which communicates whether a detector is ready to receive the trigger signal. 
The trigger message transmitted to the CRUs and detector front-ends consists of the trigger type (32 bit), the LHC orbit counter (32 bit), and bunch crossing counter (12 bit).
The CTP is read out similar to a detector and its data is merged into the continuous data stream.
The CTS readout contains information on the trigger messages sent to the detectors, the trigger input mask (specifying the active CTS trigger inputs), 
and the trigger mask (specifying the trigger conditions and the active detectors) for each bunch crossing.
The transmission of the status information of all the CRU buffers sent upstream to CTP from CRUs 
will be implemented at a later stage. 

\begin{figure}
\centering
\includegraphics[width=0.7\textwidth]{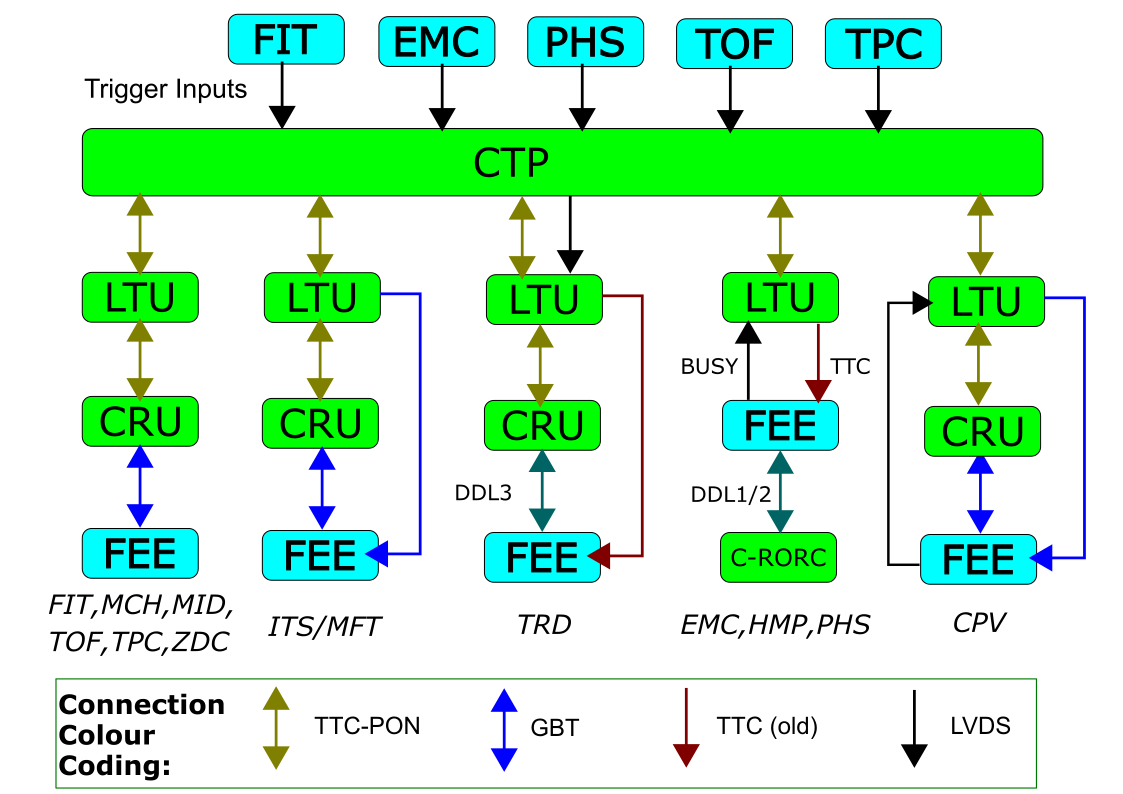}
\caption[Trigger system overview]{\label{fig:systemoverview}Trigger system overview.}
\end{figure}

A general overview of the trigger system is shown in Fig.~\ref{fig:systemoverview}.

In the continuous readout scheme, the readout rate may be downscaled by introducing periodic HBr triggers. That data rate can be adjusted by changing the ratio between HBa and HBr triggers as required.
At present, the system allows the application of a pre-defined sequence. Dynamic modification of the ratio depending on the CRU buffer status will be implemented at a later stage, once the operation of the ALICE system is fully tuned.
This functionality uses the transmission of the state of the CRU buffers in the HB acknowledge messages that are sent by the CRUs to the CTS upon reception of a HBa or HBr message. The HB acknowledge messages of all CRUs are assembled into a HB map in the CTS which is used to assess the buffer states of all CRUs and to decide on modulating the HBa/HBr rate accordingly.

\subsection{Detector Control System}
\label{sec:dcs}

The ALICE Detector Control System (DCS) ~\cite{DCS:GENERAL} ensures safe and stable operation of the experiment. Its architecture is derived from the previous versions used in Run~1 and Run~2, but significant extensions were developed and deployed for Run~3 to allow for integration of new readout electronics. The internal conditions data flow was also modified, to provide continuous streaming of conditions data in real time ~\cite{DCS:RUN3}.  

An optimized standalone DCS system is available for each detector. These systems are built by detector teams, following the guidelines prepared by the central DCS team. The central DCS  further integrates the detector DCS into one distributed system, which can be operated by a single operator.

During the design phase of the DCS great attention was given to the selection of hardware components and software tools. All systems are based on the commercial SCADA (Supervisory Control and Data Acquisition) system WINCC OA~\cite{DCS:WINCCOA} provided by Siemens. SCADA systems are based on industrial standards and are widely used to efficiently supervise processes by monitoring and controlling the devices. High level of standardization allows for deployment of common solutions  which largely simplify the development cycle and reduce deviations from operational standards adopted in ALICE:

- The CERN Joint Control Project (JCOP) provides a common framework for integration of standard devices such as power supplies, embedded logical controllers (ELMB) ~\cite{DCS:ELMB}, magnetic field sensors, etc. It is developed as a joint effort of LHC experiments.  ALICE contributes with a component for integration and control of ISEG power supplies. The JCOP framework also contains, for example, all necessary tools for integration of CERN standard protocols DIM ~\cite{DCS:DIM} and DIP ~\cite{DCS:DIP} as well as the State Management Interface (SMI++) ~\cite{DCS:SMIPP} used to model the device operation as FSMs.

- The ALICE framework further extends the JCOP framework with tools specific to ALICE such as a unified user interface, or FRED framework for integration of front-end modules.

Despite its complexity, the whole DCS can be operated by one person using a single user interface. The FSM mechanism guarantees that commands are executed in a correct order and experiment conditions (such as status of cooling or power systems) are always taken into account. Predefined states were implemented as a response to different conditions of the experiment. For example, the requirements for high voltage settings during the beam injections differ from those during data taking.

To minimize the human factor in the operation, most actions are executed by the system without the need of manual intervention. The operator specifies the desired state (for example, move ALICE to a state compatible with beam injection or magnet ramps) and the system determines and executes the corresponding sequence of commands. Direct interaction of the operator with the system is required only in case of exceptions (for example the recovery from power cut or detector trip). Almost all operations could be executed automatically, however certain checkpoints where operator response is required were introduced to ensure that sufficient attention to the operation is given by the shift crew.
 
\subsubsection{DCS computing hardware upgrades}

The core of DCS computing is installed in ALICE counting room CR\,3. Three rows of racks originally hosting the DCS cluster were replaced with 26 new racks in two rows, each equipped with water cooled doors. All DCS computers were replaced with new hardware. In total 200 new servers, mostly running WINCC OA~\cite{DCS:WINCCOA}, were installed and configured. These servers run the central distributed SCADA system and front-end control servers, and provide services required for DCS operation (DNS, BootP servers providing boot images to diskless controllers, fileservers, databases, etc.). Old servers were kept operational alongside the new servers in order to provide a smooth transition to the new hardware for the detector groups. Using this approach, the DCS maintained almost uninterrupted operation during the whole LS2 period. 

Files shared between several systems are residing on a new redundant cluster of file servers. These are also hosting installation repositories and system backups. The fileservers are mirrored in a cluster installed outside of the ALICE experiment network.

The DCS configuration data (device settings, front-end configuration, etc.) and conditions data (temperatures, currents, pressures, system states, etc.) are stored in an ORACLE database. This mission critical service is provided by a new cluster consisting of four servers and highly redundant storage. The whole database is replicated using ORACLE Active Data Guard technology (ADG)~\cite{DCS:ADG} to a twin cluster installed outside of the ALICE detector network. The ADG provides data availability and protection by mirroring data to a standby database which can replace the ALICE online cluster in case of severe failure. The standby database further provides ALICE data in read-only mode, to offload heavy load operations from the production cluster. 

All computer racks hosting the DCS cluster are equipped with switches with 10 Gigabit Ethernet uplinks to the router. This provides sufficient bandwidth for DCS operation also beyond Run~3. A total of 144 multi-mode and 48 single-mode patch panels are installed in the DCS counting room. Two high-speed (40\,Gb/s) Ethernet links connect the DCS cluster with the FLP farm. These links carry the traffic to the front-end electronics and are used for the streaming of DCS data to O$^2$ as explained below.

The DCS counting room network renewal is part of the overall O$^2$ network upgrade. The DCS network services counting rooms, control rooms, surface areas (gas system, EPN containers), and the whole cavern. The infrastructure put in operation before Run~1 reached the end of its lifetime and was entirely replaced. All routers and switches were installed in parallel to the installation and commissioning of detectors. New infrastructure was installed alongside the old one and was put in operation in phases, in order to minimize the impact on the commissioning activities of the experiment. The network upgrade process lasted two years and the old network was decommissioned recently. 

One of the key factors affecting the selection of communication busses is the distance between the servers installed on the surface and the devices in the cavern. To provide stability in a harsh environment, the Controller Area Network (CANbus)~\cite{DCS:CAN} has been adopted. It is based on a serial bus designed for robust performance over long distances.
A majority of the devices controlled by CANbus are commercial power supplies or Embedded Logical Monitoring Boards (ELMB) mainly used for environment monitoring and rack control. CANbus is also used in the ITS controls, where it provides a redundant channel for hardware access and also served the interlock control during the on-surface commissioning. All CANbus controllers, previously based on USB or PCI devices, were replaced with ANAGATE CANbus-Ethernet gateways~\cite{DCS:ANAGATE}. This change allowed to remove the dependency on a physical connection between the DCS server and CAN controller. In case of a server failure, a physical intervention on the CANbus network is no longer required.

\subsubsection{DCS software upgrades}
\label{sec:dcs_software}

\begin{figure}[h]
\centering
\includegraphics[width=0.7\textwidth]{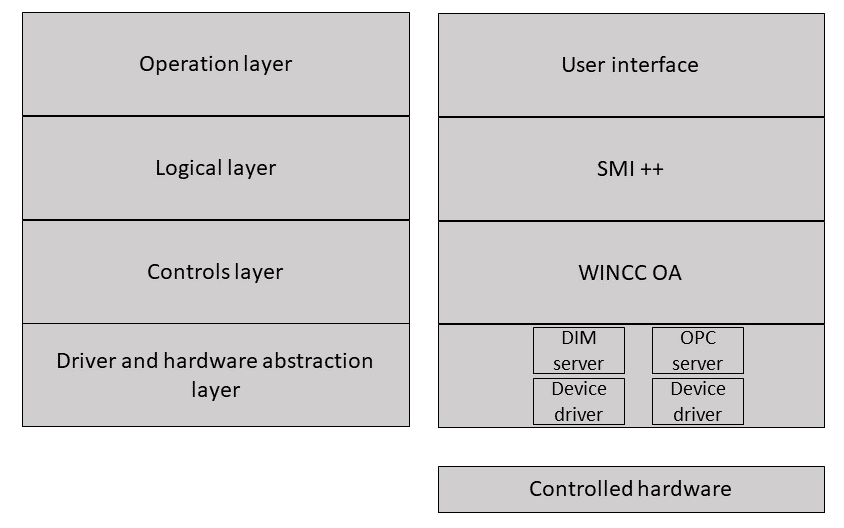}
\caption{\label{SoftwareArchitecture} DCS software architecture.}
\end{figure}

The DCS software is structured into several layers as shown in Fig.~\ref{SoftwareArchitecture}. 

The driver layer is connected directly to  controlled hardware. It provides the device specific low-level interface. To integrate the numerous hardware configurations used in ALICE, this layer also provides  hardware abstraction, which hides all device-specific details. The devices are presented to the DCS by standardized interfaces. Most commercial devices use the industrial communication standard OPC~\cite{DCS:OPC} for this purpose. An OPC server communicates with the hardware and exposes its functionality in a form of standard commands and services carrying the monitored data. 

Custom hardware modules developed for ALICE are controlled by software exposing its functionality to WINCC OA using the CERN DIM protocol.  Similar to OPC, the DIM servers convert the device specific interface to a standardized set of commands and services. Both technologies require that either an OPC or DIM client is deployed on the WINCC system, however, this component is common for all detectors. 

The controls layer implemented in WINCC OA performs the basic control and monitoring tasks. It sends commands to devices and reads back the responses. Scripted actions allow to execute procedures, based on the received response. Monitored values are in addition compared with predefined settings and in case of deviations a message can be sent to an alert system. In most cases an automated procedure can regulate the settings without any human intervention.

The logical layer encodes  detector specific operations in a form of a Finite State Machine (FSM) organized in a hierarchical way. This layer is entirely encoded in SMI++ language, which allows for definition of stable states and rules for the transitions between them. The FSM logic encodes experience gained over years of operation and defines reaction of detectors to different experiment conditions (such as beam injections, data taking, ALICE magnet ramps, etc.)

Finally, the user interface layer presents the DCS to the operator using a unified interface. All DCS functionality can be reached from a single panel in an intuitive way. Each action sent from the user interface is verified by high level scripts and protects the experiment from incorrect commands sent while detectors are not in compatible condition.

A significant part of the software upgrade concerned the communication with the devices. In the past years the widely used commercial OPC standard evolved and a new OPC UA technology~\cite{DCS:OPC} emerged. It replaces the older OPC DA technology used in Run 1 and 2. During the LS2, the new standard was tested and adopted in ALICE. Currenty, a total of 49 OPC servers control commercial hardware (power supplies, PLCs, etc.).

Upgraded front-end electronics in ALICE are read out through GBT links described in Section~\ref{sec:cru}. Those are controlled by CRUs, installed in FLP servers. The same links are used to configure, control, and monitor the detector front-end electronics. This grouping of functionality has an impact on DCS, because the FLPs are not part of the DCS domain. As a result, the DCS does not have direct control over the front-end electronics.

\begin{figure}[h]
\centering
\includegraphics[width=0.7\textwidth]{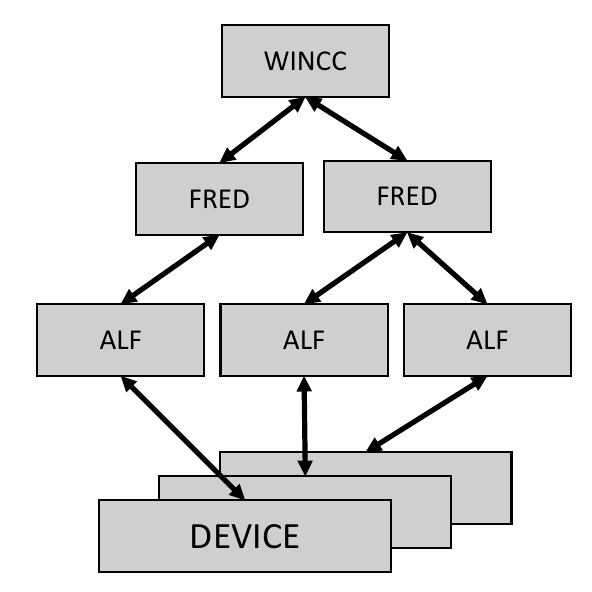}
\caption{\label{ALFRED} Access to the hardware implemented in ALF-FRED mechanism.}
\end{figure}

A client-server based architecture, named ALFRED, was developed to address the link sharing between the readout and the DCS. The low-level link access is established through ALF (ALICE Low-level Access) module and executed on the FLPs. This software is detector agnostic and transmits data received from FRED (Front-End Device server) ~\cite{DCS:FRED} to the front-end electronics. If the communication with the front-end electronics is based on the Slow Control Adapter (SCA) protocol, commands are sent through a dedicated communication channel using reserved bits in the transmitted GBT frames. For faster controls (mass configurations), the SCA channel does not provide sufficient throughput. Custom protocols, such as Single Word Transactions (SWT) are mastered by ALFRED. The data are transmitted in dedicated GBT frames in this case. The response produced by the detector front-end electronics is propagated from ALF back to FRED as illustrated in Fig.~\ref{ALFRED}.

The FRED component is a framework provided by central DCS. Detector specific code is embedded inside the FRED server framework. The framework functionality covers the ALF-to-FRED communication and synchronization. FRED also provides a communication layer to interface the system with WINCC OA. Using this architecture, a large uniformity has been achieved even between the largely heterogeneous detector front-ends. Currently eight detectors use FRED to communicate with electronics through GBT links and one detector uses FRED to control the user part of the CRU firmware.

Profiting from the flexibility of the ALFRED framework, the ITS detector developed an interlock system controlled via CANbus. A modified version of ALF (CANALF) communicates with the devices using the ANAGATE gateway instead of the CRU. The ITS electronics can also use CANbus in place of GBT as a fallback solution, bypassing the FLP. Due to reduced bandwidth of CANbus, this solution cannot fully replace the standard communication over GBT link, however, it provides the necessary level of redundancy. Using a command originating in WINCC OA, FRED can redirect communication from ALF to CANALF and continue detector operation still using the same user code.

Finally, FRED is a fully scalable and parallel framework. For large detectors, this can result in high amounts of data to be processed by a single instance of FRED. Depending on detectors size, the deployment of FRED ranges from single FRED instance on a single server, through multiple FRED instances on the same server, up to several instances distributed across multiple servers. 

\subsubsection{DCS conditions data}
\label{sec:DCS_conditions_data}

Data collected from detectors are stored by WINCC OA systems in an ORACLE database. Part of these data are used in the O$^2$ online processing. During LHC Runs~1 and 2, these data were retrieved by a dedicated process after each run (a period of stable data taking during a LHC fill). The collected data were merged with detector data acquired by the DAQ system and further processed by the offline framework. The Run~3 conditions processing expects continuous data processing, and the DCS conditions data must be provided in real time. This challenge was addressed by the newly developed ADAPOS (Alice DAtaPOint Service) system ~\cite{DCS:ADAPOS}.

The ADAPOS server collects a requested set of conditions data from individual WINCC OA systems and assembles a so-called Full Buffer Image (FBI). Each value stored in the FBI is updated whenever the WINCC OA system provides a new measurement. The FBI represents an up-to-date image of current DCS conditions. 

The ADAPOS system is able to periodically send the up-to-date FBI for further processing with an update frequency of 20\,Hz. However, most of the values do not usually change significantly, and most records in the FBI remain unchanged for longer periods of time. To save resources, ADAPOS can operate in transparent mode, where the FBI is maintained in the ADAPOS memory, but only parameters that recently changed are forwarded to the consumer. The mode selected for operarion is a hybrid one, which complements the transparent mode with a full FBI sent at regular intervals to reinforce data integrity at the point of processing.

To ensure stability and high availability, several ADAPOS servers can operate in parallel and provide full system redundancy. Implementation of ADAPOS required significant changes inside the WINCC OA system. In the standard configuration used at CERN, the communication with the ORACLE database is handled by a specialized manager (RDB manager). This connects WINCC OA directly with the database. Part of this data stream requested by ADAPOS had to be duplicated and transferred from WINCC OA to ADAPOS using a DIM protocol.

\begin{figure}[h]
\centering
\includegraphics[width=0.7\textwidth]{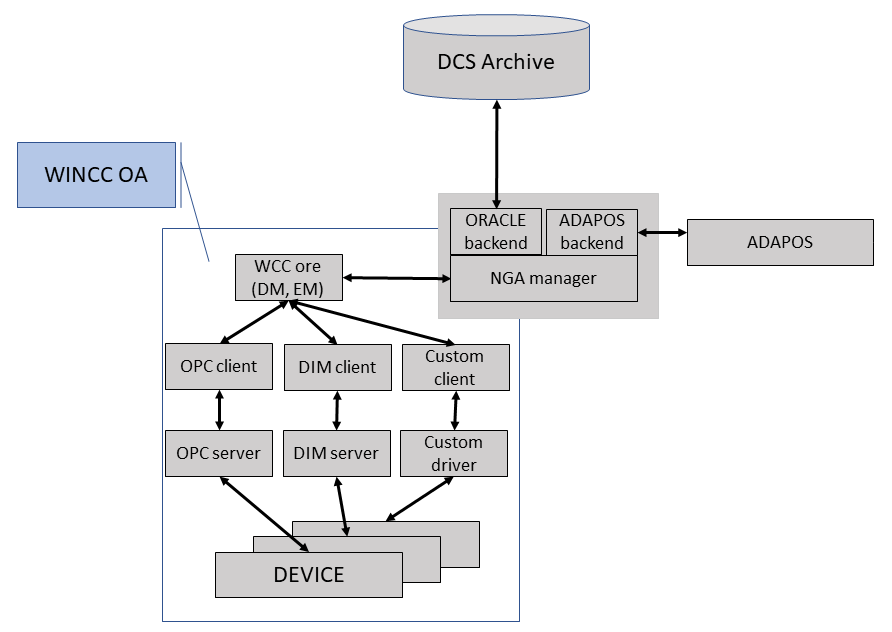}
\caption{\label{ADAPOS} DCS conditions data flow to ORACLE and ADAPOS.}
\end{figure}

A new technology developed by Siemens and further extended at CERN allowed a novel approach. The Next Generation Archive (NGA) manager plugs into the core of the system and collects data tagged for archival. The NGA can split the data stream between several backends. A CERN developed ORACLE backend handles all communication with ORACLE and fully replaces the RDB manager. The ADAPOS backend, developed in ALICE, forwards condition parameters to ADAPOS. To add new parameters to the ADAPOS data stream, the detector expert needs to tag the corresponding data, and all related configuration will be executed on the fly by the individual software components. Figure~\ref{ADAPOS} shows the data flow from WINCC OA to ORACLE and ADAPOS using the NGA manager and corresponding backends. Each WINCC OA system is built as a collection of highly specialized managers. The data flow is managed by the Data manager (DM) and Event manager (EM).

The described mechanism represents the main data flow from WINCC OA systems to ORACLE or ADAPOS, respectively. The data streaming, however, is not optimal for large data sets, such as masks, chip configurations, etc., that usually do not change during the run. A dedicated sevice called FilePush allows the injection of such data directly to the CCDB. A similar service called FilePull moves data from the CCDB to the DCS configuration database. This service was created to handle configuration data produced on the O$^2$ data processing side.  

A large part of the data is consumed by various online displays and user interfaces. This part of the data flow is covered by WINCC OA tools. Analysis and mitigation of operational issues and analysis of long term performance require retrieval of large amounts of data from the database. The online cluster in ALICE offers data recorded since 2008. Retrieval of large historical records is, however, inefficient when only WINCC OA built-in tools are used. The DCS was therefore extended by a new system, named DARMA (the DCS ARchive MAnager). 

\begin{figure}[h]
\centering
\includegraphics[width=0.7\textwidth]{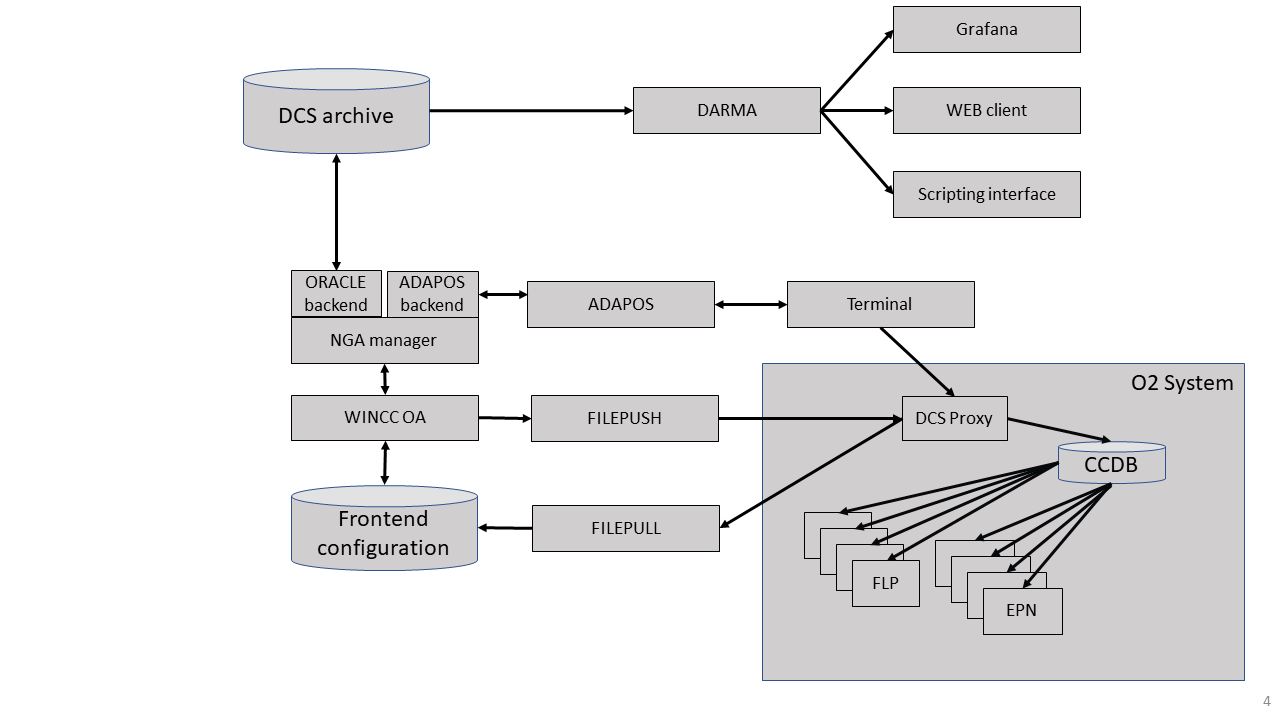}
\caption{\label{Dataflow} DCS data exchange with O$^2$ and external consumers.}
\end{figure}

DARMA is a web-based system allowing the retrieval of historical values from the database. To protect the online ORACLE cluster from possible overload, DARMA is configured to access data from the online replica of the ALICE production database. Based on ORACLE RAC ADG, this replica contains up-to-date values in real time mode. The user of DARMA can choose between a web-based GUI and a scripting interface to configure the requests. Finally, the DARMA system is extended by a plugin, allowing the Grafana system to access the DCS ORACLE archive. The full DCS data flow from/to external services is shown in Fig.~\ref{Dataflow}.

\subsubsection{DCS operator environment}

Even if the DCS provides fully automated operation and is available at all times with minimum downtime (about two or three days during the whole year for scheduled updates and backups), human supervision is maintained during detector operation. The main task of the operator is to react to anomalies indicated by the alert system and to assist the experts in mitigation. An intuitive and reliable operator environment is an important part of the DCS design. 

The complexity of the control system presents a challenge for the system designers when creating user interfaces: about one million parameters can be accessed through DCS tools and the global DCS logic is encoded in 70000 finite state machines. Central operations require large sequences of steps that need to be executed by the operator. To facilitate these tasks and protect the systems from operator errors, an ALICE DCS UI component was developed. It provides a coherent way for accessing all ALICE parameters from one user interface. Navigation is based on detector hierarchy, which makes this process intuitive. 

\begin{figure}[h]
\centering
\includegraphics[width=0.7\textwidth]{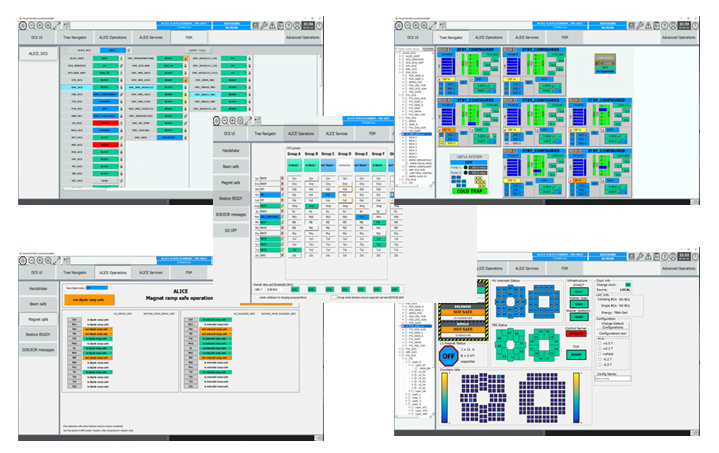}
\caption{\label{DCSUI} ALICE DCS UI with various panels used by the operators.}
\end{figure}

All controlled objects are represented in a visual pane. Navigating through the hierarchy, the main user interface displays panels for selected objects. Instructions for the use of the panel are either available in the operator manual, or are embedded in the alert instructions if an anomaly occurred. Various operator panels embedded in the common user interface are shown in Fig.~\ref{DCSUI}. 

Complex High-level operations related to detector safety, preparation of the detectors for data taking, or communication with LHC are further embedded in macro commands. The DCS operator is guided through the operations, and the procedure evaluates the status of all components in parallel and prevents human error. For example, a detector not yet fully recovered from a condition compatible with a magnet ramp operation will not allow the operator to ramp up the high voltage until the recovery steps are executed. Similar to this, a detector in a state not compatible with beam injection (‘not beam safe’) will not allow the operator to grant the injection permits to the LHC and hence blocks the injection until the detector is brought to a state safe for beam operation.

The new ALICE DCS UI component is a key operational tool. More than 300 operators were already trained for the Run~3 period, most of them being colleagues with no prior experience with system controls. Use of a single user interface component hiding as many detector specificities as possible significantly reduces the learning curve and improves the operational stability.

\section{Physics performance}
\label{sec:performance}

The physics programme motivating the upgrade was first established in the Letter of Intent~\cite{Abelevetal:2014cna} and the individual Technical Design Reports and was further expanded upon in the summary report of the heavy-ion physics programme at the LHC in Run 3 and 4 across all the LHC experiments~\cite{Citron:2018lsq}.
A precise measurement of the long-wavelength behaviour, i.e.\ the macroscopic fluid-like evolution of a high-density and high-temperature system, allows the determination of transport properties, such as the viscosity of the QGP, and provides information about the equation of state.
Probes that are sensitive to short wavelengths give access to the microscopic parton dynamics in a deconfined QGP state. 
A further goal is to study particle production in order to assess the validity of the fluid description and the role of collectivity in high-multiplicity pp collisions and in \pPb{} collisions. 
Finally, the precise measurement of nuclear parton densities over a wide $(x, Q^2)$ range is fundamental to constrain the initial conditions.

\begin{figure}[!b]
    \centering
    \includegraphics[width=.55\textwidth]{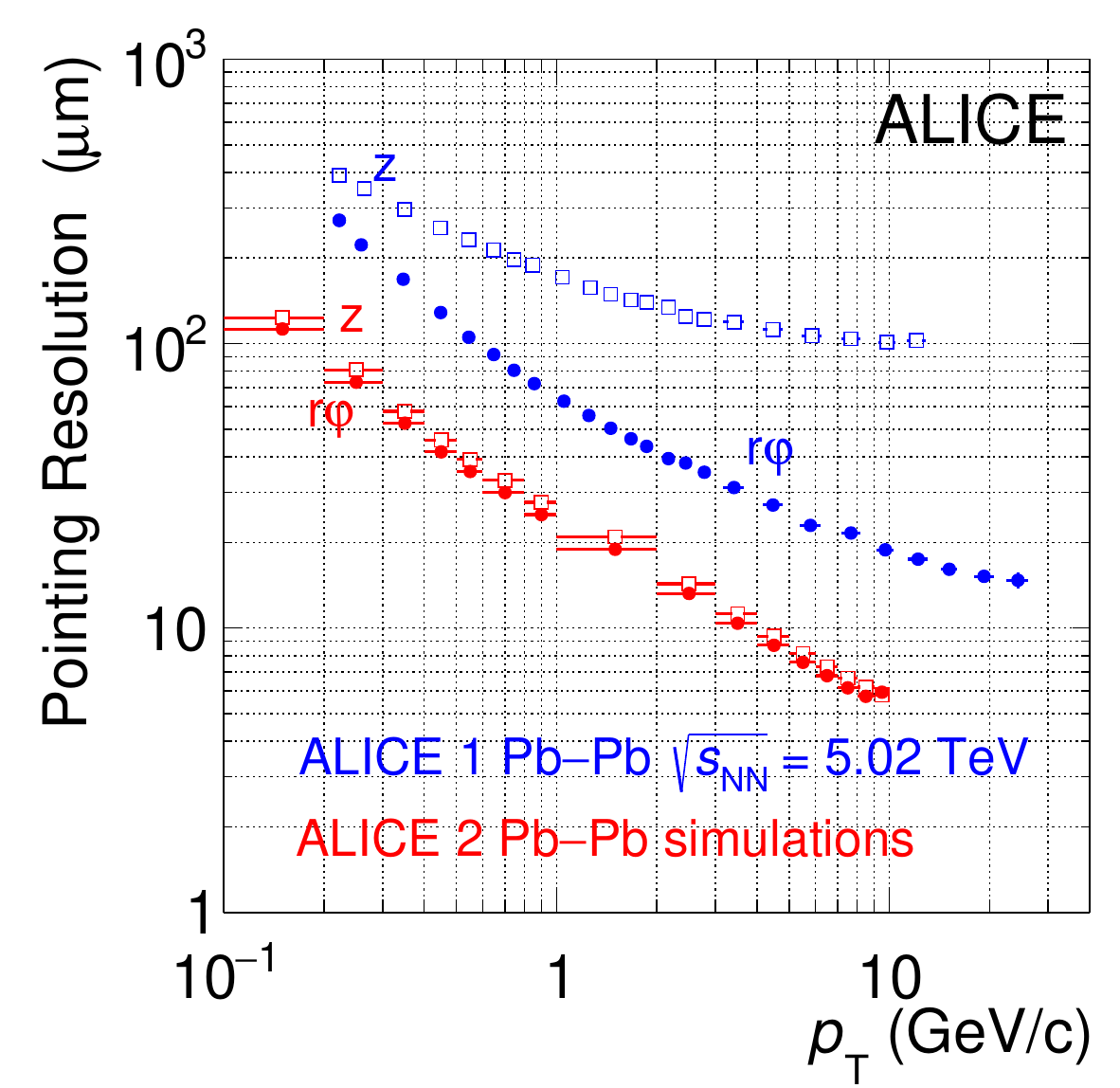}
    \caption[Impact parameter resolution]{Transverse (solid circle markers) and longitudinal (open square markers) impact parameter resolution in \PbPb{} collision data with ALICE~1 (blue points) and simulations the upgraded detector, ALICE~2 (red points).}
    \label{fig:dca_res}
  \end{figure}

\begin{figure}[!b]
    \centering
    \includegraphics[width=.45\textwidth]{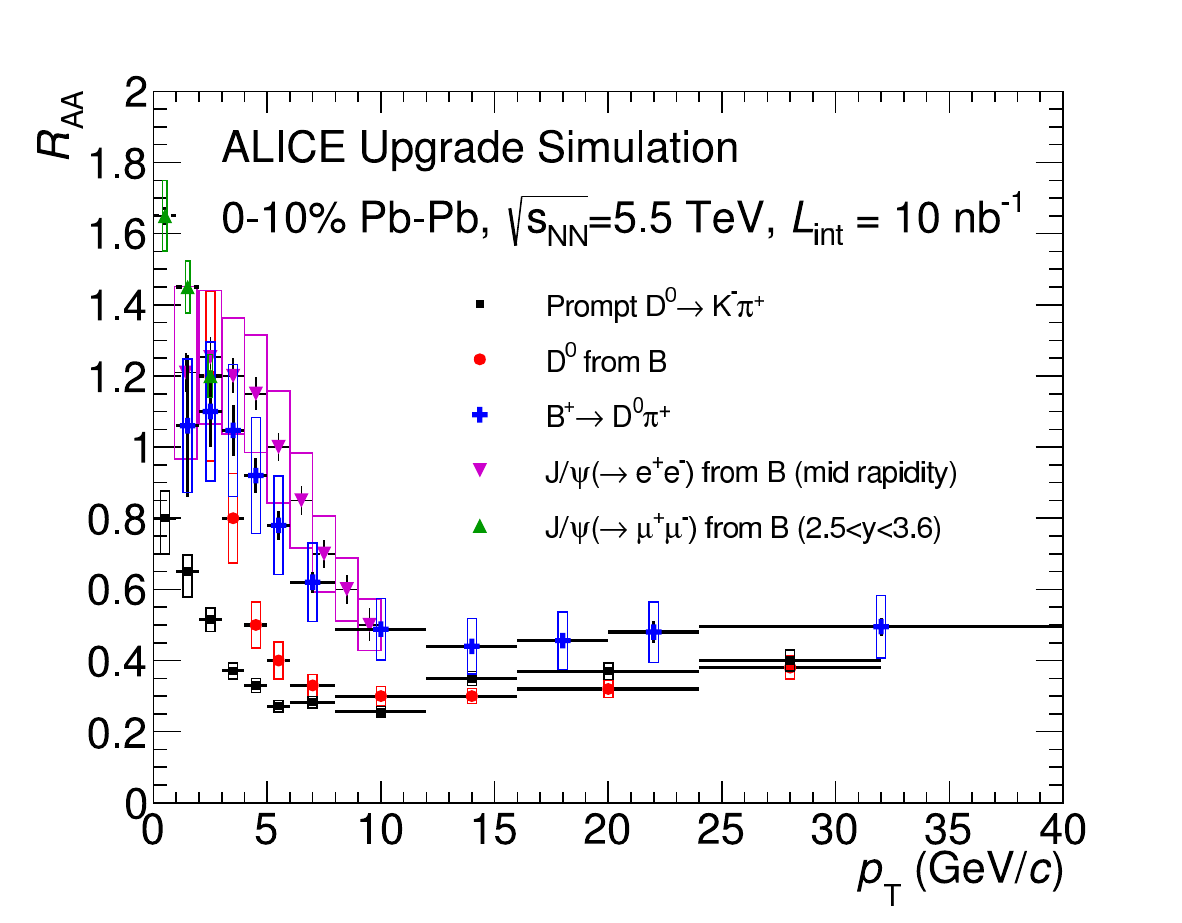}
    \includegraphics[width=.45\textwidth]{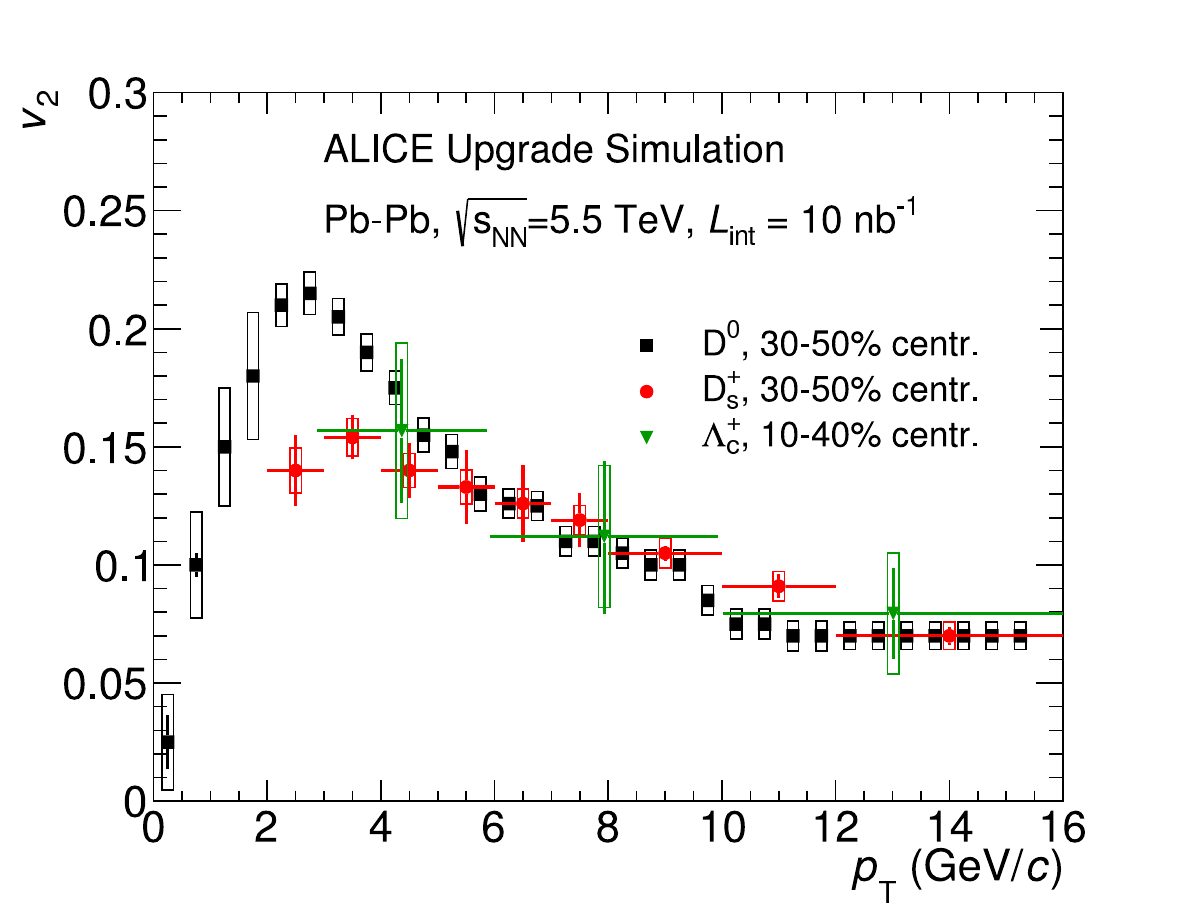}
    \caption[Projections for charm/beauty $R_\mathrm{AA}$ and $v_2$]{Performance projections for measurements of the nuclear modification factor $R_{\rm AA}$ and elliptic flow coefficient $v_{2}$ (see text for details) for charm and beauty hadrons~\cite{Citron:2018lsq}.}
    \label{fig:HF_RAA}
\end{figure}

In the following, some examples of the physics performance with the upgraded detector will be briefly highlighted. Figure~\ref{fig:dca_res} shows the resolution in the distance of closest approach of tracks to the primary vertex, in both the transverse ($r\varphi$, solid circle markers) and longitudinal ($z$, open square makers) directions in \PbPb{} collisions, as simulated with the full O$^2$ simulation framework for the upgraded detector (red points) compared to Run~2 data (blue points). The impact parameter resolution is seen to improve by factor of approximately three in the transverse direction and a factor ten or more in the longitundal direction. 

Heavy quarks are produced in hard scatterings between incoming quarks and gluons, and lose energy as they interact with the quark--gluon plasma while propagating out of the collision zone. To quantify the effect of energy loss, the transverse momentum distributions of produced hadrons are measured both in proton-proton and \PbPb{} collisions at the same centre-of-mass energy per nucleon pair. The ratio of the \pt spectra measured in \PbPb{} collisions to those in pp collisions scaled by the nuclear overlap function $T_{\rm AA}$, which is proportional to the number of binary nucleon--nucleon collisions, is then calculated:
\begin{equation}
R_{\rm AA} = \frac{\left. {\rm d}N/{\rm d}p_\mathrm{T} \right|_{\rm AA}}{ 
T_{\rm AA}\left. {\rm d}\sigma/{\rm d}p_\mathrm{T} \right|_{\rm pp}}.
\end{equation}
The ratio $R_{\rm AA}$ is called the nuclear modification factor.
The left panel of Fig.~\ref{fig:HF_RAA} shows the expected precision for measurements of the nuclear modification factor for charm and beauty hadrons, which are used to determine the mass dependence of energy loss of heavy quarks propagating through the quark--gluon plasma~\cite{Citron:2018lsq}.
The figure shows projections for the nuclear modification factor for the open charm $\mathrm{D}^0$ mesons and for measurements of beauty mesons via full hadronic reconstruction and measurements of production of non-prompt charm particles. 
The mass dependence of parton energy loss is most pronounced at transverse momenta around or below the quark mass, when the quark is moving through the QGP with a velocity that is significantly smaller than the speed of light. 
The upgraded detector provides measurements with uncertainties that are small enough to reveal the expected mass dependence in this momentum range.

The right panel of Fig.~\ref{fig:HF_RAA} shows the expected precision for measurements of the azimuthal asymmetry for charm mesons and baryons~\cite{Citron:2018lsq}, as characterised by the elliptic flow $v_{2}$, which is the second harmonic coefficient of the Fourier expansion of the distibutions of the azimuthal angles $\varphi$ of the hadrons with respect to the reaction plane angle $\Psi$:
\begin{equation}
\frac{{\rm d}N}{{\rm d}\varphi} \propto 1+2 v_2 \cos 2(\varphi - \Psi).
\end{equation}
These measurements are in particular sensitive to the diffusion of charm quarks in the QGP and the path length dependence of parton energy loss effects.

\begin{figure}
    \centering
    \includegraphics[width=.45\textwidth]{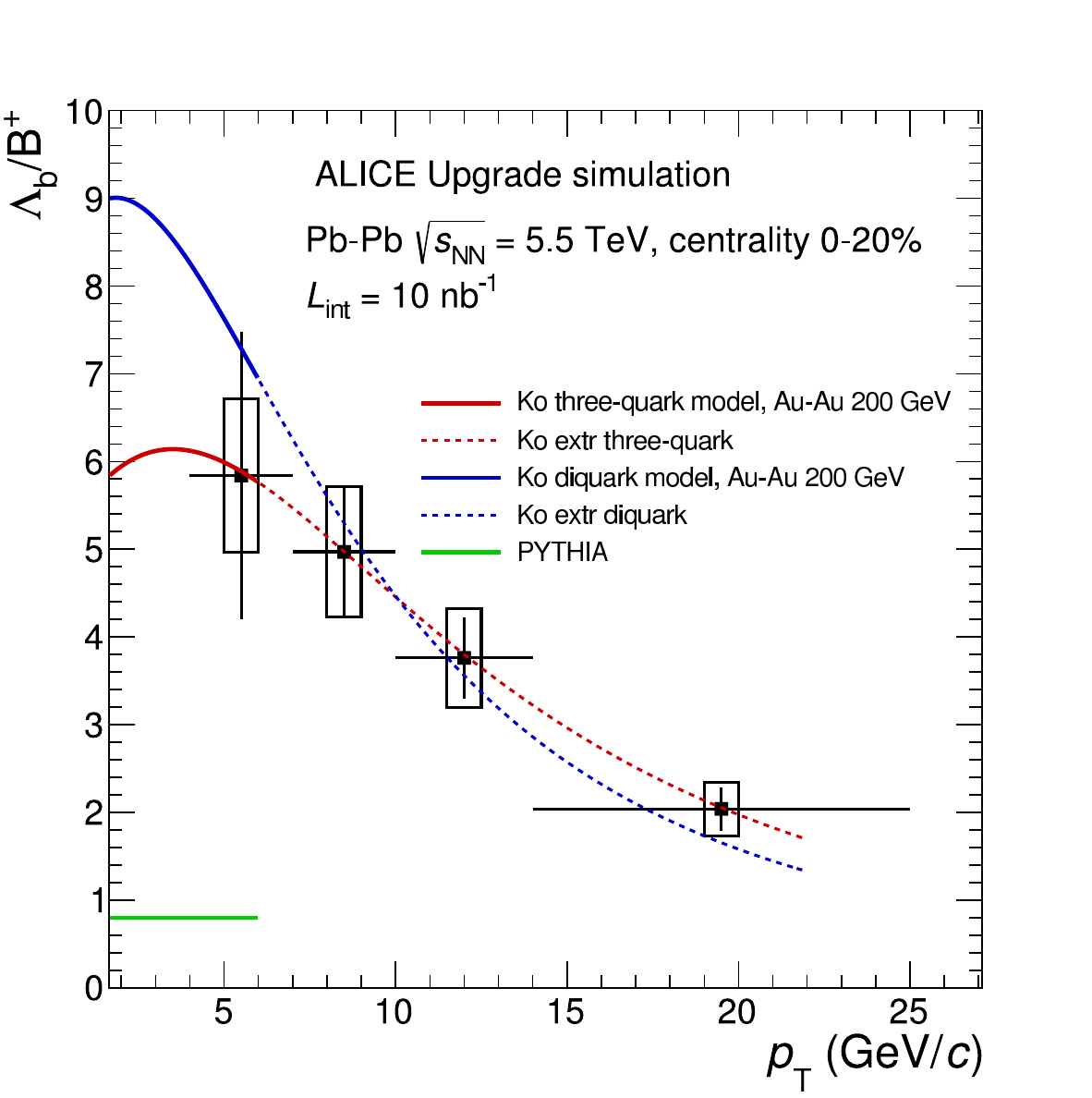}
    \caption[Beauty baryon-to-meson ratio projections]{\label{fig:BM_ratios}Performance projection for the ratio of beauty baryon and meson yields in central \PbPb{} collisions with the total expected integrated luminosity of 10 nb$^{-1}$ (at full magnetic field) for Run 3 and 4. From~\cite{Citron:2018lsq}.}
\end{figure}
Figure~\ref{fig:BM_ratios} shows the expected performance for measurements of the beauty baryon-to-meson ratio. These measurements probe the mechanisms for meson and baryon formation. For example, hadronisation via quark coalescence (red and blue curve) is expected to lead to an increased production rate of beauty baryons relative to that of the $\mathrm{B}^+$ meson~\cite{Oh:2009zj,Plumari:2017ntm,He:2019vgs}. Performing such measurements for the first time in the beauty sector will enhance the understanding of hadronisation mechanisms.

\begin{figure}
    \centering
    \includegraphics[width=.45\textwidth]{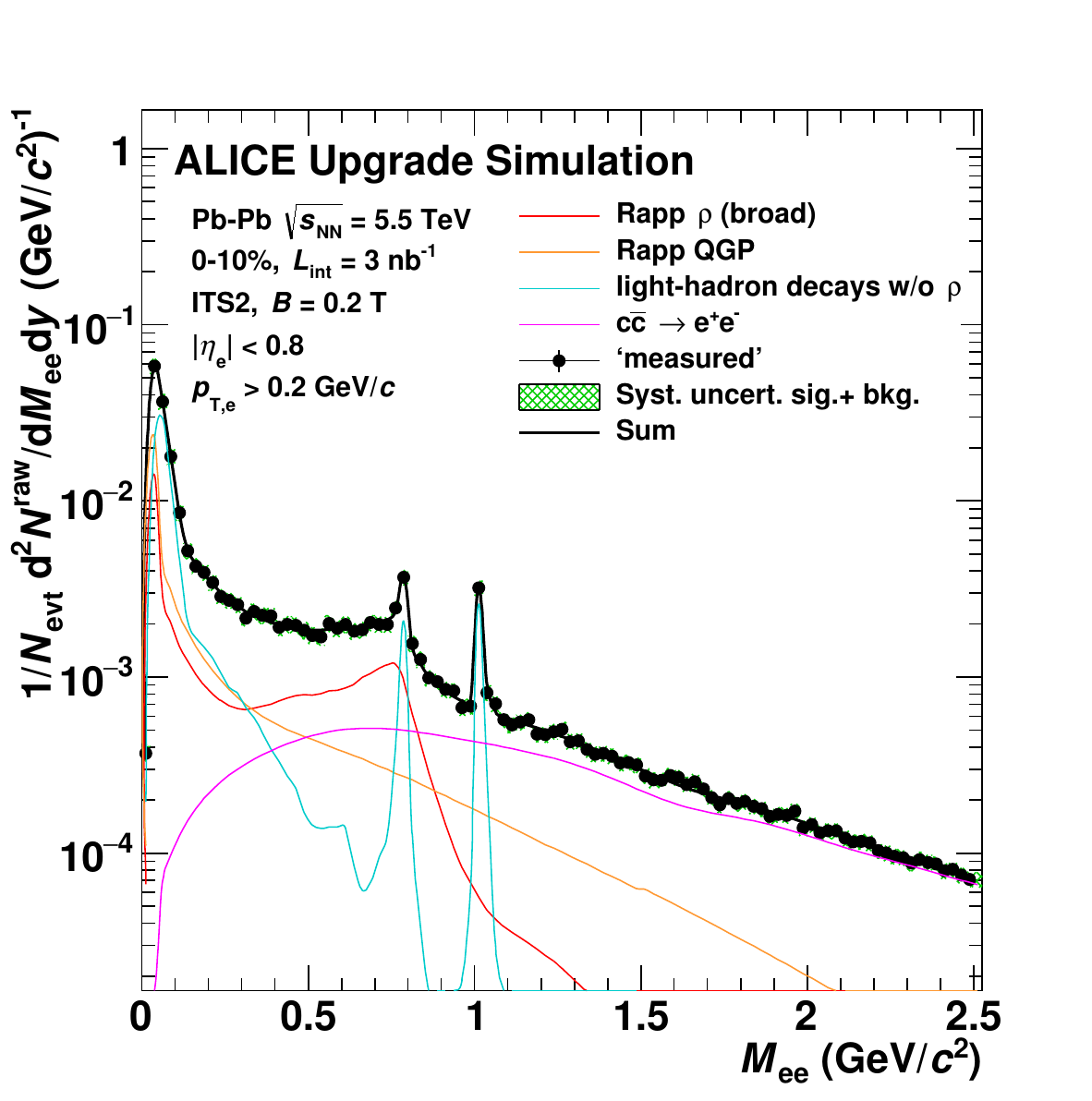}
    \includegraphics[width=.45\textwidth]{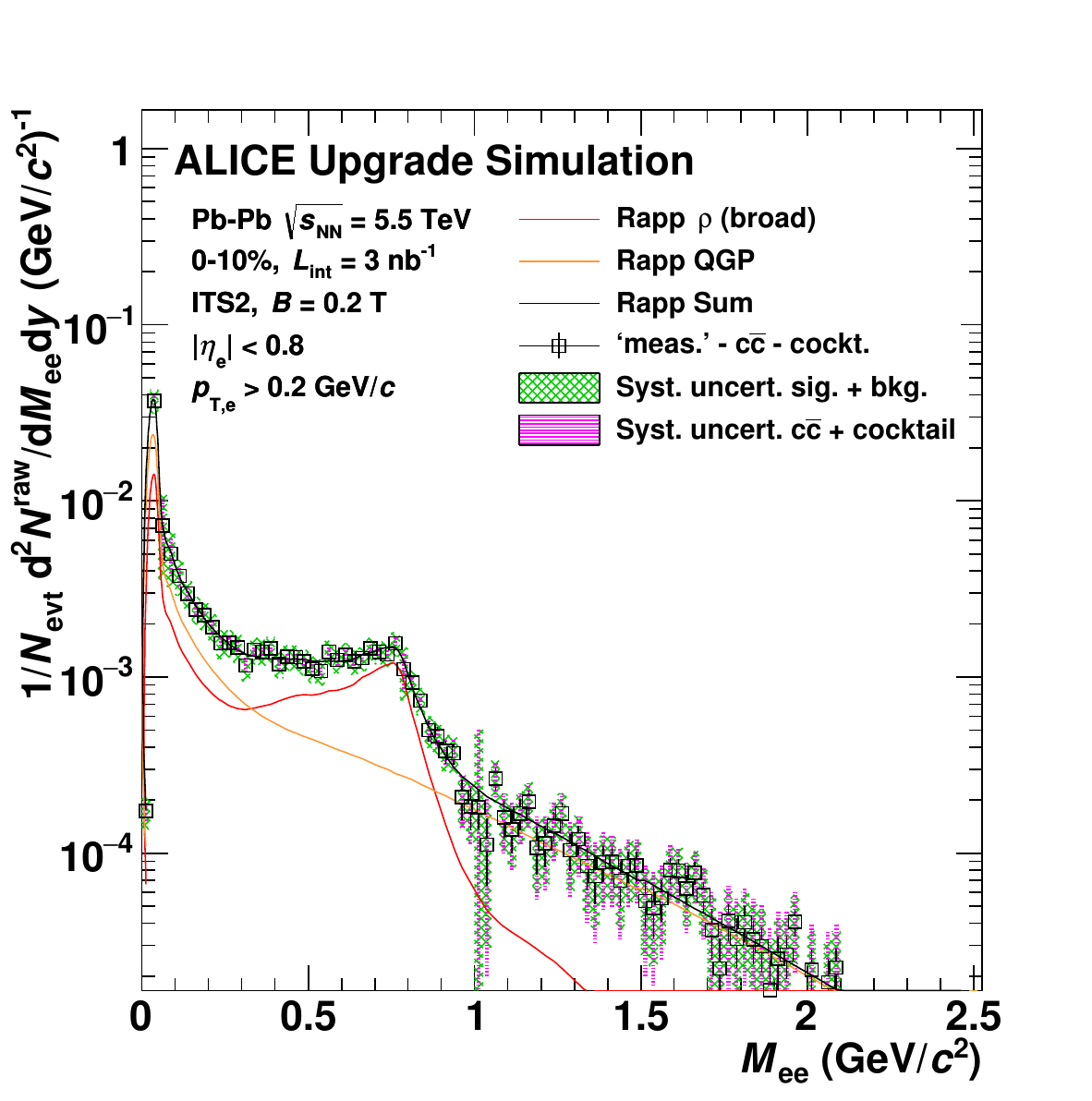}
    \caption[Di-electron performance]{\label{fig:dilepton_perf}Projection of the invariant mass distribution of electron-positron pairs before (left panel) and after (right panel) subtraction of the contribution from decays of light and heavy flavour mesons, except the $\rho$ meson. From~\cite{Citron:2018lsq}.}
\end{figure}

Figure~\ref{fig:dilepton_perf} shows the expected precision of the measurement of electron-positron pair production. The left panel shows the mass distribution of all pairs, and the right panel shows the result after subtracting the calculated contributions from the decays of light hadrons, except the $\rho$ meson and heavy-flavour decays. The result in the right panel is sensitive to the modification of the $\rho$ meson spectral function in the high-density collision system (indicated by the red line), as well as thermal emission (orange line) from all stages of the collision, including the QGP phase~\cite{Rapp:2013nxa}, which provides a unique window on the system before hadronisation. 

A further upgrade of the inner tracking system is planned to further improve the capabiltites of ALICE for measurements of electron-positron pairs as well as heavy-flavour measuremnts. To achieve this, the three innermost layers of the ITS will be replaced with wafer-scale silicon sensors that are bent into a cylindrical shape, supported by carbon foam. More details can be found in the ITS3 Letter of Intent~\cite{Musa:2703140}.

\begin{figure}
    \centering
    \includegraphics[width=.5\textwidth]{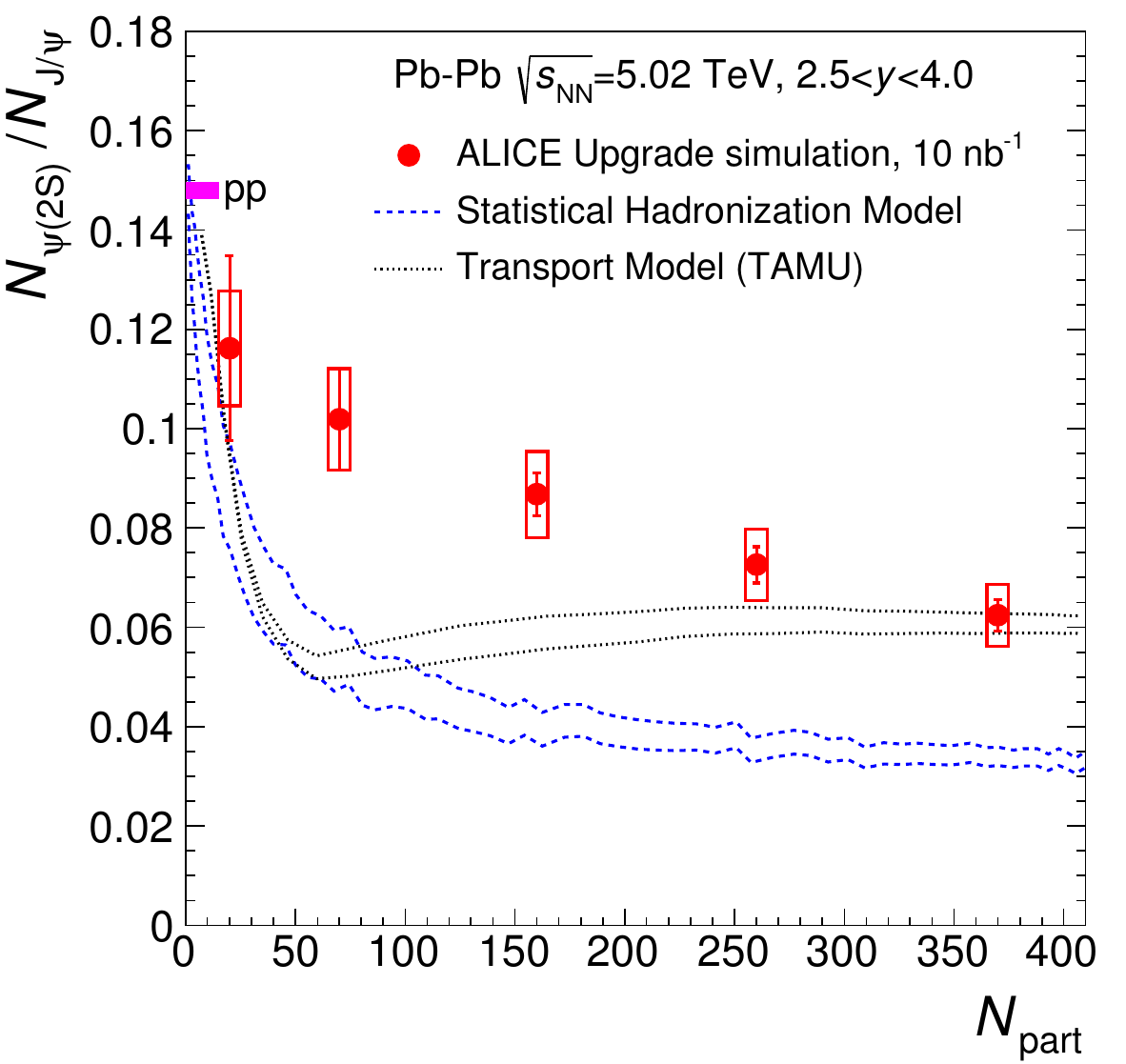}
    \caption[Performance projection for $\psi(2S)$ and $J/\psi$]{\label{fig:quarkonia}Projected yield ratio of $\psi$(2S) and J$/\psi$ in pp and \PbPb{} collisions, measured with the MFT and muon arm~\cite{Abelevetal:2014cna}. Model calculations from~\cite{Andronic:2017pug,Du:2015wha}.}
\end{figure}

Figure~\ref{fig:quarkonia} shows the expected performance for measurements of the production of the two main charmonium states with the MFT and the muon detectors. The MFT provides additional background rejection of non-prompt muons and an improved momentum resolution, which make a precise measurement of the production of the $\psi$(2S) possible. The production rates of J$/\psi$ and $\psi$(2S) are compared to gain further insight into the mechanisms for quarkonium dissociation and regeneration in the QGP.

\begin{figure}
    \centering
    \includegraphics[width=.6\textwidth]{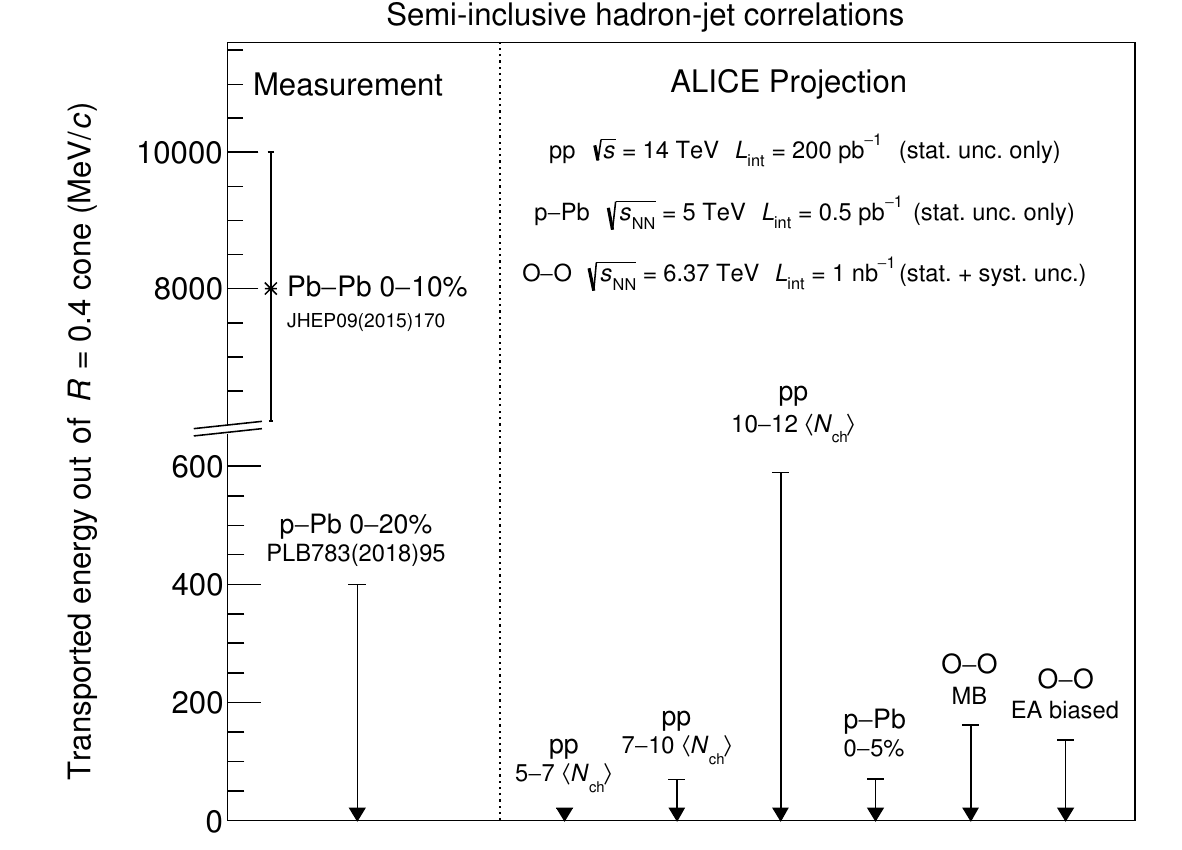}
    \caption[Energy loss from recoil jet yield suppression]{\label{fig:hjet}Projected limits on jet energy loss from the measurement of the yields of jets recoiling from high-\pt{} trigger particles in different collision systems. From~\cite{ALICE-PUBLIC-2021-004}.     }
\end{figure}
Another important part of the physics program is the investigation of several effects that have been observed in high-multiplicity collisions of protons that are reminiscent of heavy-ion collisions~\cite{ALICE-PUBLIC-2020-005}, such as the increased production yields of multi-strange baryons with respect to pions and azimuthal asymmetry of produced particles. In particular, these observations raise the question whether quark--gluon plasma is formed in high-multiplicity pp and p--Pb collisions. Parton energy loss is a distinctive signature of QGP formation that was yet observed in small collision systems. The search for this effect will continue with the much larger pp and p--Pb samples of Run 3 and also using a short pilot run with oxygen--oxygen collisions, which provides similar multiplicities but larger geometrical size. Figure~\ref{fig:hjet} shows the projected sensitivity limits on energy transported outside the jet cone based on the measurement of jet yield recoiling from high-\pt{} particles in various smaller collision systems~\cite{ALICE-PUBLIC-2021-004} compared to the values determined from existing measurements in \pPb{} and \PbPb collisions~\cite{ALICE:2017svf,ALICE:2015mdb}. The large data sample of proton--proton collisions of Runs 3 and 4 (with a target integrated luminosity of 200~pb$^{-1}$)  will also enable unique studies of perturbative QCD effects, such as the dead cone in gluon radiation off heavy quarks, of the residual strong interaction among hadrons pairs, including multistrange hadrons and light nuclei, and of hypernuclei production~\cite{ALICE-PUBLIC-2020-005}.

\section{Conclusions and outlook}
\label{chap:concl}

After the successful conclusion of data taking in the LHC Runs 1 and 2, the ALICE detector has undergone a major upgrade in order to enable new or more precise measurements in Runs 3 and 4.
The inner tracking system was completely replaced and is now fully instrumented with silicon pixel sensors, which provide a much better pointing resolution and support continuous readout. The muon forward tracker uses the same pixel sensors and provides for the first time precise pointing information for forward-rapidity muons reconstructed in the muon chambers and the muon identifier.
The time projection chamber is now read out using new detectors based on gas electron multiplication foils that reduce ion backflow and enable continuous readout at much higher interaction rates. 
The new fast interaction trigger system provides the interaction trigger, as well as multiplicity measurements and time measurements for offline analysis.
All other detector systems have undergone consolidation work and several systems have been equipped with upgraded readout electronics and/or new firmware to increase the readout speed or support continuous readout.

All upgrades were completed on schedule and have been commissioned with standalone runs as well as global commissioning runs in 2021 and 2022. The systems have been tested with pilot beams in October 2021 before the official start of Run~3 and are being operated successfully for routine data taking with proton-proton collisions at $\sqrt{s}=\SI{13.6}{\tera\eV}$ since July 2022. High-rate tests are being performed to qualify the systems for the high data rates in \PbPb collisions. The upgraded detectors will strongly extend the physics reach of the experiment, in particular in the sector of open and hidden heavy-flavour probes, measurements of thermal radiation from the quark--gluon plasma, as well as measurements of light nuclear states, final-state interactions of hadrons, and the internal structure of jets.

Two additional upgrades are in preparation for Long Shutdown 3 (2026--2028), with the goal of further enhancing the physics reach of the experiment in Run 4: new inner layers for the inner tracking system (ITS3 ~\cite{Musa:2703140}) and a forward calorimeter with high-granularity readout (FoCal~\cite{ALICECollaboration:2719928}). 

The ITS3 upgrade consists in the replacement of the three innermost layers of the ITS2 with three ultra-light truly-cylindrical layers made with curved large-area MAPS sensors~\cite{ALICEITSproject:2021kcd}. The innermost layer will have a radius of 18~mm and will surround a new beam pipe with reduced radius (16~mm) and thickness. The pointing resolution will be better than that of the ITS2 by a factor of two up to a transverse momentum of 5~GeV/$c$, reaching down to about \SI{12}{\um} at $\pt=1$~GeV/$c$. The low-$p_{\rm T}$ tracking efficiency will also improve. The ITS3 will strongly enhance the low-mass dielectron, heavy-flavour meson, and baryon production measurements.

The FoCal consists of an electromagnetic calorimeter with high-granularity readout for optimal separation of direct-photon showers from those of neutral pions at forward pseudorapidity ($3.4 < \eta < 5.8$), coupled to a hadronic calorimeter for additional hadron rejection. The required granularity is achieved using a combination of MAPS silicon pixel readout planes and silicon pad readout planes. The main physics goal of the FoCal is the study of gluon parton distribution functions in the lead nucleus at Bjorken $x$ values down to $10^{-6}$ using the nuclear modification factor of forward direct photons with transverse momentum $2<\pt<20$~GeV/$c$ in p--Pb collisions.  

For the LHC Runs 5 and 6, a completely new apparatus, named ALICE~3, is proposed~\cite{ALICE:2022wwr}.
The ALICE~3 detector consists of a vertexing and tracking system with unique pointing resolution over a large pseudorapidity range ($-4<\eta<+4$), complemented by multiple sub-detector systems for particle identification, including silicon time-of-flight layers, a ring-imaging Cherenkov detector with high-resolution readout, a muon identification system, and an electromagnetic calorimeter.
Unprecedented pointing resolution of \SI{10}{\um} at $\pt= 200$~MeV/$c$ at midrapidity in both the transverse and longitudinal directions can be achieved by placing the first layers as close as possible to the interaction point, on a retractable structure to leave sufficient aperture for the beams at injection energy.
This next-generation apparatus will, on the one hand, enable novel studies of the QGP and, on the other hand, open up important physics opportunities in other areas of QCD and beyond.
The main new studies in the QGP sector focus on low-$\pt$ heavy-flavour production, including beauty hadrons, multi-charm baryons and charm--charm correlations, as well as on precise multi-differential measurements of dielectron emission to probe the mechanism of chiral-symmetry restoration and the time-evolution of the QGP temperature.
Besides QGP studies, ALICE~3 can uniquely contribute to hadronic physics, for example with femtoscopic studies of the interaction potentials between charm mesons and searches for nuclei with charm, to fundamental physics, with tests of the Low theorem for ultra-soft photon emission, and to searches for physics beyond the Standard Model, for example in the sector of axion-like particles and the anomalous magnetic moment of $\tau$ leptons. 
The programme aims to collect an integrated luminosity of about 35~nb$^{-1}$ with Pb--Pb collisions and 18~fb$^{-1}$ with pp collisions at top LHC energy. The potential to further increase the luminosity for ion running in the LHC by using smaller ions (e.g.\,$^{84}$Kr or $^{128}$Xe), as well as further runs with small collision systems to explore the approach to thermal equilibrium, are being explored.

\newenvironment{acknowledgement}{\relax}{\relax}
\begin{acknowledgement}
\section*{Acknowledgements}

The ALICE Collaboration would like to thank all its engineers and technicians for their invaluable contributions to the construction of the experiment and the CERN accelerator teams for the outstanding performance of the LHC complex.
The ALICE Collaboration gratefully acknowledges the resources and support provided by all Grid centres and the Worldwide LHC Computing Grid (WLCG) collaboration.
The ALICE Collaboration acknowledges the following funding agencies for their support in building and running the ALICE detector:
A. I. Alikhanyan National Science Laboratory (Yerevan Physics Institute) Foundation (ANSL), State Committee of Science and World Federation of Scientists (WFS), Armenia;
Austrian Academy of Sciences, Austrian Science Fund (FWF): [M 2467-N36] and Nationalstiftung f\"{u}r Forschung, Technologie und Entwicklung, Austria;
Ministry of Communications and High Technologies, National Nuclear Research Center, Azerbaijan;
Conselho Nacional de Desenvolvimento Cient\'{\i}fico e Tecnol\'{o}gico (CNPq), Financiadora de Estudos e Projetos (Finep), Funda\c{c}\~{a}o de Amparo \`{a} Pesquisa do Estado de S\~{a}o Paulo (FAPESP) and Universidade Federal do Rio Grande do Sul (UFRGS), Brazil;
Bulgarian Ministry of Education and Science, within the National Roadmap for Research Infrastructures 2020-2027 (object CERN), Bulgaria;
Ministry of Education of China (MOEC) , Ministry of Science \& Technology of China (MSTC) and National Natural Science Foundation of China (NSFC), China;
Ministry of Science and Education and Croatian Science Foundation, Croatia;
Centro de Aplicaciones Tecnol\'{o}gicas y Desarrollo Nuclear (CEADEN), Cubaenerg\'{\i}a, Cuba;
Ministry of Education, Youth and Sports of the Czech Republic, Czech Republic;
The Danish Council for Independent Research | Natural Sciences, the VILLUM FONDEN and Danish National Research Foundation (DNRF), Denmark;
Helsinki Institute of Physics (HIP), Finland;
Commissariat \`{a} l'Energie Atomique (CEA) and Institut National de Physique Nucl\'{e}aire et de Physique des Particules (IN2P3) and Centre National de la Recherche Scientifique (CNRS), France;
Bundesministerium f\"{u}r Bildung und Forschung (BMBF) and GSI Helmholtzzentrum f\"{u}r Schwerionenforschung GmbH, Germany;
General Secretariat for Research and Technology, Ministry of Education, Research and Religions, Greece;
National Research, Development and Innovation Office, Hungary;
Department of Atomic Energy Government of India (DAE), Department of Science and Technology, Government of India (DST), University Grants Commission, Government of India (UGC) and Council of Scientific and Industrial Research (CSIR), India;
National Research and Innovation Agency - BRIN, Indonesia;
Istituto Nazionale di Fisica Nucleare (INFN), Italy;
Japanese Ministry of Education, Culture, Sports, Science and Technology (MEXT) and Japan Society for the Promotion of Science (JSPS) KAKENHI, Japan;
Consejo Nacional de Ciencia (CONACYT) y Tecnolog\'{i}a, through Fondo de Cooperaci\'{o}n Internacional en Ciencia y Tecnolog\'{i}a (FONCICYT) and Direcci\'{o}n General de Asuntos del Personal Academico (DGAPA), Mexico;
Nederlandse Organisatie voor Wetenschappelijk Onderzoek (NWO), Netherlands;
The Research Council of Norway, Norway;
Commission on Science and Technology for Sustainable Development in the South (COMSATS), Pakistan;
Pontificia Universidad Cat\'{o}lica del Per\'{u}, Peru;
Ministry of Education and Science, National Science Centre and WUT ID-UB, Poland;
Korea Institute of Science and Technology Information and National Research Foundation of Korea (NRF), Republic of Korea;
Ministry of Education and Scientific Research, Institute of Atomic Physics, Ministry of Research and Innovation and Institute of Atomic Physics and University Politehnica of Bucharest, Romania;
Ministry of Education, Science, Research and Sport of the Slovak Republic, Slovakia;
National Research Foundation of South Africa, South Africa;
Swedish Research Council (VR) and Knut \& Alice Wallenberg Foundation (KAW), Sweden;
European Organization for Nuclear Research, Switzerland;
Suranaree University of Technology (SUT), National Science and Technology Development Agency (NSTDA) and National Science, Research and Innovation Fund (NSRF via PMU-B B05F650021), Thailand;
Turkish Energy, Nuclear and Mineral Research Agency (TENMAK), Turkey;
National Academy of  Sciences of Ukraine, Ukraine;
Science and Technology Facilities Council (STFC), United Kingdom;
National Science Foundation of the United States of America (NSF) and United States Department of Energy, Office of Nuclear Physics (DOE NP), United States of America.
In addition, individual groups or members have received support from:
European Research Council, Strong 2020 - Horizon 2020, Marie Sk\l{}odowska Curie (grant nos. 950692, 824093, 896850), European Union;
Academy of Finland (Center of Excellence in Quark Matter) (grant nos. 346327, 346328), Finland;
Programa de Apoyos para la Superaci\'{o}n del Personal Acad\'{e}mico, UNAM, Mexico.

\end{acknowledgement}

\bibliographystyle{utphys}   
\bibliography{bibliography}

\newpage
\appendix

%
%

\section{The ALICE Collaboration}
\label{app:collab}
\begin{flushleft} 
\small

S.~Acharya\,\orcidlink{0000-0002-9213-5329}\,$^{\rm 127}$, 
R.~Acosta Hernandez$^{\rm 111}$, , 
D.~Adamov\'{a}\,\orcidlink{0000-0002-0504-7428}\,$^{\rm 86}$, 
A.~Adler$^{\rm 70}$, 
J.~Adolfsson\,\orcidlink{0000-0001-5651-4025}\,$^{\rm 75}$, 
D.~Agguiaro$^{\rm 54}$, 
G.~Aglieri Rinella\,\orcidlink{0000-0002-9611-3696}\,$^{\rm 32}$, 
M.~Agnello\,\orcidlink{0000-0002-0760-5075}\,$^{\rm 29}$, 
F.~Agnese\,\orcidlink{0000-0003-2806-6709}\,$^{\rm 129}$, 
N.~Agrawal\,\orcidlink{0000-0003-0348-9836}\,$^{\rm 51}$
% ...

\section*{Affiliation Notes}

$^{\rm I}$ Deceased\\
$^{\rm II}$ Also at: Max-Planck-Institut f\"{u}r Physik, Munich, Germany\\
$^{\rm III}$ Also at: Italian National Agency for New Technologies, Energy and Sustainable Economic Development (ENEA), Bologna, Italy\\
$^{\rm IV}$ Also at: Dipartimento DET del Politecnico di Torino, Turin, Italy\\
$^{\rm V}$ Also at: Department of Applied Physics, Aligarh Muslim University, Aligarh, India\\
$^{\rm VI}$ Also at: Institute of Theoretical Physics, University of Wroclaw, Poland\\
$^{\rm VII}$ Also at: An institution covered by a cooperation agreement with CERN\\

\section*{Collaboration Institutes}

$^{1}$ A.I. Alikhanyan National Science Laboratory (Yerevan Physics Institute) Foundation, Yerevan, Armenia\\
$^{2}$ AGH University of Krakow, Cracow, Poland\\
$^{3}$ Bogolyubov Institute for Theoretical Physics, National Academy of Sciences of Ukraine, Kiev, Ukraine\\
$^{4}$ Bose Institute, Department of Physics  and Centre for Astroparticle Physics and Space Science (CAPSS), Kolkata, India\\
$^{5}$ California Polytechnic State University, San Luis Obispo, California, United States\\
$^{6}$ Central China Normal University, Wuhan, China\\
$^{7}$ Centro de Aplicaciones Tecnol\'{o}gicas y Desarrollo Nuclear (CEADEN), Havana, Cuba\\
$^{8}$ Centro de Investigaci\'{o}n y de Estudios Avanzados (CINVESTAV), Mexico City and M\'{e}rida, Mexico\\
$^{9}$ Chicago State University, Chicago, Illinois, United States\\
$^{10}$ China Institute of Atomic Energy, Beijing, China\\
$^{11}$ Chungbuk National University, Cheongju, Republic of Korea\\
$^{12}$ Comenius University Bratislava, Faculty of Mathematics, Physics and Informatics, Bratislava, Slovak Republic\\
$^{13}$ COMSATS University Islamabad, Islamabad, Pakistan\\
$^{14}$ Creighton University, Omaha, Nebraska, United States\\
$^{15}$ Department of Physics, Aligarh Muslim University, Aligarh, India\\
$^{16}$ Department of Physics, Pusan National University, Pusan, Republic of Korea\\
$^{17}$ Department of Physics, Sejong University, Seoul, Republic of Korea\\
$^{18}$ Department of Physics, University of California, Berkeley, California, United States\\
$^{19}$ Department of Physics, University of Oslo, Oslo, Norway\\
$^{20}$ Department of Physics and Technology, University of Bergen, Bergen, Norway\\
$^{21}$ Dipartimento di Fisica, Universit\`{a} di Pavia, Pavia, Italy\\
$^{22}$ Dipartimento di Fisica dell'Universit\`{a} and Sezione INFN, Cagliari, Italy\\
$^{23}$ Dipartimento di Fisica dell'Universit\`{a} and Sezione INFN, Trieste, Italy\\
$^{24}$ Dipartimento di Fisica dell'Universit\`{a} and Sezione INFN, Turin, Italy\\
$^{25}$ Dipartimento di Fisica e Astronomia dell'Universit\`{a} and Sezione INFN, Bologna, Italy\\
$^{26}$ Dipartimento di Fisica e Astronomia dell'Universit\`{a} and Sezione INFN, Catania, Italy\\
$^{27}$ Dipartimento di Fisica e Astronomia dell'Universit\`{a} and Sezione INFN, Padova, Italy\\
$^{28}$ Dipartimento di Fisica `E.R.~Caianiello' dell'Universit\`{a} and Gruppo Collegato INFN, Salerno, Italy\\
$^{29}$ Dipartimento DISAT del Politecnico and Sezione INFN, Turin, Italy\\
$^{30}$ Dipartimento di Scienze MIFT, Universit\`{a} di Messina, Messina, Italy\\
$^{31}$ Dipartimento Interateneo di Fisica `M.~Merlin' and Sezione INFN, Bari, Italy\\
$^{32}$ European Organization for Nuclear Research (CERN), Geneva, Switzerland\\
$^{33}$ Faculty of Electrical Engineering, Mechanical Engineering and Naval Architecture, University of Split, Split, Croatia\\
$^{34}$ Faculty of Engineering and Science, Western Norway University of Applied Sciences, Bergen, Norway\\
$^{35}$ Faculty of Nuclear Sciences and Physical Engineering, Czech Technical University in Prague, Prague, Czech Republic\\
$^{36}$ Faculty of Physics, Sofia University, Sofia, Bulgaria\\
$^{37}$ Faculty of Science, P.J.~\v{S}af\'{a}rik University, Ko\v{s}ice, Slovak Republic\\
$^{38}$ Frankfurt Institute for Advanced Studies, Johann Wolfgang Goethe-Universit\"{a}t Frankfurt, Frankfurt, Germany\\
$^{39}$ Fudan University, Shanghai, China\\
$^{40}$ Gangneung-Wonju National University, Gangneung, Republic of Korea\\
$^{41}$ Gauhati University, Department of Physics, Guwahati, India\\
$^{42}$ Helmholtz-Institut f\"{u}r Strahlen- und Kernphysik, Rheinische Friedrich-Wilhelms-Universit\"{a}t Bonn, Bonn, Germany\\
$^{43}$ Helsinki Institute of Physics (HIP), Helsinki, Finland\\
$^{44}$ High Energy Physics Group,  Universidad Aut\'{o}noma de Puebla, Puebla, Mexico\\
$^{45}$ Horia Hulubei National Institute of Physics and Nuclear Engineering, Bucharest, Romania\\
$^{46}$ Hubei University of Technology, Wuhan, China\\
$^{47}$ Indian Institute of Technology Bombay (IIT), Mumbai, India\\
$^{48}$ Indian Institute of Technology Indore, Indore, India\\
$^{49}$ INFN, Laboratori Nazionali di Frascati, Frascati, Italy\\
$^{50}$ INFN, Sezione di Bari, Bari, Italy\\
$^{51}$ INFN, Sezione di Bologna, Bologna, Italy\\
$^{52}$ INFN, Sezione di Cagliari, Cagliari, Italy\\
$^{53}$ INFN, Sezione di Catania, Catania, Italy\\
$^{54}$ INFN, Sezione di Padova, Padova, Italy\\
$^{55}$ INFN, Sezione di Pavia, Pavia, Italy\\
$^{56}$ INFN, Sezione di Torino, Turin, Italy\\
$^{57}$ INFN, Sezione di Trieste, Trieste, Italy\\
$^{58}$ Inha University, Incheon, Republic of Korea\\
$^{59}$ Institute for Gravitational and Subatomic Physics (GRASP), Utrecht University/Nikhef, Utrecht, Netherlands\\
$^{60}$ Institute of Experimental Physics, Slovak Academy of Sciences, Ko\v{s}ice, Slovak Republic\\
$^{61}$ Institute of Physics, Homi Bhabha National Institute, Bhubaneswar, India\\
$^{62}$ Institute of Physics of the Czech Academy of Sciences, Prague, Czech Republic\\
$^{63}$ Institute of Space Science (ISS), Bucharest, Romania\\
$^{64}$ Institut f\"{u}r Kernphysik, Johann Wolfgang Goethe-Universit\"{a}t Frankfurt, Frankfurt, Germany\\
$^{65}$ Instituto de Ciencias Nucleares, Universidad Nacional Aut\'{o}noma de M\'{e}xico, Mexico City, Mexico\\
$^{66}$ Instituto de F\'{i}sica, Universidade Federal do Rio Grande do Sul (UFRGS), Porto Alegre, Brazil\\
$^{67}$ Instituto de F\'{\i}sica, Universidad Nacional Aut\'{o}noma de M\'{e}xico, Mexico City, Mexico\\
$^{68}$ iThemba LABS, National Research Foundation, Somerset West, South Africa\\
$^{69}$ Jeonbuk National University, Jeonju, Republic of Korea\\
$^{70}$ Johann-Wolfgang-Goethe Universit\"{a}t Frankfurt Institut f\"{u}r Informatik, Fachbereich Informatik und Mathematik, Frankfurt, Germany\\
$^{71}$ Korea Institute of Science and Technology Information, Daejeon, Republic of Korea\\
$^{72}$ KTO Karatay University, Konya, Turkey\\
$^{73}$ Laboratoire de Physique Subatomique et de Cosmologie, Universit\'{e} Grenoble-Alpes, CNRS-IN2P3, Grenoble, France\\
$^{74}$ Lawrence Berkeley National Laboratory, Berkeley, California, United States\\
$^{75}$ Lund University Department of Physics, Division of Particle Physics, Lund, Sweden\\
$^{76}$ Nagasaki Institute of Applied Science, Nagasaki, Japan\\
$^{77}$ Nara Women{'}s University (NWU), Nara, Japan\\
$^{78}$ National and Kapodistrian University of Athens, School of Science, Department of Physics , Athens, Greece\\
$^{79}$ National Centre for Nuclear Research, Warsaw, Poland\\
$^{80}$ National Institute of Science Education and Research, Homi Bhabha National Institute, Jatni, India\\
$^{81}$ National Nuclear Research Center, Baku, Azerbaijan\\
$^{82}$ National Research and Innovation Agency - BRIN, Jakarta, Indonesia\\
$^{83}$ Niels Bohr Institute, University of Copenhagen, Copenhagen, Denmark\\
$^{84}$ Nikhef, National institute for subatomic physics, Amsterdam, Netherlands\\
$^{85}$ Nuclear Physics Group, STFC Daresbury Laboratory, Daresbury, United Kingdom\\
$^{86}$ Nuclear Physics Institute of the Czech Academy of Sciences, Husinec-\v{R}e\v{z}, Czech Republic\\
$^{87}$ Oak Ridge National Laboratory, Oak Ridge, Tennessee, United States\\
$^{88}$ Ohio State University, Columbus, Ohio, United States\\
$^{89}$ Physics department, Faculty of science, University of Zagreb, Zagreb, Croatia\\
$^{90}$ Physics Department, Panjab University, Chandigarh, India\\
$^{91}$ Physics Department, University of Jammu, Jammu, India\\
$^{92}$ Physics Program and International Institute for Sustainability with Knotted Chiral Meta Matter (SKCM2), Hiroshima University, Hiroshima, Japan\\
$^{93}$ Physikalisches Institut, Eberhard-Karls-Universit\"{a}t T\"{u}bingen, T\"{u}bingen, Germany\\
$^{94}$ Physikalisches Institut, Ruprecht-Karls-Universit\"{a}t Heidelberg, Heidelberg, Germany\\
$^{95}$ Physik Department, Technische Universit\"{a}t M\"{u}nchen, Munich, Germany\\
$^{96}$ PINSTECH, Islamabad, Pakistan\\
$^{97}$ Politecnico di Bari and Sezione INFN, Bari, Italy\\
$^{98}$ Research Division and ExtreMe Matter Institute EMMI, GSI Helmholtzzentrum f\"ur Schwerionenforschung GmbH, Darmstadt, Germany\\
$^{99}$ Saga University, Saga, Japan\\
$^{100}$ Saha Institute of Nuclear Physics, Homi Bhabha National Institute, Kolkata, India\\
$^{101}$ School of Physics and Astronomy, University of Birmingham, Birmingham, United Kingdom\\
$^{102}$ Secci\'{o}n F\'{\i}sica, Departamento de Ciencias, Pontificia Universidad Cat\'{o}lica del Per\'{u}, Lima, Peru\\
$^{103}$ Stefan Meyer Institut f\"{u}r Subatomare Physik (SMI), Vienna, Austria\\
$^{104}$ SUBATECH, IMT Atlantique, Nantes Universit\'{e}, CNRS-IN2P3, Nantes, France\\
$^{105}$ Sungkyunkwan University, Suwon City, Republic of Korea\\
$^{106}$ Suranaree University of Technology, Nakhon Ratchasima, Thailand\\
$^{107}$ Technical University of Ko\v{s}ice, Ko\v{s}ice, Slovak Republic\\
$^{108}$ The Henryk Niewodniczanski Institute of Nuclear Physics, Polish Academy of Sciences, Cracow, Poland\\
$^{109}$ The University of Texas at Austin, Austin, Texas, United States\\
$^{110}$ Universidad Aut\'{o}noma de Sinaloa, Culiac\'{a}n, Mexico\\
$^{111}$ Universidade de S\~{a}o Paulo (USP), S\~{a}o Paulo, Brazil\\
$^{112}$ Universidade Estadual de Campinas (UNICAMP), Campinas, Brazil\\
$^{113}$ Universidade Federal do ABC, Santo Andre, Brazil\\
$^{114}$ University of Cape Town, Cape Town, South Africa\\
$^{115}$ University of Houston, Houston, Texas, United States\\
$^{116}$ University of Jyv\"{a}skyl\"{a}, Jyv\"{a}skyl\"{a}, Finland\\
$^{117}$ University of Kansas, Lawrence, Kansas, United States\\
$^{118}$ University of Liverpool, Liverpool, United Kingdom\\
$^{119}$ University of Malta, Msida, Malta\\
$^{120}$ University of Science and Technology of China, Hefei, China\\
$^{121}$ University of South-Eastern Norway, Kongsberg, Norway\\
$^{122}$ University of Tennessee, Knoxville, Tennessee, United States\\
$^{123}$ University of the Witwatersrand, Johannesburg, South Africa\\
$^{124}$ University of Tokyo, Tokyo, Japan\\
$^{125}$ University of Tsukuba, Tsukuba, Japan\\
$^{126}$ University Politehnica of Bucharest, Bucharest, Romania\\
$^{127}$ Universit\'{e} Clermont Auvergne, CNRS/IN2P3, LPC, Clermont-Ferrand, France\\
$^{128}$ Universit\'{e} de Lyon, CNRS/IN2P3, Institut de Physique des 2 Infinis de Lyon, Lyon, France\\
$^{129}$ Universit\'{e} de Strasbourg, CNRS, IPHC UMR 7178, F-67000 Strasbourg, France, Strasbourg, France\\
$^{130}$ Universit\'{e} Paris-Saclay, Centre d'Etudes de Saclay (CEA), IRFU, D\'{e}partment de Physique Nucl\'{e}aire (DPhN), Saclay, France\\
$^{131}$ Universit\'{e}  Paris-Saclay, CNRS/IN2P3, IJCLab, Orsay, France\\
$^{132}$ Universit\`{a} degli Studi di Foggia, Foggia, Italy\\
$^{133}$ Universit\`{a} del Piemonte Orientale, Vercelli, Italy\\
$^{134}$ Universit\`{a} di Brescia, Brescia, Italy\\
$^{135}$ Variable Energy Cyclotron Centre, Homi Bhabha National Institute, Kolkata, India\\
$^{136}$ Warsaw University of Technology, Warsaw, Poland\\
$^{137}$ Wayne State University, Detroit, Michigan, United States\\
$^{138}$ Westf\"{a}lische Wilhelms-Universit\"{a}t M\"{u}nster, Institut f\"{u}r Kernphysik, M\"{u}nster, Germany\\
$^{139}$ Wigner Research Centre for Physics, Budapest, Hungary\\
$^{140}$ Yale University, New Haven, Connecticut, United States\\
$^{141}$ Yonsei University, Seoul, Republic of Korea\\
$^{142}$  Zentrum  f\"{u}r Technologie und Transfer (ZTT), Worms, Germany\\
$^{143}$ Affiliated with an institute covered by a cooperation agreement with CERN\\
$^{144}$ Affiliated with an international laboratory covered by a cooperation agreement with CERN.\\

\end{flushleft} 

\end{document}